\documentclass[iop,numberedappendix]{emulateapj}

\pdfoutput=1
\usepackage[linktoc=all]{hyperref}
\usepackage{graphicx}
\usepackage{color}
\usepackage{transparent}
\usepackage{rotating}
\usepackage[squaren, Gray, mediumspace]{SIunits}
\usepackage{MnSymbol}
\graphicspath{{.}}
\usepackage[outdir=./]{epstopdf}
\usepackage{subfigure}
\usepackage{mathtools}

\emergencystretch=\maxdimen
\slugcomment{Draft 2018 March 22 -- Accepted to ApJ}

\shorttitle{CARLA {\it HST} Results}
\shortauthors{Noirot et al.}

\begin{document}

\title{{\it HST} Grism Confirmation of $16$ Structures at $1.4<\MakeLowercase{z}<2.8$\\
from the Clusters Around Radio-Loud AGN (CARLA) Survey}

\author{Ga\"el Noirot\altaffilmark{1,2,3}, Daniel Stern\altaffilmark{3}, Simona Mei\altaffilmark{1,2,4}, Dominika Wylezalek\altaffilmark{5,6}, Elizabeth A. Cooke\altaffilmark{7}, Carlos De Breuck\altaffilmark{6}, Audrey Galametz\altaffilmark{8}, Nina A. Hatch\altaffilmark{9}, Jo\"el Vernet\altaffilmark{6}, Mark Brodwin\altaffilmark{10}, Peter Eisenhardt\altaffilmark{3}, Anthony H. Gonzalez\altaffilmark{11}, Matt Jarvis\altaffilmark{12,13}, Alessandro Rettura\altaffilmark{14}, Nick Seymour\altaffilmark{15}, S. A. Stanford\altaffilmark{16}
}

\altaffiltext{1}{Universit\'e Paris-Diderot Paris VII, Sorbonne Paris Cit\'e, 75205 Paris Cedex 13, France}
\altaffiltext{2}{LERMA, Observatoire de Paris, PSL Research University, CNRS, Sorbonne Universit\'es, UPMC Univ. Paris 06, 75014 Paris, France}
\altaffiltext{3}{Jet Propulsion Laboratory, California Institute of Technology, 4800 Oak Grove Drive, Pasadena, CA 91109, USA}
\altaffiltext{4}{California Institute of Technology, Pasadena, CA 91125, USA}
\altaffiltext{5}{Johns Hopkins University, Zanvyl Krieger School of Arts \& Sciences, 3400 N. Charles St, Baltimore, MD 21218, USA}
\altaffiltext{6}{European Southern Observatory, Karl-Schwarzschildstrasse 2, 85748 Garching, Germany}
\altaffiltext{7}{Centre for Extragalactic Astronomy, Department of Physics, Durham University, South Road, Durham DH1 3LE, UK}
\altaffiltext{8}{Max-Planck-Institut f\"ur extraterrestrische Physik, Giessenbachstrasse 1, 85748 Garching, Germany}
\altaffiltext{9}{School of Physics and Astronomy, University of Nottingham, University Park, Nottingham NG7 2RD, UK}
\altaffiltext{10}{Department of Physics, University of Missouri, 5110 Rockhill Road, Kansas City, MO 64110, USA}
\altaffiltext{11}{Department of Astronomy, University of Florida, Gainesville, FL 32611-2055, USA}
\altaffiltext{12}{Oxford Astrophysics, Department of Physics, Keble Road, Oxford OX1 3RH, UK}
\altaffiltext{13}{Department of Physics, University of the Western Cape, Bellville 7535, South Africa}
\altaffiltext{14}{IPAC, Caltech, KS 314-6, 1200 E. California Blvd, Pasadena, CA 91125, USA}
\altaffiltext{15}{International Center for Radio Astronomy Research, Curtin University, GPO Box U1987, 6102 Perth, Australia}
\altaffiltext{16}{Department of Physics, University of California, One Shields Avenue, Davis, CA 95616, USA}

\begin{abstract}

We report spectroscopic results from our 40-orbit {\it Hubble Space Telescope} slitless grism spectroscopy program observing the 20 densest CARLA candidate galaxy clusters at $1.4<z<2.8$. These candidate rich structures, among the richest and most distant known, were identified on the basis of $[3.6]-[4.5]$ color from a $408$~hr multi-cycle {\it Spitzer} program targeting $420$ distant radio-loud AGN. We report the spectroscopic confirmation of $16$ distant structures at $1.4<z<2.8$ associated with the targeted powerful high-redshift radio-loud AGN. We also report the serendipitous discovery and spectroscopic confirmation of seven additional structures at $0.87<z<2.12$ not associated with the targeted radio-loud AGN. We find that $10^{10} - 10^{11}\,M_{\odot}$ {member galaxies of our confirmed CARLA structures} form significantly fewer stars than their field counterparts at all redshifts within $1.4\leq z\leq 2$. We also observe higher star-forming activity in the {structure cores} up to $z=2$, finding similar trends as cluster surveys at slightly lower redshifts ($1.0<z<1.5$). By design, our efficient strategy of obtaining just two grism orbits per {field} only obtains spectroscopic confirmation of emission-line galaxies. Deeper spectroscopy will be required to study the population of evolved, massive galaxies in these (forming) clusters.  {Lacking multi-band coverage of the fields, we adopt a very conservative approach of calling all confirmations ``structures'', although we note that a number of features are consistent with some of them being bona fide galaxy clusters.} Together, this survey represents a unique and large homogenous sample of spectroscopically confirmed structures at high redshifts, {potentially} more than doubling the census of confirmed, massive clusters at $z>1.4$.

\end{abstract}


\section{Introduction}\label{sec:intro}

The last decade has seen an explosion in our understanding of the
early stages of galaxy cluster formation, 
largely built upon samples of distant galaxy clusters identified
from the Infrared Array Camera (IRAC; \citealp{Fazio04}) on the
{\it Spitzer Space Telescope}.  Mid-infrared observations provide
an incredibly sensitive tool for identifying massive galaxies at
high redshift.  For stellar populations formed at high redshift,
negative $k$-corrections provide a nearly constant $4.5\, \mu$m
flux density over a wide redshift range, while red $[3.6]-[4.5]$
colors provide an effective means of isolating galaxies at $z > 1.3$
(e.g., \citealp{Papovich08}).  
Indeed, the sensitivity required to detect $L^\star$ galaxies at $z \sim 0.7$ with mid-infrared observations, which {\it Spitzer} reaches in $< 2$~min integrations, is sufficient to detect such galaxies to $z \ga 2$.
Several large projects are 
exploiting {mid-infrared data} to search for high-redshift galaxy
clusters using {\it Spitzer} and {\it WISE}
(e.g., \citealp{Eisenhardt08}; \citealp{Papovich10}; \citealp{Galametz10,
Galametz13}; \citealp{Stanford12, Stanford14}; \citealp{Muzzin13};
\citealp{Wylezalek14}; \citealp{Rettura14}; \citealp{Brodwin15,
Brodwin16}; \citealp{PaternoMahler17}). Such work has, for example, proved useful for {\it
(i)} measuring the galaxy cluster autocorrelation function out to
$z \sim 1.5$, which provides a measure of the typical galaxy cluster
mass (\citealp{Brodwin07}), {\it (ii)} targeted cosmological surveys
for $z > 1$ Type~Ia SNe in dust-free environments (\citealp{Suzuki12}),
{\it (iii)} probing evolution in the $\sigma - T_x$ correlation
(\citealp{Brodwin11}), {\it (iv)} using the rest-frame near-infrared
luminosity function and rest-frame optical colors to probe the
formation epoch of cluster galaxies (\citealp{Mancone10, Mancone12};
\citealp{Snyder12}, \citealp{Wylezalek14}; \citealp{Cooke15};
\citealp{Nantais16}), {\it (v)} probing the role of AGN feedback
in forming clusters (\citealp{Galametz09}; \citealp{Martini13}),
{\it (vi)} leading cosmological investigations based on the incidence of
massive, high-redshift clusters (\citealp{Brodwin12};
\citealp{Gonzalez12}), and {\it (vii)} probing the dependency of
galaxy quenching on environment (\citealp{Muzzin14}, \citealp{Nantais17}).

A significant challenge, however, is to confirm these massive,
distant cluster candidates, particularly at the highest redshifts
which probe the earliest stages of their evolution. At $z \sim
1.4$, the strong [\ion{O}{2}] and D4000 features shift to infrared
wavelengths where ground-based spectroscopic follow-up is more
challenging, particularly for absorption-line redshifts. Near-infrared grism spectroscopy using the Wide Field Camera 3 (WFC3) onboard the {\it Hubble Space Telescope} ({\it HST}) provides impressive sensitivity
for studying galaxy clusters at high redshift. Courtesy of the
multiplexing advantages of slitless spectroscopy combined with the
low, absorption-free background provided from space, grism data
is sufficient to confirm clusters out to $z \sim 2$.  Indeed, the
majority of the most distant clusters currently known were selected
on the basis of {\it Spitzer} mid-infrared data and confirmed with {\it
Hubble} grism spectroscopy. This includes clusters at 
$z = 1.75$
(\citealp{Stanford12}), $z = 1.80$ (\citealp{Newman14}), $z = 1.89$
(\citealp{Zeimann12}), and our own recent confirmation of {one evolved cluster and one younger forming structure at $z = 2.00$ and $z=1.99$, respectively} (\citealp{Noirot16};
hereafter N16).

Much work on distant clusters is based on blind field searches, which
provide both a strength and a weakness.  Field surveys have a
simpler selection function for measuring the growth of structure
through cluster counts.  However, field surveys --- both in the
mid-infrared and at other wavelengths (e.g., Sunyaev-Zel'dovich or
X-ray surveys; \citealp{Bleem15};
\citealp{Tozzi15}) --- find few massive clusters at the highest
redshifts ($z \ga 1.5$).  Many key galaxy cluster studies do not
require knowledge of the cluster space density (e.g., \citealp{Krick08};
\citealp{Stern10}; \citealp{Rettura11}; \citealp{Suzuki12}), and
therefore targeted searches for high-redshift galaxy clusters have
many advantages.

We report on a comprehensive, 40-orbit {\it Hubble} program using
the unique near-infrared grism capabilities of the WFC3 to attempt confirmation
of the 20 richest $z > 1.4$ galaxy cluster candidates identified
from our 408-hr {\it Warm Spitzer} survey of 420 radio-loud AGN
across the full sky, Clusters Around Radio-Loud AGN (CARLA; \citealp{Wylezalek13, Wylezalek14}).  With IRAC exposures of $\sim 1$~hr per {field},
CARLA reaches $10\sigma$ depths of $m^* + 2$ out to $z \ga 2$.
Extensive literature reaching back 50 years shows that powerful
radio-loud AGN preferentially reside in overdense environments
(e.g., \citealp{Matthews64}).  We identified sources with
red mid-infrared colors ($[3.6]-[4.5] > -0.1$; AB), which are primarily
expected to be galaxies at $z \ga 1.3$ (e.g., \citealp{Wylezalek13}).  Considering the surface density of sources with such colors
as selected from the 1 deg$^2$ {\it Spitzer} UKIDSS Ultra-Deep
Survey (SpUDS, PI: J. Dunlop), we find that the average blind field
contains $\langle \Sigma_{\rm SpUDS} \rangle = 8.3 \pm 1.6$ red
{\it Spitzer} galaxies arcmin$^{-2}$.  In contrast, 92.4\% of the
CARLA fields are denser than this SpUDS average density.  
Approximately $10\%$ of the CARLA fields are as rich or richer than twice the
average SpUDS density of red galaxies (i.e., $\Sigma \geq 16.3$
arcmin$^{-2}$), compared to 0.7\% of SpUDS fields.
Indeed, many of the CARLA
fields --- and all the fields reported here --- are richer than
the highest density region in SpUDS, the $z = 1.62$ cluster reported
by \cite{Papovich10} and \cite{Tanaka10}.  
This {\it Hubble} program
enhances the census of confirmed {\em rich} {structures} at $z > 1.4$
by a factor of several, identifying systems at a time when clusters
are believed to be actively forming.

The paper is structured as follows. Section~\ref{sec:obs} reports
on the observations, while Section~\ref{sec:red} briefly summarizes the
data analysis; we refer the reader to N16 for a detailed description
of the data processing.  Section~\ref{sec:res} reports on some
initial results from the rest-frame optical spectroscopy, including
cluster membership (\S~\ref{sec:memb}), cluster star formation rates
(SFRs; \S~\ref{sec:sfr}), and a detailed description of several
newly confirmed structures of particular interest
(\S~\ref{sec:clinterest}).  Section~\ref{sec:discussion} investigates
statistical correlations probed by the data, including
comparison to other cluster surveys (\S~\ref{sec:highzcomp}), typical H$\alpha$/[\ion{O}{3}] line ratios at high redshift
(\S~\ref{sec:ratios}), and how SFR depends on galaxy mass
(\S~\ref{sec:sfrm}) and distance from the cluster center
(\S~\ref{sec:sfrr}).  We summarize our results in Section~\ref{sec:con}.
Throughout, we use AB magnitudes and we
adopt the concordance cosmology, $\Omega_{\rm M} = 0.3$, $\Omega_\Lambda
= 0.7$ and $H_0 = 70\, {\rm\,km\,s^{-1}\,Mpc^{-1}}$.\\

\section{Observations}\label{sec:obs}

\begin{deluxetable*}{lccc}  
\tablewidth{500pt}
\tablecolumns{4}
\tablecaption{{\it HST} WFC3 Observations \label{table:obs}}
\tablehead{   
  \colhead{Field} &
  \colhead{UT Date} &
  \colhead{Position Angle\tablenotemark{a} (degrees)} &
  \colhead{F140W/G141 Exp. Time (s)}
  }
\startdata
CARLA~J0116$-$2052 &	2014 Dec 06 &	$-$70 &	512/2012	 \\
				     &	2015 Oct 25 &	$-$100 &	512/2012	 \\
CARLA~J0800+4029 &	2014 Nov 03 &	+125 &	537/2062	 \\
				 &	2014 Nov 07 &	+162 &	537/2062	 \\
CARLA~J0958$-$2904 &	2015 Oct 13 &	+124 &	512/2012	 \\
 				     &	2015 Dec 02 &	+161 &	512/2012	 \\
CARLA~J1017+6116 &	2014 Dec 22\tablenotemark{b} &	+104 &	512/2262	 \\
 				 &	2015 Feb 05 &	+76 &	512/2262	 \\
				 &	2015 Nov 22 &	+173 &	512/2262	 \\
CARLA~J1018+0530 &	2015 Mar 14 &	$-$30 &	487/2012	 \\
 				 &	2015 Mar 14 &	$-$10 &	487/2012	 \\
CARLA~J1052+0806 &	2015 May 14 &	$-$8 &	487/2012	 \\
 				 &	2015 May 17 &	$-$28 &	487/2012	 \\
6CSS1054+4459 &	2014 Dec 26 &	+122 &	537/2062  \\
			    &	2015 May 15 &	$-$19 &	537/2062  \\
CARLA~J1103+3449 &	2014 Dec 25 &	+167 &	537/2012	 \\
 				 &	2016 Mar 03 &	+45 &	537/2012	 \\
CARLA~J1129+0951 &	2015 Apr 19 &	$-$20 &	487/2012	 \\
				 &	2015 Apr 19 &	0        &	487/2012	 \\
CARLA~J1131$-$2705 &	2014 Nov 12 &	+131 &	512/2012	 \\
 				     &	2016 Jan 31 &	$-$159 &	512/2012	 \\
CARLA~J1300+4009 &	2015 Jul 04 &	$-$31 &	537/2062	 \\
				 &	2015 Oct 30 &	$-$160 &	537/2062	 \\
J1317+3925 		&	2015 Jun 30 &	$-$25 &	512/2062	 \\
 				 &	2016 Feb 29 &	+111 &	512/2062	 \\
CARLA~J1358+5752 &	2015 Feb 03 &	+142 &	537/2212	 \\  
				 &	2016 Feb 17 &	+111 &	537/2212	 \\  
CARLA~J1510+5958 &	2015 May 18 &	+21     &	537/2212	 \\
 				 &	2015 Jul 04   &	$-$9 &	537/2212	 \\
J1515+2133	    &	2015 Apr 08 &	+99 &	512/2012  \\
			    &	2015 Apr 13 &	+79 &	512/2012  \\
CARLA~J1753+6310 &	2015 Jul 02 &	+45 &	512/2262	\\
 				 &	2015 Aug 21 &	$-$25 &	512/2262	\\
CARLA~J2039$-$2514 &	2014 Oct 14 &	$-$62 &	512/2012	 \\
 				     &	2014 Oct 14 &	$-$39 &	512/2012	 \\
CARLA~J2227$-$2705 &	2014 Oct 30 &	$-$50 &	512/2012	 \\
				     &	2015 Jul 26 &	+160 &	512/2012	 \\
TNR 2254+1857  &	2014 Nov 05 &	$-$32 &	512/2012  \\
  			    &	2014 Nov 09 &	$-$72 &	512/2012  \\
CARLA~J2355$-$0002 &	2014 Oct 22 &	$-$46 &	487/2012	 \\
				     &	2015 Jul 13 &	+97 &	462/2012	 \\
\enddata
\tablenotetext{a}{East of north.}
\tablenotetext{b}{Failed. Re-observed on UT 2015 Nov 22.}
\end{deluxetable*}

Our {\it HST} program consists of WFC3/F140W imaging and WFC3/G141
grism spectroscopy of the 20 densest CARLA cluster candidates,
selected from 420 radio-loud AGN at $z > 1.3$ observed
in our {\it Spitzer} program (\citealp{Wylezalek13}). The {\it HST} fields are
$4\sigma$ to $8\sigma$ overdense in {\it Spitzer} color-selected
sources (i.e, selected to have mid-infrared colors consistent with
$z>1.3$) compared to the mean SpUDS density of similarly selected
sources (\citealp{Wylezalek14}). The targeted radio-loud AGN (RLAGN) at the center
of the fields observed with {\it HST} cover the redshift range
$1.37<z<2.80$, with a median redshift $\tilde{z} =1.655$.
Ten fields are associated with high-redshift radio galaxies (HzRGs, type-2 RLAGN),
and the other ten with radio-loud quasars (RLQs, type-1 RLAGN). The 20 fields
were observed between 2014 October and 2016 April (Program ID:
13740) with a 2-visit per field strategy, using different orientations
to mitigate contamination from overlapping spectra.  For each visit,
we obtained $0.5$ ksec F140W imaging and $2$ ksec G141 grism
spectroscopy, divided into four dithered blocks of exposures, with
the direct images taken just after the grism exposures to enable
wavelength calibration of the spectra based on source position.
Table~\ref{table:obs} lists the observation dates, orientation
angles, and exposure times.  We refer to confirmed structures by
their CARLA names, whereas we refer to unconfirmed structures by
their radio-loud AGN target names.

Each image covers a field of view of $2 \times 2.3$ $\rm arcmin^2$
at a sampling of $0.13$ arc$\sec~\rm pix^{-1}$. The G141 grism
covers the wavelength range $1.08 - 1.70\, \micron$ with a throughput
$> 10\%$ at low spectral resolution, $R \equiv \lambda / \Delta \lambda
= 130$. This grism was chosen to enable identification of strong
spectroscopic features at the redshifts of our cluster candidates
($1.4 < z < 2.8$), namely H$\alpha$ at $0.65 < z < 1.59$, [\ion{O}{3}]
at $1.16 < z < 2.40$, H$\beta$ at $1.22 < z < 2.50$, and [\ion{O}{2}]
at $1.90 < z < 3.56$. Our efficient strategy of two orbits per field
represents shallow observations, {only allowing} us to spectroscopically
confirm star-forming galaxies with their strong, narrow emission
lines. Specifically, this strategy of shallow observations does
not provide spectroscopic confirmation of the important, but more
challenging, population of passive galaxies in the distant clusters.\\

\section{Data Processing}\label{sec:red}

Data reduction mainly follows the steps described in N16. In short,
we first combine the individual F140W exposures with {\tt aXe} (v2.2.4;
\citealp{Kümmel09,Kümmel11}) to create deep drizzled, cosmic ray
cleaned, direct images of each field, on which we perform source
extraction using SExtractor (\citealp{BertinArnouts96}). Using {spectrally} empty
sky regions determined from the SExtractor catalogs, we then subtract
the sky from the grism images using {\tt aXe} and appropriate configuration
files. Individual two-dimensional spectra in each field and visit are then
extracted from the grism data based on source position and sizes.
For each spectrum  we also determine contamination estimates from
neighboring objects using the {\tt aXe} Gaussian contamination model.

We then use an internally developed Graphical User Interface (GUI) to facilitate spectral and source characterization. We determine source redshifts and emission line fluxes with the python version of {\ttfamily mpfit} using the same method as described in N16, now implemented within our GUI. 
See N16 for a detailed, in-depth description of the data processing methodology and implementation.\\

\section{Results}\label{sec:res}

\begin{deluxetable*}{lccccccc}  
\tablewidth{500pt}
\tablecolumns{8}
\tablecaption{CARLA {\it HST} Results  \label{table:res}}
\tablehead{   
  \multicolumn{1}{c}{Field} &
  \multicolumn{1}{c}{$\sigma_{\rm IRAC}$\tablenotemark{a}} &
  \multicolumn{1}{c}{$z_{\rm RLAGN}$} &
  \multicolumn{1}{c}{$\left<z\right>_{\rm cl}$\tablenotemark{b}} &
  \multicolumn{1}{c}{$\tilde{z}_{\rm cl}$\tablenotemark{c}} &
    \multicolumn{3}{c}{$\#z_{cl}$/$\#z_{tot}$\tablenotemark{d}} \\
 \cline{6-8}
 & \colhead{\mbox{}}
 & \colhead{\mbox{}}
 & \colhead{\mbox{}}
 & \colhead{\mbox{}}
 & \colhead{{\it HST}}
 & \colhead{{IRAC}}
 & \colhead{{$[3.6]-[4.5]>-0.1$}}
 }
\startdata
\cutinhead{CONFIRMED CARLA STRUCTURES}
CARLA~J0116$-$2052 &	5.14 &	1.417 &	1.425 &	1.430 &	12/31 &	9/22	&	7/14 \\
CARLA~J0800+4029 &	6.38 &	2.004 &	1.986 &	1.986 &	10/26 &	5/18	&	5/14\\
CARLA~J0958$-$2904 &	5.00 &	1.411 &	1.392 &	1.396 &	8/23 &	5/12	&	5/9 \\
CARLA~J1017+6116 &	6.67 &	2.80	  &	2.801 &	2.801 &	7/41 &	3/20	&	3/10 \\
CARLA~J1018+0530 &	5.00 &	1.949 &	1.952 &	1.953 &	8/26 &	5/18	&	4/8 \\
CARLA~J1052+0806 &	4.71 &	1.641 &	1.646 &	1.648 &	6/40 &	1/17	&	1/9 \\
CARLA~J1103+3449 &	6.38 &	1.444 &	1.442 &	1.443 &	8/26 &	4/15	&	4/10 \\
CARLA~J1129+0951 &	6.33 &	1.520 &	1.528 &	1.531 &	12/39 &	4/16	&	4/7 \\
CARLA~J1131$-$2705 &	4.38 &	1.444 &	1.446 &	1.445 &	9/36 &	6/23	&	6/18 \\
CARLA~J1300+4009 &	4.86 &	1.669 &	1.675 &	1.676 &	8/28 &	2/10	&	2/7 \\
CARLA~J1358+5752 &	6.24 &	1.370 &	1.368 &	1.373 &	14/48 &	8/23	&	8/17 \\  
CARLA~J1510+5958 &	5.62 &	1.719 &	1.725 &	1.719 &	6/46 &	5/24	&	5/14 \\
CARLA~J1753+6310 &	4.52 &	1.576 &	1.582 &	1.581 &	5/35 &	1/8	&	1/7 \\
CARLA~J2039$-$2514 &	8.00 &	1.997 &	1.999 &	2.000 &	9/30 &	3/14	&	3/10 \\
CARLA~J2227$-$2705 &	5.29 &	1.684 &	1.692 &	1.686 &	7/51 &	2/28	&	2/10 \\
CARLA~J2355$-$0002 &	5.62 &	1.487 &	1.490 &	1.489 &	12/44 &	7/28	&	7/14 \\
\cutinhead{UNCONFIRMED CARLA STRUCTURES\tablenotemark{e}}
6CSS1054+4459 &	4.67 &	2.573 &	(2.566) &	(2.566) &	(2)/38 &	(0)/17	&	(0)/7 \\
J1317+3925 	    &	4.86 &	1.569 &	(1.574) &	(1.569) &	(3)/39 &	(3)/22	&	(3)/10 \\
J1515+2133	    &	4.24 &	2.249 &	(2.262) &	(2.262) &	(2)/35 &	(1)/22	&	(1)/10 \\
TNR 2254+1857  &	5.62 &	2.164 &	(2.159) &	(2.157) &	(3)/25 &	(1)/14	&	(1)/10 \\
\cutinhead{SERENDIPITOUS DISCOVERIES}
CARLA-Ser~J1017+6116 &	\nodata	&	\nodata	   &	1.235 &	1.234 &	5/41 &	3/20	&	1/10 \\
CARLA-Ser~J1317+3925 	&	\nodata 	&	\nodata 	   &	1.467 &	1.465 &	8/39 &	3/22	&	3/10 \\
CARLA-Ser~J1510+5958 &	\nodata	&	\nodata	   &	0.875 &	0.876 &	6/46 &	1/24	&	0/14 \\
CARLA-Ser2~J1510+5958  &	\nodata	&	\nodata	   &	0.977 &	0.976 &	7/46 &	5/24	&	0/14 \\
CARLA-Ser~J1753+6310 &	\nodata	&	\nodata        &	2.117 &	2.117 &	6/35 &	0/8	&	0/7\\
CARLA-Ser~J2227$-$2705 &	\nodata	&	\nodata        &	1.355 &	1.358 &	10/51 &	3/28	&	1/10\\
CARLA-Ser2~J2227$-$2705 &	\nodata	&	\nodata   &	1.477 &	1.478 &	6/51 &	2/28	&	2/10\\

\enddata
\tablenotetext{a}{Overdensity significance of color-selected sources above the field value (\citealp{Wylezalek14}).}
\tablenotetext{b}{Mean redshift of structure members.}
\tablenotetext{c}{Median redshift of structure members.}
\tablenotetext{d}{Number of confirmed structure members ($\#z_{cl}$) compared to the total number of sources for which we measure a redshift ($\#z_{tot}$). The first column corresponds to sources detected in our {\it HST}/F140W imaging, the second column to {secure} sources detected in our {\it Spitzer}/IRAC imaging, and the third column to sources passing our {\it Spitzer}/IRAC color-selection criterion.}
\tablenotetext{e}{For unconfirmed structures, we show in parenthesis mean and median redshifts and source numbers based on the few confirmed sources at the RLAGN redshifts.}
\end{deluxetable*}

All spectra were initially analyzed by GN, with secondary assessments
provided by co-authors.  In our final catalogs, we provide the
{fitted redshifts and consensus quality flags}, described in 
\S~\ref{sec:zflag}. Table~\ref{table:res} shows the spectroscopic
results, and spectra of all confirmed members are shown  
in Appendix \ref{app:spectra}. Among the 20 fields, 16 are confirmed structures
associated with the RLAGN, while we fail to confirm four cluster candidates
(at $z=1.57, 2.16, 2.25$, and $2.57$). In five of the 20 fields, we also identify seven background or foreground structures. Four (seven) of the 16 confirmed
structures have 12 (9) or more confirmed star-forming members. All four
non-confirmed clusters have at least one source confirmed at the
RLAGN redshift in addition to the RLAGN itself\footnote{Note that these non-confirmed clusters also have the potential to host low star-forming or quiescent populations that we cannot identify with our shallow grism observations.}. We provide a catalog of all spectroscopically identified sources in the online material. We describe the content of this catalog in Appendix \ref{app:catalog}.

\subsection{Redshift Quality Flags}\label{sec:zflag}
We use the same redshift qualities as in N16: A, B$^{+}$, and B$^{-}$. In short, we have three proxies for redshift determination: emission lines, {\it Spitzer}/IRAC colors, and the RLAGN prior redshifts. We only consider strong emission lines characteristic of star formation: H$\alpha$, [\ion{O}{3}], H$\beta$, and [\ion{O}{2}]. When {two or more emission lines are securely} detected, we assign quality A redshifts. {These redshifts are considered to be very secure.} 
When {only one strong line is detected we also use line non-detections to disentangle between possible identifications and reject uncertain cases. Additionally, if} a source also possesses a secure {\it Spitzer}/IRAC counterpart with a mid-infrared color that disentangles between possible identifications (see \S~\ref{sec:intro}), the redshift quality becomes B$^{+}$. {Such sources are considered to have relatively secure redshifts.} In the absence of a secure {\it Spitzer}/IRAC identification, the redshift quality flag is B$^{-}$. {Such a situation could happen either because the source is undetected by IRAC or because an IRAC-detected source is a blend of several sources detected in the {\it HST} imaging. The redshift quality flag is also B$^{-}$ when a source is securely detected by IRAC but the mid-infrared color alone leaves an ambiguous redshift identification --- e.g., a red IRAC source with an isolated emission line at $1.6\,\micron$ could potentially be, based on the IRAC color, H$\alpha$ at $z=1.44$ or [\ion{O}{3}] at $z=2.20$ (or [\ion{O}{2}] at yet higher redshift).} {Quality B$^{-}$ redshift are considered likely correct, albeit with the potential for some mis-identifications. Note that recurrent identification of strong line(s) at the same wavelengths as the strong emission lines of the targeted RLAGN {provides additional strong support that} B identifications {are} robust.}  
{{Without deep multi-band coverage of the majority of the fields, robust photometric redshifts are not possible.}}  
Overall, we identify $308$ quality A, $181$ quality B$^{+}$, and $218$ quality B$^{-}$ redshifts, in the range $0.38 < z < 2.85$. Among them, two-thirds ($473/707$) are at $z>1.3$.\\

\subsection{Membership Definition}\label{sec:memb}
As in N16, we {adapt} the \cite{Eisenhardt08} criteria  
to define a spectroscopically confirmed galaxy cluster. {{Developed} to confirm $z>1$ clusters}, the \cite{Eisenhardt08} criteria require at least five galaxies within a physical radius of 2~Mpc whose spectroscopic redshifts are confined to lie within $\pm 2000(1+\left<z_{\rm spec}\right>) \rm~km~\sec^{-1}$. {The physical radius of our confirmations is here constrained by our {\it HST} field of views, which probe $\sim 4\times$ smaller radii that the 2~Mpc criterion. Our adopted criteria therefore require at least five galaxies within {our {\it HST} field of views ($\sim 0.5$~Mpc physical radius)} whose spectroscopic redshifts are confined to lie within $\pm 2000 \rm~km~\sec^{-1}$.} As emphasized in N16, the \cite{Eisenhardt08} definition was designed for ground-based spectroscopic surveys and, alone, may also identify groups, protoclusters, sheets and filaments.  
As such, these criteria are imperfect, but have the advantage of providing a clearly stated and easily measurable threshold. Ideal criteria for defining robust clusters would likely require additional multi-wavelength data such as extended X-ray detections, Sunyaev-Zel'dovich decrements,
and/or weak-lensing total mass measurements.  We are pursuing such observations, but they are beyond the scope of the results presented herein. We note, however, that {\it (i)} our {\it HST} observations probe $16\times$ smaller areas than the area considered by the \cite{Eisenhardt08} criteria, {\it (ii)} the color-selected (i.e., $z > 1.3$) cluster member candidates are, on average, {highly} concentrated around the targeted RLAGN (\citealp{Wylezalek13}) and {\it (iii)} the overdensities of these candidates reach $4\sigma$ to $8\sigma$ above the field value for all $20$ fields (\citealp{Wylezalek14}). Therefore, 
the fields confirmed herein have additional supporting properties that argue for them being associated with rich clusters and protoclusters. 

{In an attempt to better assess the status of our confirmations, we investigate their spectroscopic overdensity significance using the similarly deep 3D-HST field survey (e.g., \citealp{Momcheva16}) and compare them to what is expected from numerical simulations (\citealp{Cautun14}). In Appendix~\ref{app:specsig}, we describe three classes of confirmation: {\it (i)} highly probable confirmed clusters (HPC), {\it (ii)} probable confirmed clusters (PC), and {\it (iii)} confirmed galaxy concentrations (CGC). Three of our confirmed CARLA cluster candidates fall in the first category (HPC) while the rest ($13/16$) fall in the second category (PC). This analysis suggests that, albeit imperfect, our confirmation criteria are robust and likely {suffer minimal contamination from} groups, sheets and filaments. For simplicity, and lacking additional multi-wavelength coverage of the fields, we however adopt a very conservative approach of calling all spectroscopic confirmations ``structures'' and refer the reader to Appendix~\ref{app:specsig} for a more detailed analysis. We {did obtain} optical wavelength observations for two fields confirmed herein (CARLA~J2039-2514 and CARLA~J1753+6310), revealing the presence of two red sequences populated by passive galaxies (see N16 and \citealp{Cooke16}, respectively). Based on these additional observations, CARLA~J2039-2514 and CARLA~J1753+6310 are therefore consistent with being evolved galaxy clusters at $z=2.0$ and $z=1.6$, respectively.} 
We also {emphasize} that our 2-orbit per field strategy only confirms star-forming galaxies, 
and not the typically dominant passive galaxy population present in evolved galaxy clusters. 
Among the 16 confirmed CARLA structures associated with the targeted RLAGN, we identify $63$ quality A, $14$ quality B$^{+}$, and $64$ quality B$^{-}$ members in the range $1.34 < z < 2.82$, with an average (median) of $9$ (8) emission line members per confirmed {structure}. {In Appendix~\ref{app:velspadist}, we show the redshift/velocity and spatial distributions of all {\it HST} confirmed members of the $16$ confirmed CARLA cluster candidates and briefly discuss characteristics of these structures.}\\

\subsection{Line Fluxes}

\begin{figure*}
\includegraphics[scale=0.48]{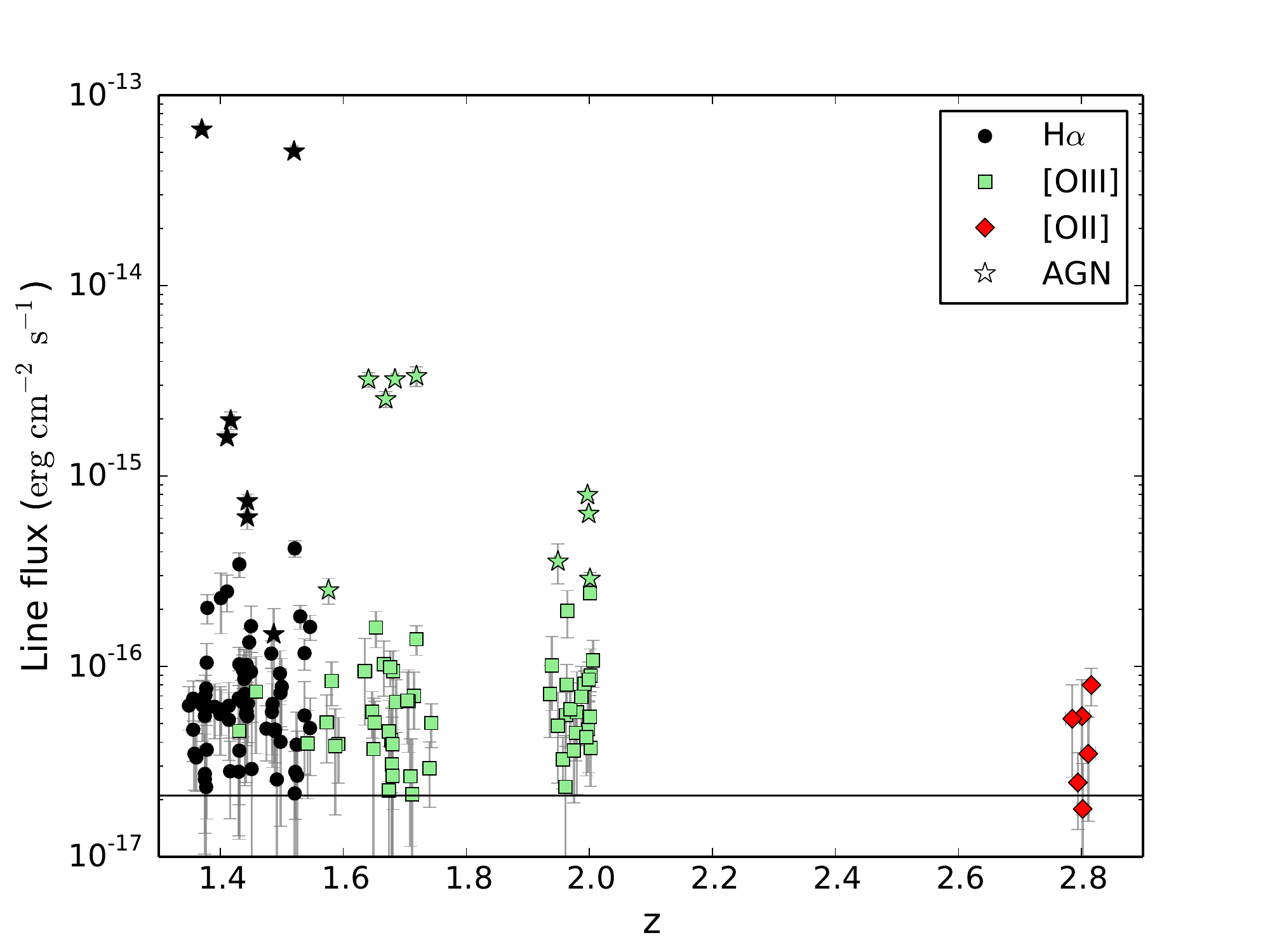}\hspace{-0.6cm}
\includegraphics[scale=0.48]{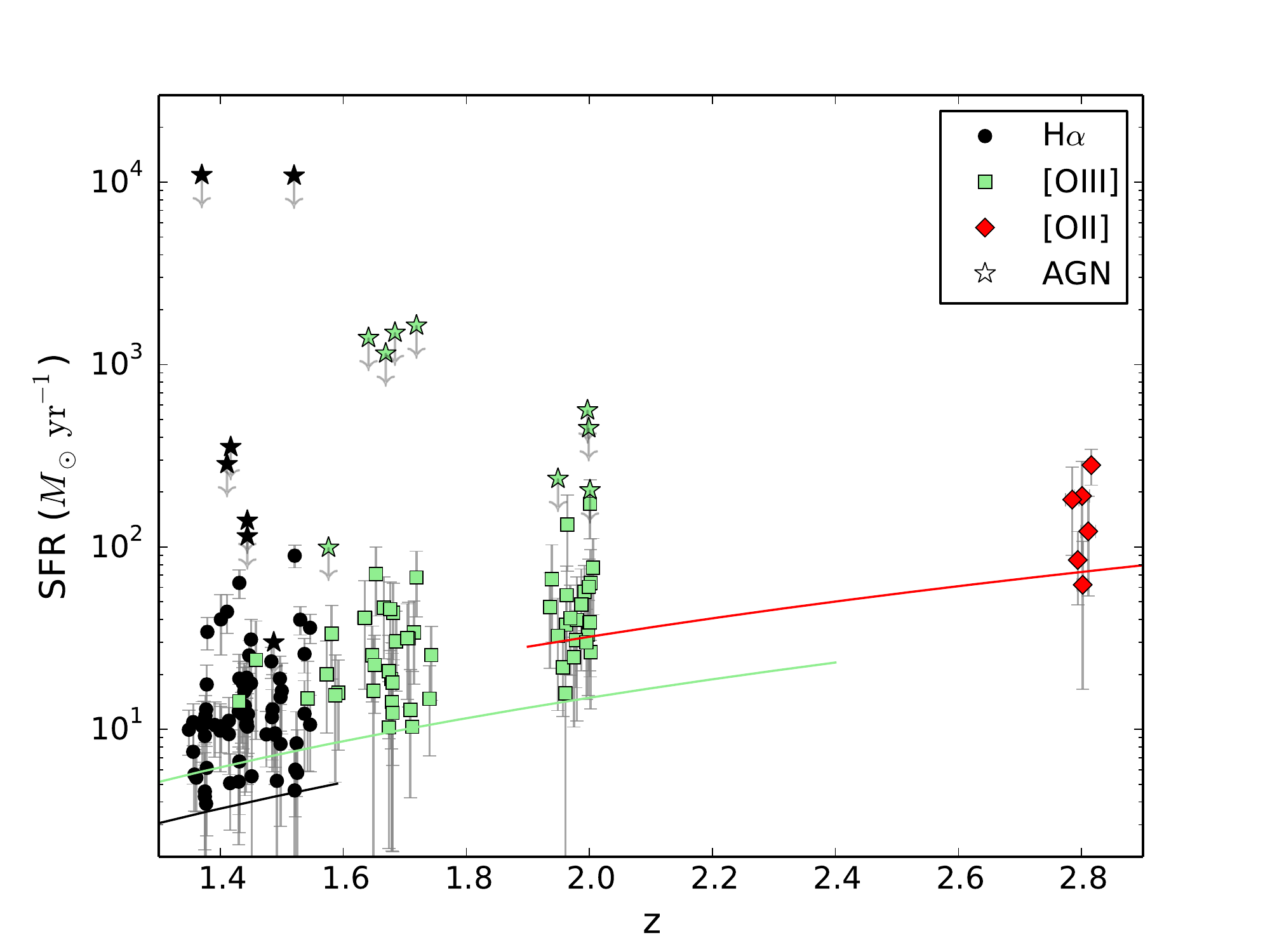}
\caption[Fluxes \& SFRs]{Observed fluxes (left) and {inferred} SFRs (right) of confirmed members against source redshift. We show one emission line per confirmed source with the following priority: H$\alpha$ (black cirlcles), [\ion{O}{3}] (light green squares), [\ion{O}{2}] (red diamonds). Stars represent the targeted RLAGN and identified member AGN. SFRs for AGN are likely overestimated and therefore considered upper limits. The solid line in the left panel is the 3D-HST $3\sigma$ line detection limit, which used a similar observering strategy to the program
reported here, and solid lines in the right panel are corresponding SFRs color-coded according to the emission line, and spanning the redshift range for which they are visible in the G141 grism.}
\label{fig:fluxsfrz}
\end{figure*}

The left panel of Figure \ref{fig:fluxsfrz} shows the emission line
fluxes of confirmed members as a function of redshift. For members
with multiple-line identifications, we show only one line with the
following priority order: H$\alpha$, [\ion{O}{3}], and then
[\ion{O}{2}]. In our low resolution grism data, H$\alpha$
and the [\ion{N}{2}]$\lambda6548, 6584\rm\,\AA$ doublet are blended.
Similar to other teams using WFC3 grism data (e.g., 3D-HST,
\citealp{Fumagalli12}, \citealp{Momcheva16}; the WISP survey,
\citealp{Colbert13}), our reported H$\alpha$ fluxes include the
[\ion{N}{2}] contributions.  However, we consider [\ion{N}{2}]
contributions when estimating H$\alpha$ star formation rates,
assuming a typical [\ion{N}{2}]/H$\alpha$ ratio of $0.3$ (see
\S~\ref{sec:sfr}). The [\ion{O}{3}]$\lambda4959, 5007\rm\,\AA$
doublet is also unresolved in our grism data, but we fit and measure
line fluxes of both emission lines in our fitting procedure (see
N16 for details) and only refer to [\ion{O}{3}]$\lambda5007\rm\,\AA$
when using the generic [\ion{O}{3}] denomination. Finally, the
[\ion{O}{2}]$\lambda3727, 3729\rm\,\AA$ doublet is also blended in
our grism spectroscopy.  However, we do not distinguish between the
two lines when referring to [\ion{O}{2}], and our [\ion{O}{2}]
fluxes include the contribution of both lines, fitted as a single
Gaussian.  

In N16 we determined a $2.5\times 10^{-17}\, \rm erg\, cm^{-2}\, s^{-1}$ line detection limit, with some scatter in the range $(1.2-4.0)\times 10^{-17}\, \rm erg\, cm^{-2}\, s^{-1}$. This limit varies somewhat depending on the specific observation and grism exposure time. 
The 3D-HST survey had a similar 2-orbit depth per WFC3 G141 observation, and determined a $3\sigma$ emission line flux limit of $2.1\times 10^{-17}\, \rm erg\, cm^{-2}\, s^{-1}$ (\citealp{Momcheva16}). This limit, shown as the solid black line in the left panel of Figure \ref{fig:fluxsfrz}, is consistent with our measured fluxes.
Overall, fluxes are in the range $(0.2 - 4.2) \times 10^{-16}\, \rm erg\, cm^{-2}\, s^{-1}$ for star-forming members and in the range $(1.5 - 661) \times 10^{-16}\, \rm erg\, cm^{-2}\, s^{-1}$ for the RLAGN.\\

\subsection{Star Formation Rates}\label{sec:sfr}
The right panel of Figure \ref{fig:fluxsfrz} shows the SFRs of confirmed members as a function of redshift. We measure SFRs based on the following lines in priority order: H$\alpha$, [\ion{O}{3}], and [\ion{O}{2}]. We use the \cite{Kennicutt83} relation, ${\rm SFR} = (8.9 \times 10^{-42}{\rm\, erg^{-1} \, sec}) \times L({\rm H}\alpha) \, M_{\odot}\,\rm yr^{-1}$, to convert H$\alpha$ luminosities to SFRs. As noted earlier, H$\alpha$ and the [\ion{N}{2}]$\lambda6548, 6584\rm\,\AA$ doublet are blended in our low resolution grism data. The typical [\ion{N}{2}]$\lambda6584\rm\,\AA$/H$\alpha$ ratio for local star-forming galaxies is in the range $10^{-1.5}-10^{-0.3}$ (e.g., \citealp{Brinchmann04}). \cite{Shapley05} found similar values at higher redshifts from a sample of star-forming galaxies at $1.0<z<1.5$, and determined an average ratio of $0.25$. [\ion{N}{2}]$\lambda6548\rm\,\AA$ is fainter than [\ion{N}{2}]$\lambda6584\rm\,\AA$, with a ratio of [\ion{N}{2}]$\lambda6548$/[\ion{N}{2}]$\lambda6584 = 1/3$ predicted theoretically (e.g., \citealp{Osterbrock&Ferland06}) and confirmed empirically (e.g., \citealp{Boselli13}). We therefore correct the H$\alpha$ SFRs from [\ion{N}{2}] flux contributions using [\ion{N}{2}]/H$\alpha = 0.3$. When estimating SFRs, the WISP survey and 3D-HST teams corrected their H$\alpha$ fluxes for [\ion{N}{2}] contributions using a [\ion{N}{2}]$\lambda6584$/H$\alpha$ ratio of $0.25$ (\citealp{Atek10}, \citealp{Fumagalli12}, respectively), {similar to the} correction used here. {This ratio typically shows a $\leq0.1\,\rm dex$ scatter (e.g., \citealp{Shapley05}), which we also add in quadrature to the H$\alpha$ flux errors to estimate H$\alpha$ SFR uncertainties.} 

When H$\alpha$ is not available, we estimate SFRs based on the [\ion{O}{3}] fluxes. As in N16, we assume a crude H$\alpha$/[\ion{O}{3}] ratio of unity and estimate SFRs using the same \cite{Kennicutt83} relation as before. {Typically, H$\alpha$/[\ion{O}{3}] ratios show a $0.2-0.5\,\rm dex$ scatter (e.g., \citealp{Mehta15}, \citealp{Suzuki16}). We therefore add a {typical} $0.35\,\rm dex$ scatter in quadrature to the [\ion{O}{3}] flux errors to estimate [\ion{O}{3}] SFR uncertainties.} We investigate H$\alpha$/[\ion{O}{3}] line ratios in \S~\ref{sec:ratios} from member galaxies where both lines are measured. When only [\ion{O}{2}] is available, we measure SFRs using the \cite{Kennicutt98} relation, ${\rm SFR} = (1.4 \times 10^{-41}{\rm\, erg^{-1} \, sec}) \times L($[\ion{O}{2}]$) \, M_{\odot}\,\rm yr^{-1}$.  

As in N16, and as done by other teams (e.g., \citealp{Zeimann12},
\citealp{Newman14}), we crudely correct our observed fluxes for
dust extinction assuming a constant dust attenuation in the $V$-band
of $1$~mag (typical of star forming galaxies; e.g., \citealp{Kewley04}, \citealp{Sobral12}),
and using the \cite{Calzetti00} extinction curve with $R_V = 4.05$
(see N16 for details). Unfortunately, we cannot accurately
evaluate the amount of dust in our galaxies without additional,
longer wavelength data. We further discuss dust contributions
in \S~\ref{sec:sfrm}.

The right panel of Figure \ref{fig:fluxsfrz} shows the line detection limit shown in the left panel converted to SFRs. 
Overall, SFRs are in the range $4 - 280 \, M_{\odot}\,\rm yr^{-1}$ (excluding the RLAGN). Due to AGN contamination, SFR values for (RL)AGN are likely overestimated, and are therefore considered upper limits. The median (mean) star-forming structure member SFR is $11\, (16) \, M_{\odot}\,\rm yr^{-1}$ based on H$\alpha$ (i.e., $1.36<z<1.59$), $31\, (37) \, M_{\odot}\,\rm yr^{-1}$ based on [\ion{O}{3}] (i.e., $1.59<z<2.02$), and $178\, (157) \, M_{\odot}\,\rm yr^{-1}$ based on [\ion{O}{2}] (CARLA~J1017+6116 at $z=2.8$). 
For similar limiting fluxes, we are limited to confirming sources with higher SFRs at higher redshifts, as expected. These numbers also clearly demonstrate the added difficulty in confirming clusters at higher redshifts, as only the galaxies with the highest SFRs are confirmed in our shallow grism data.\\

\subsection{Stellar Masses}\label{sec:mass}
We determine the stellar masses of sources with {\it Spitzer} detections from their {\it Spitzer}/IRAC $3.6$ and $4.5\, \micron$ fluxes. We scale the fluxes with the python version of {\ttfamily EZGAL} (\citealp{ManconeGonzalez12}) to \cite{BruzualCharlot03} stellar population synthesis models using a \cite{Chabrier03} initial mass function (IMF), single stellar population (SSP), and a $z_f = 4.5$ formation redshift.
We normalize the models to match the best-fit CARLA $z=1.45$ luminosity function for a Schechter parametrization with $\alpha = 1$ (\citealp{Wylezalek14}). Using the CARLA luminosity function at $z=2.05$ instead does not significantly change the results.  
Using this methodology, structure members with {\it Spitzer}/IRAC detections have stellar masses in the range $0.6 \times 10^{10} - 3.4\times10^{11}\, M_{\odot}$, with a median (mean) stellar mass of $4.1\, (7.2)\times 10^{10}\, M_{\odot}$. We derive masses in the range $10^{12}-10^{13}\, M_{\odot}$ for a few RLAGN, though these masses are likely overestimated due to AGN contributions to the spectral energy distributions (e.g., \citealp{Drouart12}) and are therefore upper limits. For structure members without {\it Spitzer}/IRAC detections, we determine upper limits based on the {\it Spitzer}/IRAC $[4.5]$ depths (\citealp{Wylezalek14}); typically, these sources have masses $<10^{10}\, M_{\odot}$.

{To investigate {the robustness of our masses}, we also estimate stellar masses using models comprised of {varying} contributions between an SSP and a population of exponentially decaying SFRs. \cite{Wylezalek14} showed that models with up to an $80\%$ contribution of the star-forming population (with an $e$-folding time $\tau=1\,\rm Gyr$) were still in good agreement with the empirical evolution of $m^{\star}_{\rm CARLA}$, while higher contributions of the star-forming population shows disagreement, especially in the highest redshift bins. In the extreme case of an $80\%$ contribution of the star-forming population, we find $\sim20\%$ lower stellar masses in the range $1.4<z<2.0$ compared to an SSP model for our {\it Spitzer}-detected member galaxies. We discuss how this can bias the SFR-mass relation in Section \ref{sec:sfrm}.}\\

\subsection{Clusters of Interest}\label{sec:clinterest}
 
We next discuss several structures and bona fide clusters of interest, including ones with supporting  
data in the literature (CARLA~J2355$-$0002, CARLA~J1753+6310, CARLA~J0800+4029, and CARLA~J2039$-$2514), and the most distant confirmed structure in our sample  
(CARLA~J1017+6116, at $z=2.8$).\\

\subsubsection{CARLA~J2355$-$0002}\label{sec:carla2355-0002}
\begin{figure}
\hspace{-0.7cm}\includegraphics[scale=0.54]{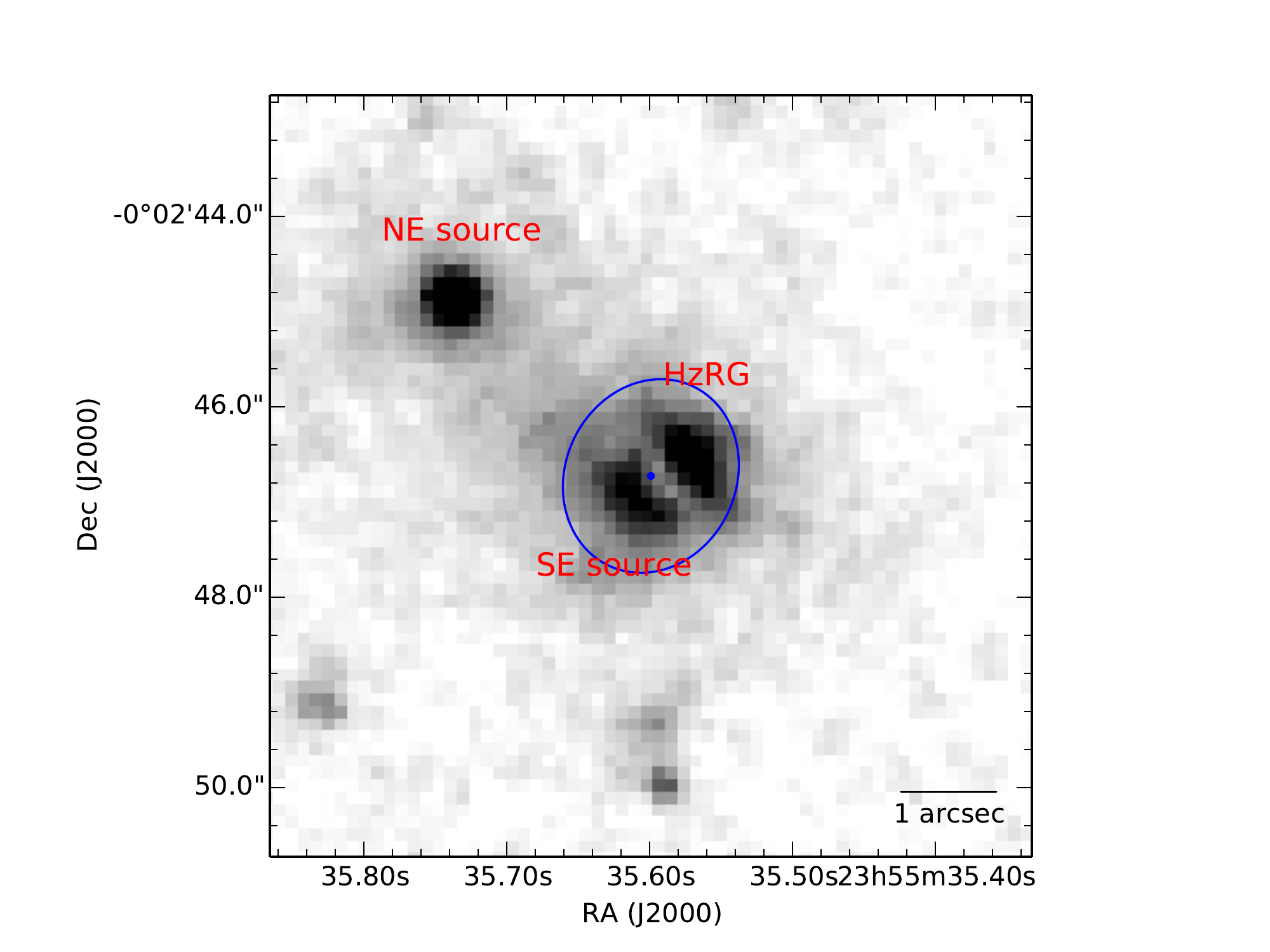}
\caption[TXS2353-003]{Close-up view of TXS2353$-$003, the targeted HzRG of CARLA~J2355$-$0002, with North up and East to the left. The WFC3 F140W imaging data reveals two sources at the Sextractor position of the HzRG (blue ellipse). Based on the G141 grism spectroscopy, we identify the HzRG as the NW source, and the SE source as a potential companion. The two sources are separated by less than $0.7\arcsec$. A third source, in the top left corner, is located $\sim 3\arcsec$ to the NE.}
\label{fig:TXSdirim}
\end{figure}

\cite{DeBreuck01} reported the redshift of the targeted HzRG of this field, TXS2353$-$003, as $z=2.587\pm0.003$ from long-slit spectroscopic observations with the ESO 3.6m EFOSC1 instrument. They reported detection of Ly$\alpha$,  
\ion{N}{5}$\lambda1240\,\rm\AA$, and \ion{C}{4}$\lambda1549\,\rm\AA$ emission. We do not find emission lines consistent with this redshift in our grism spectroscopy. In both orientations, we identify two strong emission lines consistent with H$\alpha$ and [\ion{O}{3}] at $z=1.500\pm0.006$. A similar redshift, consistent with our measurement, was independently reported in \cite{Collet15}, who studied the radio-jet and gas properties of this HzRG 
as part of a larger study on the warm ionized gas properties of $50$ HzRGs at $z\ga2$. 
With moderate-resolution VLT/SINFONI near-infrared spectroscopy ($R=1500$), they measured a redshift of $z=1.487$ for TXS2353$-$003 based on H$\alpha$ and [\ion{N}{2}]$\lambda\lambda6548,6583$ emission. Given the higher spectral resolution of their data compared to our low-resolution grism data, we adopt this value as the redshift of the HzRG. {Note that the redshift discrepancy with our grism measurement is due to the Sextractor source centroid being offset from the emission line region (see last paragraph of this section and Fig.~\ref{fig:TXSdirim}).}  \cite{Collet15} do not identify any lines consistent with 
$z\simeq2.59$.

\begin{figure}
\hspace{-0.7cm}\includegraphics[scale=0.54]{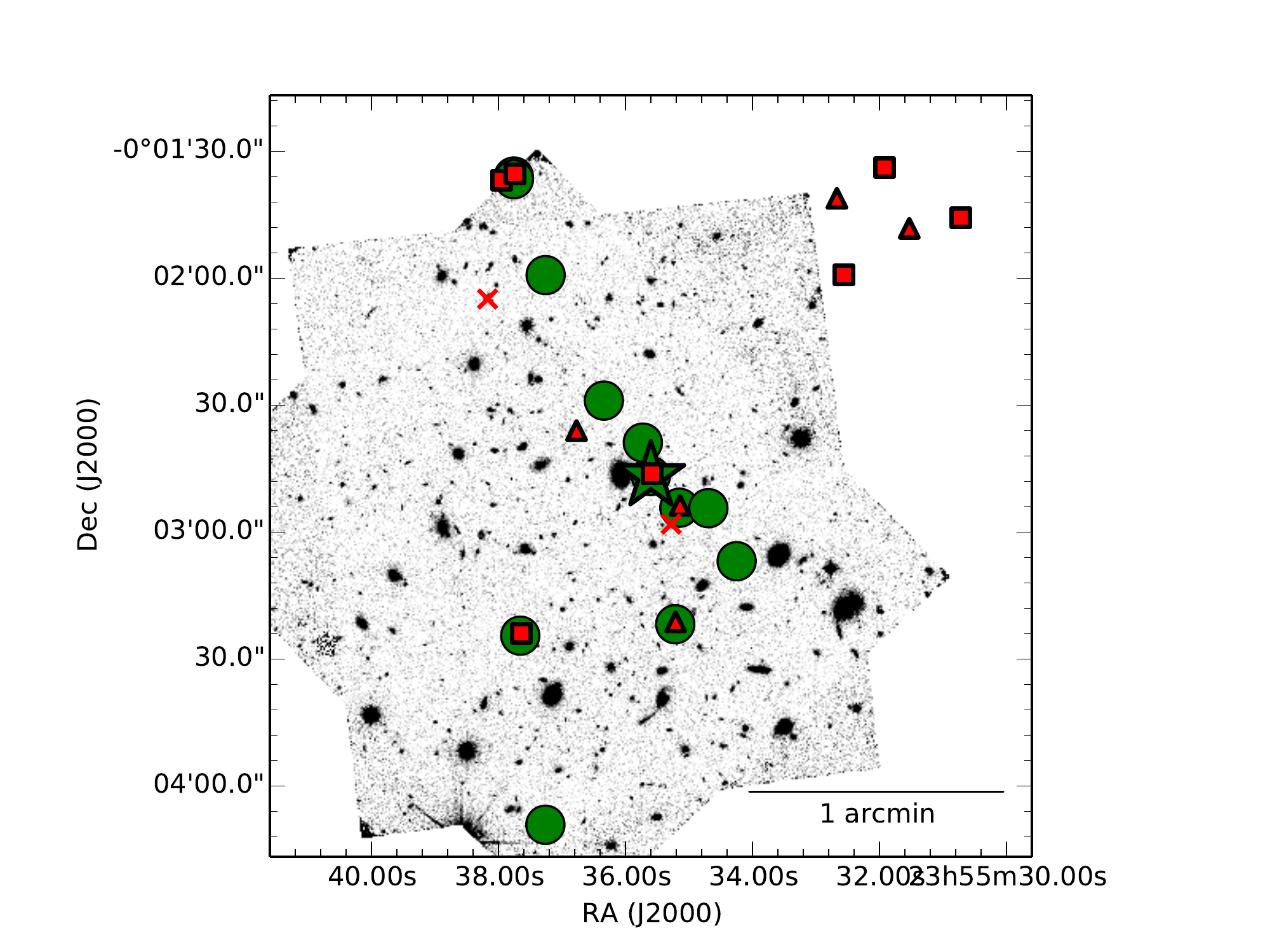}
\caption[CARLA~J2355$-$0002 Spatial Distribution]{Spatial distribution of CARLA~J2355$-$0002 members identified in our grism data (green circles), and H$\alpha$ $z\sim1.5$ emitters from \cite{Collet15} (red markers). North is up and East is to the left. The green star represents the target RLAGN of this field, TXS2353$-$003. We highlight the high-confidence H$\alpha$ emitters with red squares, the low-confidence emitters with red triangles, and additional low-confidence emitters with red crosses (private communication).}
\label{fig:TXSspadist}
\end{figure}

We only identify one source showing a tentative emission line at
$\lambda \sim 13,380\, \rm \AA$ which could be consistent with
[\ion{O}{2}] at $z\simeq2.59$. However, this identification is very
uncertain and therefore not included in our catalog. {\it HST} grism
data of this field does, however, identify twelve emission line
galaxies at a median redshift $\tilde{z}_{\rm cl} = 1.489$,
out of a total of $46$ redshift identifications. We identify both
H$\alpha$ and [\ion{O}{3}] emission for all twelve members (i.e.,
all redshifts are therefore of quality A). \cite{Collet15} also
obtained H$\alpha$ narrow-band imaging of the field around TXS
J2355$-$002 with the VLT/ISAAC $1.64\micron$ narrow-band filter.
They identified six high-confidence H$\alpha$ emitters at $z=1.5$
(down to a $3\sigma$ flux of $7\times10^{-17}\rm\,erg\, cm^{-2}\,
s^{-1}$) in addition to the HzRG, and five lower-confidence emitters.
In Figure~\ref{fig:TXSspadist} we show the spatial distribution of the grism confirmed members (green markers) and H$\alpha$ emitters (red markers) in the field.
Four high-confidence and two lower-confidence emitters fall outside
of our {\it HST} field of view (red squares and red triangles in Fig.~\ref{fig:TXSspadist}, respectively).   
Considering the narrow-band emitters within the {\it HST} field, all three remaining
high-confidence $z\sim 1.5$ H$\alpha$ emitters (which includes the
HzRG) correspond to independently confirmed members from our {\it
HST} analysis. One of these H$\alpha$ emitters actually corresponds
to two confirmed {\it HST} members separated by less than one
arcsecond. Of the three remaining low-confidence $z\sim 1.5$
H$\alpha$ emitters, all but one is independently confirmed from our
{\it HST} analysis; the unconfirmed member shows bright continuum
in our grism data, but with no indication of emission lines consistent
with $z\sim 1.5$. \cite{Collet15} also identified two additional
low-confidence $z\sim 1.5$ H$\alpha$ emitters, unpublished in
their paper (private communication; red crosses in Fig. ~\ref{fig:TXSspadist}). One of these sources
corresponds to four sources within a $1\arcsec$ radius aperture in
our {\it HST} imaging. One of these {\it HST} sources is confirmed
at $z=1.463\pm0.006$ (quality A), but is not considered a member
of CARLA~J2355$-$0002 due to its velocity shift of $\sim2500\,\rm
km\,sec^{-1}$ from the mean structure redshift, higher than our
threshold of $2000\,\rm km\,sec^{-1}$. The other three sources of
this group do not show notable spectral features. The other unpublished
low-confidence emitter was not confirmed in our initial {\it HST} analysis.
However, {\it a posteriori}, we identify tentative H$\alpha$ and H$\beta$
emission lines in both grism orientations consistent with $z=1.5$,
without {the detection of [\ion{O}{3}]}. This suggests that this source is
likely an additional structure member.  However, the identification
is uncertain at the current depth of our {\it HST} grism
data, and therefore is not included in our catalog.

\cite{Collet15} also reported peculiar radio-jet and gas properties for TXS2353$-$003, finding many similarities with brightest cluster galaxies in low-redshift cool core clusters. Specifically, they found a large ($\sim 90\deg$) offset between the position angle of the radio jets and the warm ionized gas, and a large radio size of $328\rm\,kpc$. This suggests no direct interaction between the radio jets and the galaxy gas. They also identified a possible companion to the HzRG, $3 \arcsec$ to the North-East, with no emission lines in the SINFONI spectroscopic data. We also detect this source in the {\it HST} data (see Fig. \ref{fig:TXSdirim}), similarly detecting only continuum without emission lines in the grism data. However, the approximately two times better {\it HST}/WFC3 spatial resolution compared to SINFONI allows us to identify two sources at the position of the HzRG, with visually complex morphologies. As seen in Fig. \ref{fig:TXSdirim}, the two sources are separated by less than $0.7\arcsec$, and show distinct spectral traces in the grism data (see Panel d in Fig. \ref{fig:J2355-0002spectra} in Appendix \ref{app:spectra}). We identify faint continuum with strong H$\alpha$ and [\ion{O}{3}] emission lines for the North-Western source, which we identify as the HzRG. We identify bright continuum with no clear emission lines for the South-Eastern source.  
\cite{Collet15} argued that the misalignment of the gas and jet could be explained by gas supply from a satellite galaxy.  
They however also argued that the potential companion $3\arcsec$ to the NE is unlikely
a perturbing satellite. The other source we identify less than $0.7\arcsec$ to the SE of the HzRG, unresolved in their imaging, seems a better candidate to be associated to the HzRG and responsible for the jet-gas misalignment.\\

\subsubsection{CARLA~J1753+6310}\label{sec:carlaJ1753+6310}
Based on William Herschel Telescope (WHT) ISIS spectroscopic data, \cite{Lacy99} measured a tentative redshift of $z=1.96$ for this source based on uncertain identification of a \ion{He}{2} emission line and Ly$\alpha$ absorption. We do not find any emission lines consistent with this redshift in our grism data. 
\cite{Cooke16} 
studied the broadband spectral energy distributions of candidate members of this distant galaxy cluster, CARLA~J1753+6310 (CARLA~J1753+6311 in their paper) by combining the CARLA {\it Spitzer}/IRAC observations with deep optical imaging data.
They also obtained Keck/LRIS 
optical spectroscopic data for the RLAGN and measured a redshift of $z=1.576$ based on identification of a single [\ion{O}{2}] emission line. 
In our grism data, we detect bright continuum in both orientations and strong emission at $13,063\rm\,\AA$ and $12,993\rm\,\AA$ in the first and second orientations, respectively (see Panel b in Fig. \ref{fig:J1753+6310spectra} in Appendix \ref{app:spectra}). The emission line is spatially offset from the continuum, by approximately $+0.5\arcsec$ in the first orientation and approximately $-0.5\arcsec$ in the second orientation, while the continuum is aligned with the source centroid from SExtractor in both orientations. This suggests the presence of a line-emitting region located less than an arcsecond to the West of the SExtractor source centroid. With this close separation we consider the system as a single galaxy, the HzRG, at the redshift measured 
with Keck. 
The emission lines in our grism data are consistent  
with [\ion{O}{3}] at 
similar redshift\footnote{We initially measured $z = 1.595 \pm 0.004$ in the first orientation and $z = 1.609 \pm 0.004$ in the second orientation, under the assumption that the emission line regions were associated with the SExtractor source centroid. Adjusting the wavelength calibration based on the apparent small spatial offset should slightly lower the measured redshift, bringing it into agreement with the redshift measured by \cite{Cooke16}.}.
H$\alpha$ unfortunately falls outside of the grism range at this redshift, but we assign a quality A to the HzRG redshift based on the complementary {Keck detections of} [\ion{O}{2}] and H$\alpha$ lines {reported} by \cite{Cooke16} and A.~Rettura et al.~(in prep), respectively.

We identify five emission line members at a median redshift of $\tilde{z}_{\rm cl}=1.581$. These identifications are based on [\ion{O}{3}] emission only, as H$\alpha$ falls outside of the grism range at this redshift. With additional Keck/MOSFIRE observations of the field, we confirm a combined total of $8$ emission line members at $\langle z \rangle$=1.582 (A.~Rettura et al., in prep.). Among these $8$ sources, three are new members confirmed outside the {\it HST} field of view with the MOSFIRE spectroscopy based on H$\alpha$ emission. Excluding the HzRG which was re-observed as well, A.~Rettura et al.~(in prep.) also present complementary H$\alpha$ emission for {\it HST} source $\#619$, which was already confirmed in our grism data based on [\ion{O}{3}] emission. We therefore assign a quality A redshift to this source.

\cite{Cooke16} 
demonstrate that CARLA~J1753+6310 is a mature cluster at high-redshift. They identify a red sequence population dominated by passive galaxies (i.e., $80\%$ of the red sequence galaxies have broadband colors indicative of a passive population). They show that half the cluster galaxies in the core are quiescent, as compared to only $16\%$ of field galaxies of similar mass and redshift. The relatively small number of confirmed star-forming members in our grism data is therefore consistent with this picture of an evolved cluster largely comprised of passive members with minimal star formation.

The {\it HST} grism data of this field also identify a group of {six} emission line sources at a median redshift of $\tilde{z}_{\rm cl}=2.117$, within $\pm1000\,\rm km\,sec^{-1}$. This structure, serendipitously discovered, is comprised of {one} quality A and five quality B$^{-}$ sources. None of the sources have secure {\it Spitzer}/IRAC detections. {The quality A source was} identified on the basis of [\ion{O}{3}] and [\ion{O}{2}] emissions, while the quality B$^{-}$ sources were identified based solely on [\ion{O}{3}] emission. 
This system is the second highest redshift confirmed structure in our sample. A.~Rettura et al.~(in prep.) present this structure in more detail.
This discovery, as well as the serendipitous discovery of three additional $z>1.3$ structures not associated with the targeted RLAGN (in the fields of CARLA~J1317+3925 and CARLA~J2227$-$2705; see Table \ref{table:res}), suggests that a fraction of the {\it Spitzer}/IRAC overdensities are enhanced by line-of-sight projections 
--- though we note that all the confirmed members of the background structure identified in this field are actually below the CARLA IRAC detection limit. Potentially, passive, massive members of this background structure are in the CARLA sample, but were not confirmed in the grism spectroscopy. {Section \ref{sec:ser} discusses serendipitous structures identified in our grism data.}\\

\subsubsection{CARLA~J0800+4029 \& CARLA~J2039$-$2514}
CARLA~J0800+4029 and CARLA~J2039$-$2514 are two of the highest-redshift confirmed structures associated with the targeted RLAGN, at $\tilde{z}_{\rm cl}=1.986$ and $\tilde{z}_{\rm cl}=2.000$, respectively. These were two of the first fields observed by our {\it HST} program, and the grism results are presented in N16. Based on additional optical data, we show in N16 that CARLA~J2039$-$2514 possesses a red-sequence population of passive galaxies. This suggests that CARLA~J2039$-$2514 is a bona fide $z=2$ galaxy cluster, while our analysis suggests that CARLA~J0800+4029 is a younger forming cluster at similar epoch. We also show in N16 that the targeted HzRG of CARLA~J2039$-$2514 is kinematically complex with two components, likely a dual AGN, separated by $3\,\rm kpc$ and confirmed at the same redshift. This complex morphology, together with similar findings for the HzRGs of CARLA~J2355$-$0002, CARLA~J1753+6310 and several other fields (e.g., the targeted RLAGN MRC0955$-$288,  
6CE1100+3505,  
MRC1128$-$268,  
J1510+5958) 
is consistent with type-2 RLAGN often being found in close merging systems (e.g., \citealp{Chiaberge15}). We redirect the reader to N16 for further details concerning CARLA~J0800+4029 and CARLA~J2039$-$2514, such as redshift distributions, mass estimates, color-magnitude relation (CMR) analysis, and comparison to other individual high-redshift clusters.\\

\subsubsection{CARLA~J1017+6116}
CARLA~J1017+6116 is the highest redshift confirmed structure in our CARLA {\it HST} sample. We identify {seven} emission line members at $\tilde{z}_{\rm cl} =2.801$. At this redshift, H$\alpha$, [\ion{O}{3}], and H$\beta$ fall outside the G141 grism wavelength range, and identifications are therefore based solely on [\ion{O}{2}] emission. We however do not identify [\ion{O}{2}] emission in the grism data of the targeted QSO of this field, but additional strong [\ion{O}{3}]$\lambda4363\,\rm\AA$, H$\delta$, and [\ion{Ne}{3}]$\lambda3869\,\rm\AA$, at $z = 2.80$ (quality A). {Excluding the targeted QSO}, two confirmed members ({\it HST} sources $\#124$ and $\#382$) have {\it Spitzer}/IRAC detections with mid-infrared colors consistent with $z>1.3$. This simple color criterion alone, however, does not disentangle between identifying the detected lines as [\ion{O}{3}] or [\ion{O}{2}]. 
{These two} sources have some of the reddest {\it Spitzer}/IRAC colors measured in our sample, with $([3.6] - [4.5])_{\rm AB}$ colors of $0.4$ {mag for source $\#124$} and $0.7$ mag {for source $\#382$}, consistent with very distant galaxies. We {conservatively} assign quality B$^{-}$ redshifts to these two sources, as the color information alone also allows for an [\ion{O}{3}] identification. The redshifts of the other five sources are based on the detection of single emission lines, and are therefore of quality B$^{-}$ as well.

We also serendipitously identify another structure in the grism data of this field. We confirm five emission line sources at a median redshift of $\tilde{z}_{\rm cl}=1.234$. Three of these sources have both multiple-line identifications of H$\alpha$ and [\ion{O}{3}], as well as {\it Spitzer}/IRAC detections. The other two are identified based solely on H$\alpha$.\\

\subsection{Serendipitous Discoveries}\label{sec:ser}

\begin{figure*}
\centering
\includegraphics[scale=0.54]{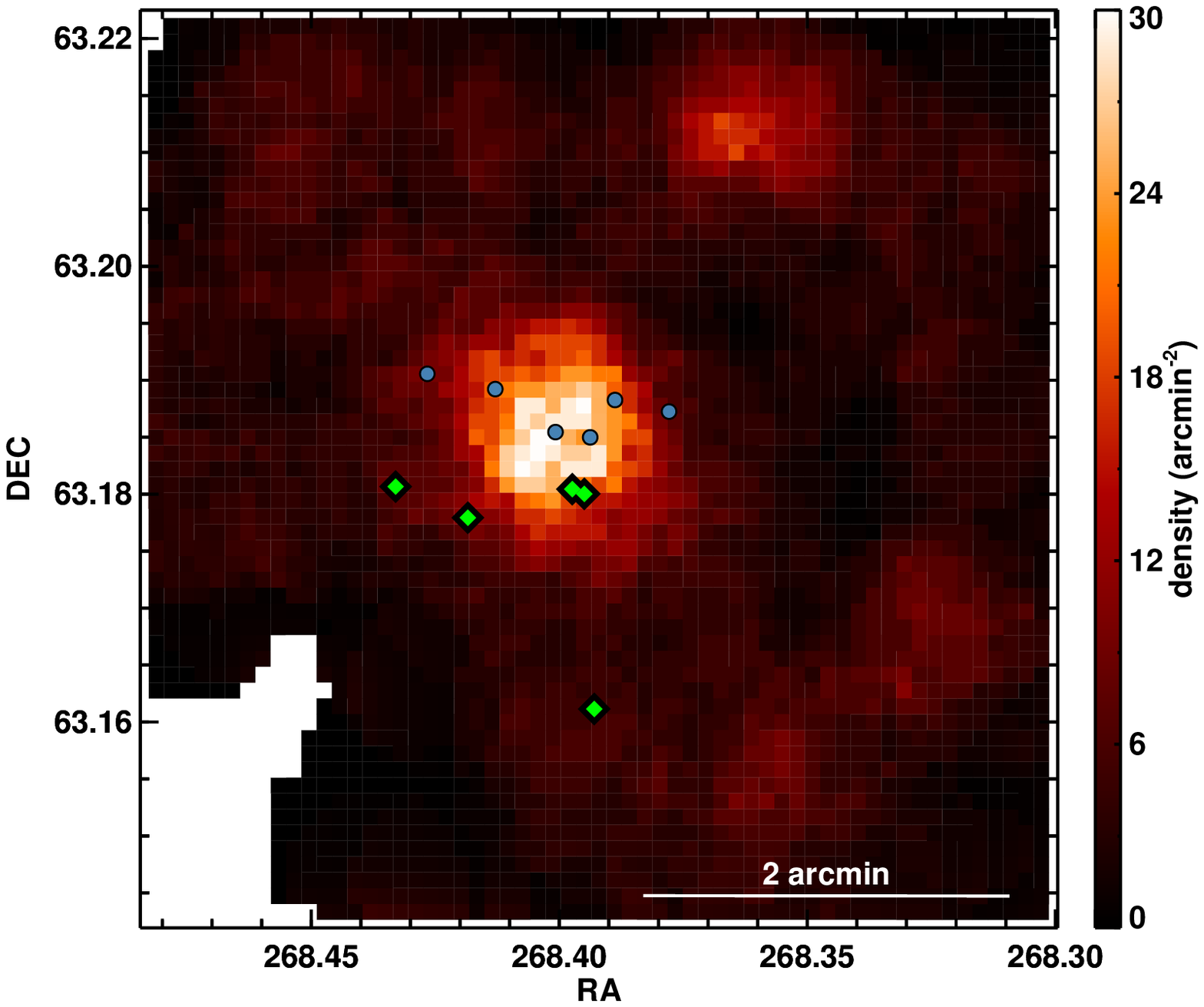}\vspace{-1.5cm}\\
\includegraphics[scale=0.54]{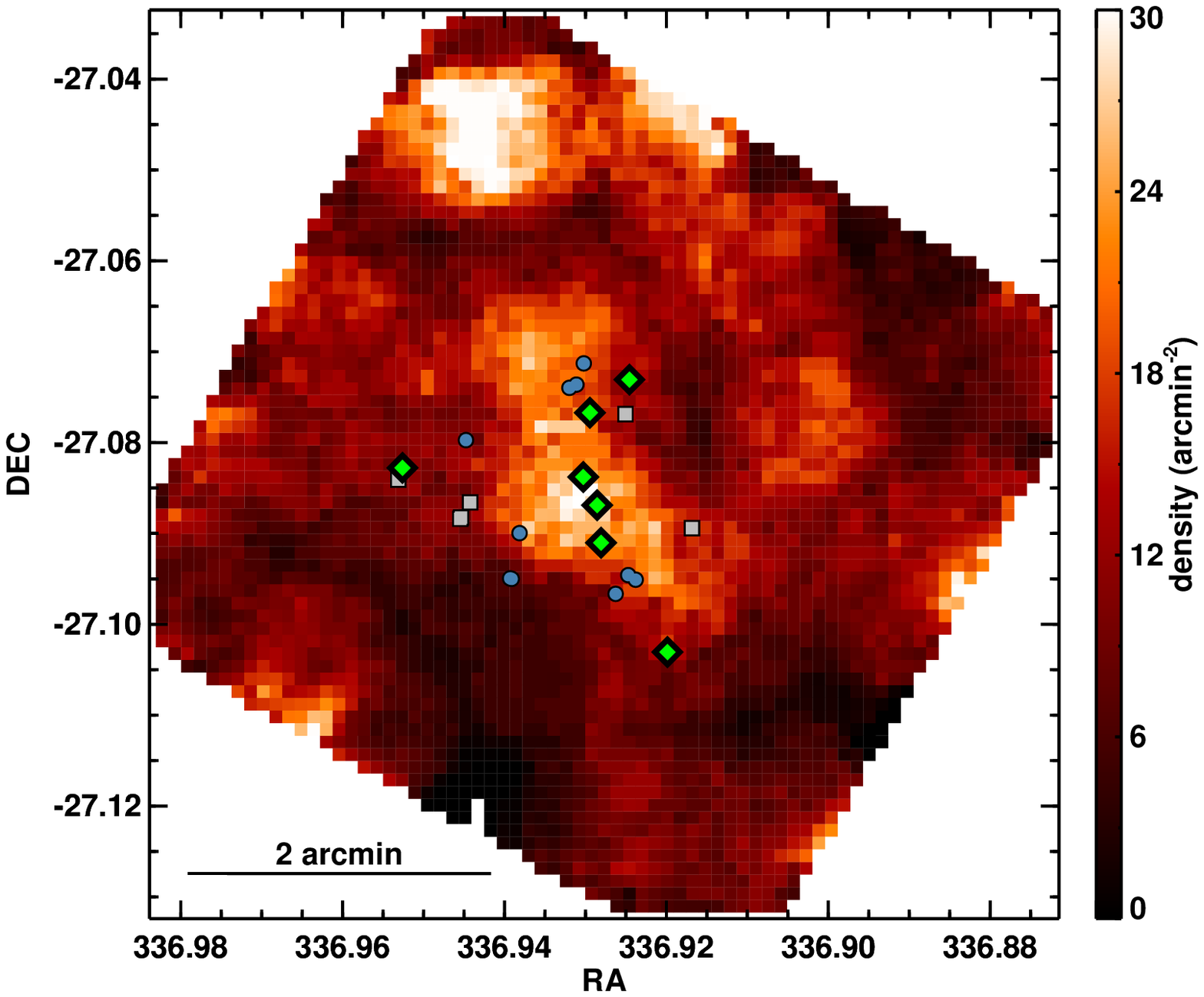}\hspace{-0.4cm}
\includegraphics[scale=0.54]{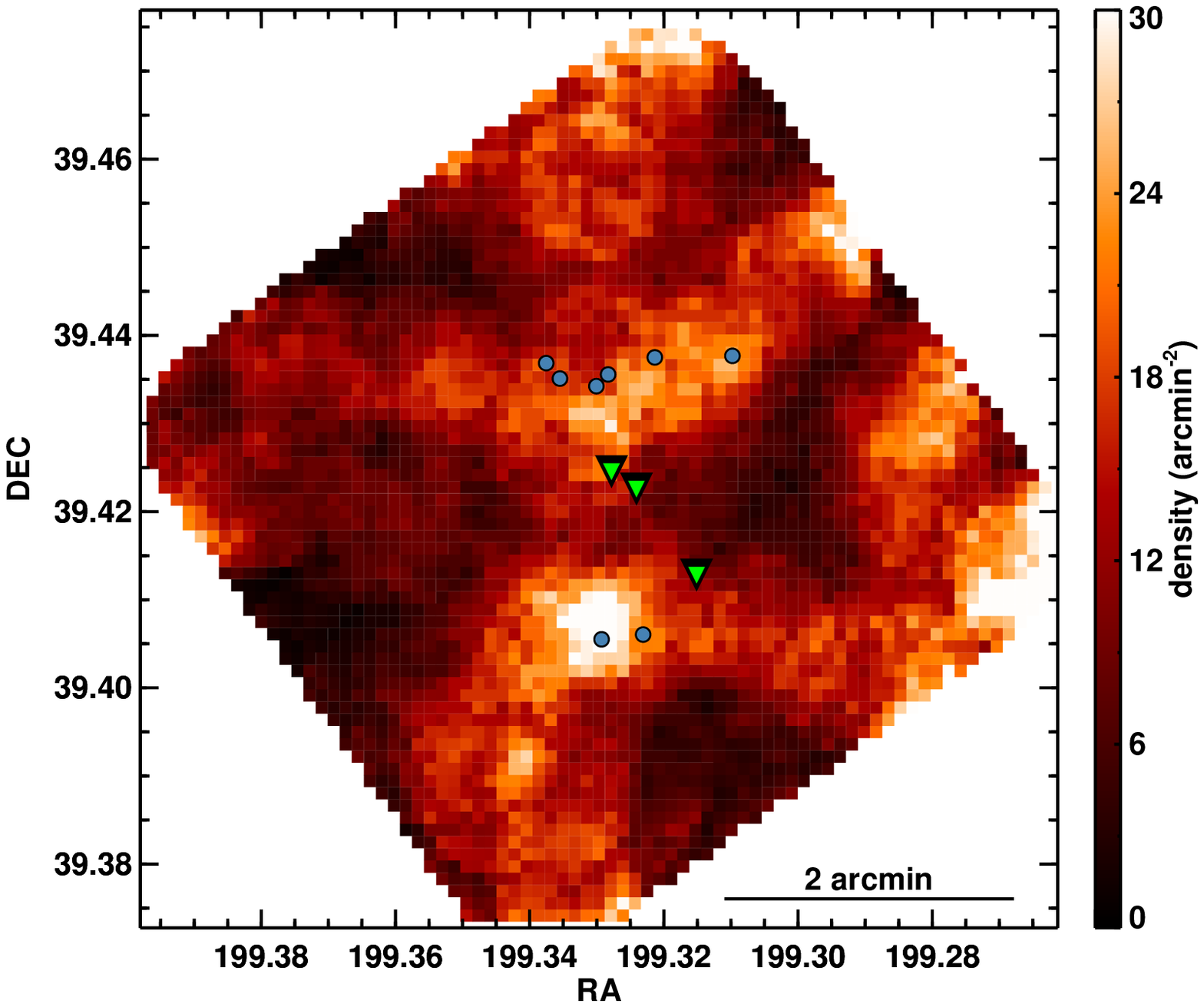}\vspace{-0.4cm}
\caption[Density Maps]{Density maps, in terms of {Spitzer}/IRAC color-selected sources, of the three fields containing $z>1.3$ serendipitous structures. Top panel: field of CARLA~J1753+6310. Bottom left panel: field of CARLA~J2227-2705. Bottom right panel: field around RLAGN J1317+3925. In each field, confirmed CARLA members are shown with green diamonds, and foreground or background members of serendipitous structures with blue circles, and also light-blue squares if a second serendipitous structure is present in the field (e.g., bottom left panel). Galaxies at similar redshift as the targeted RLAGN J1317+3925 in this unconfirmed CARLA field (bottom right panel) are shown with green triangles (RLAGN included). We masked regions of poor coverage, edges, and bright stars in our {\it Spitzer}/IRAC imaging and measured the number density of color-selected sources within 30 arcsec radius apertures around each 5 arcsec pixel. The {\it HST} field of view ($2\times2.3\rm\, arcmin^{2}$) covers only the central region of each {\it Spitzer}/IRAC image.}
\label{fig:densitymaps}
\end{figure*}

In addition to the $16$ confirmed CARLA structures associated with the targeted RLAGN of our program, we identify a total of seven serendipitous structures among five of the twenty {\it HST} fields. We refer to these {serendipitous structures in the CARLA fields by the nomenclature ``CARLA-Ser\textbf{\it X}~Jhhmm$\pm$ddmm" where \textbf{\it X} is an integer, starting at 2 and in redshift order, if two or more serendipitous structures are identified in the same CARLA field, and where Jhhmm$\pm$ddmm are the CARLA coordinates of the respective field}. As noted earlier, a fraction of the {\it Spitzer}/IRAC overdensities might be enhanced by line-of-sight projections. In the last column of Table \ref{table:res}, we show the number of structure members which satisfy our {\it Spitzer}/IRAC color-criterion among the total number of sources with a measured redshift satisfying the same criterion. When compared to the second-to-last column, which shows the same numbers for {\it Spitzer}/IRAC detected sources, we see that  
only one of the {9 {\it Spitzer}-detected} confirmed members of the $z<1.3$ serendipitous structures has a color matching our $z>1.3$ selection criterion.
This corroborates the goodness of our selection criterion and further suggests that the overdensity significance ($\sigma_{\rm IRAC}$, second column in Table \ref{table:res}) of the fields containing the three $z<1.3$ serendipitous structures is {minimally} contaminated by these foreground structures.
The serendipitous structures at $z>1.3$, however, are more likely to enhance the CARLA overdensities since a large fraction of $z>1.3$ galaxies detected by {\it Spitzer} should satisfy our color-selection criterion. We discover four serendipitous structures at $z>1.3$, including two within the same field (CARLA~J2227$-$2705). {CARLA-Ser~J1753+6310}, discovered at $\tilde{z}_{cl} = 2.117$ in the field of CARLA~J1753+6310, is comprised of {six} faint, star-forming members, none of which are detected in our {\it Spitzer}/IRAC data. The members of this serendipitous structure, however, lie close to the core of CARLA~J1753+6310 as seen in the density map of that field in the top panel of Figure \ref{fig:densitymaps}. We therefore cannot rule out a contribution of potential {CARLA-Ser~J1753+6310} members to the CARLA overdensity of this field.
The two serendipitous structures identified in the field of CARLA~J2227$-$2705 appear not to be contributing significantly to the CARLA overdensity in this field, as shown by their member spatial distribution overlaid on the CARLA density map in the bottom left panel of Figure \ref{fig:densitymaps}. We see that CARLA members align with the density of red sources whereas the serendipitous members are less clearly associated.
We do not confirm a CARLA structure associated with the targeted RLAGN J1317+3925 at $z=1.569$. In the grism data of this field, we only identify two additional sources at the redshift of the RLAGN, as shown in Table \ref{table:res}. We however serendipitously identify {CARLA-Ser~J1317+3925}, a foreground structure at $\tilde{z}_{cl} = 1.465$ comprised of $8$ galaxies. Unlike the two serendipitous structures in the field of CARLA~J2227$-$2705, {CARLA-Ser~J1317+3925} members spatially align with the overdensity of red sources as shown in the density map of that field in the bottom right panel of Figure \ref{fig:densitymaps}. On the other hand, the two sources at the RLAGN redshift do not appear associated with the overdensity. This suggests that the CARLA overdensity in this field is potentially associated with {CARLA-Ser~J1317+3925}. This structure is therefore likely a ``spurious" detection of an overdense structure, and not associated with the targeted RLAGN.

To investigate whether we find a larger number of serendipitous structures than expected, we calculate the number of sources in the 3D-HST survey (\citealp{Momcheva16}) which belong to a structure as defined in Section \ref{sec:memb}. {To compare to our observations and redshift determinations, we select 3D-HST sources which have usable grism redshifts ($use\_zgrism = 1$), and have any combinations of identified H$\alpha$, [\ion{O}{3}], or [\ion{O}{2}] emission lines above our limiting fluxes ($2.5\times10^{-17}\rm\,erg\,cm^{-2}\,\sec^{-1}$) and falling within the wavelength range covered by the G141 grism ($\lambda = 1.08 - 1.7\rm\,\micron$).} Using $1$ arcmin and $\pm 2000\, {\rm km}\, {\rm s}^{-1}$ thresholds, we find that $7\%$ of {such selected} galaxies in 3D-HST are members of structures containing at least five confirmed members. In contrast, the comparable number from our similarly deep CARLA program is $27\%$, with three-quarters ($20\%$) associated with the RLAGN structures and one-quarter ($7\%$) associated with serendipitous structures. {This is consistent with our observations probing biased, rich, environments. Assuming no strong confirmation bias, }
the number of serendipitous structures identified in the CARLA program is  
consistent with expectations based on 3D-HST.\\

\section{Discussion}\label{sec:discussion}

\subsection{High-Redshift Clusters}\label{sec:highzcomp}

\begin{figure*}
\hspace{-0.5cm}\includegraphics[scale=1]{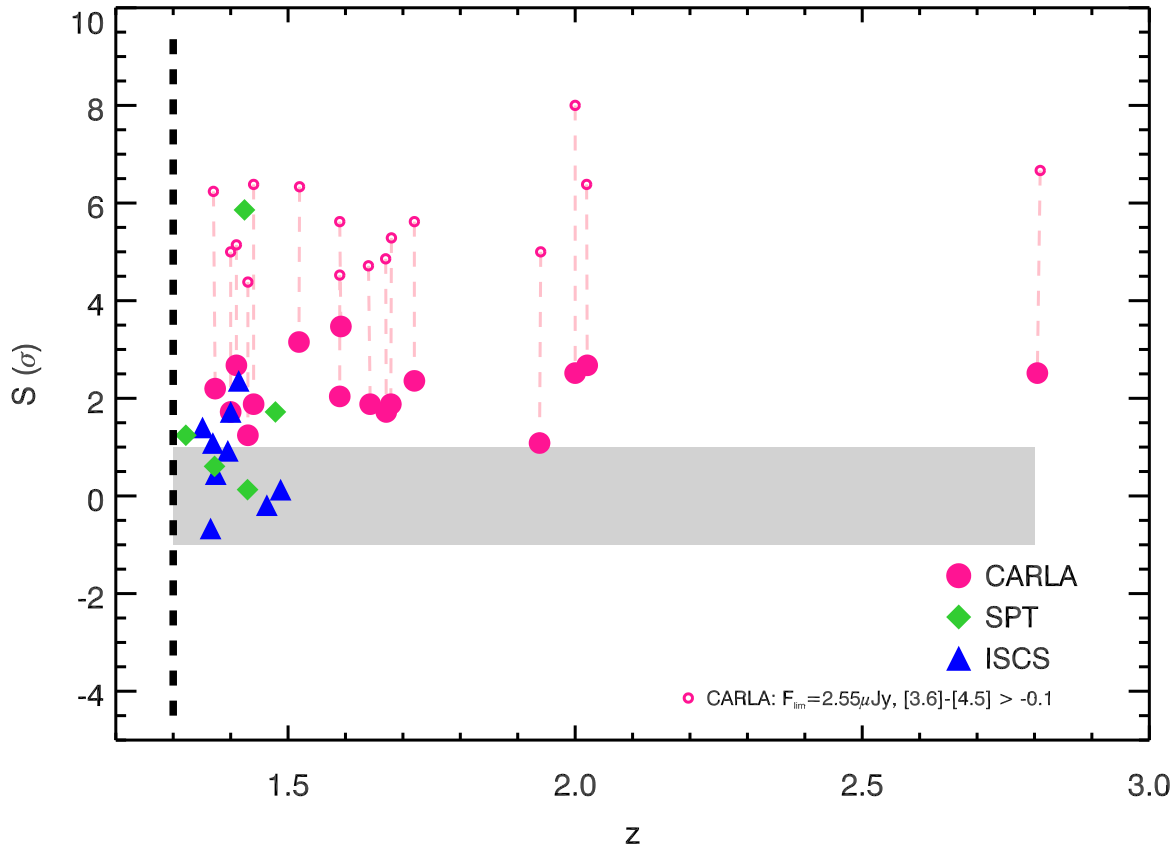}
\caption[Overdensities vs. $z$]{
Overdensity significance $S$ versus redshift for spectroscopically confirmed SPT and ISCS clusters at $z >1.3$ and our spectroscopically confirmed CARLA clusters. The horizontal grey box shows the $\pm 1\sigma$ confidence region around $S = 0$.  
We apply the same color cut of $[3.6]-[4.5]>-0.1$ to SPT and ISCS clusters as for the CARLA fields and show the overdensity significance for all three samples at the same limiting flux density of $F_{\rm{lim}} = 10\,\mu$Jy (filled green diamonds, filled blue triangles and filled pink circles, respectively). 
Although selected in different ways, the CARLA fields are similar or richer than SPT and ISCS clusters at comparable redshifts. For completeness, we also show the CARLA overdensity significance $S$ at the original limiting flux density of the CARLA IRAC observations, $F_{\rm{lim}} = 2.55\,\mu$Jy (pink open circles).}
\label{fig:sigvsz}
\end{figure*}

\begin{figure*}
\centering
\includegraphics[scale=0.65]{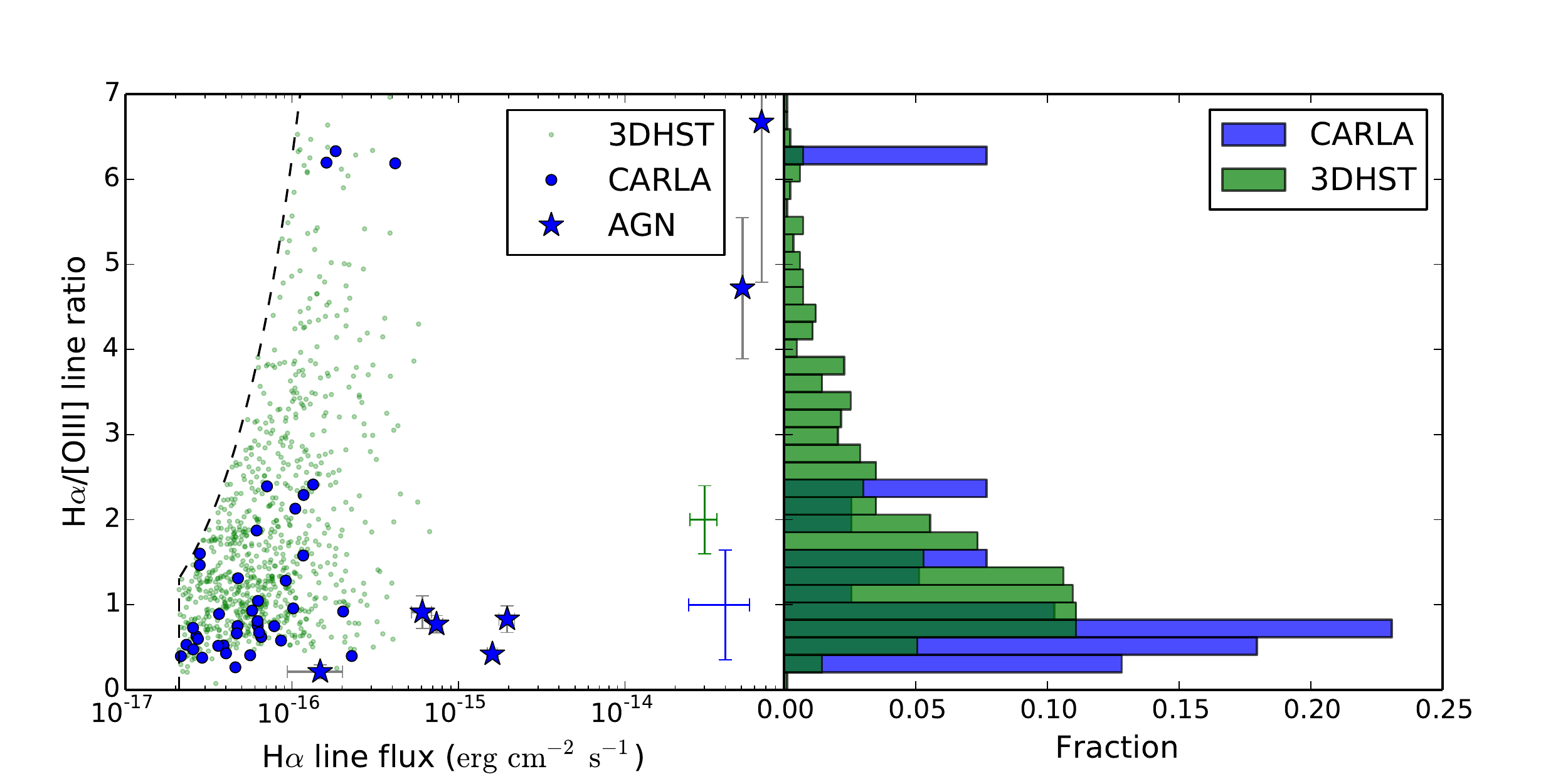}
\caption[Line Ratios]{The left panel shows H$\alpha$/[\ion{O}{3}] line ratios of confirmed members (solid blue circles) against H$\alpha$ fluxes, compared to 3D-HST (small green dots). Stars represent the targeted RLAGN. {In the bottom right corner of the left panel, we also show the typical error bars of star-forming CARLA members and 3D-HST sources in blue and green, respectively. We show individual error bars for our RLAGN. The dashed line represents the H$\alpha$/[\ion{O}{3}] line-ratio lower limit derived from the 3D-HST line detection limit.} The right panel shows the ratio distributions (normalized), in blue for our confirmed structures (excluding AGN), and in green for 3D-HST (overlaid on the blue histogram).}
\label{fig:lineratio}
\end{figure*}

Our shallow 2-orbit per field strategy proves to be efficient at confirming {\it Spitzer} color-selected high-redshift structures, with an $80\%$ confirmation rate (16/20 CARLA confirmations). Our grism confirmation method additionally allowed us to serendipitously discover and confirm other high-redshift structures not associated with the targeted RLAGN.

In the following section, we compare the spectroscopically confirmed high redshift CARLA clusters with other high redshift massive clusters. As laid out in the introduction, finding and confirming high redshift galaxy clusters is non trivial. Although much progress has been made in the past decade in finding and characterizing galaxy clusters at $z >1$, confirmed galaxy clusters at even higher redshift ($z \gtrsim 1.3$) remain rare, making a direct comparison to the confirmed CARLA clusters challenging.

The South Pole Telescope (SPT) Sunyaev-Zel'dovich (SZ) survey (\citealp{Reichardt13}; \citealp{Bleem15}) discovered a few tens of high redshift galaxy clusters at $z  > 1$, with about a third of those at $z > 1.3$ \citep{Bleem15}. In contrast to CARLA clusters, the SZ-selected SPT clusters naturally represent a mass-selected sample and the structures belong to the most massive structures known at this epoch. Another very successful program, the IRAC Shallow Cluster Survey (ISCS, \citealp{Eisenhardt08}), identified $335$ galaxy cluster and group candidates, including $106$ at $z > 1$, using a $4.5\, \micron$-selected sample of objects in the $8.5\deg^{2}$ Bo\"{o}tes field. These systems are typically less massive clusters as they were selected from a relatively small survey area. The ISCS clusters were identified as three-dimensional overdensities using a wavelet algorithm, based on photometric redshift probability distributions. Similar to CARLA clusters, ISCS clusters therefore form an IRAC-based galaxy richness-selected sample, although the exact details of the selection differ from the CARLA selection. \citet{Brodwin13} report on the high central SFRs in $16$ spectroscopically confirmed ISCS clusters at $1 < z < 1.5$.

For all spectroscopically confirmed $z>1.3$ SPT and ISCS clusters with existing IRAC data, we generate IRAC1 and IRAC2 source catalogs using the same procedures and parameters as for the CARLA catalogs (\citealp{Wylezalek13,Wylezalek14}).  
We use the same color cut of $[3.6]-[4.5]> -0.1$ as for the CARLA fields (\citealp{Wylezalek13}). Similar to \cite{Wylezalek13,Wylezalek14}, we then measure the density of the IRAC-selected sources, $\Sigma_{\rm{x}}$, within an $1\arcmin$ radius aperture around each cluster center. SPT cluster centers have been re-calculated using the IRAC photometry to calculate the centroid of the galaxy distribution from the IRAC-selected sources (Gonzalez et al.~in prep., Wylezalek et al.~in prep.). ISCS cluster centers were determined based on the wavelet analysis from weighted galaxy density maps (\citealp{Eisenhardt08}). 

To account for the different depths between the SPT, ISCS and CARLA {\it Spitzer}/IRAC observations, we apply a common flux density cut of $F_{\rm{lim}} = 10\,\mu$Jy (see \citealp{Wylezalek13} for details on estimating limiting flux densities) to the SPT, ISCS and CARLA fields, shallower than the original depth of the CARLA survey ($F_{\rm{lim}} = 2.55\,\mu$Jy). For completeness, we also show the CARLA overdensity significance at the original depth of the CARLA survey, $F_{\rm{lim}} = 2.55\,\mu$Jy (pink open circles in Figure \ref{fig:sigvsz}). To derive density estimates for a blind field, $\Sigma_{\rm{background}}$, and the corresponding standard deviation to the blind field density distribution, $\sigma_{\rm{background}}$, we apply the same criteria to a distribution of random locations from SpUDS. We then estimate the significance $S$ of the overdensity of IRAC-selected sources in the fields using: 

\begin{equation} S =\frac{ \Sigma_{x} - \Sigma_{\rm{background, x}}}{\sigma_{\rm{background, x}}}
\end{equation}
where $x$ denotes that we repeat this calculation for the SPT, ISCS and CARLA fields with the corresponding background estimates, respectively. 

In Figure \ref{fig:sigvsz}, we compare the IRAC overdensity significances, $S$, of SPT and ISCS clusters to our confirmed CARLA clusters.  
While CARLA, SPT and ISCS clusters have been selected in different ways, CARLA clusters tend to show a similar or even higher galaxy richness than SPT and ISCS clusters although the SPT clusters are expected to represent the most massive and virialized systems among the three samples. It is beyond the scope of this paper to evaluate how and if galaxy richness can be used to estimate the total mass of a galaxy cluster (see \citealp{Rettura17} in this respect). Though, this simple analysis shows that our CARLA fields (the most overdense among the full CARLA sample, by selection) are comparable to confirmed massive, evolved systems in terms of overdensity of {\it Spitzer}/IRAC color-selected galaxies. The strength of the CARLA survey is to find similarly significant overdensities at $z > 1.5$. {We however caution the reader that the non-evolving flux cuts used here preclude direct comparison of the different redshift bins with each other. Investigation of evolutionary trends --and whether the current sample of 16 spectroscopically confirmed structures represent an evolutionary sequence-- is therefore beyond the scope of the current analysis.}\\

\begin{figure*}
\centering
\includegraphics[scale=0.8]{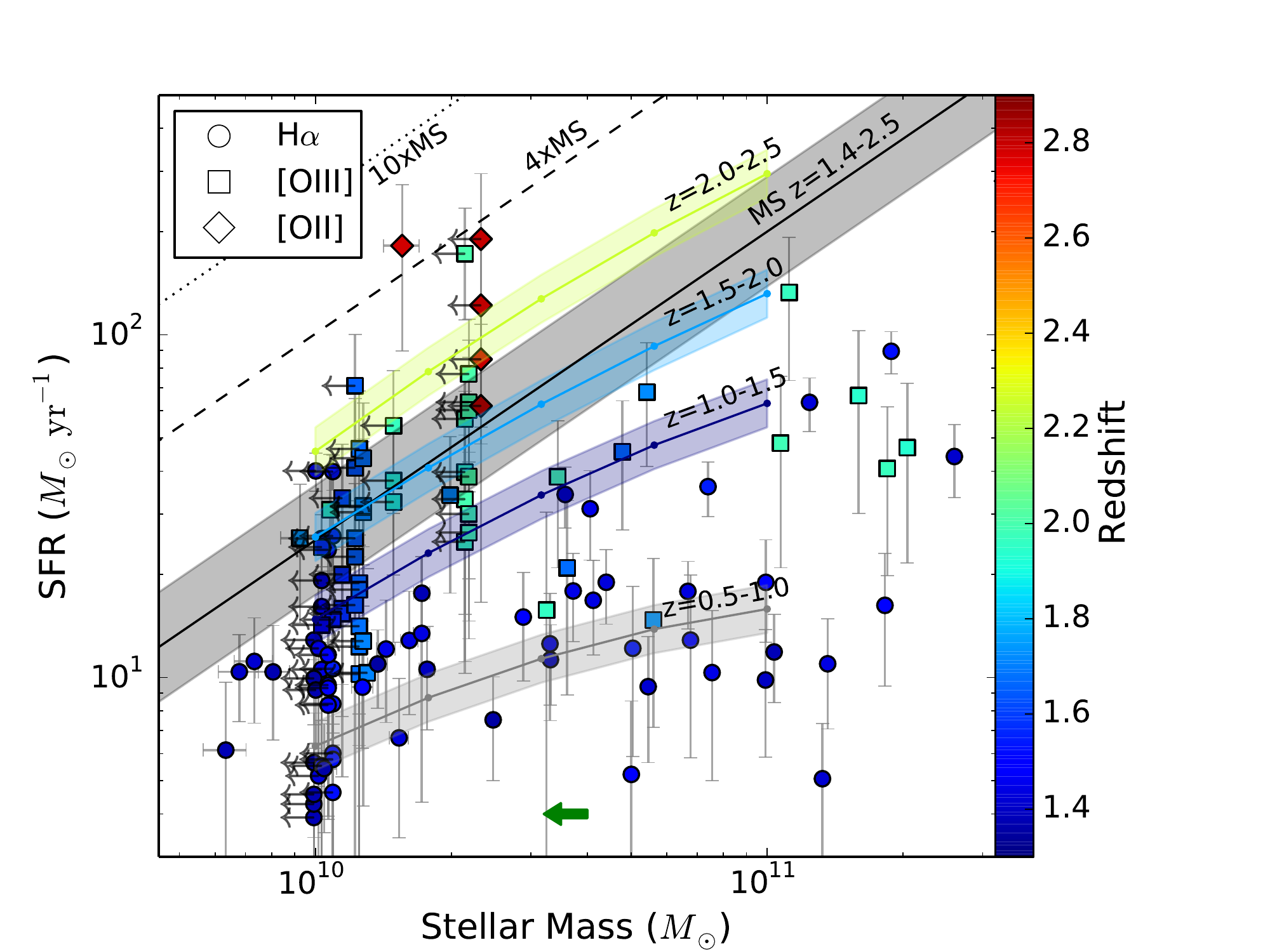}
\caption[SFRs vs. M]{SFR of members as a function of galaxy stellar mass compared to the \cite{Daddi07} $z=1.4-2.5$ star-forming main-sequence (black line), and the main-sequences derived in \cite{Whitaker14} in the ranges $z=0.5-1.0, 1.0-1.5, 1.5-2.0,$ and $2.0-2.5$ (colored curves). The dashed and dotted lines represent $4\times$ and $10\times$ the \cite{Daddi07} main-sequence, respectively. SFRs derived from H$\alpha$, [\ion{O}{3}], and [\ion{O}{2}] are shown with circles, squares and diamonds, respectively. All sources are color-coded according to redshift.  
We show upper limits on the stellar masses (leftward arrows) of sources without {\it Spitzer}/IRAC detection. The leftward green arrow represents the shift in stellar mass when using a combination of SSP ($20\%$) and decaying star-forming population ($80\%$) instead of an SSP-only model (see \S~\ref{sec:mass}).  
See \S~\ref{sec:sfrm} for details on the confidence regions around the field main-sequences.}
\label{fig:sfrvsm}
\end{figure*}

\subsection{Line Ratios}\label{sec:ratios}

In Figure \ref{fig:lineratio}, we show the H$\alpha$/[\ion{O}{3}] line ratios of confirmed members against H$\alpha$ flux, compared to field values from the 3D-HST survey (\citealp{Momcheva16}). We select sources from 3D-HST {with usable grism redshifts ($use\_zgrism=1$) in the range} where we observe both H$\alpha$ and [\ion{O}{3}] (i.e., $1.37<z<1.59$) and use 
their $3\sigma$ emission line flux limit of $2.1\times 10^{-17}\, \rm erg\, cm^{-2}\, s^{-1}$. We assume a 1:3.2 [\ion{O}{3}] doublet ratio (see N16) to convert the 3D-HST [\ion{O}{3}] doublet fluxes into [\ion{O}{3}]$\lambda 5007\, \rm \AA$ fluxes. As noted earlier, the H$\alpha$ fluxes include the [\ion{N}{2}] contribution. 
We find a median of $0.76$ for star-forming CARLA members (i.e., excluding AGN), slightly lower than unity, and lower than the 3D-HST median of $1.43$ for sources within $0.15 < \rm H\alpha/$[\ion{O}{3}]$\mbox{}< 7$. Our results suggest that, on average, [\ion{O}{3}] is brighter than H$\alpha$ at these redshifts for cluster galaxies. This is consistent with a decreasing ratio with redshift, where H$\alpha$ is typically brighter than [\ion{O}{3}] at lower redshifts (e.g., \citealp{Ly07}, \citealp{Dominguez13}). This would additionally be consistent with elevated AGN activity in high-redshift (proto)clusters compared to the field, as seen in other works (e.g., \citealp{Galametz09}), enhancing [\ion{O}{3}] line fluxes via AGN photoionization.

Five sources {from our grism survey} have high ratios, in the range $4 < \rm H\alpha/$[\ion{O}{3}]$\mbox{}< 7$. 
Two of these sources are the central QSOs of CARLA~J1358+5752  
and CARLA~J1129+0951. We identify and measure broad and strong H$\alpha$, and weaker [\ion{O}{3}] and H$\beta$ fluxes for the 
two QSOs. Unless due to intrinsic QSO line properties, the high H$\alpha$/H$\beta$ ratios ($4.0$, 
and $5.6$, respectively) suggest the presence of intrinsic dust extinction that affects the H$\alpha$/[\ion{O}{3}] ratios; where the typical Balmer-decrement (i.e., H$\alpha$/H$\beta$ ratio) is $3.37$ for low-redshift radio-loud QSOs with negligible dust extinction (\citealp{Dong08}). The three other sources with high H$\alpha$/[\ion{O}{3}] ratios are star-forming members of CARLA~J1129+0951. We detect H$\beta$ emission for one of the three sources, for an H$\alpha$/H$\beta$ ratio of $5.2$; where the intrinsic (i.e., dust-free) ratio is $2.98$ for Case B recombination (\citealp{Osterbrock89}). We only detect [\ion{O}{3}] to the level of our detection limit and a $\sim6\times$ stronger H$\alpha$ flux for the other two sources. In both cases, this suggests that the high H$\alpha$/[\ion{O}{3}] ratios are due to dust-attenuated [\ion{O}{3}] emission. Another possibility for the latter case, would be a very strong [\ion{N}{2}] contribution to the H$\alpha$ flux.
Two sources have low H$\alpha$/[\ion{O}{3}] ratios, with values $< 0.3$. One source is the central HzRG of CARLA~J2355$-$0002, and the other is a star-forming member of this structure.
Using sources with {\it Spitzer} detections we do not find a dependence of the line ratio with galaxy stellar mass.\\

\subsection{SFR vs. Mass}\label{sec:sfrm}

In Figure~\ref{fig:sfrvsm}, we show the SFRs of members as a function
of galaxy stellar mass, where masses are determined as described in \S~\ref{sec:mass}. {The leftward green arrow represents the typical shift in stellar mass ($20\%$) when using a sum of SSP and star-forming decaying models contributing $20\%$ and $80\%$ {of the mass}, respectively, instead of using an SSP-only model, as described in \S~\ref{sec:mass}. Note that this shift does not significantly change our results.  
We also show uncertainties in the SSP-derived stellar masses, propagated from uncertainties in {\it Spitzer}/IRAC flux measurements. Most symbols have sizes similar to that of these uncertainties}.

As noted in \S~\ref{sec:sfr}, depending on the available emission lines, we estimate SFRs with the following priority order: H$\alpha$, [\ion{O}{3}], and [\ion{O}{2}] (highlighted by circles, squares, and diamonds in Fig.~\ref{fig:sfrvsm}, respectively).
To compare the SFR-$M_{\star}$ relation of our confirmed members with field values, we show the star-forming main-sequence of $1.4<z<2.5$ field galaxies
established in \cite{Daddi07} (solid black line in Fig.~\ref{fig:sfrvsm}). The dashed and dotted lines represent $4\times$ and $10\times$ the main-sequence, respectively.
We also show the \cite{Whitaker14} main-sequences for $(10^{10} - 10^{11})\,M_{\odot}$
galaxies, comparable to our mass estimates of {\it Spitzer}-detected members. We show the main-sequences derived for the redshift ranges $z=0.5-1.0, 1.0-1.5, 1.5-2.0,$ and $2.0-2.5$ (grey, blue, cyan, and yellow
curves in Fig. \ref{fig:sfrvsm}, respectively). The grey confidence region around the \cite{Daddi07} $z\sim2$ main-sequence corresponds to the semi-interquartile range ($0.16$ dex) of their SFR-$M_{\star}$ distribution. The confidence regions around the \cite{Whitaker14} main-sequences represent the typical error ($\sim0.07$ dex) in their SFR median stacks used to derive the polynomial fits of the main-sequences in the range $(10^{10} - 10^{11})\,M_{\odot}$, as in Table 2 of their paper. Using the limiting
{\it Spitzer}/IRAC fluxes determined in \cite{Wylezalek14}, we
determine upper limits on the galaxy masses of non-{\it Spitzer}-detected
confirmed members. We highlight these sources with leftward
arrows in Figure \ref{fig:sfrvsm}.

We find that {\it Spitzer}-detected confirmed members, with stellar masses typically
above $10^{10}\, M_{\odot}$, are located under the main-sequence of their {corresponding} redshift bins up to $z=2$. Indeed, {\it Spitzer}-detected $z\sim1.5$ members (highlighted by dark blue markers in Fig.~\ref{fig:sfrvsm}) are located below the field $z\sim1.5$ star-forming main-sequence, and tend to better agree with the lower-redshift (i.e., $z=0.5-1.0$) star-forming main-sequence of \cite{Whitaker14}. This would be consistent with an accelerated galactic evolution in overdense environments. Similarly, {\it Spitzer}-detected $z\sim2$ members are located below the $z\sim2$ \cite{Daddi07} and \cite{Whitaker14} main-sequences. 
For many of the
lower mass galaxies, we only have upper limits on their masses and are
therefore unable to confidently address where they reside relative to
the main-sequence. 
On the other hand, at all redshifts, the more massive (i.e., {\it
Spitzer}-detected) member galaxies, with stellar masses typically
above $10^{10}\, M_{\odot}$, form very few stars for their mass,
as shown by their location under their respective star-forming main-sequences of
field galaxies. This suggests that these are evolved galaxies,
that have already undergone a major episode of star formation.

These results, however, are dependent on
the robustness of the SFR estimates. We might be underestimating SFRs
of dustier member galaxies since our calculations assume a constant
dust attenuation of $1$~mag in the $V$-band for all galaxies. Additional
longer wavelength data will be required to fully investigate how
dust affects our results. However, we emphasize that our treatment is
identical to the approach adopted in previous analyses (e.g., \citealp{Zeimann12}, \citealp{Newman14}), and that this value is typical of star-forming galaxies
(e.g., \citealp{Kewley04}, \citealp{Sobral12}).  Even adopting $2$~mag of dust attenuation
in the $V$-band, we still find that $> 75\%$ of {\it Spitzer}-detected
star-forming members are located below the \cite{Daddi07} 
main-sequence. 
{Alternatively, we fit a second order polynomial to the subsample of $10^{10} - 10^{11}\, M_{\odot}$ H$\alpha$ members and compute the amount of dust attenuation A$_{V}$, as a function of $M_{\star}$, required to reproduce the \cite{Whitaker14} main-sequence at $z=1.5$. We find that the required dust attenuation A$_V$ is monotonically increasing with stellar mass and ranges from $2\,\rm mag$ at $10^{10} \, M_{\odot}$ to $4\,\rm mag$ at $10^{11}\, M_{\odot}$. {These values are significantly higher than typical dust attenuation for massive galaxies, both at low redshifts and in color-selected massive galaxies and comparable redshift to the CARLA sample. Specifically,} using a sample of $90,000$ star-forming galaxies from Data Release 7 of the Sloan Digital Sky Survey (\citealp{Abazajian09}), \cite{GarnBest10} find H$\alpha$ dust attenuations ranging from $0.91\,\rm mag$ for $10^{10} \, M_{\odot}$ galaxies to $1.70\,\rm mag$ at $10^{11} \, M_{\odot}$ with an average $1\sigma$ width of the distribution of $\sim0.3\,\rm mag$. This corresponds to A$_V$ attenuations of $1-2\,\rm mag$, respectively. \cite{Kashino13,Kashino14} find similar dust attenuations from a sample of $271$ $sBzK$-selected star-forming galaxies at $1.4<z<1.7$ from the COSMOS field (\citealp{McCracken10}).}
Our analysis therefore supports the result that {massive ($10^{10} - 10^{11}\, M_{\odot}$)} confirmed
members of these distant structures did indeed form earlier than field
galaxies at similar redshift, and reside below the main sequence of
star-forming galaxies at their redshift, {unless significantly dust-obscured}.\\

\subsection{SFR vs. Radius}\label{sec:sfrr}

\begin{figure*}
\hspace{4.6cm}\includegraphics[scale=0.46]{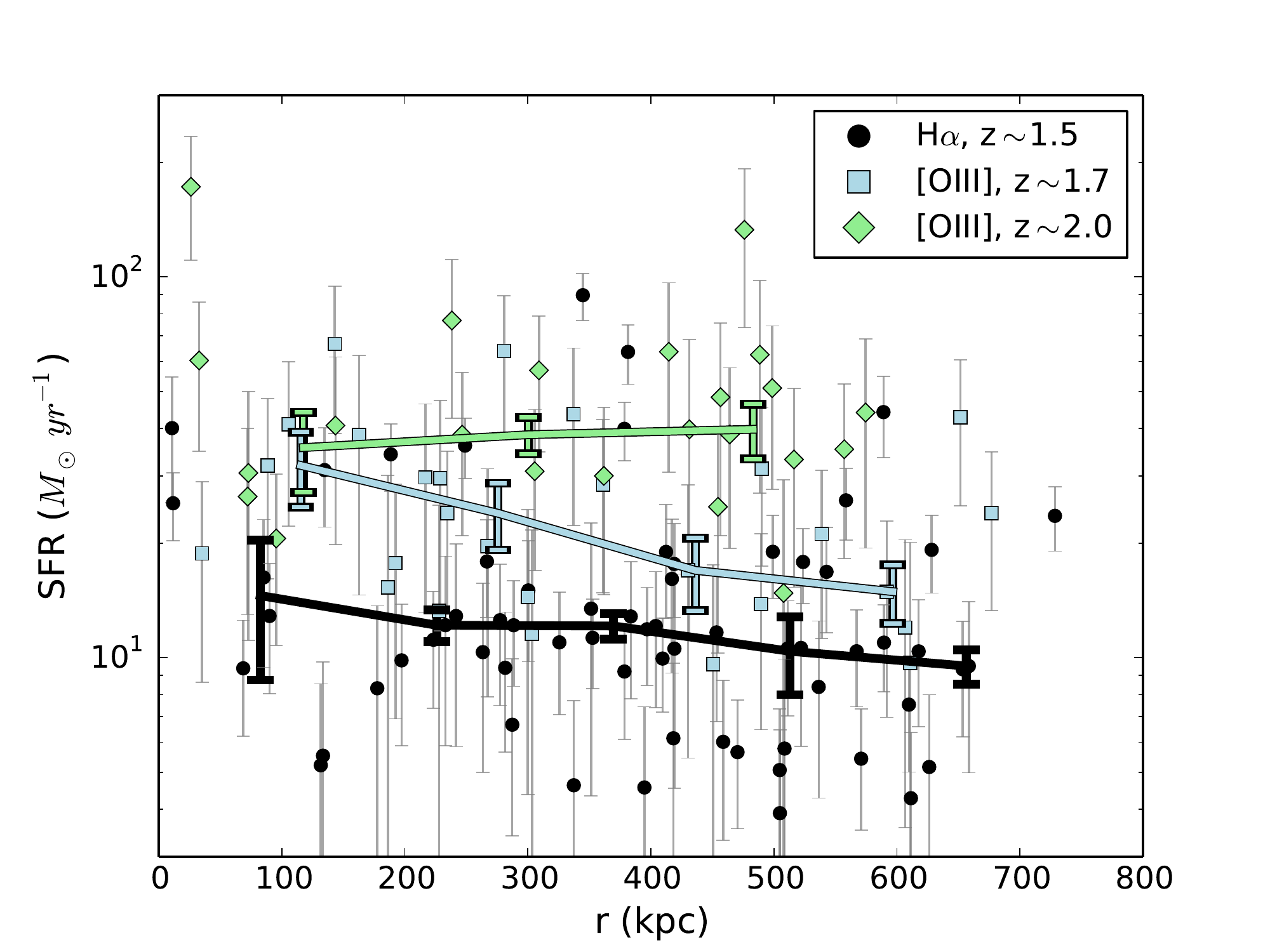}\\
\includegraphics[scale=0.46]{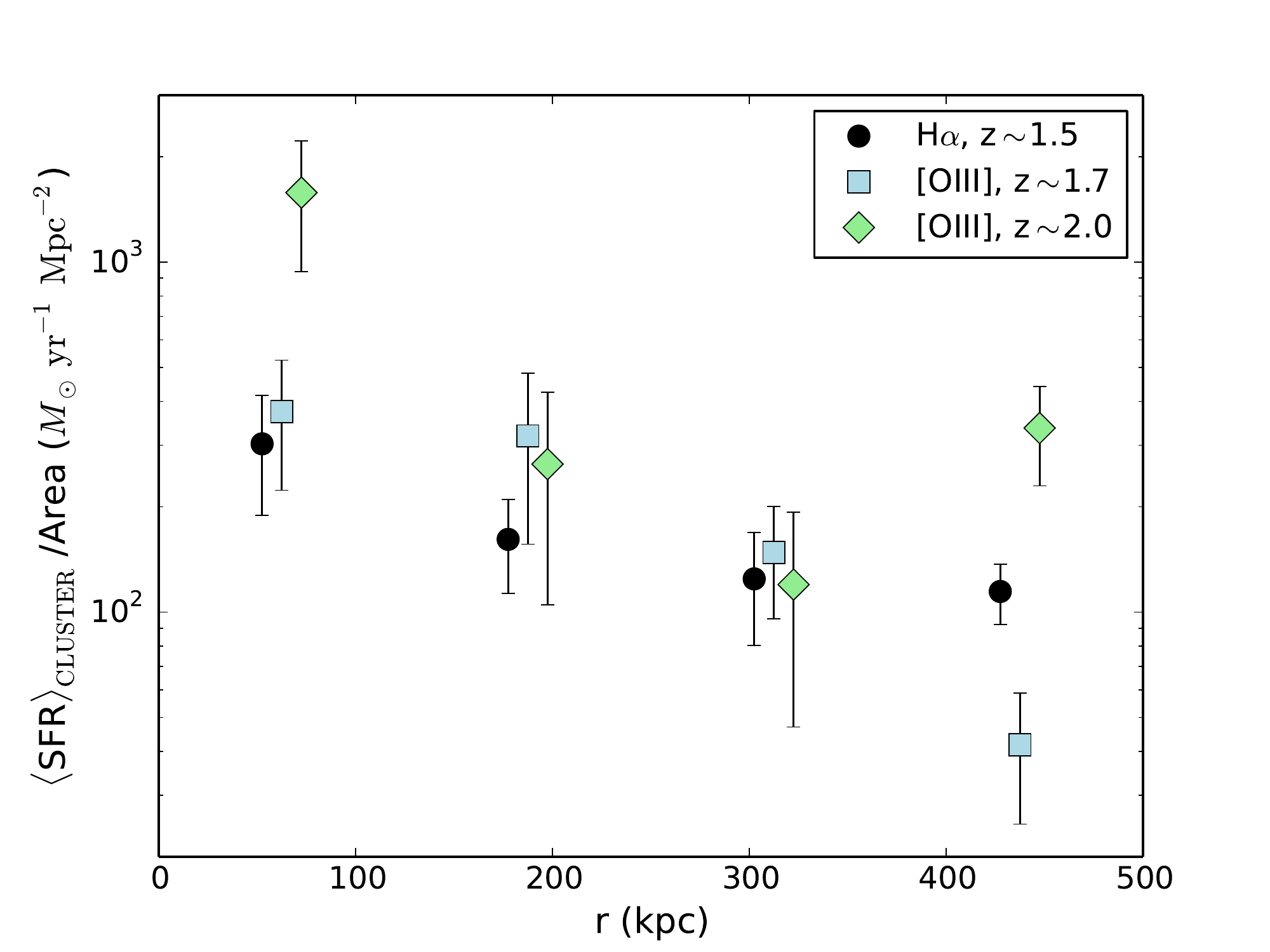}
\includegraphics[scale=0.46]{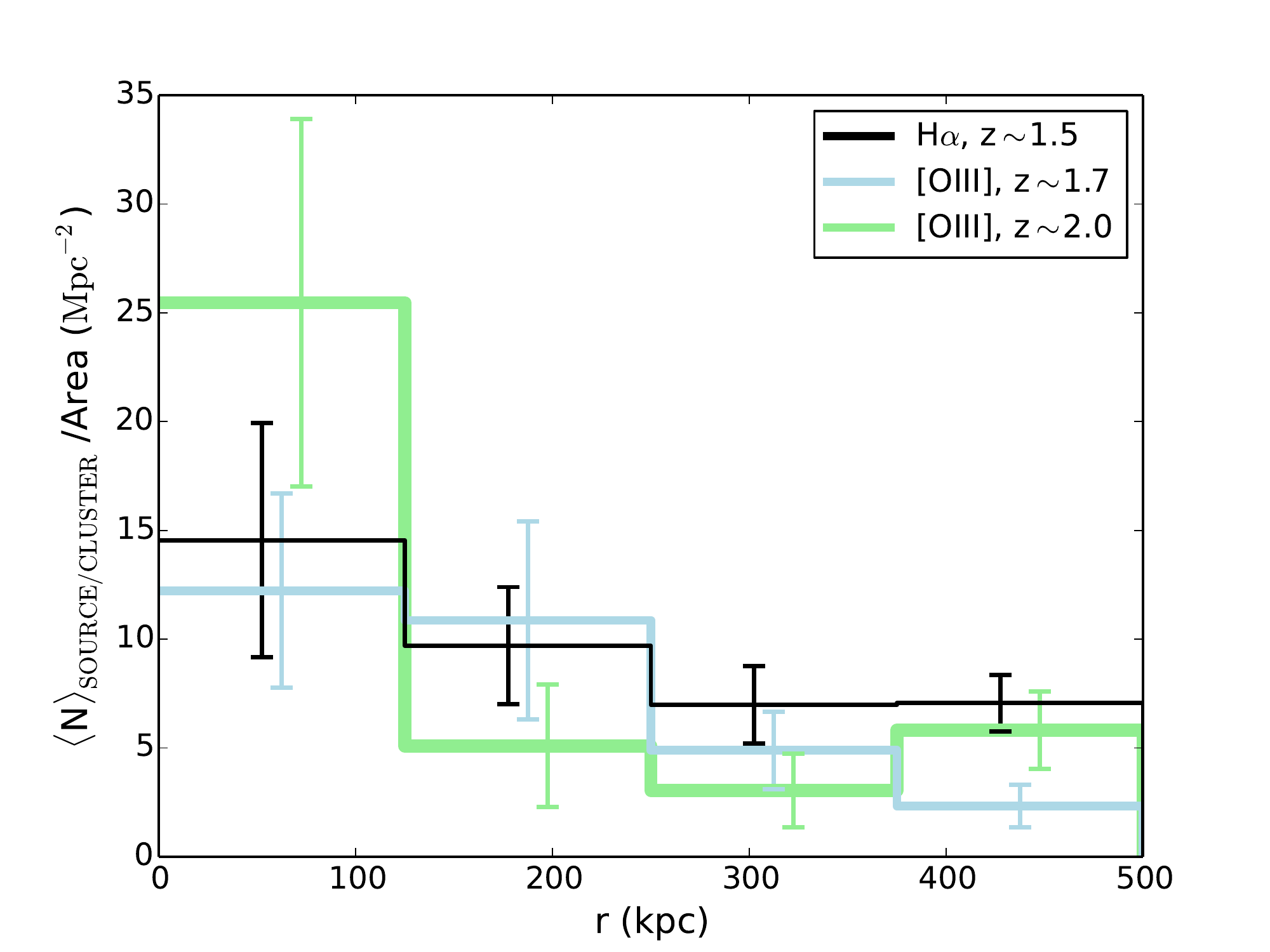}
\caption[SFRs vs. R]{Top panel: source SFR (in $M_{\odot}\,\rm yr^{-1}$) as a function of distance from the targeted RLAGN. {We show running medians as solid lines. Error bars on the medians represent the $68\%$ confidence intervals determined via $10,000$ bootstrap resampling.} Bottom left panel: average SFR density per structure (in $M_{\odot}\,\rm yr^{-1}\, Mpc^{-2}$) as a function of distance from the targeted RLAGN. {For each radial bin, data points are slightly offset from each other for clarity.} Bottom right panel: histogram of the average source number density per structure (in $\rm Mpc^{-2}$). Members are divided in three groups consisting of H$\alpha$-based SFR members at $z\sim1.5$, [\ion{O}{3}]-based SFR members at $z\sim1.7$, and [\ion{O}{3}]-based SFR members at $z\sim2.0$ (black circles, light-blue squares, and light-green diamonds, respectively). For the two bottom panels, values are derived within bins of $125\,\rm kpc$ from the RLAGN and error bars represent the error in the mean.}
\label{fig:sfrvsr}
\end{figure*}

In the top panel of Figure \ref{fig:sfrvsr}, we show {individual} SFRs of members {and corresponding running medians} as a function of physical distance from the targeted RLAGN. We divided members into three redshift groups. The first group, represented by solid black circles, corresponds to members of the seven {confirmed CARLA structures} at $z\sim1.5$ for which the SFRs are based on H$\alpha$, and limited to $>4\,M_{\odot}\,\rm yr^{-1}$. The second group, represented by solid light-blue squares, includes members of the five [\ion{O}{3}]-based SFR {structures} at $z\sim1.7$, which have limiting SFRs $>10\,M_{\odot}\,\rm yr^{-1}$. The third group, represented by solid light-green squares, is comprised of the three [\ion{O}{3}]-based SFR {structures} at $z\sim2.0$ (with SFRs $>20\,M_{\odot}\,\rm yr^{-1}$; see Fig.~\ref{fig:fluxsfrz}). {We caution the reader that the use of different SFR tracers, each also having different limiting SFRs, precludes direct comparison of the three different redshift bins.}  
Despite large scatter, we observe a general trend of slightly decreasing source SFR with distance from the targeted RLAGN for members of the H$\alpha$ SFR-based group and for members of the [\ion{O}{3}] SFR-based group at $z\sim1.7$ (top panel Fig.~\ref{fig:sfrvsr}). {Confirmed members of the [\ion{O}{3}] SFR-based group at $z\sim2.0$, limited to higher ($>20\,M_{\odot}\,\rm yr^{-1}$) SFRs, do not exhibit such behavior. Higher number statistics and deeper observations to identify lower star-forming members would therefore be required to infer the radial SFR trend of individual sources at $z\sim2$. At close proximity to the RLAGN ($<50\,\rm kpc$), we see a steep increase of member galaxy SFRs (for the H$\alpha$ SFR-based group and the [\ion{O}{3}] SFR-based group at $z\sim2.0$). This could be consistent with RLAGN nuclear activity triggering the star-formation of satellite galaxies.}

In the bottom left panel of Figure \ref{fig:sfrvsr}, we show the average SFR density {per structure} (in $M_{\odot}\,\rm yr^{-1}\, Mpc^{-2}$) as a function of distance from the targeted RLAGN. Values are all derived within bins of $125\,\rm kpc$ from the RLAGN, and error bars represent the error in the mean. We observe a clear decreasing trend with distance from the targeted RLAGN for all SFR tracers and redshift bins, {except the last radial bin of the [\ion{O}{3}] SFR-based group at $z\sim2.0$ (light-green diamonds) which exhibits an increase of SFR density}. This is directly correlated to the average number density of sources with distance from the RLAGN. In the bottom right panel of Figure \ref{fig:sfrvsr}, we indeed observe a general higher concentration of star-forming members at close proximity to the RLAGN, {and also a slightly increased concentration of sources within $r = 375 - 500\,\rm kpc$ compared to $r= 125 - 375\,\rm kpc$ for the [\ion{O}{3}] SFR-based group at $z\sim2.0$}. Even though the RLAGN may not always reside at the center of the structures, this result supports that it is the case on average, as also shown in \cite{Wylezalek13}. {These results also show that, on average, most of the star-forming budget in our confirmed CARLA structures is enclosed within their densest, central regions.}
 
Overall, these trends are consistent with \cite{Brodwin13} who studied $16$ confirmed clusters from the {\it Spitzer}/ISCS (\citealp{Eisenhardt08}) in the range $1.0 < z < 1.5$ and found similar trends for a subsample of clusters in the range $1.37 < z < 1.50$, while lower redshift clusters showed 
lower SFR in clusters cores, ubiquitous at $z=0$ (see also, e.g., \citealp{Tran10}). {Together, this suggests that cluster cores are the preferred sites of star formation at these epochs.}\\

\section{Summary}\label{sec:con}

We conclude the following from our 40-orbit {\it HST}/WFC3 F140W
and G141 follow-up observations on the 20 densest CARLA candidate
clusters at $1.4<z<2.8$.

\begin{itemize}
\item[1.] We spectroscopically confirm $16$ {\it Spitzer} color-selected
distant structures associated with the targeted RLAGN, including
three at $z \sim 2$ (CARLA~J1018+0530, CARLA~J0800+4029 and CARLA~J2039-2514) and one at $z=2.8$ (CARLA~J1017+6116). 
These
structures are among the most distant {confirmed cluster candidates} currently known. We
identify $143$ members in these $16$ structures, with an average of
$9$ members per confirmed structure. We also serendipitously discover and confirm
seven other structures at $0.87 < z < 2.12$ not associated with the
targeted RLAGN of our program.

\item[2.] With just two orbits of {\it HST} imaging and grism
spectroscopy per field, we confirm emission-line sources down to
$\sim 2.5 \times 10^{-17}\, \rm erg\, cm^{-2}\, s^{-1}$, corresponding
to limiting SFRs in the range $4 - 70\, M_{\odot} \, \rm yr^{-1}$, depending
on the identified emission line and galaxy redshift.

\item[3.] We show that these newly confirmed CARLA structures at $1.4
< z < 2.8$ are comparable or slightly richer in mid-infrared overdensity to spectroscopically confirmed massive
SPT and ISCS clusters at $1.3 < z < 1.5$, implying similar galaxy richness
despite being at much higher redshift.

\item[4.] We find that {massive ($10^{10} - 10^{11}\,M_{\odot}$)} confirmed structure members, all showing
evidence of star formation, reside below the star-forming main-sequence
of galaxies at their redshift. {Unless significantly dust-obscured,} this implies that these galaxies
in rich environments underwent a significant episode of star formation
prior to the epoch that we are now observing them.

\item [5.] We find that the density of star-forming galaxies rises
sharply at smaller {radii from the central RLAGN}, implying that most
of the star-forming budget is enclosed within the densest regions of these
structures. This trend is consistent with the results of \citet{Brodwin13},
who studied a sample of mid-infrared selected clusters at slightly
lower redshift, and is consistent with an overall reversal of the
SFR-density relation at higher redshifts.

\item[6.] {Comparing spectroscopically confirmed member densities to expectations from field {observations} and numerical simulations, we classify, in Appendix~\ref{app:specsig}, our confirmed structures into three classes: {\it (i)} highly probable confirmed clusters (HPC), {\it (ii)} probable confirmed clusters (PC), and {\it (iii)} confirmed galaxy concentrations (CGC). Our analysis classifies three confirmed CARLA structures in the HPC category, while all other confirmed CARLA structures ($13/16$) are classified as probable confirmed clusters.}

\item[7.] Finally, these results highlight both the strengths and
weaknesses of shallow {\it HST} grism spectroscopy for confirming
high-redshift galaxy clusters. With an extremely efficient strategy
of just 2-orbits per field, we are able to confirm most of these
candidate high-redshift clusters and demonstrate the robust
selection efficiency of the CARLA {\it Spitzer} imaging program.
However, by design, this strategy fails to confirm member
galaxies lacking star formation and therefore fails to identify the
interesting sample of massive, evolved, early-type galaxies
in these distant structures. A deeper grism program would be able
to achieve such science.\\

\end{itemize}

\acknowledgments
{We thank the referee for useful comments and suggestions which improved the paper.} We thank Nicole Nesvadba for kindly sharing data concerning the field around TXS2353$-$003. This work is based on observations made with the NASA/ESA {\it Hubble Space
Telescope}, obtained at the Space Telescope Science Institute, which
is operated by the Association of Universities for Research in
Astronomy, Inc., under NASA contract NAS 5-26555. These observations
are associated with program \#GO-13740.  Support for program
\#GO-13740 was provided by NASA through a grant from the Space
Telescope Science Institute, which is operated by the Association of
Universities for Research in Astronomy, Inc., under NASA contract NAS
5-26555. This work is also based in part on observations made with the {\it Spitzer Space Telescope}, which is operated by the Jet Propulsion Laboratory, California Institute of Technology, under a contract with NASA. This work is also based in part on observations made with the $200$-inch Hale Telescope, Palomar Observatory, operated by the California Institute of Technology. The work of P.E.~and D.S.~was carried out at the Jet Propulsion Laboratory, California Institute of Technology, under a contract with NASA. S.M.~acknowledges financial support from the Institut Universitaire de France (IUF), of which she is senior member. D.W.~acknowledges support by Akbari-Mack Postdoctoral Fellowship. E.A.C.~acknowledges support from the ERC Advanced Investigator program DUSTYGAL $321334$. N.A.H.~acknowledges support from STFC through an Ernest Rutherford Fellowship.\\

{\it Facilities:} \facility{HST (WFC3; STScI)}, \facility{Spitzer (IRAC; JPL/Caltech)}, \facility{Palomar (DBSP; Caltech)}.\\

\appendix
\section{Spectroscopic Catalog}\label{app:catalog}
We describe in Table~\ref{table:catalog} the content of the spectroscopic catalog available in the online material. This catalog gathers spectroscopic information on all sources with a measured redshift in our 20 {\it HST} fields.\\
\begin{deluxetable}{lll}  
\tablewidth{500pt}
\tablecolumns{3}
\tablecaption{Spectroscopic Catalog Content\label{table:catalog}}
\tablehead{	
  \colhead{\#} &
  \colhead{Column Label} &
  \colhead{Description}
  }
\startdata
col1&	Field&		CARLA names for all fields where a CARLA confirmed structure is present. Target RLAGN name otherwise, \\
&			&		except for CARLA-Ser~J1317+3925 (only serendipitous structure within a field with no confirmed CARLA structure).\\
col2&	ID&			Unique source ID among each field.\\
col3&	RA&			Right Ascension (J$2000$), in degrees.\\
col4&	Dec&		Declination (J$2000$), in degrees.\\
col5&	F140W&		Magnitude (AB).\\
col6&	F140W\_ERR&	Magnitude uncertainty (AB).\\
col7&	z&			Redshift.\\
col8&	z\_ERR&		Redshift uncertainty.\\
col9&	Q&			Redshift quality.\\
col10&	f\_Ha&		H$\alpha$ flux ($\rm erg\,cm^{-2}\,sec^{-1}$). H$\alpha$ fluxes include [\ion{N}{2}] contributions.\\
col11	&	f\_Ha\_ERR&	H$\alpha$ flux uncertainty ($\rm erg\,cm^{-2}\,sec^{-1}$).\\
col12&	f\_Hb&		H$\beta$ flux ($\rm erg\,cm^{-2}\,sec^{-1}$)\\
col13&	f\_Hb\_ERR&	H$\beta$ flux uncertainty ($\rm erg\,cm^{-2}\,sec^{-1}$)\\
col14&	f\_OIII&		[\ion{O}{3}]$\lambda5007$ flux ($\rm erg\,cm^{-2}\,sec^{-1}$)\\
col15&	f\_OIII\_ERR&	[\ion{O}{3}]$\lambda5007$ flux uncertainty ($\rm erg\,cm^{-2}\,sec^{-1}$)\\
col16&	f\_OII&		[\ion{O}{2}] flux ($\rm erg\,cm^{-2}\,sec^{-1}$)\\
col17&	f\_OII\_ERR&	[\ion{O}{2}] flux uncertainty ($\rm erg\,cm^{-2}\,sec^{-1}$)
\enddata
\tablecomments{The catalog is published in its entirety in the machine-readable format.
      Meta-data are shown here for guidance regarding its form and content.}
\end{deluxetable}\\

\section{Significance of the Spectroscopic Confirmations}\label{app:specsig}
{In this Appendix we evaluate the significance of our CARLA {spectroscopic overdensities} ({both targeted and} serendipitous). This analysis provides an additional flag to better assess the status of our confirmations and non-confirmations.}

{As already mentioned, in \cite{Wylezalek13,Wylezalek14} we derived an estimate of the {\it Spitzer}-selected galaxy overdensities for all CARLA fields and compared them to the average distribution of similarly selected galaxies from SpuDS. We {measured} a field galaxy overdensity of $9.6\pm2.1$ galaxies per arcmin$^{2}$ (\citealp{Wylezalek14}).
The fields of our $16$ confirmed structures (CARLA clusters) have galaxy overdensity significances $>4.4\sigma$. This means that, based on the {\it Spitzer}-selected galaxy overdensities alone, there is an almost zero probability that our confirmed structures are simply field galaxy overdensities.
As mentioned before, our detections can still be clusters, groups, filaments or sky superpositions which can not be distinguished in {\it Spitzer} colors alone. We therefore measure the overdensity of spectroscopically confirmed members in our CARLA fields and compare to expected values in the field (using 3D-HST) and numerical simulations (using \citealp{Cautun14}) in an attempt to understand whether our confirmations are in better agreement with being nodes of the cosmic web (i.e., virialized clusters) or filaments.}

\begin{figure}
\includegraphics[scale=0.46]{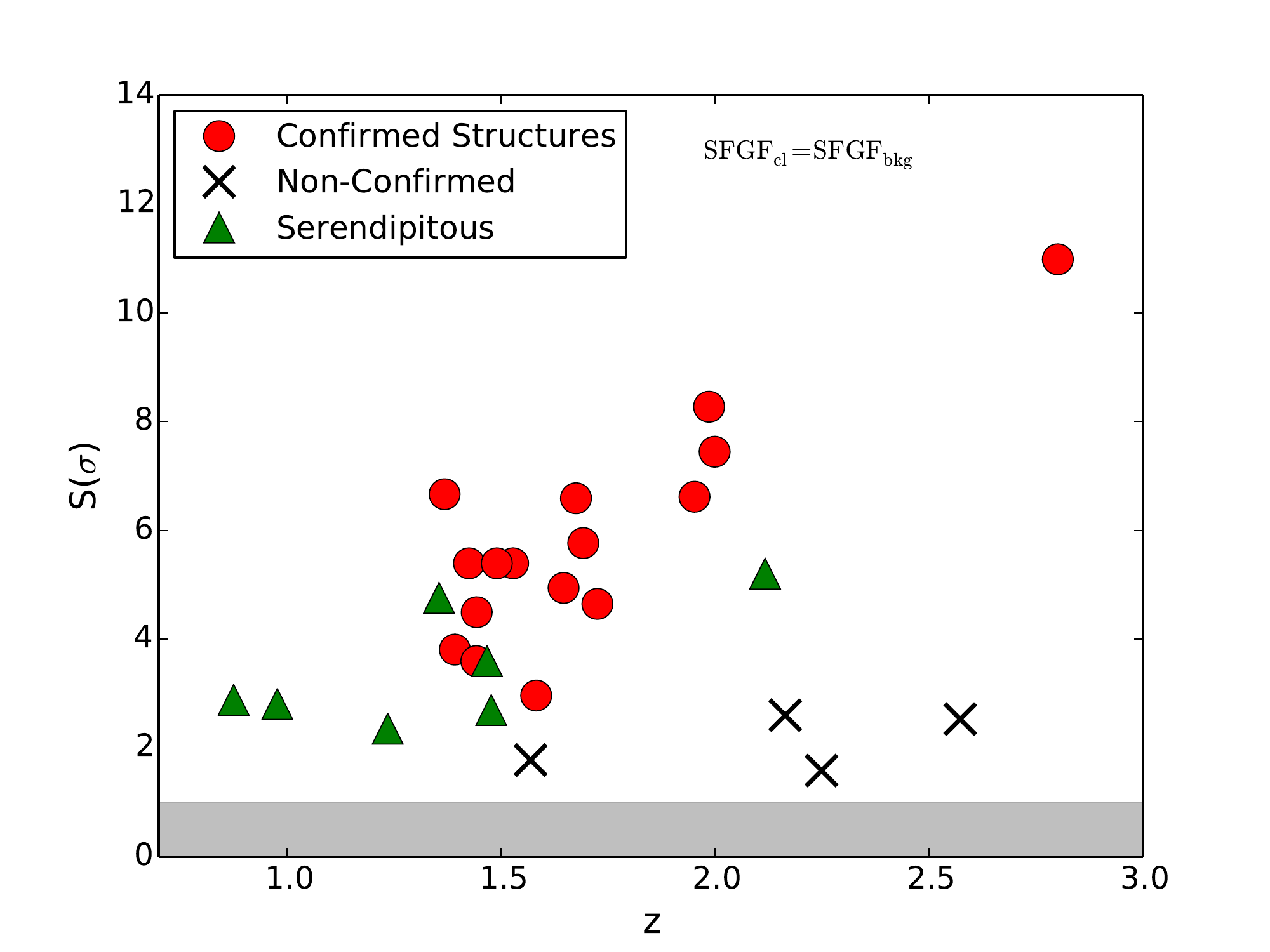}
\vspace{0.2in}
\includegraphics[scale=0.46]{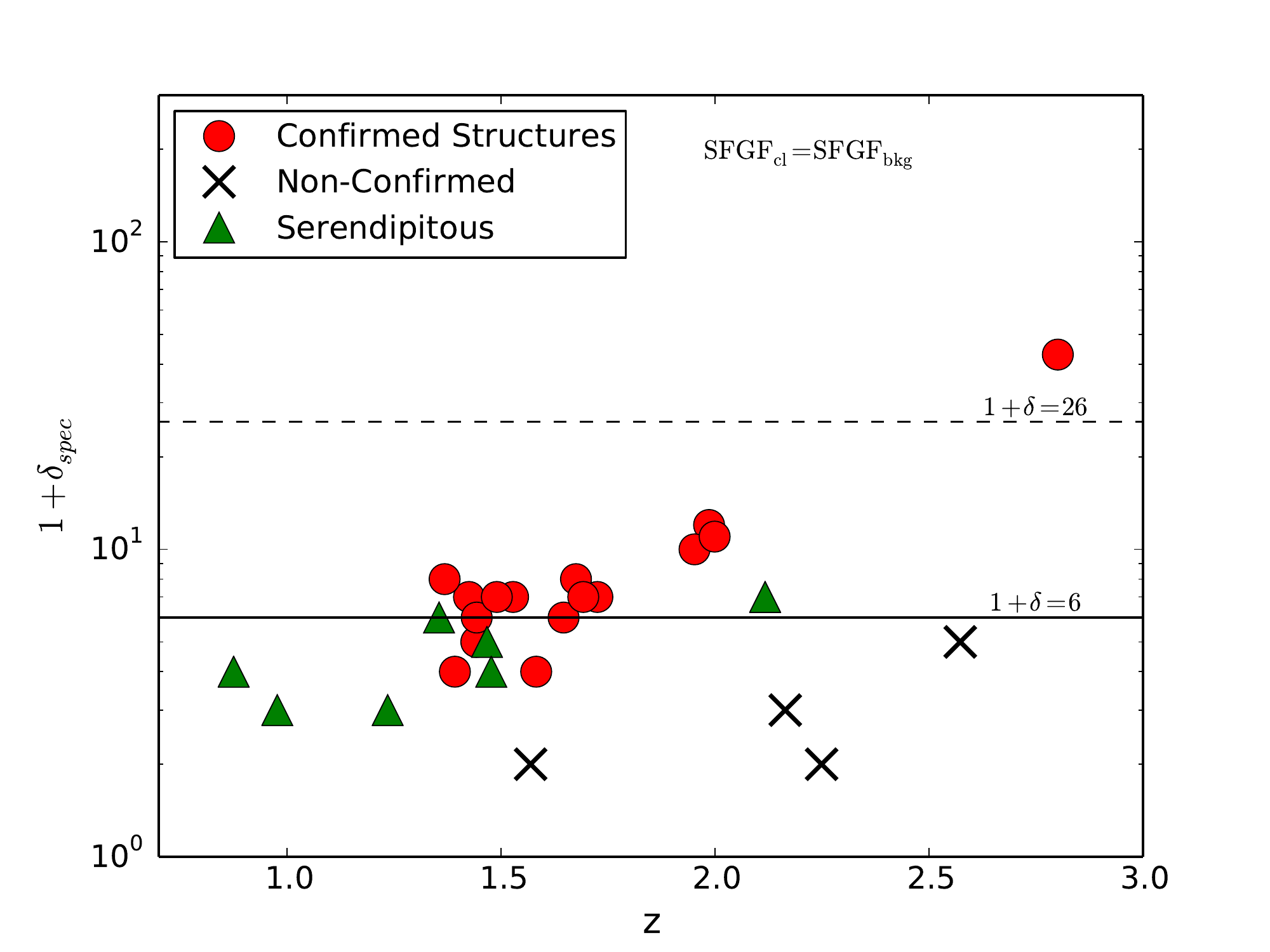}
\includegraphics[scale=0.44]{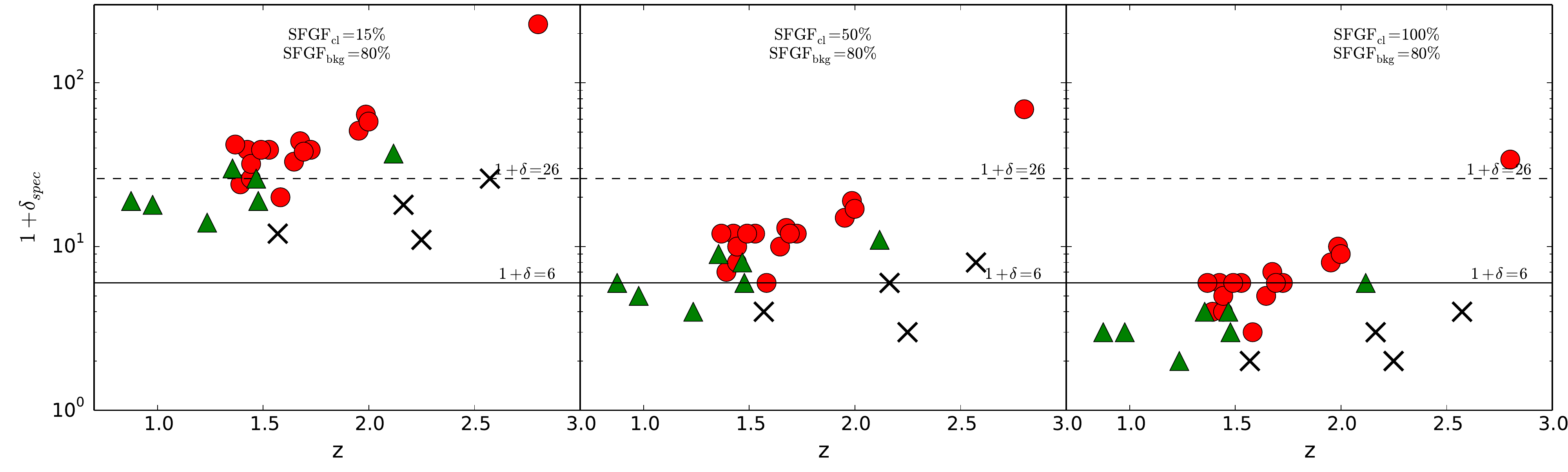}
\caption[Spectroscopic Significance]{Top left: overdensity significance (in $\sigma$) of spectroscopic members of our confirmed CARLA and serendipitous structures and non-confirmed CARLA cluster candidates, shown as red circles, green triangles, and black crosses, respectively. The horizontal grey area represents the $1\sigma$ standard deviation of the field value, derived from 3D-HST as described in the text. Top right: galaxy contrast $(1 + \delta_{spec})$ of the same fields.  Both top panels assume identical star-forming galaxy fractions in the field and cluster environments. Bottom: galaxy contrasts for three cases of star-forming galaxy fractions in cluster environments: $\rm SFGF_{\rm cl} = 15\%$, $50\%,$ and $100\%$, from left to right, respectively, and with a field value of $\rm SFGF_{\rm bkg} = 80\%$. In all galaxy contrast panels, the solid and dashed lines represent the $(1+\delta) = 6$ and $(1+\delta) = 26$ thresholds from \cite{Cautun14}, respectively.}
\label{fig:specsig}
\end{figure}

{As previously, we select 3D-HST sources which have usable grism redshifts ($use\_zgrism = 1$), and have any combinations of identified H$\alpha$, [\ion{O}{3}], or [\ion{O}{2}] emission lines above our limiting fluxes ($2.5\times10^{-17}\rm\,erg\,cm^{-2}\,\sec^{-1}$) and falling within the wavelength range covered by the G141 grism ($\lambda = 1.08 - 1.7\rm\,\micron$). We scan these sources in the redshift range $z=0.7-3.0$ with a step of $\pm 2000\, {\rm km}\, {\rm s}^{-1}$, corresponding to our adopted definition. For each redshift bin, we measure the number density of galaxies using randomly distributed non-overlapping $1\arcmin$ radius apertures. We then stack the density distributions within steps of $0.1$ in redshift and fit Gaussians to the distributions to obtain the $1\sigma$ standard deviation of the spectroscopic field galaxy density (in $\rm arcmin^{-2}$) at these redshifts. We then compare confirmed member densities in our CARLA fields to the field values at the appropriate redshifts. The top left panel of Figure \ref{fig:specsig} shows the overdensity significance (in $\sigma$) of spectroscopic members of our confirmed CARLA and serendipitous structures and non-confirmed CARLA cluster candidates, shown as red circles, green triangles, and black crosses, respectively.}

{We also measure galaxy contrasts $\delta_{spec} = \frac{N_{members}-{N_{3DHST}}}{N_{3DHST}}$ ({where $N_{members}$ and $N_{3DHST}$ are measured} in galaxies per arcmin$^{2}$) which we can compare to numerical simulations (e.g., \citealp{Cautun14}).  
Following a similar approach as in \cite{Mei15}, we compare the $(1+\delta) = \frac{\rho}{\rho_{bkg}}$ galaxy contrast measurements of \cite{Cautun14} to our $(1+\delta_{spec})$. \cite{Cautun14} measured the probability density function (PDF) of contrasts $(1+\delta)$ belonging to a void, a wall, a filament or a node (virialized region) of the cosmic web. Following \cite{Cautun14}, we define, for this analysis, clusters and cluster progenitors as virialized regions at all redshifts. Numerical simulations predict that the haloes that are most probably accreted in clusters with masses $M > 10^{14} M_\odot$ at $z<0.5$ have masses $M \ge 5 \times 10^{13} M_\odot$ at $z>1.5$ (e.g., \citealp{Chiang13}, \citealp{Cautun14}).}

{Following \cite{Cautun14}, overdensities within $1\sigma$ from the mean of the node PDF have {contrasts of} $26 \leq (1+\delta) \leq 507$ and a probability of $\la 15\%$ to be filaments, while those with {contrasts of} $6 \leq (1+\delta) < 26$ are within $1\sigma - 2\sigma$ from the mean of the node PDF and have a probability of $\la 40\%$ to be filaments. 
Overdensities with {contrasts of} $(1+\delta) < 6$ are more than $2\sigma$ from the mean of the node PDF; their probability to be nodes of the cosmic web is $\la 10\%$ and they also have a probability of $\sim 40\%-50\%$ to be filaments. 
We find galaxy contrasts $(1+\delta_{spec}) \geq 6$ {for $13$ of our $16$} confirmed CARLA cluster candidates, including one (CARLA~J1017+6116) with a contrast higher than $26$ ($42$; see top right panel of Fig.~\ref{fig:specsig}). CARLA~J1753+6310, showing a strong red sequence, dominated by a passive population (\citealp{Cooke16}) and for which we confirm five star-forming members in our ${\it HST}$ grism data, is among the {three confirmed} CARLA structures with $(1+\delta_{spec}) < 6$. {This illustrates} that the different star-formation histories in the field and cluster environments can significantly affect our spectroscopic overdensity significances and galaxy contrasts. We therefore estimate different contributions of the star-forming population from the literature to scale our galaxy contrasts. Also note that the higher grism confusion in crowded (cluster) environments compared to the field tends to underestimate our measurements of spectroscopic significance and galaxy contrast.}

{The fraction of star-forming galaxies in the field is typically $\sim 80\%$ at $z>1$ (e.g., \citealp{Darvish16}). The fraction of star-forming galaxies in clusters within a region of radius $0.5$~Mpc ($\sim 1\arcmin$) is typically $\sim 10-20\%$ and $\sim 40-50\%$ at $ 1 \leq z \leq 1.4$ and $ z > 1.4 $, respectively (e.g., \citealp{Brodwin13}). The exact percentage depends on the evolutionary status of the cluster. In fact, clusters dominated by passive populations are found up to $z\sim2$ (e.g., \citealp{Newman14}, \citealp{Cooke16}).
To quantify how different star-forming galaxy fractions in clusters and the field ($\rm SFGF_{cl}$ and $\rm SFGF_{bkg}$, respectively) change our classification, we scale $(1+\delta_{spec})$ by $\frac{{\rm SFGF}_{bkg}}{{\rm SFGF}_{cl}}$ for three different cases: {\it (i)} $\rm SFGF_{cl} = 15\%$, which is typical of evolved clusters, {\it (ii)} $\rm SFGF_{cl} = 50\%$, which corresponds to typical clusters at $z\sim 1.5$ (e.g., \citealp{Brodwin13}), and {\it (iii)} $\rm SFGF_{cl} = 100\%$, which corresponds to the most extreme case in which all galaxies are star-forming galaxies. We show our galaxy contrasts for these three cases in the bottom panels of Figure~\ref{fig:specsig}.} 

{In the case $\rm SFGF_{cl} = 15\%$ (Fig.~\ref{fig:specsig}, bottom left panel), {14 of our 16} confirmed CARLA {structures} are within $1\sigma$ from the mean of the \cite{Cautun14} node PDF (i.e., they have $(1+\delta_{spec})\geq26$). Three of the serendipitous confirmations are also within $1\sigma$ of the node PDF, while four are within $2\sigma$ (i.e., they have $6 \leq (1+\delta_{spec}) < 26$). 
In the case $\rm SFGF_{cl} = 50\%$ (Fig.~\ref{fig:specsig}, bottom middle panel), the only confirmed CARLA {structure} within $1\sigma$ from the mean of the node PDF is CARLA~J1017+6116, confirmed at $z=2.8$ with $8$ star-forming members. {With a contrast of $(1+\delta_{spec}) = 78$}, \cite{Cautun14} predicts a probability larger than $\sim 60\%$ that this is a node of the cosmic web and $<10\%$ to be a filament. In this scenario, all other confirmed CARLA {structures} and five serendipitous structures  
have $6 \leq (1+\delta_{spec} )< 26$. 
In the extreme case of $\rm SFGF_{cl} = 100\%$ (Fig.~\ref{fig:specsig}, bottom right panel), again only the $z=2.8$ confirmed structure is within $1\sigma$ from the mean of the node PDF. In this last scenario, eleven confirmed CARLA {structures} and one serendipitous have $6 \leq (1+\delta_{spec} )< 26$; all other sources have $(1+\delta_{spec} )< 6$.}

{In Table~\ref{table:specsig}, we show for all 20 fields the spectroscopic overdensity significance ($\sigma_{spec}$), the galaxy contrast ($1+\delta_{spec}$) for the three star-forming {galaxy} fractions discussed here, and a robustness flag. This flag is comprised of three integers, each corresponding to one of the three star-forming {galaxy} fraction cases, and each ranging from one to three, with $1$, $2$, and $3$ corresponding to fields which have galaxy contrasts $(1+\delta_{spec})$ $<1\sigma$, $= 1\sigma - 2\sigma$, and $>2\sigma$ from the mean of the node PDF, respectively. For example, a field with flag $111$ is within $1\sigma$ from the mean of the node PDF (i.e., $26 \leq (1+\delta_{spec}) \leq 507$) for $\rm SFGF_{cl}$ of $15\%$, $50\%$, and $100\%$, in this sequence.
We also add a {flag}, not available to all fields, indicating the presence of a clearly defined red sequence as from \cite{Cooke16} and N16.}

{Using this analysis, we consider confirmed structures (confirmed CARLA structures and serendipitous discoveries) with at least two flag components equal to $1$ as highly probable confirmed clusters {or protoclusters} (HPC). We consider confirmed structures with at least two flag components equal or less than $2$ as probable confirmed clusters or protoclusters (PC). For any flag combination, we consider confirmed structures with clear red sequences as highly probable confirmed clusters or protoclusters (HPC). We consider all other confirmed structures as confirmed galaxy concentrations (CGC). We show the classifications for our confirmed structures in Table~\ref{table:specsig}. 
We have three confirmed CARLA structures classified as {HPC}, while the rest ($13/16$) are classified as {PC}. Among serendipitous discoveries, $5/7$ are classified as {PC}, while the other two are classified as {CGC}. We do not classify unconfirmed CARLA structures but show their flags in Table~\ref{table:specsig} for completeness. 
We will refine these considerations when additional multi-wavelength imaging and spectroscopy will be available to us on these fields. For simplicity, in the main body of this paper, we adopt the very conservative approach to call all spectroscopic confirmations ``structures".}\\

\begin{deluxetable*}{lccccccccc}  
\tablewidth{500pt}
\tablecolumns{10}
\tablecaption{CARLA {\it HST} Spectroscopic Significance and Galaxy Contrasts \label{table:specsig}}
\tablehead{   
  \multicolumn{1}{c}{Field} &
  \multicolumn{1}{c}{$\tilde{z}_{\rm cl}$} &  
  \multicolumn{1}{c}{$\sigma_{\rm IRAC}$\tablenotemark{a}} &
  \multicolumn{1}{c}{$\sigma_{spec}$\tablenotemark{b}} &
  \multicolumn{3}{c}{$(1+\delta_{spec})$\tablenotemark{c}} &
  \multicolumn{1}{c}{Flag\tablenotemark{d}} &
    \multicolumn{1}{c}{RS\tablenotemark{e}} &
 \multicolumn{1}{c}{Class\tablenotemark{f}} \\
 \cline{5-7}
 & \colhead{\mbox{}}
 & \colhead{\mbox{}}
 & \colhead{\mbox{}}
 & \colhead{{$\rm SFGF_{cl} = 15\%$}}
 & \colhead{{$\rm SFGF_{cl} = 50\%$}}
 & \colhead{{$\rm SFGF_{cl} = 100\%$}}
 & \colhead{\mbox{}}
 & \colhead{\mbox{}}
 & \colhead{\mbox{}}
 }
\startdata
\cutinhead{CONFIRMED CARLA STRUCTURES}
CARLA~J0116$-$2052 &	1.430&	5.14 &	5.54 &	39 	&	12 	&	6	 &	122	&	\nodata 	& PC\\
CARLA~J0800+4029 &	1.986&	6.38 &	8.91 &	64 	&	19 	&	10	 &	122	&	0		& PC\\
CARLA~J0958$-$2904 &	1.396&	5.00 &	3.71 &	24 	&	7 	&	4	 &	223	&	\nodata 	& PC\\
CARLA~J1017+6116 &	2.801&	6.67 &	12.26&	228 	&	69 	&	34	 &	111	&	\nodata 	& HPC\\
CARLA~J1018+0530 &	1.953&	5.00 &	7.13 &	51 	&	15 	&	8	 &	122	&	\nodata 	& PC\\
CARLA~J1052+0806 &	1.648&	4.71 &	6.19 &	33 	&	10 	&	5	 &	123	&	\nodata 	& PC\\
CARLA~J1103+3449 &	1.443&	6.38 &	3.69 &	26 	&	8 	&	4	 &	123	&	\nodata 	& PC\\
CARLA~J1129+0951 &	1.531&	6.33 &	5.54 &	39 	&	12	 &	6	 &	122	&	\nodata 	& PC\\
CARLA~J1131$-$2705 &	1.445&	4.38 &	4.62 &	32 	&	10 	&	5	 &	123	&	\nodata 	& PC\\
CARLA~J1300+4009 &	1.676&	4.86 &	7.07 &	44 	&	13 	&	7	 &	122	&	\nodata 	& PC\\
CARLA~J1358+5752 &	1.373&	6.24 &	6.49 &	42 	&	12 	&	6 	&	122	&	\nodata 	& PC\\  
CARLA~J1510+5958 &	1.719&	5.62 &	4.52 &	39 	&	12	 &	6	 &	122	&	\nodata 	& PC\\
CARLA~J1753+6310 &	1.581&	4.52 &	2.90 &	20 	&	6 	&	3 	&	223	&	1 		& HPC\\
CARLA~J2039$-$2514 &	2.000&	8.00 &	8.02 &	58 	&	17 	&	9	 &	122	&	1 		& HPC\\
CARLA~J2227$-$2705 &	1.686&	5.29 &	6.19 &	38 	&	12	 &	6	 &	122	&	\nodata 	& PC\\
CARLA~J2355$-$0002 &	1.489&	5.62 &	5.54 &	39 	&	12	 &	6	 &	122	&	\nodata 	& PC\\
\cutinhead{UNCONFIRMED CARLA STRUCTURES\tablenotemark{g}}
6CSS1054+4459 	&	(2.566)&	4.67 &	2.27 &	26 	&	8	 &	4 	&	(123)	&	\nodata 	& \nodata\\
J1317+3925 		&	(1.569)&	4.86 &	1.74 &	12 	&	4 	&	2	 &	(233)	&	\nodata 	& \nodata \\
J1515+2133	    	&	(2.262)&	4.24 &	1.63 &	11 	&	3 	&	2	 &	(233)	&	\nodata 	& \nodata \\
TNR 2254+1857  	&	(2.157)&	5.62 &	3.15 &	18 	&	6 	&	3	 &	(223)	&	\nodata 	& \nodata \\
\cutinhead{SERENDIPITOUS DISCOVERIES}
CARLA-Ser~J1017+6116 		&	1.234&	\nodata	&	2.17	   &	14 	&	4 	&	2	 &	233	&	\nodata 	& CGC\\
CARLA-Ser~J1317+3925 		&	1.465&	\nodata 	&	3.69 	   &	26 	&	8	 &	4	 &	123	&	\nodata 	& PC\\
CARLA-Ser~J1510+5958 		&	0.875&	\nodata	&	2.69	   &	19 	&	6	 &	3	 &	223	&	\nodata	& PC \\
CARLA-Ser2~J1510+5958 	&	0.976&	\nodata	&	2.36	   &	18 	&	5 	&	3	 &	233	&	\nodata 	& CGC\\
CARLA-Ser~J1753+6310 		&	2.117&	\nodata	&	5.51     &	37 	&	11	 &	6 	&	122	&	\nodata 	& PC\\
CARLA-Ser~J2227$-$2705 	&	1.357&	\nodata	&	4.64     &	30 	&	9	 &	4	&	123	&	\nodata 	& PC\\
CARLA-Ser2~J2227$-$2705 	&	1.478&	\nodata	&	2.77     &	19 	&	6	 &	3	 &	223	&	\nodata 	& PC\\
\enddata
\tablenotetext{a}{Overdensity significance (in $\sigma$) of color-selected sources above the field value (\citealp{Wylezalek14}).}
\tablenotetext{b}{Overdensity significance (in $\sigma$) of spectroscopically confirmed sources above the field value. The values derived assume identical star-forming galaxy fractions in the field and cluster environments.}
\tablenotetext{c}{Galaxy contrast of spectroscopically confirmed sources for three cases of star-forming galaxy fractions in clusters; respectively $15\%$, $50\%$, and $100\%$, with a constant fraction of $80\%$ in the field.}
\tablenotetext{d}{Robustness flag, as described in the text.}
\tablenotetext{e}{{Flag} indicating if the confirmed structure exhibits a tight red sequence (0: False, 1: True)}
\tablenotetext{f}{Confirmed structure classification following our analysis. HPC: highly probable confirmed cluster, PC: probable confirmed cluster, CGC: confirmed galaxy concentration.}
\tablenotetext{g}{For unconfirmed structures, overdensity significances and galaxy contrasts are derived based on the few ($<5$) confirmed sources at the RLAGN redshifts. }
\end{deluxetable*}

\section{Velocity Scatter and Spatial Distribution of Confirmed Members}\label{app:velspadist}
{In this Appendix we show and briefly discuss the redshift/velocity scatter and spatial distribution of confirmed members in our $16$ confirmed CARLA cluster candidates (additional members of CARLA~J1753+6311 confirmed with Keck spectroscopy are not shown here; see A.~Rettura et al., in prep). CARLA confirmations are sorted by increasing scatter in redshift space (standard deviation). CARLA~J2039$-$2514 and CARLA~J0800+4029, already published in N16, are reproduced here for consistency.}\\

\begin{figure*}
{%
\setlength{\fboxsep}{0pt}%
\setlength{\fboxrule}{1pt}%
\fbox{\includegraphics[scale=0.38]{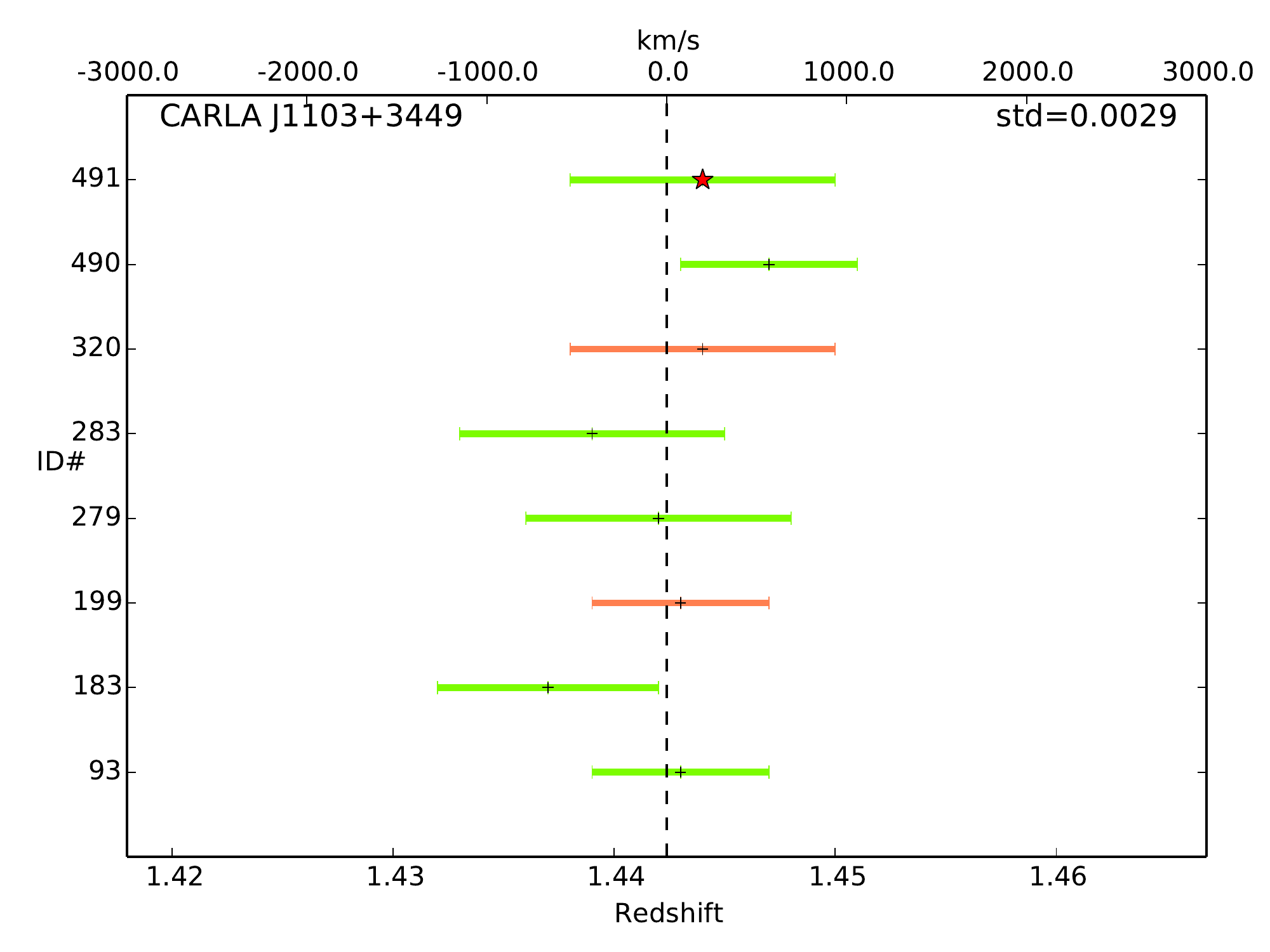} \hfill \includegraphics[scale=0.38]{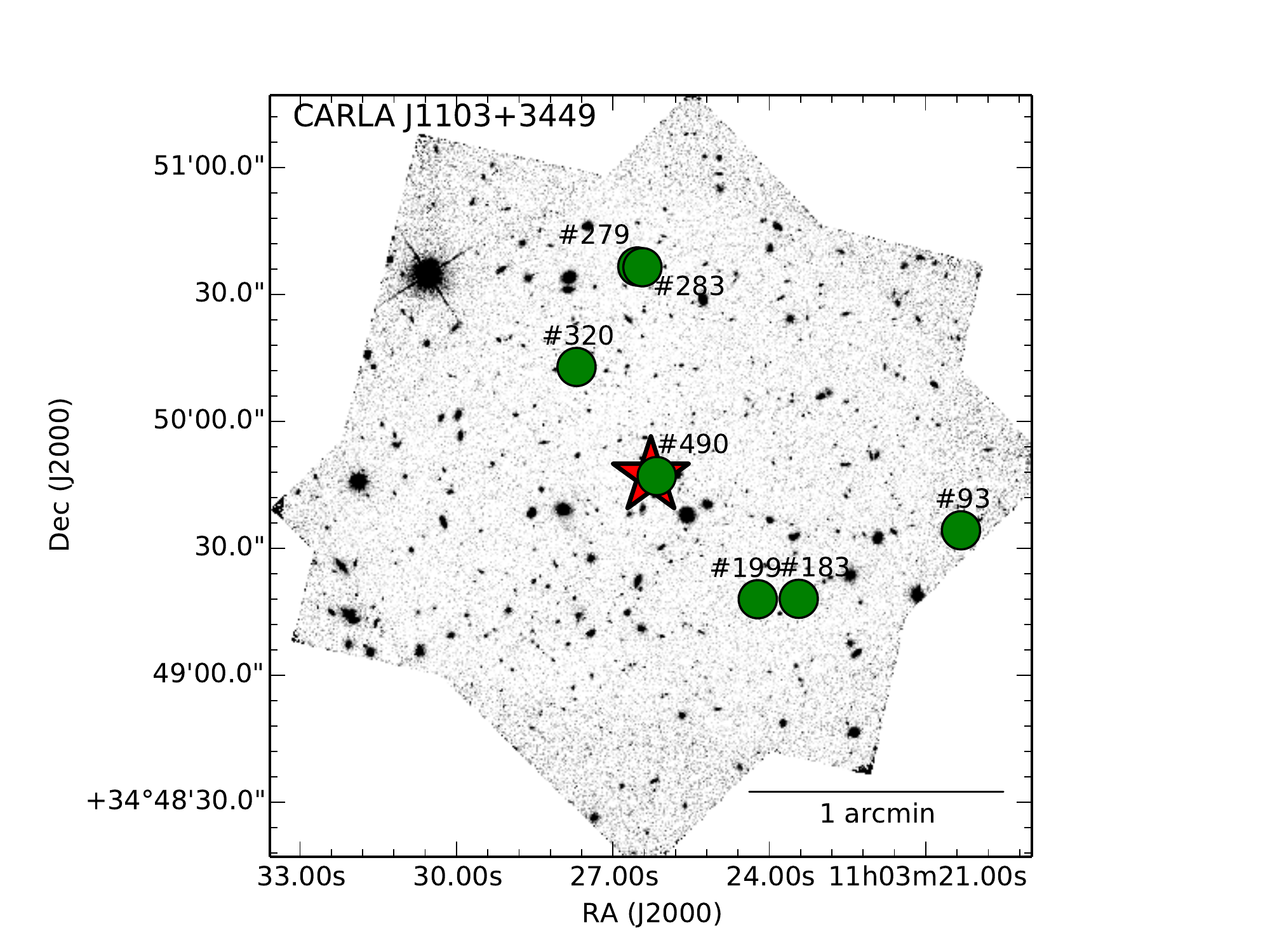} \mbox{(a)}}%
}\\%
{%
\setlength{\fboxsep}{0pt}%
\setlength{\fboxrule}{1pt}%
\fbox{\includegraphics[scale=0.38]{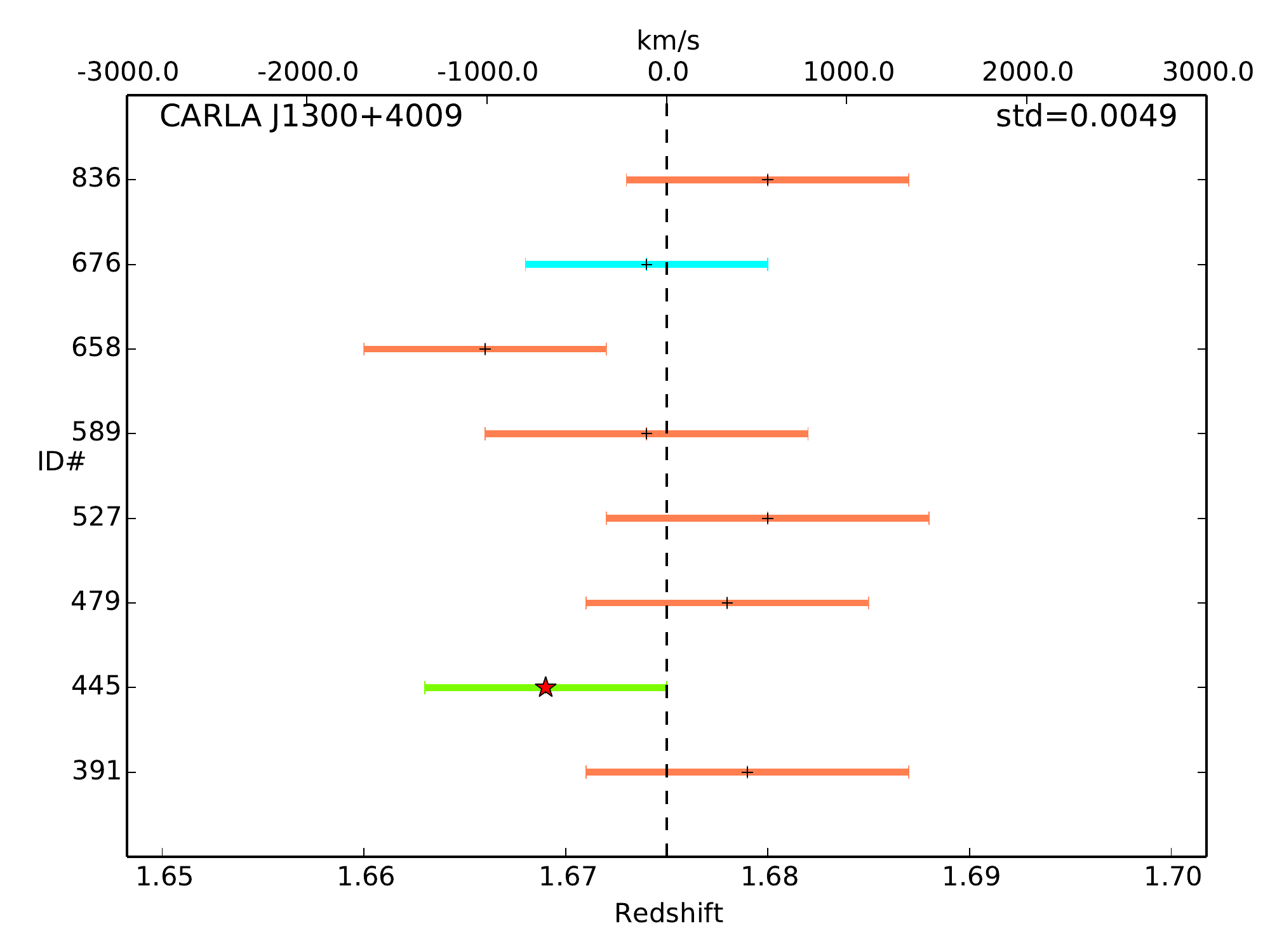} \hfill \includegraphics[scale=0.38]{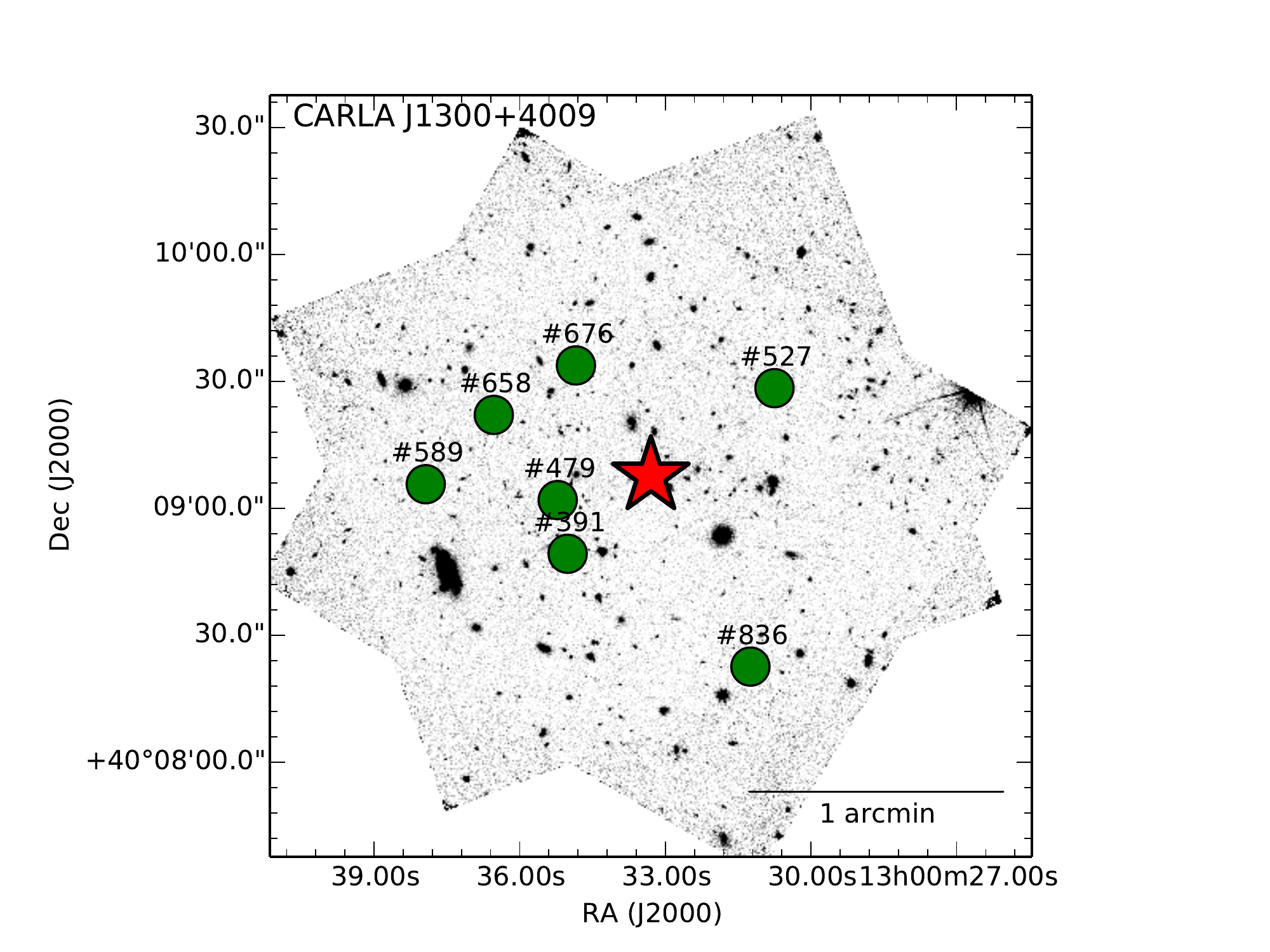} \mbox{(b)}}%
}\\%
{%
\setlength{\fboxsep}{0pt}%
\setlength{\fboxrule}{1pt}%
\fbox{\includegraphics[scale=0.38]{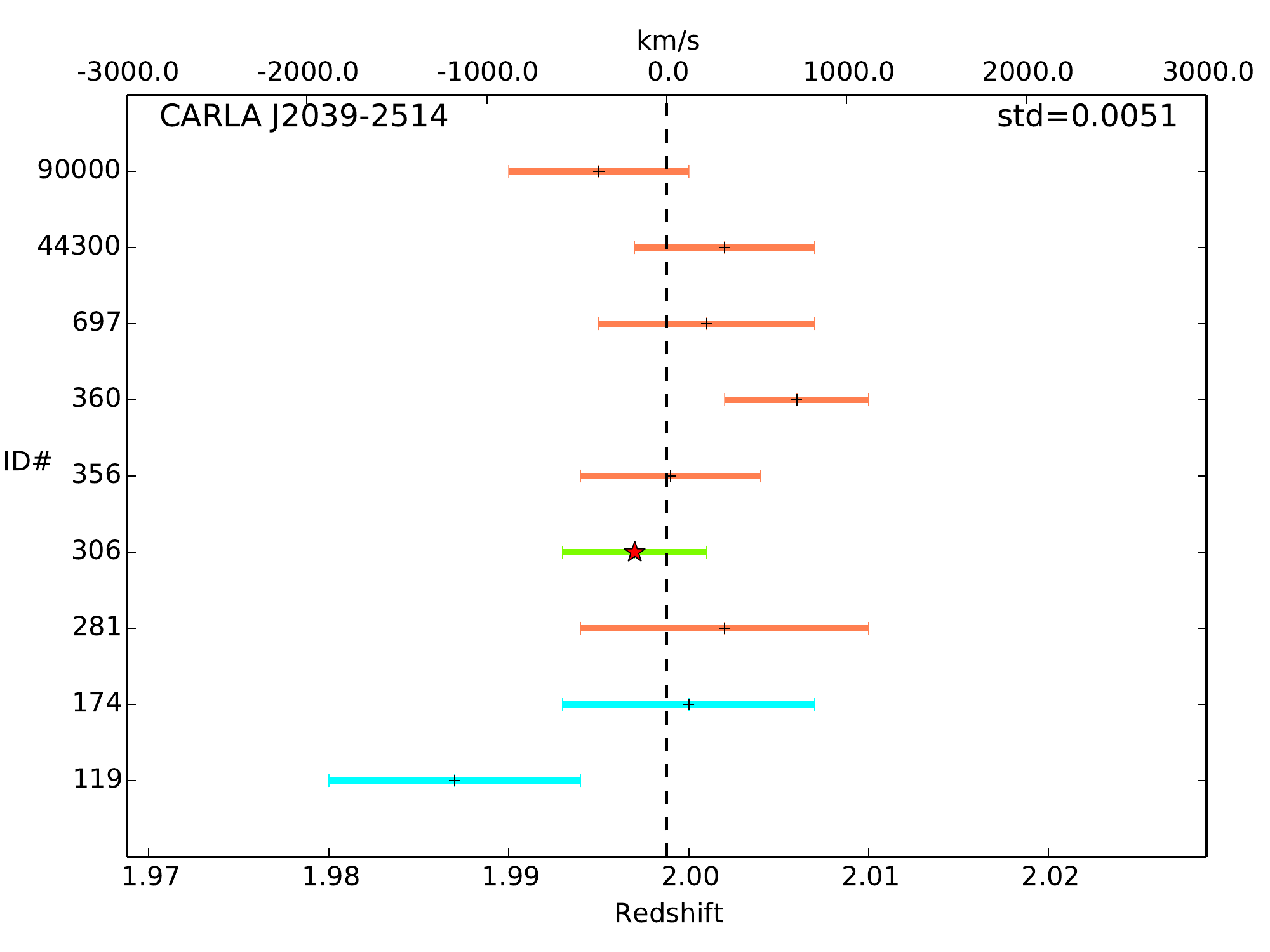} \hfill \includegraphics[scale=0.38]{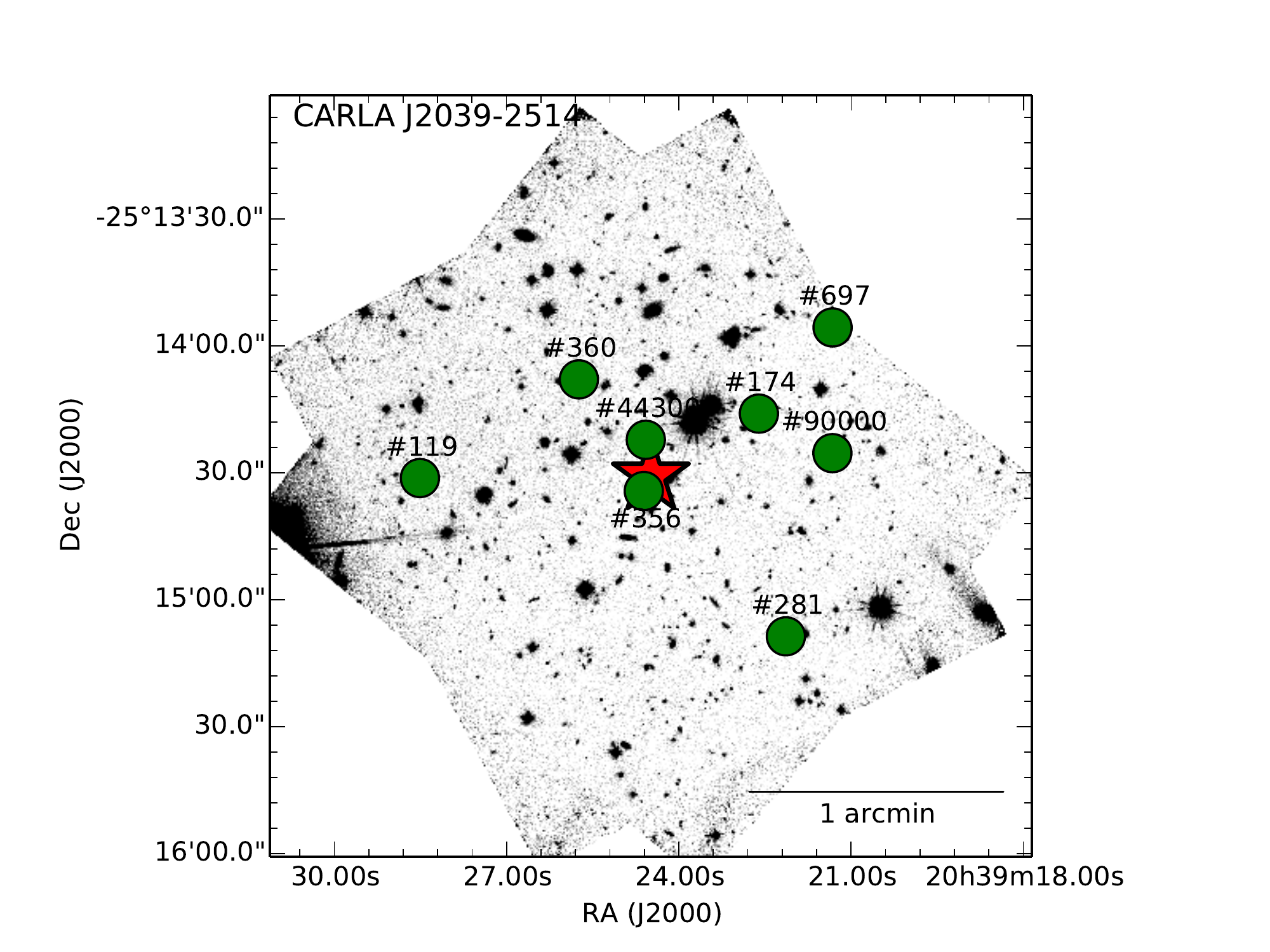} \mbox{(c)}}%
}\\%
{%
\setlength{\fboxsep}{0pt}%
\setlength{\fboxrule}{1pt}%
\fbox{\includegraphics[scale=0.38]{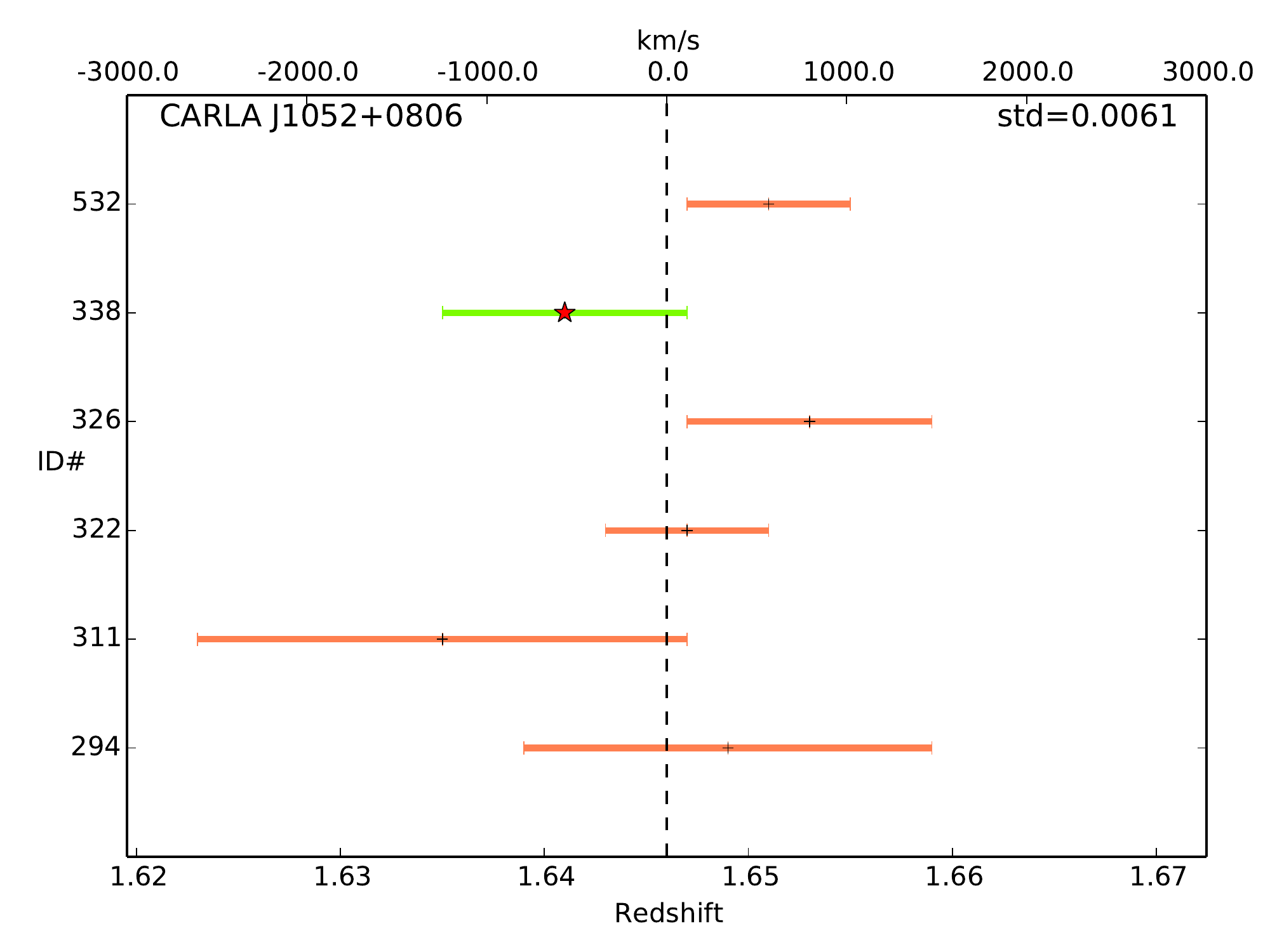} \hfill \includegraphics[scale=0.38]{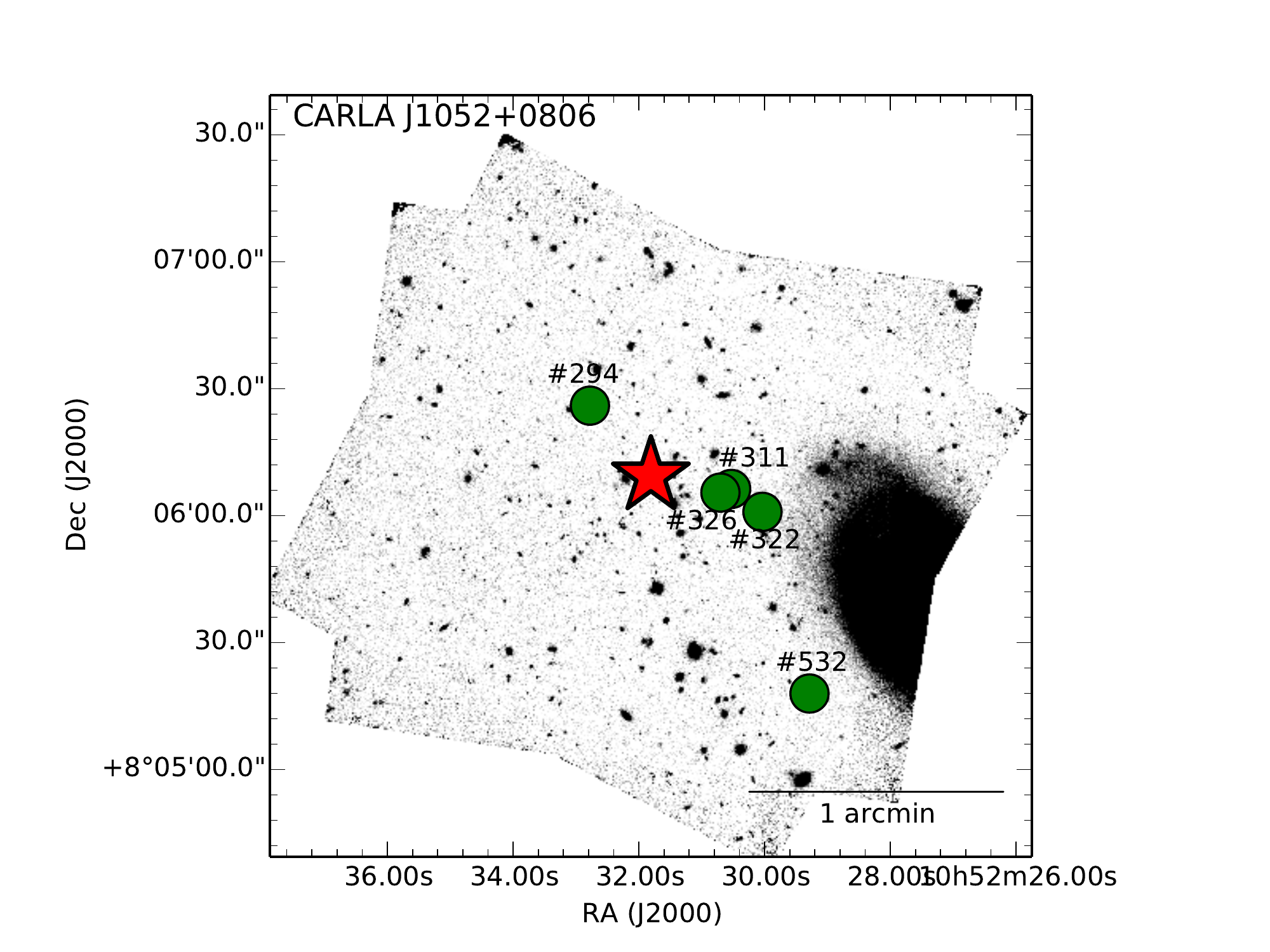} \mbox{(d)}}%
}\\%
\caption[Velocity scatter spatial distribution]{Left: redshift/velocity distributions of confirmed CARLA cluster members. Green, cyan and orange lines represent quality A, B$^{+}$, and B$^{-}$ redshifts, respectively. Vertical dashed lines represent structure mean redshifts. All left panels are scaled to the same velocity range. Right: spatial distributions of structure members; North is up and East is to the left. In all panels, red stars represent the RLAGN.}
\label{fig:velspadist}
\end{figure*}

{{\it CARLA~J1103+3449}, Fig.~\ref{fig:velspadist}, Panel (a) -- This confirmed CARLA cluster candidate is comprised of $8$ members {at $z=1.44$}, six being of quality A with both H$\alpha$ and [\ion{O}{3}] detected. All members are confirmed within $\pm1000\,\rm km\,s^{-1}$ and strongly peak at the same redshift as the RLAGN, which could argue against a very massive virialized cluster of larger velocity scatter/dispersion. We however determined a $123$ flag for this confirmed structure (see Appendix \ref{app:specsig}), which, following our analysis in Appendix \ref{app:specsig}, would tend to support that CARLA~J1103+3449, {classified as PC}, is probably a node of the cosmic web, as opposed to a filament. The RLAGN appears to possess at least one confirmed satellite member ($\#490$).}

{{\it CARLA~J1300+4009} (b) -- Given its redshift, [\ion{O}{3}] was the only strong line available in our grism data for confirming members of this CARLA cluster candidate. The $8$ confirmed members have again very low velocity scatter, for a combined flag $122$. {Classified as PC}, this would tend to support that CARLA~J1300+4009 is most probably a node of the cosmic web at $z=1.68$.}

{{\it CARLA~J2039$-$2514} (c) -- We {previously presented} CARLA~J2039$-$2514 in detail in N16 and showed that this $z=2.00$ confirmed structure is already comprised of a red-sequence of passive candidate members, consistent with being a bona fide galaxy cluster. With a flag of $122$, CARLA~J2039$-$2514, comprised of 9 confirmed members including a dual AGN {and classified as HPC}, appears as a very strong candidate for being a massive node of the cosmic web at $z=2$. The (dual-AGN) RLAGN seem to possess at least two confirmed satellite members ($\#356$ and $\#44300$).}

{{\it CARLA~J1052+0806} (d) -- This confirmed structure shows one of the densest cores, among our 20 CARLA {\it HST} fields, of red color-selected galaxies which only exhibit continuum in our grism data, without detection of emission lines. Due to proximity, their spectra are additionally contaminating each other. These sources are very promising targets for deeper spectroscopic observations and additional multi-wavelength follow-ups to reveal and study passive populations already present in a cluster core at $z=1.65$. A flag of $123$ {classifies CARLA~J1052+0806 as PC}. A bright nearby galaxy, falling on the West side of our {\it HST} field of view, additionally contaminated a significant fraction of our grism data in this field, likely hindering confirmation of additional star-forming members of CARLA~J1052+0806.}

{{\it CARLA~J1753+6310} (e) -- We {previously presented} CARLA~J1753+6310 in Section \ref{sec:carlaJ1753+6310}; see that section for details. We only confirm five members in our grism data, while additional Keck observations confirm an additional three members (A.~Rettura et al., in prep). CARLA~J1753+6310 has a $223$ flag, and \cite{Cooke16} demonstrated that this structure possesses a strong red-sequence of passive candidate members and that its core is dominated by passive galaxies. CARLA~J1753+6310, {classified as HPC}, therefore appears as a very strong candidate for being a massive node of the cosmic web at $z=1.58$. The RLAGN likely possesses one confirmed satellite member ($\#470$) at least.}

{{\it CARLA~J2355$-$0002} (f) -- We {previously presented} CARLA~J2355$-$0002 in Section \ref{sec:carla2355-0002}; see that section for details. With a total of $12$ quality-A-only members at $z=1.49$, CARLA~J2355$-$0002 appears as one of the most convincing spectroscopic confirmation from our {\it HST} data alone. Supporting data (\citealp{Collet15}) additionally identifies $z\sim1.5$ H$\alpha$ emitters outside our {\it HST} field of view. CARLA~J2355$-$0002 has a $122$ flag and its velocity distribution shows some scatter. This tends to support that CARLA~J2355$-$0002, {classified as PC}, is probably a massive virialized cluster caught at an epoch of star formation. Confirmed (star-forming) members interestingly appear to be aligned in the NNE-SSW direction. Detailed study of CARLA~J2355$-$0002 therefore appears very promising with respect to investigating the interplay between AGN nuclear activity, large-scale star formation and the infall of gas from the cosmic web. The RLAGN seems to possess multiple confirmed satellite members.}

{{\it CARLA~J1131$-$2705} (g) -- CARLA~J1131$-$2705 is comprised of $9$ members {at $z=1.45$}, including four quality A and four quality B$^{+}$ sources. Confirmed members are well peaked around the redshift of the RLAGN, with a scatter similar to CARLA~J2355$-$0002. This confirmed structure has a $123$ flag. Together, this tends to support that CARLA~J1131$-$2705, {classified as PC} is a probable node of the cosmic web. One member (\#442) appears as a probable satellite of the RLAGN.}

{{\it CARLA~J0116$-$2052} (h) -- We confirm $12$ members {at $z=1.43$}, including six quality A sources, in the grism data of CARLA~J0116$-$2052. As opposed to previously discussed confirmed structures, confirmed members of CARLA~J0116$-$2052 appear to form two peaks around its mean redshift and are predominantly located to the North of the RLAGN. This might suggest that the RLAGN is not located at the center {of the structure}. With a flag of $122$, our analysis however supports that this confirmed structure is a probable node of the cosmic web ({it is classified as PC}).}

{{\it CARLA~J1017+6116} (i) -- At a redshift of $z=2.80$, CARLA~J1017+6116 is the highest redshift confirmed structure among our 20 fields. With {seven} confirmed members at this redshift, CARLA~J1017+6116 also possesses the highest spectroscopic overdensity significance and galaxy contrast of our sample. Confirmed members appear well distributed around the RLAGN redshift. At this high redshift, we however relied on the sole identification of the relatively weak [\ion{O}{2}] emission line to confirm members of this structure. Additional spectroscopy would be required to obtain more robust redshift qualities on these sources. With a $111$ flag, our analysis suggests that CARLA~J1017+6116 is a highly probable node of the cosmic web ({i.e., it is classified as HPC}). Note that {\it Spitzer}/IRAC and {\it HST}/F140W flux measurements of source $\#124$ are contaminated by a bright nearby source and a diffraction spike and should be considered as upper limits only.}

{{\it CARLA~J1358+5752} (j) -- CARLA~J1358+5752 is the structure with the highest number of confirmed members in our data, with $14$ confirmations {at $z=1.37$} including $10$ quality A sources. Five confirmed sources appear slightly blueshifted compared to the rest of the members. This could suggest infalling galaxies in the structure potential. Together with a $122$ flag, CARLA~J1358+5752, {classified as PC}, appears as a probable node of the cosmic web. One source ($\#414$) appears as a probable satellite of the RLAGN.}

{{\it CARLA~J1129+0951} (k) -- We confirm $12$ members {at $z=1.53$}, including $8$ quality A sources, in the grism data of CARLA~J1129+0951. The velocity distribution is again double peaked, with five members slightly redshifted compared to the bulk of confirmed members including the RLAGN. Deeper, higher-resolution spectroscopy will be required to understand whether they are infalling galaxies. This structure has a $122$ flag, {classifying CARLA~J1129+0951 as PC.}}

{{\it CARLA~J1018+0530} (l) -- Together with CARLA~J2039$-$2514 and CARLA~J0800+4029, CARLA~J1018+0530 is one of the three confirmed structures at $z=2$ among our 20 fields. CARLA~J1018+0530 interestingly shows a concentration of bright continuum-only sources in the NE with respect to the RLAGN. This group shows hint of potential intra-cluster light, while narrow-band imaging available for this field suggests that sources of this group have photometric redshifts consistent with that of the RLAGN. A posteriori tentative detection of an emission line consistent with [\ion{O}{3}] at the redshift of the RLAGN for one source in this group tends to support this claim.  {We will present these data in a forthcoming paper.} With a $122$ flag, CARLA~J1018+0530 {is classified as PC}. Two sources ($\#354$ and $\#466$) appear as probable satellites of the RLAGN.}

{{\it CARLA~J1510+5958} (m) -- CARLA~J1510+5958 is comprised of six confirmed members, mainly of B qualities due to the redshift ($z=1.72$) of the target. This allows identification of a sole strong emission line ([\ion{O}{3}]) in our grism data. Two sources appear significantly redshifted ($>1000\,\rm km\,s^{-1}$) compared to the other confirmed members. With a $122$ flag, our analysis {classifies} CARLA~J1510+5958 {as PC}. One source ($\#771$) appears as a probable satellite of the RLAGN.}

{{\it CARLA~J0958$-$2904} (n) -- CARLA~J0958$-$2904 is confirmed with $8$ members {at $z=1.40$} including five quality A sources. Among our confirmed CARLA clusters, CARLA~J0958$-$2904 shows the most redshifted RLAGN relative to the mean redshift of the structure. Unless due to e.g., wavelength mis-calibration, this could suggest that the targeted RLAGN of CARLA~J0958$-$2904 might not reside at the center of the structure. Source $\#405$ appears as a likely satellite of the RLAGN. With a $223$ flag, CARLA~J0958$-$2904 {is classified as PC}. Source $\# 405$ appears as a probable satellite of the RLAGN.}

{{\it CARLA~J2227$-$2705} (o) -- CARLA~J0958$-$2904, confirmed with $7$ members at $z=1.69$, shows a double-peaked velocity distribution, which could potentially suggest that confirmed members are not yet well stabilized in the structure potential. With a $122$ flag, our analysis {classifies} CARLA~J2227$-$2705 {as PC}.}

{{\it CARLA~J0800+4029} (p) -- We {previously presented} CARLA~J0800+4029 in detail in N16. In N16, we showed that CARLA~J0800+4029 does not exhibit a clear red-sequence of bright ($\rm F140W <23\, mag$) candidate members, as opposed to CARLA~J2039$-$2514, and is likely a younger forming structure than CARLA~J2039$-$2514, which was confirmed at a similar redshift ($z=2$). This would tend to better argue in favor of {the large velocity scatter of CARLA~J0800+4029 being} consistent with infalling members in the structure potential rather than showing evidence for a massive stabilized cluster. Additional data, however, is required to better characterize the evolutionary status of this confirmed structure. With a flag of $122$, our analysis {classifies} CARLA~J0800+4029 {as PC}. Source $\# 372$ appears as a probable satellite of the RLAGN.}\\

\begin{figure*}
{%
\setlength{\fboxsep}{0pt}%
\setlength{\fboxrule}{1pt}%
\fbox{\includegraphics[scale=0.38]{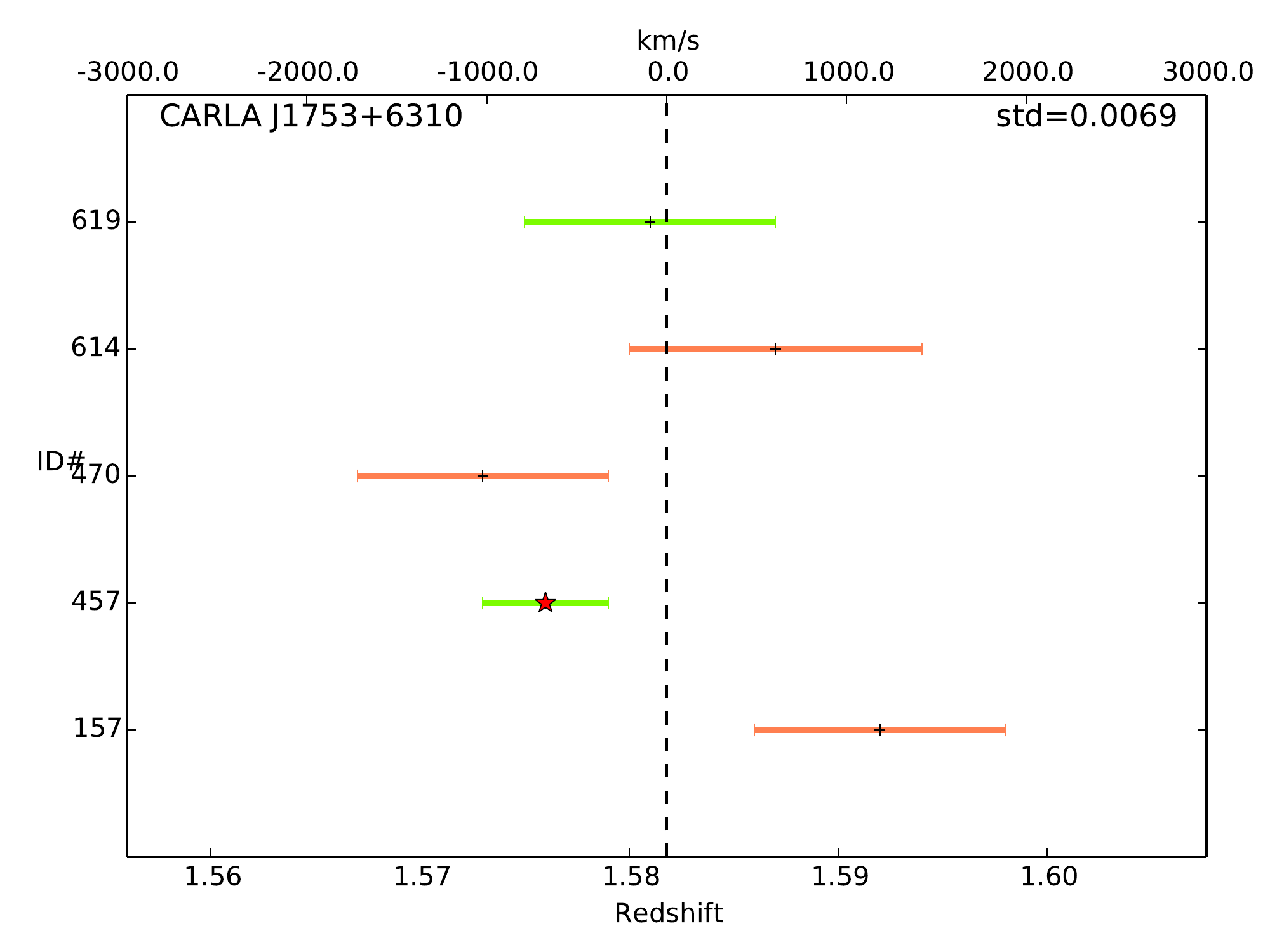} \hfill \includegraphics[scale=0.38]{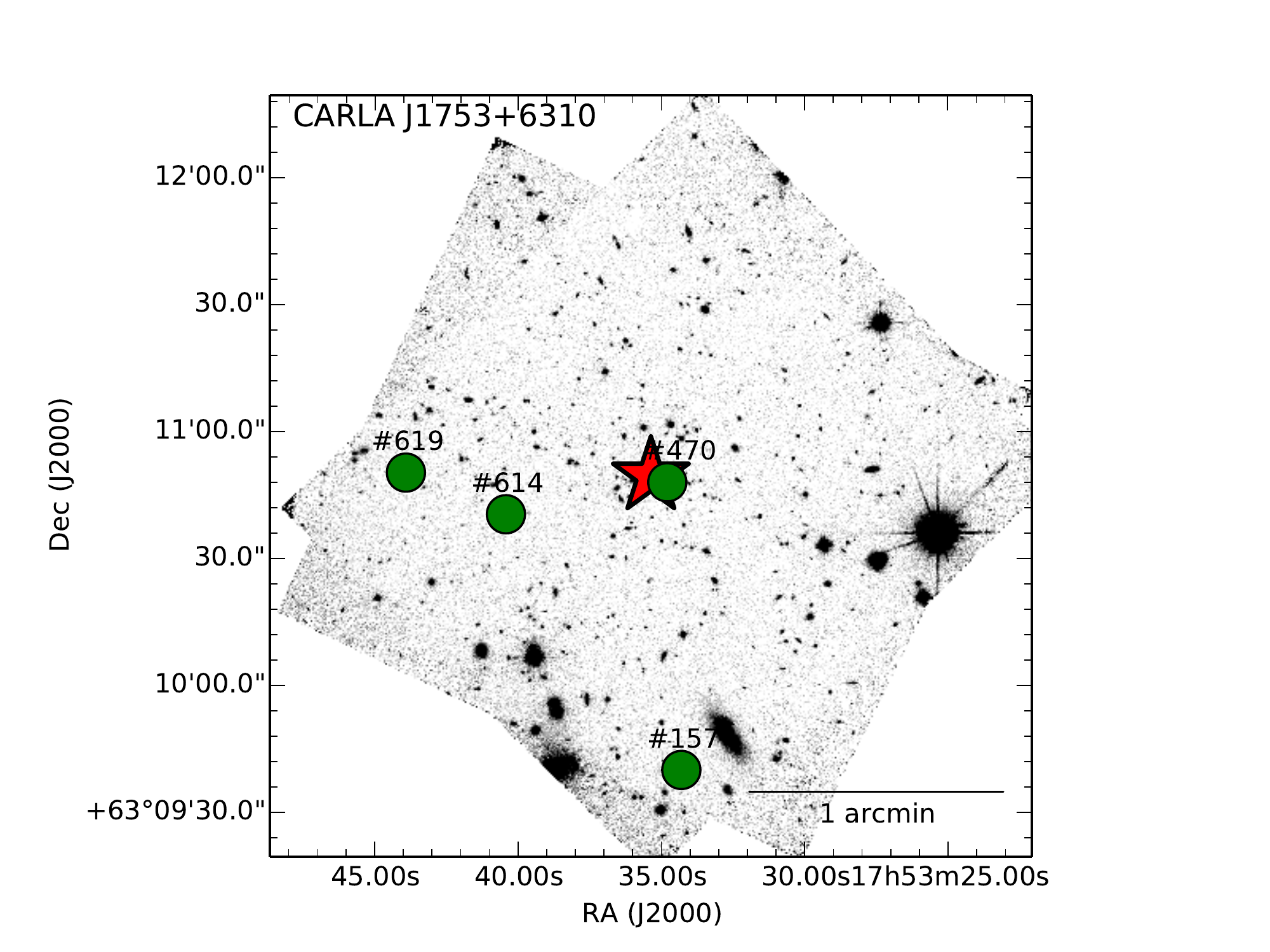} \mbox{(e)}}%
}\\%
{%
\setlength{\fboxsep}{0pt}%
\setlength{\fboxrule}{1pt}%
\fbox{\includegraphics[scale=0.38]{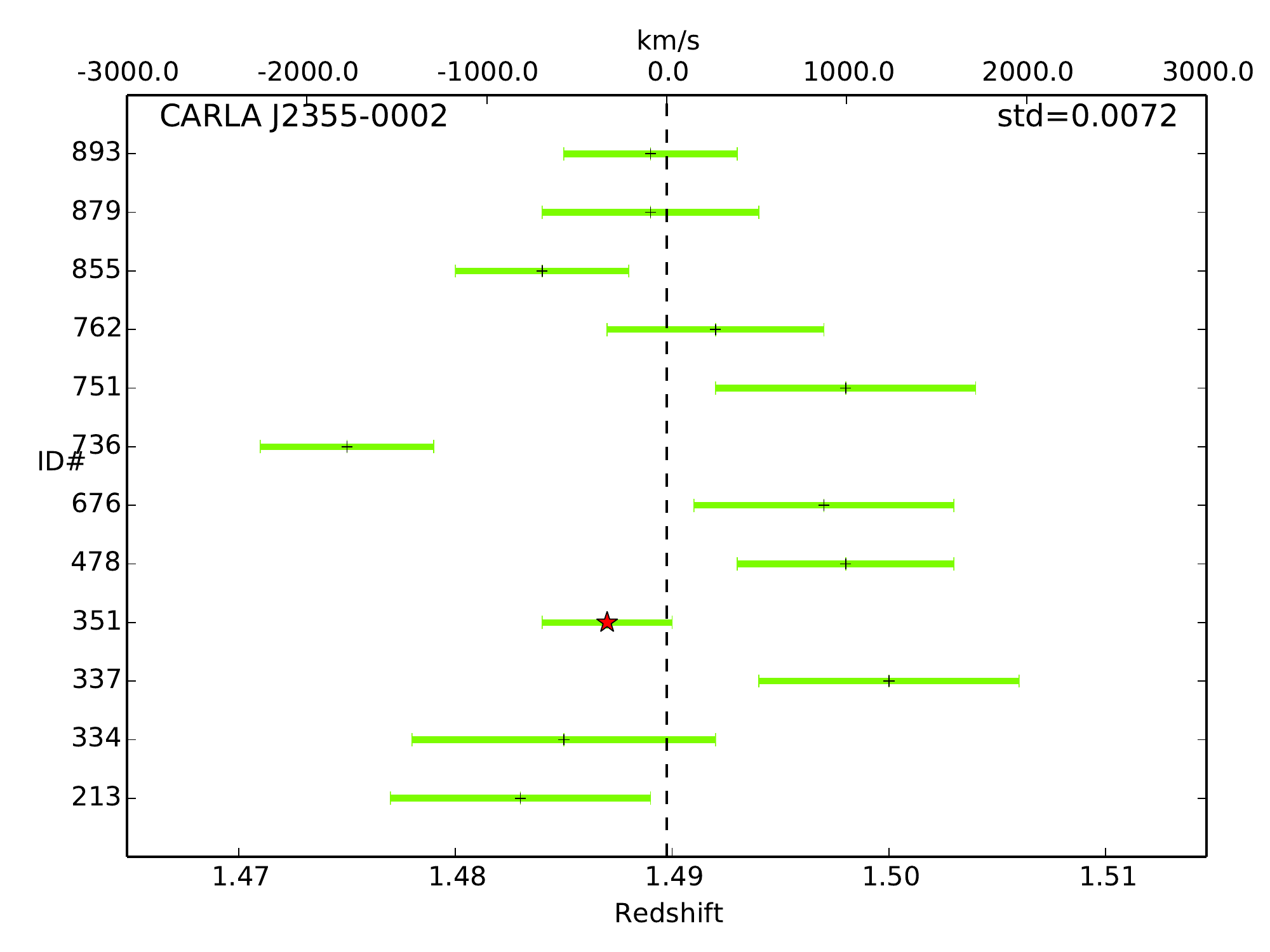} \hfill \includegraphics[scale=0.38]{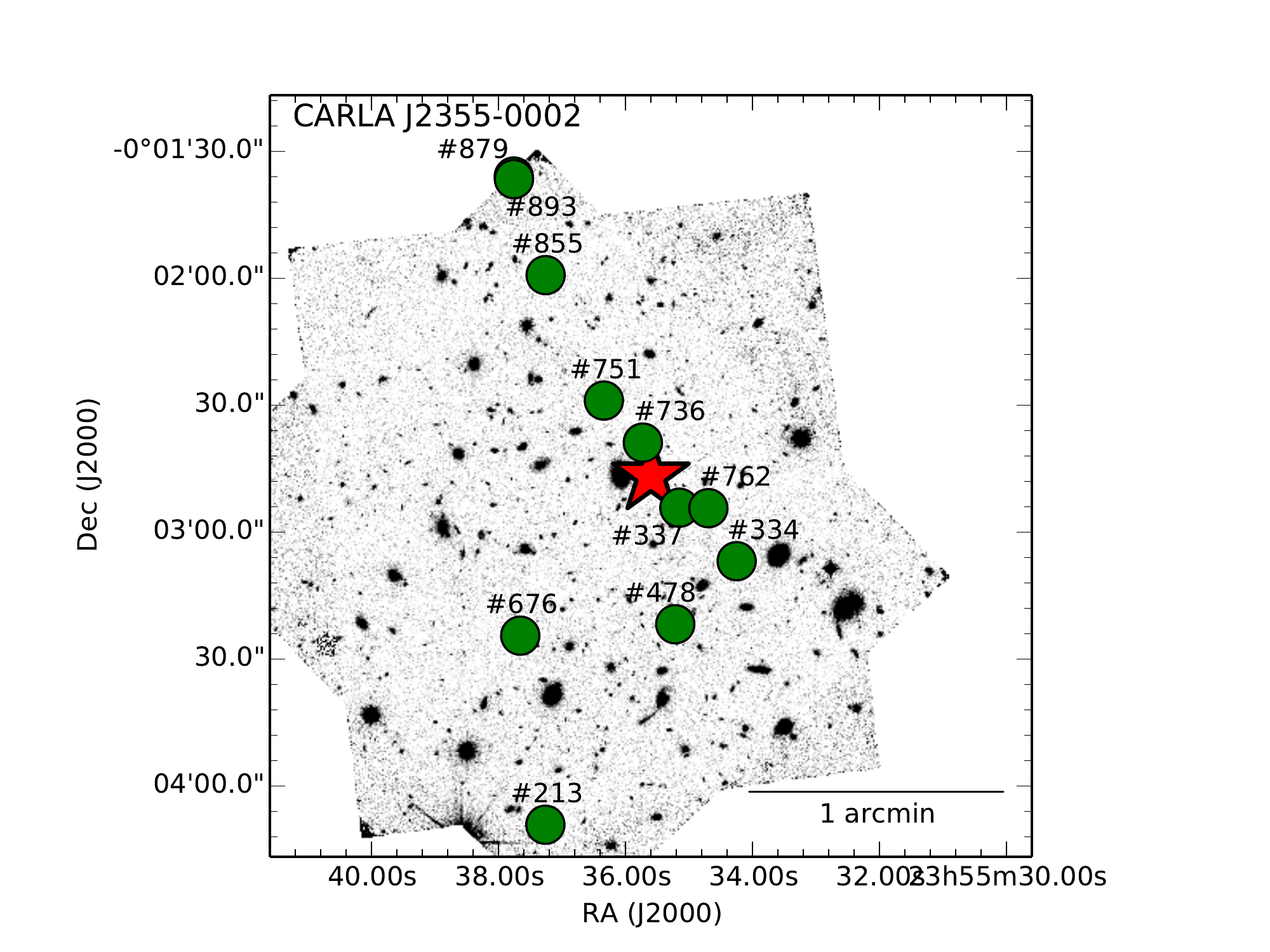} \mbox{(f)}}%
}\\%
{%
\setlength{\fboxsep}{0pt}%
\setlength{\fboxrule}{1pt}%
\fbox{\includegraphics[scale=0.38]{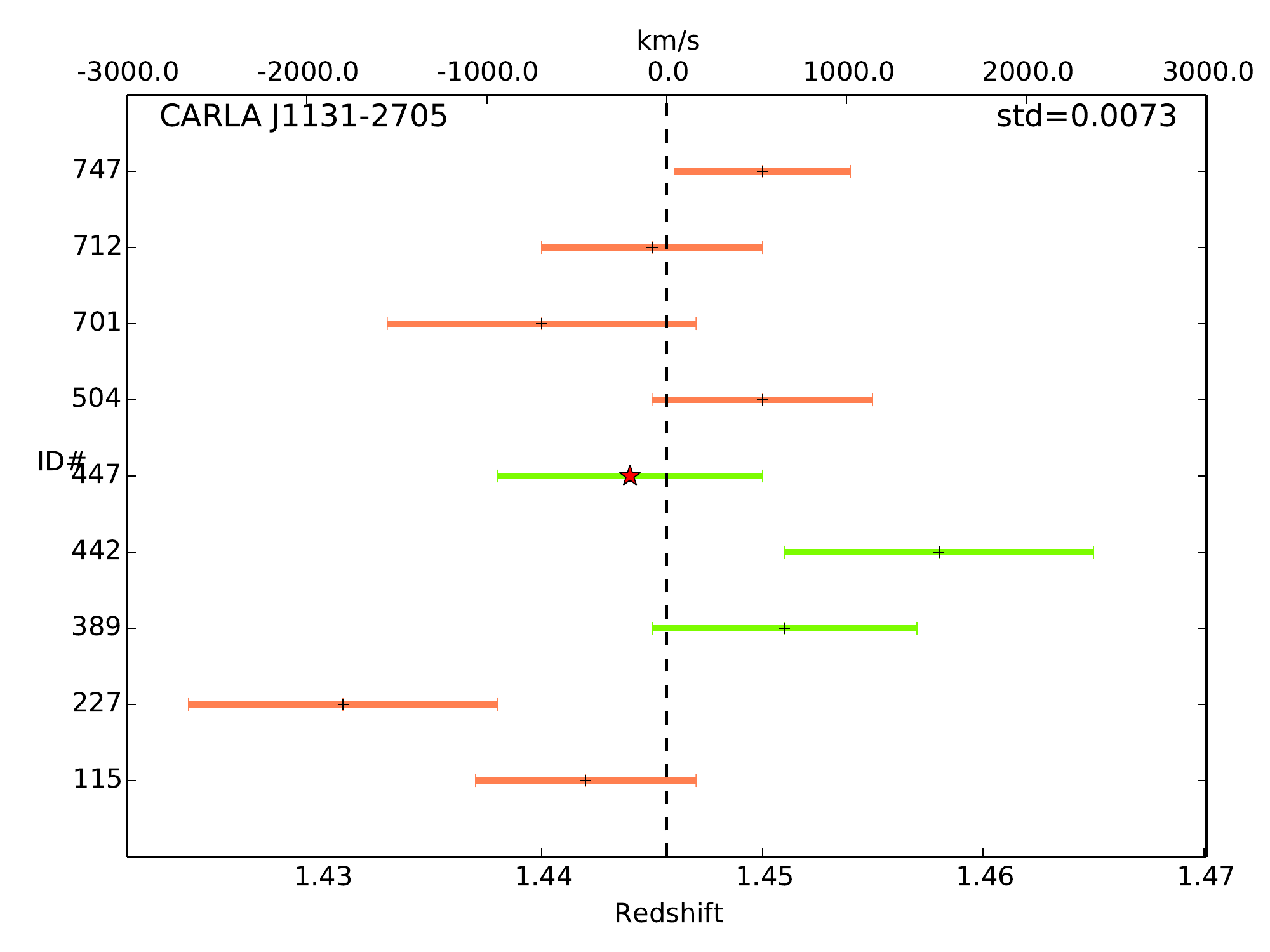} \hfill \includegraphics[scale=0.38]{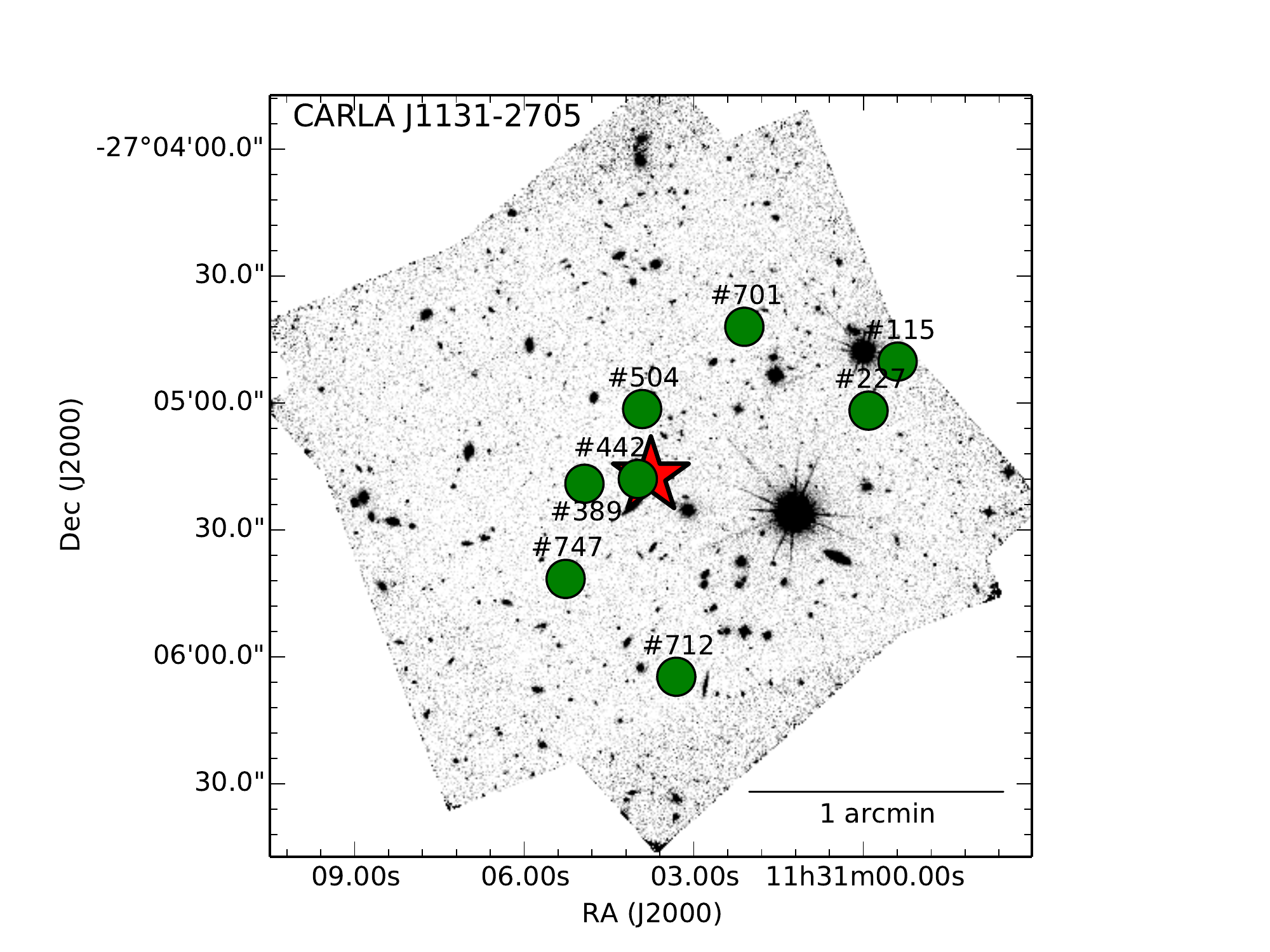} \mbox{(g)}}%
}\\%
{%
\setlength{\fboxsep}{0pt}%
\setlength{\fboxrule}{1pt}%
\fbox{\includegraphics[scale=0.38]{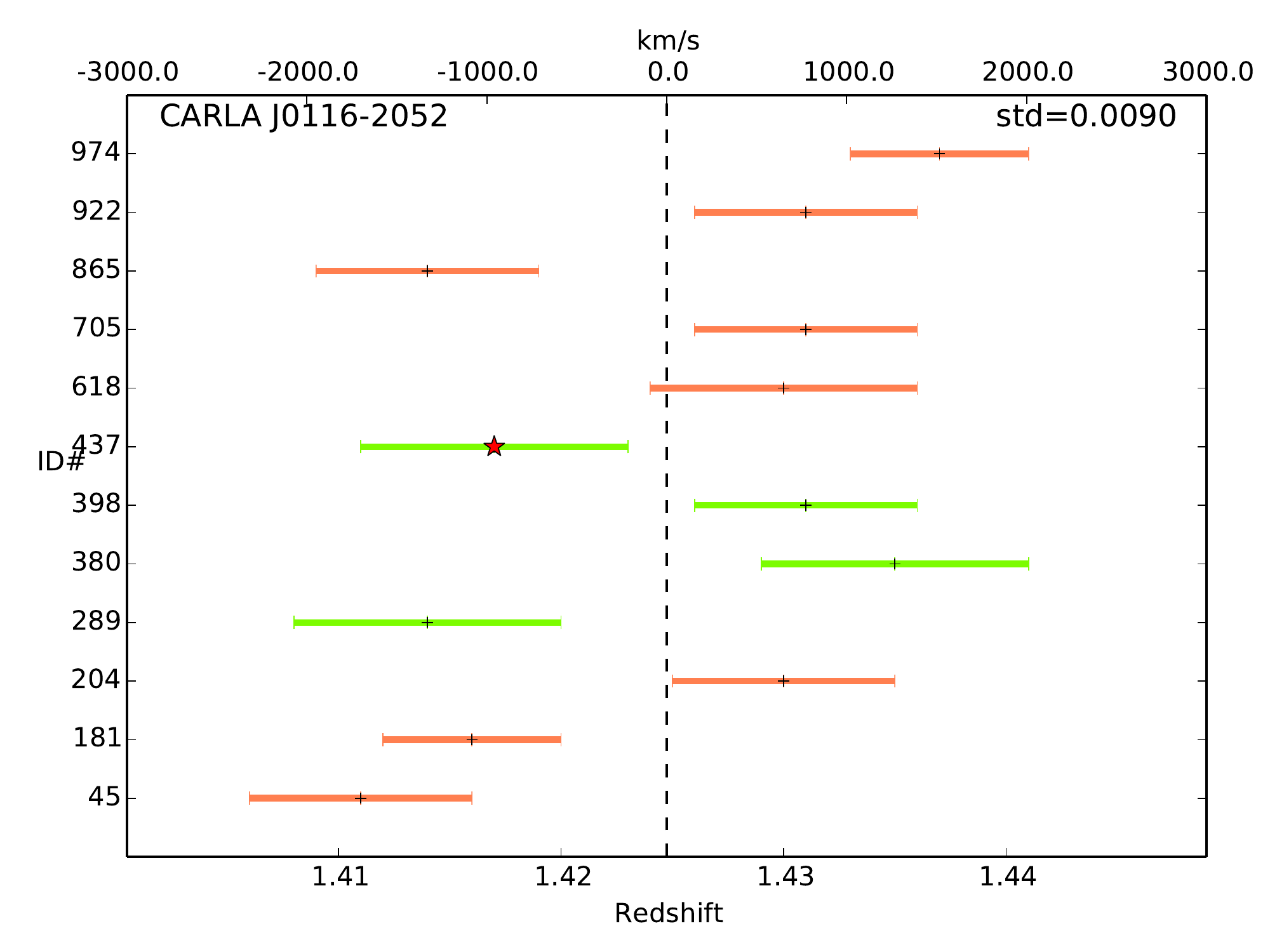} \hfill \includegraphics[scale=0.38]{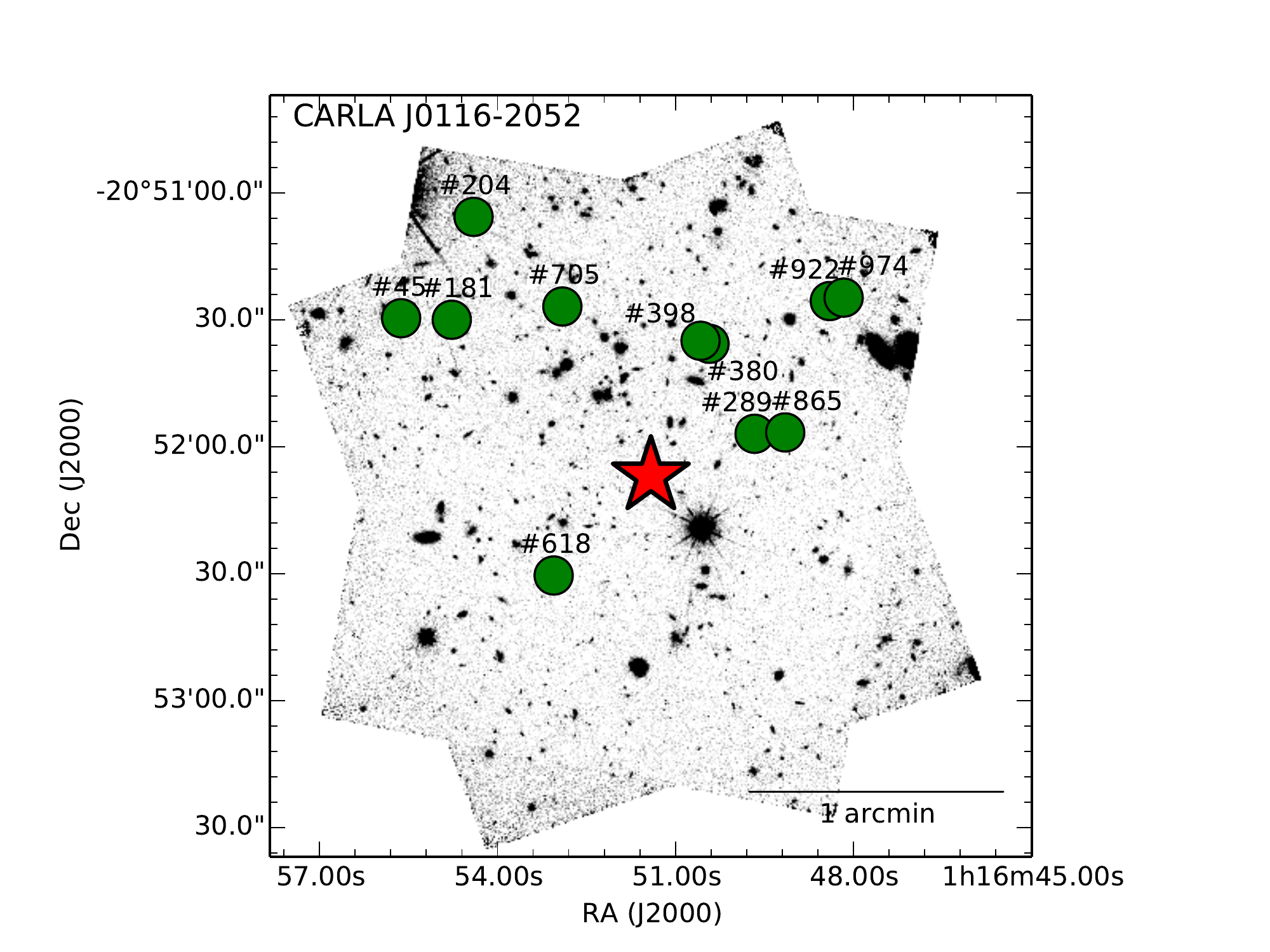} \mbox{(h)}}%
}\\%
\textbf{\mbox{}\\ Figure \ref{fig:velspadist}} --- Continued.
\end{figure*}

\begin{figure*}
{%
\setlength{\fboxsep}{0pt}%
\setlength{\fboxrule}{1pt}%
\fbox{\includegraphics[scale=0.38]{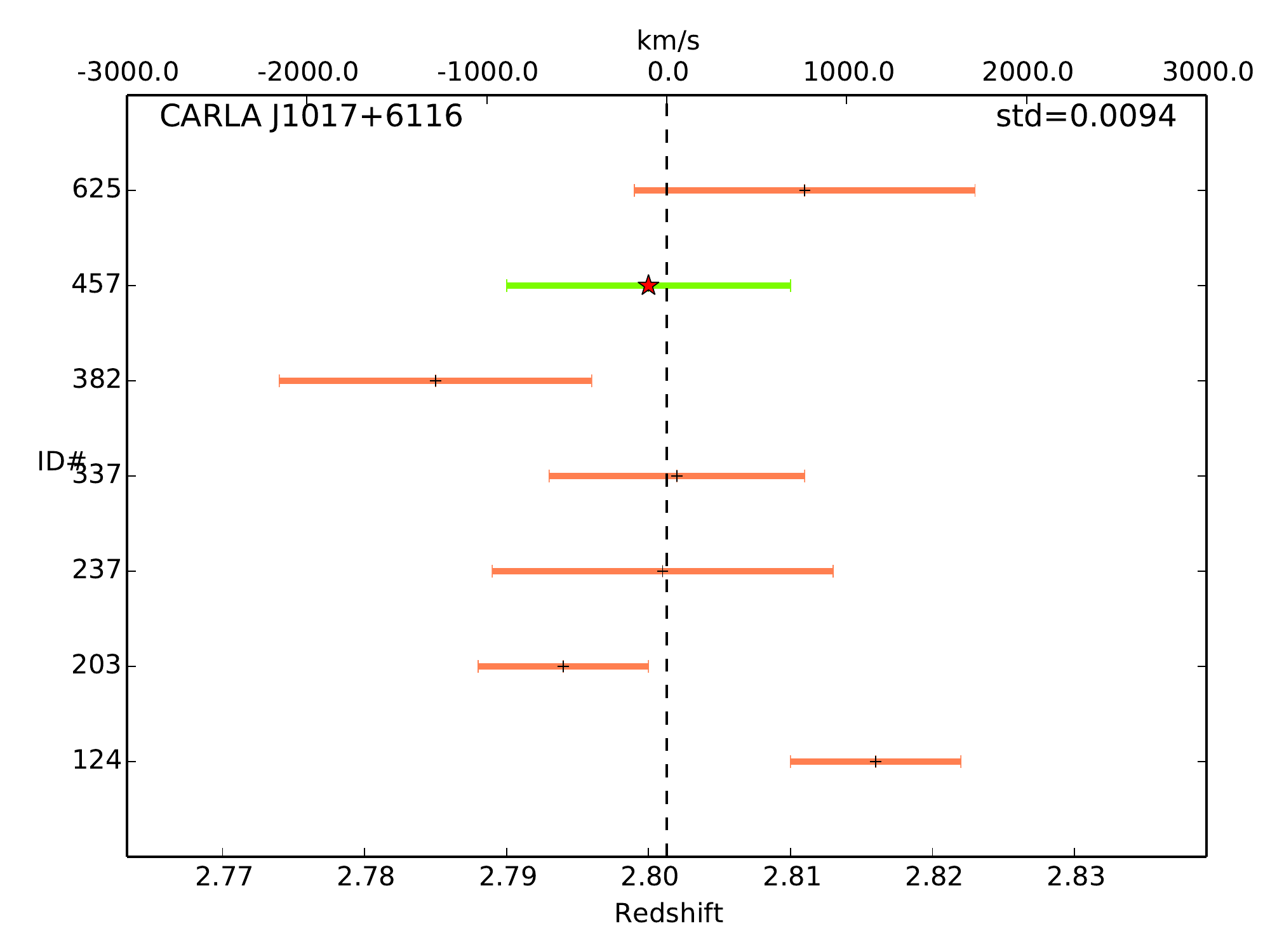} \hfill \includegraphics[scale=0.38]{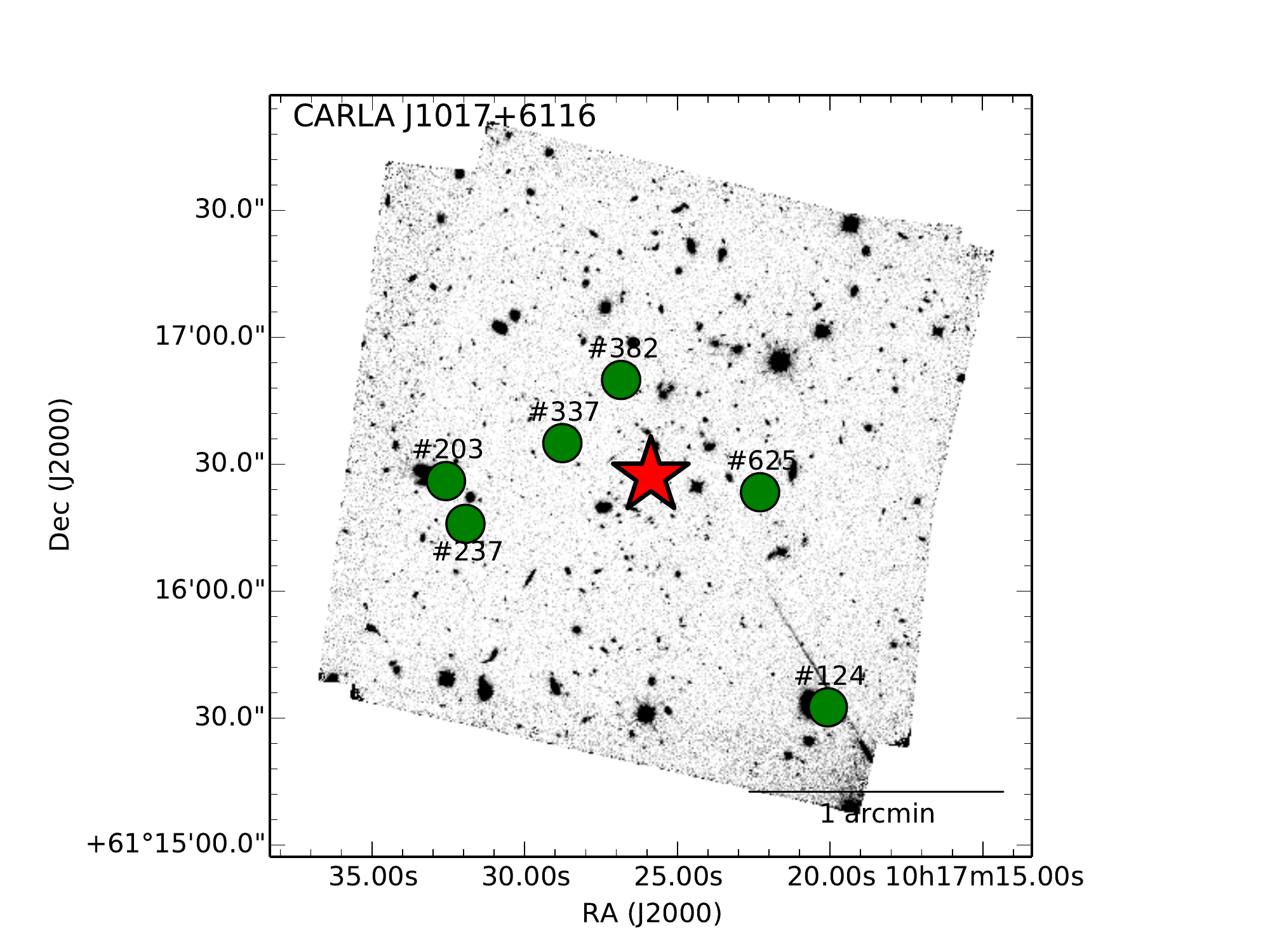} \mbox{(i)}}%
}\\%
{%
\setlength{\fboxsep}{0pt}%
\setlength{\fboxrule}{1pt}%
\fbox{\includegraphics[scale=0.38]{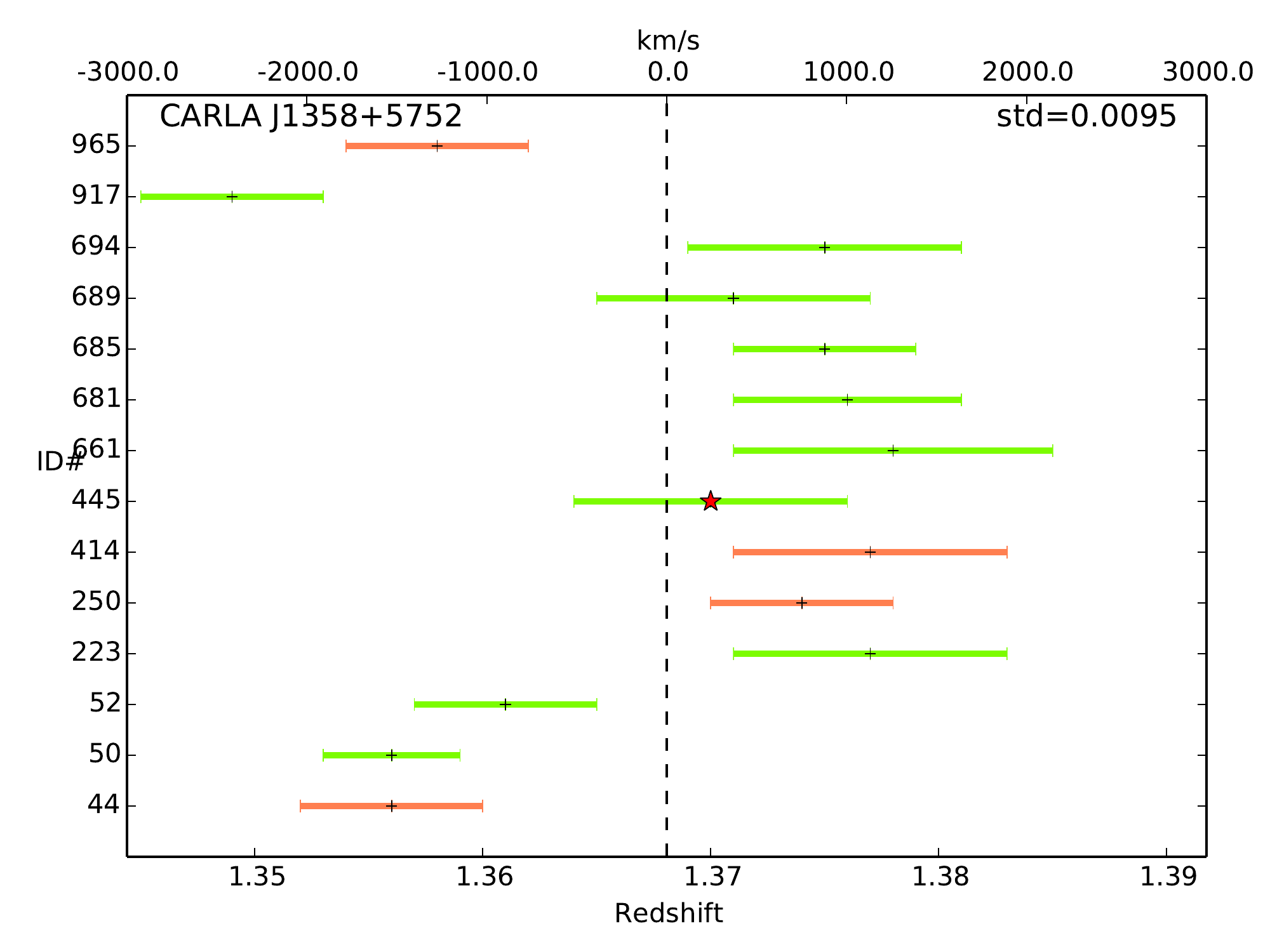} \hfill \includegraphics[scale=0.38]{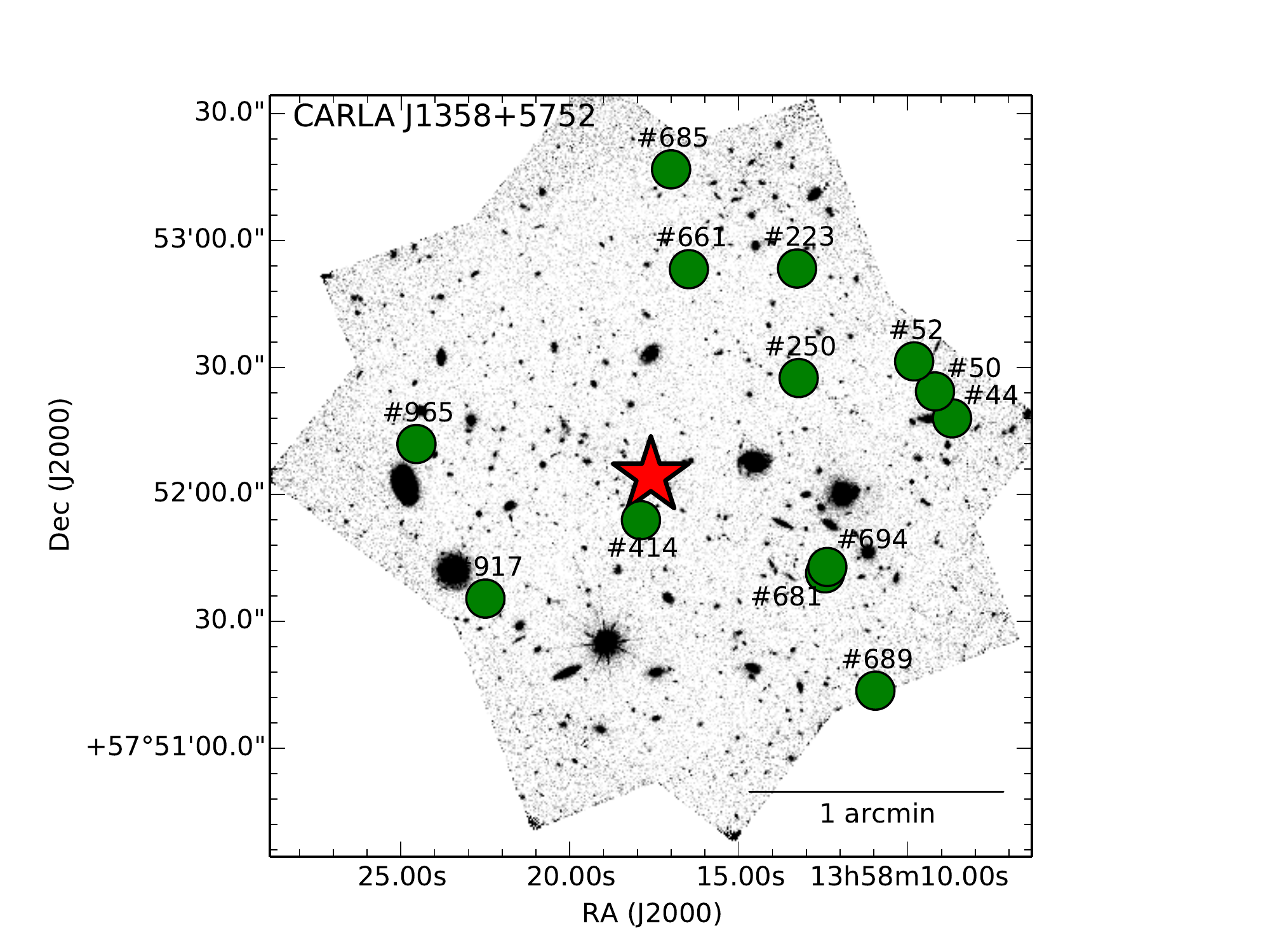} \mbox{(j)}}%
}\\%
{%
\setlength{\fboxsep}{0pt}%
\setlength{\fboxrule}{1pt}%
\fbox{\includegraphics[scale=0.38]{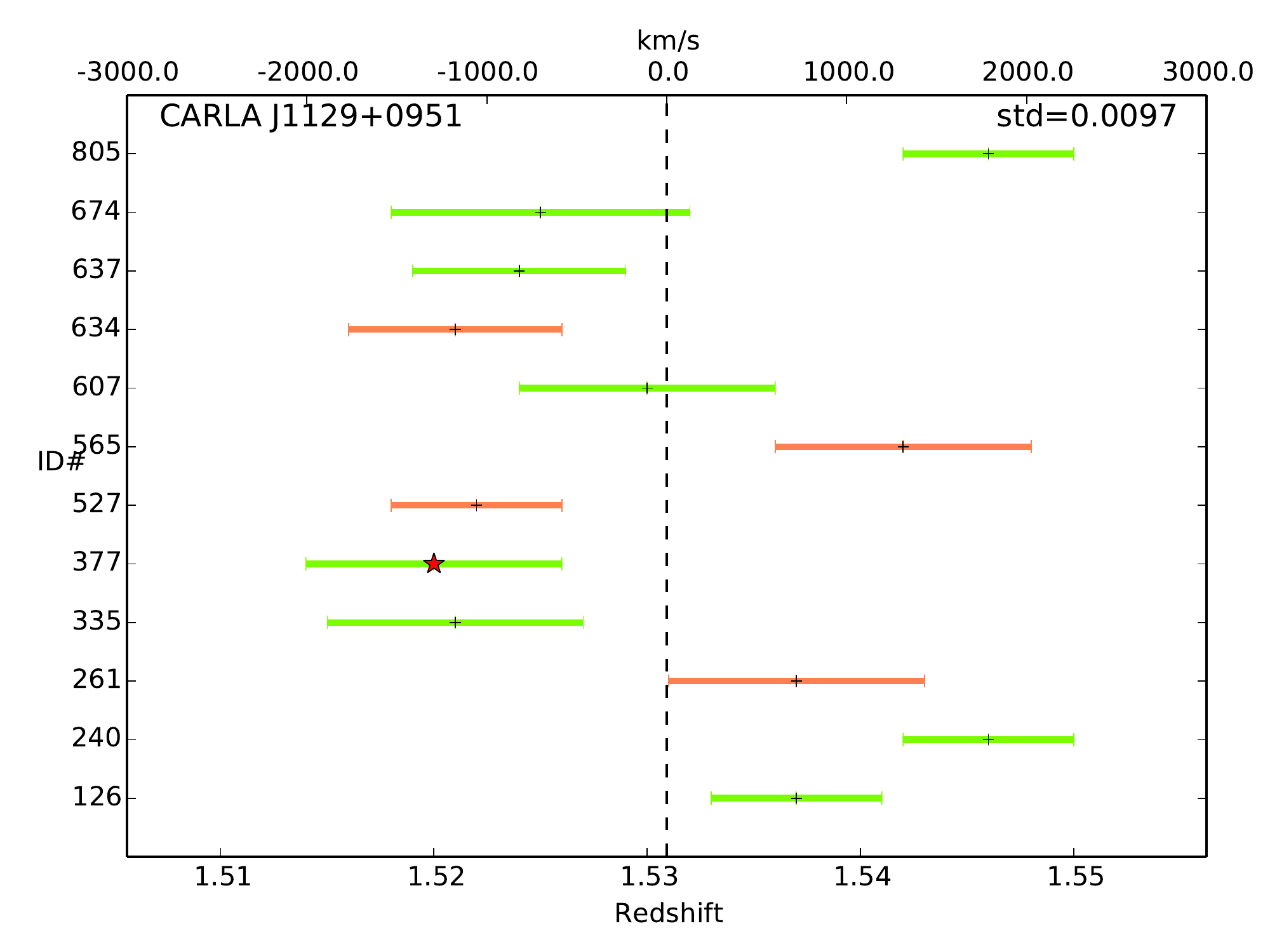} \hfill \includegraphics[scale=0.38]{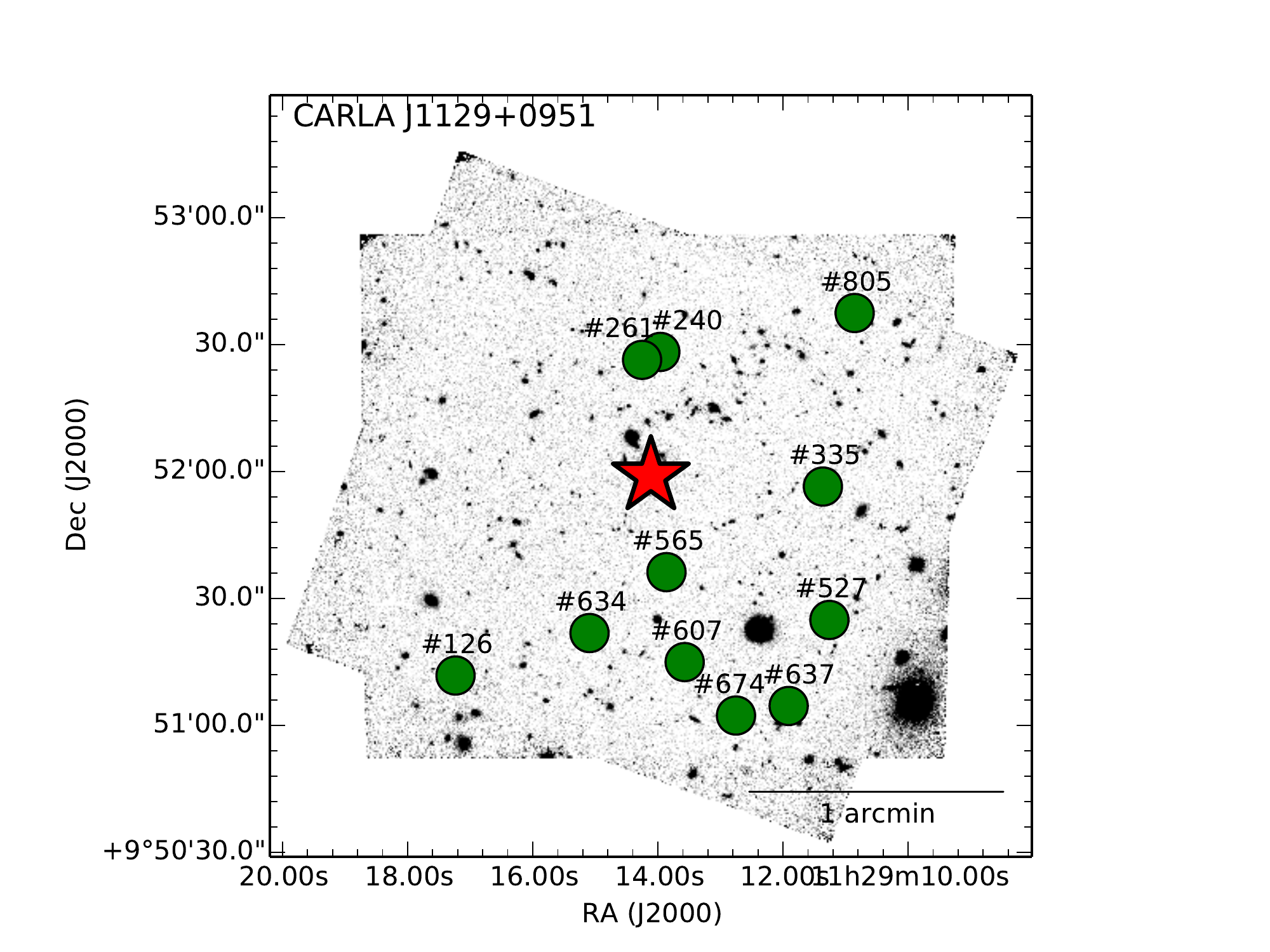} \mbox{(k)}}%
}\\%
{%
\setlength{\fboxsep}{0pt}%
\setlength{\fboxrule}{1pt}%
\fbox{\includegraphics[scale=0.38]{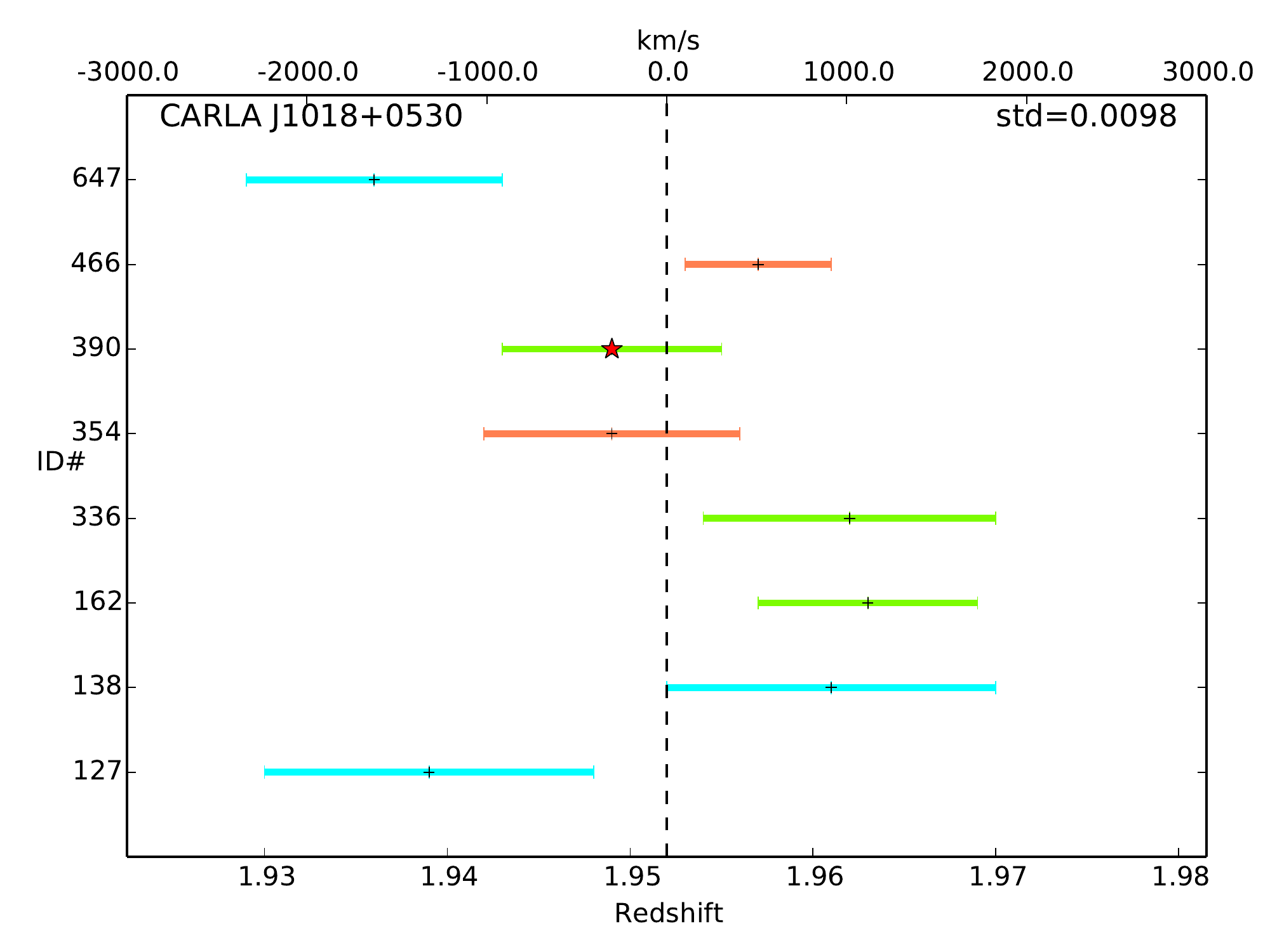} \hfill \includegraphics[scale=0.38]{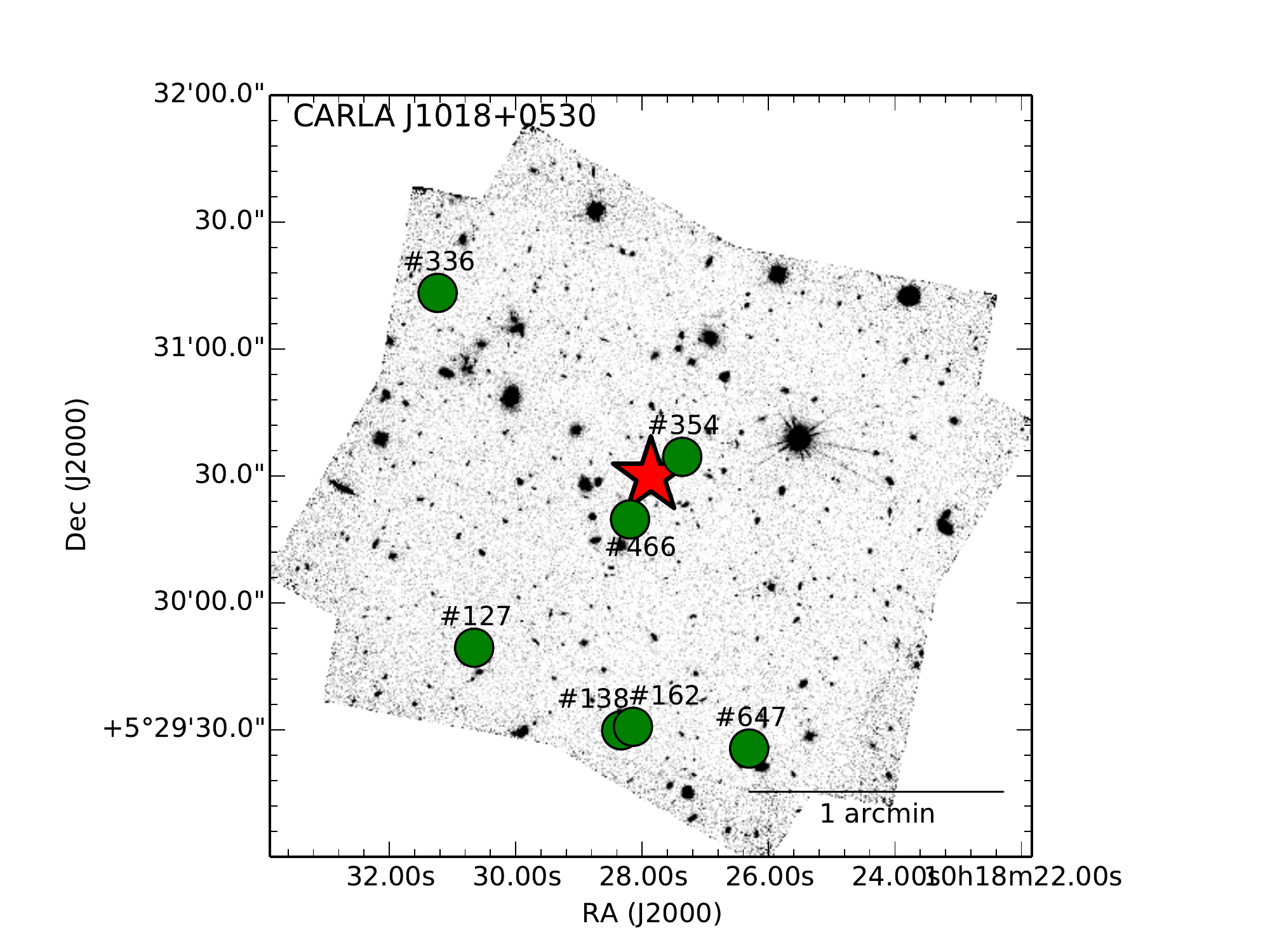} \mbox{(l)}}%
}\\%
\textbf{\mbox{}\\ Figure \ref{fig:velspadist}} --- Continued.
\end{figure*}

\begin{figure*}
{%
\setlength{\fboxsep}{0pt}%
\setlength{\fboxrule}{1pt}%
\fbox{\includegraphics[scale=0.38]{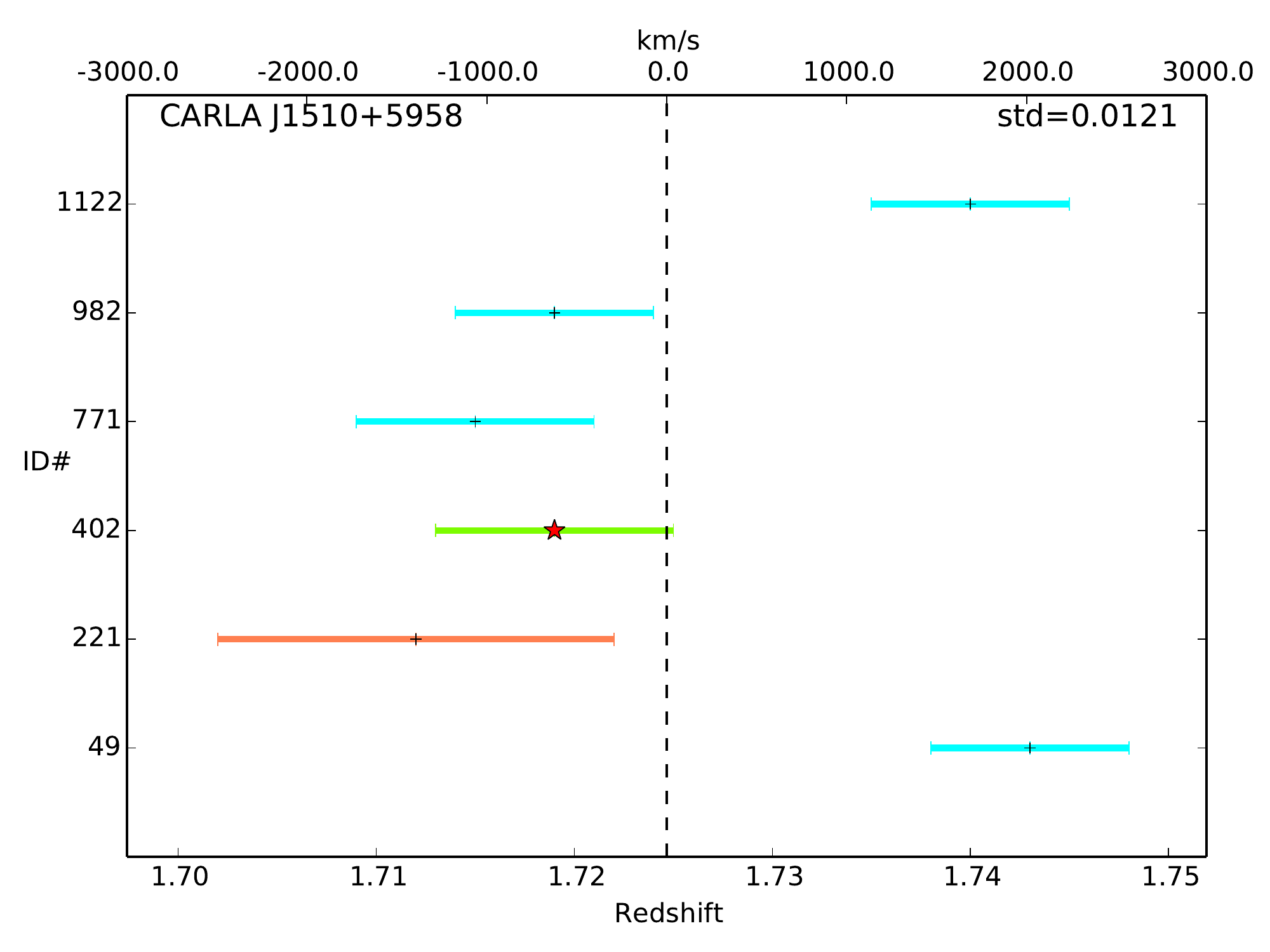} \hfill \includegraphics[scale=0.38]{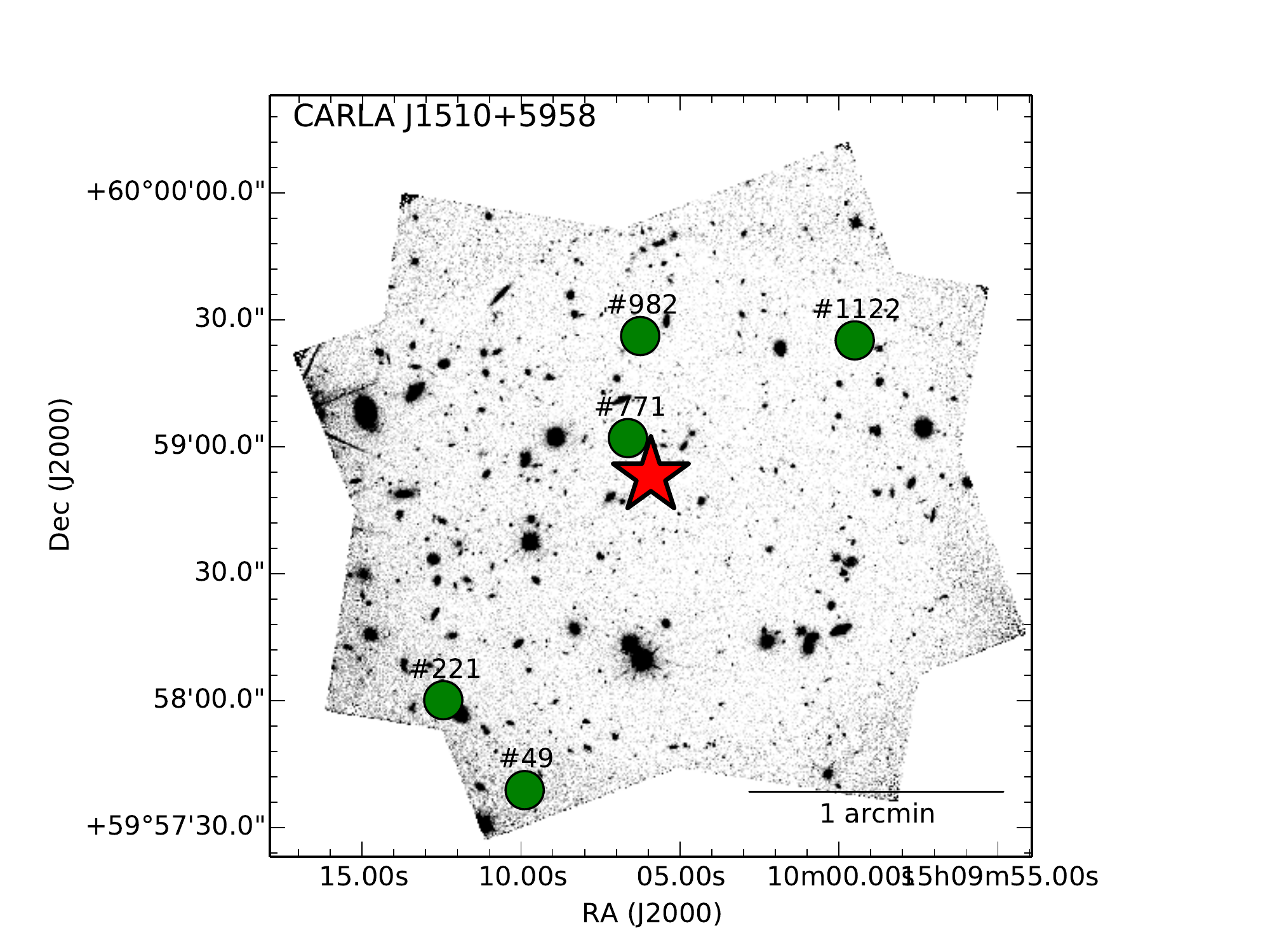} \mbox{(m)}}%
}\\%
{%
\setlength{\fboxsep}{0pt}%
\setlength{\fboxrule}{1pt}%
\fbox{\includegraphics[scale=0.38]{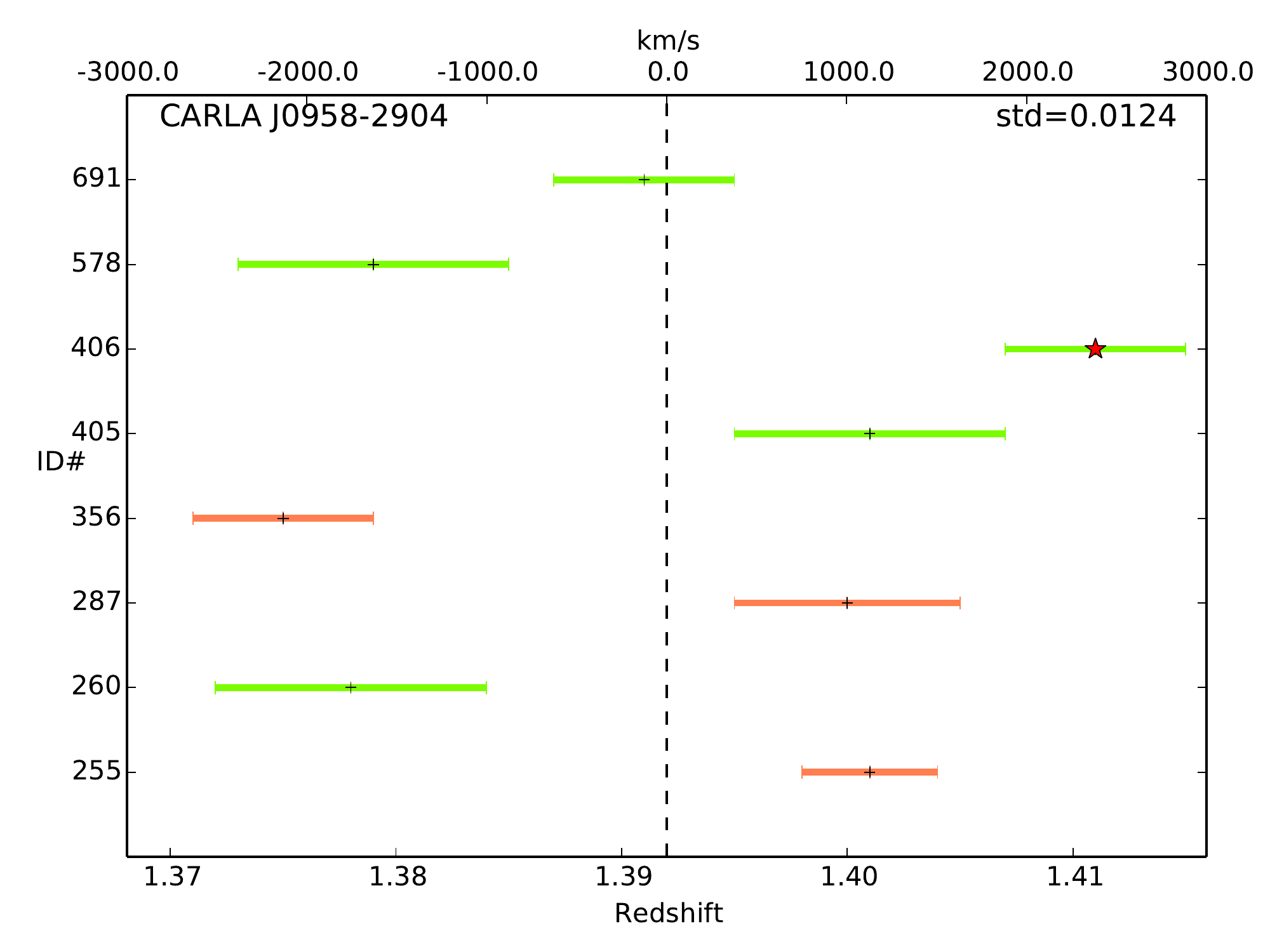} \hfill \includegraphics[scale=0.38]{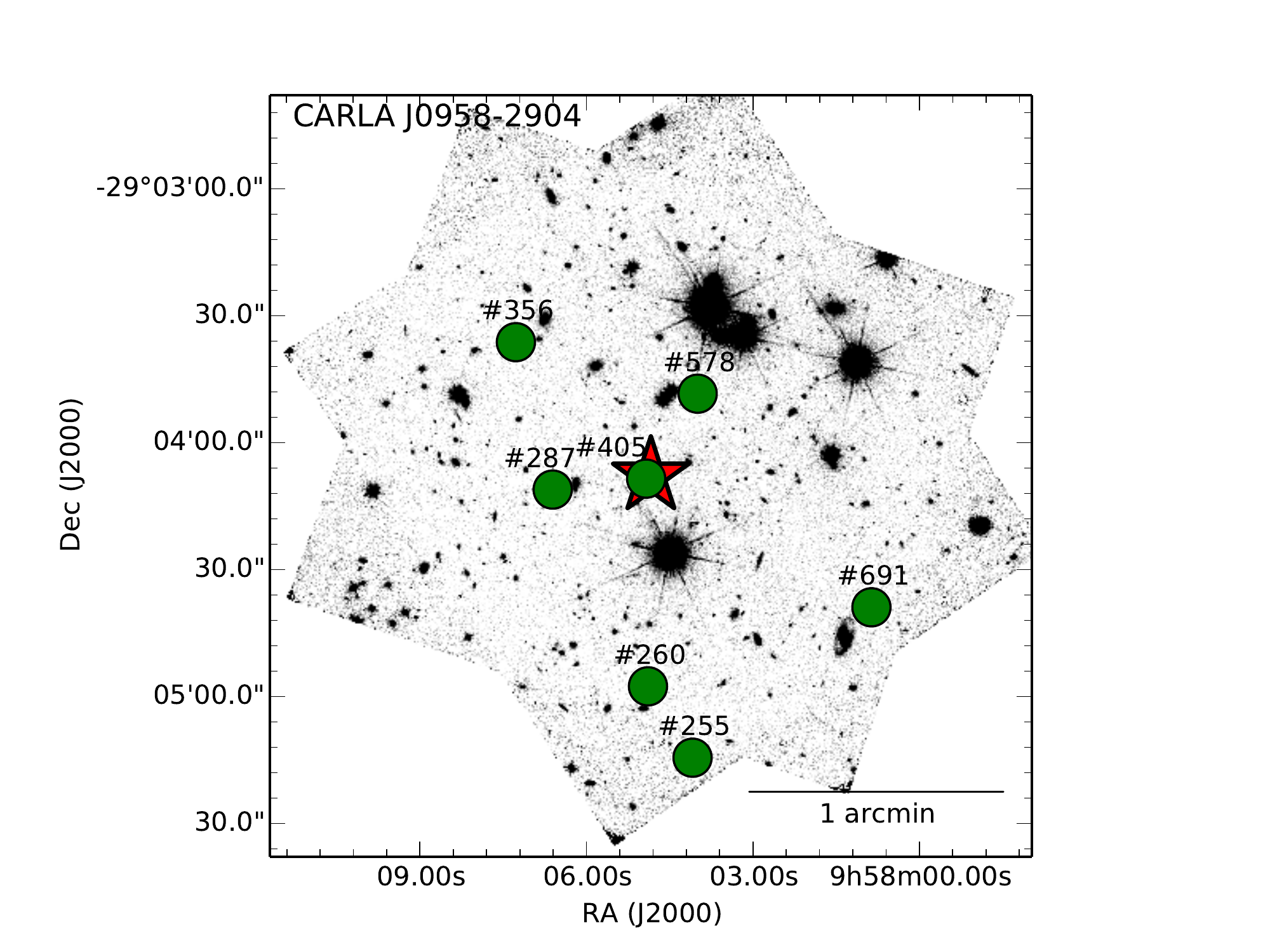} \mbox{(n)}}%
}\\%
{%
\setlength{\fboxsep}{0pt}%
\setlength{\fboxrule}{1pt}%
\fbox{\includegraphics[scale=0.38]{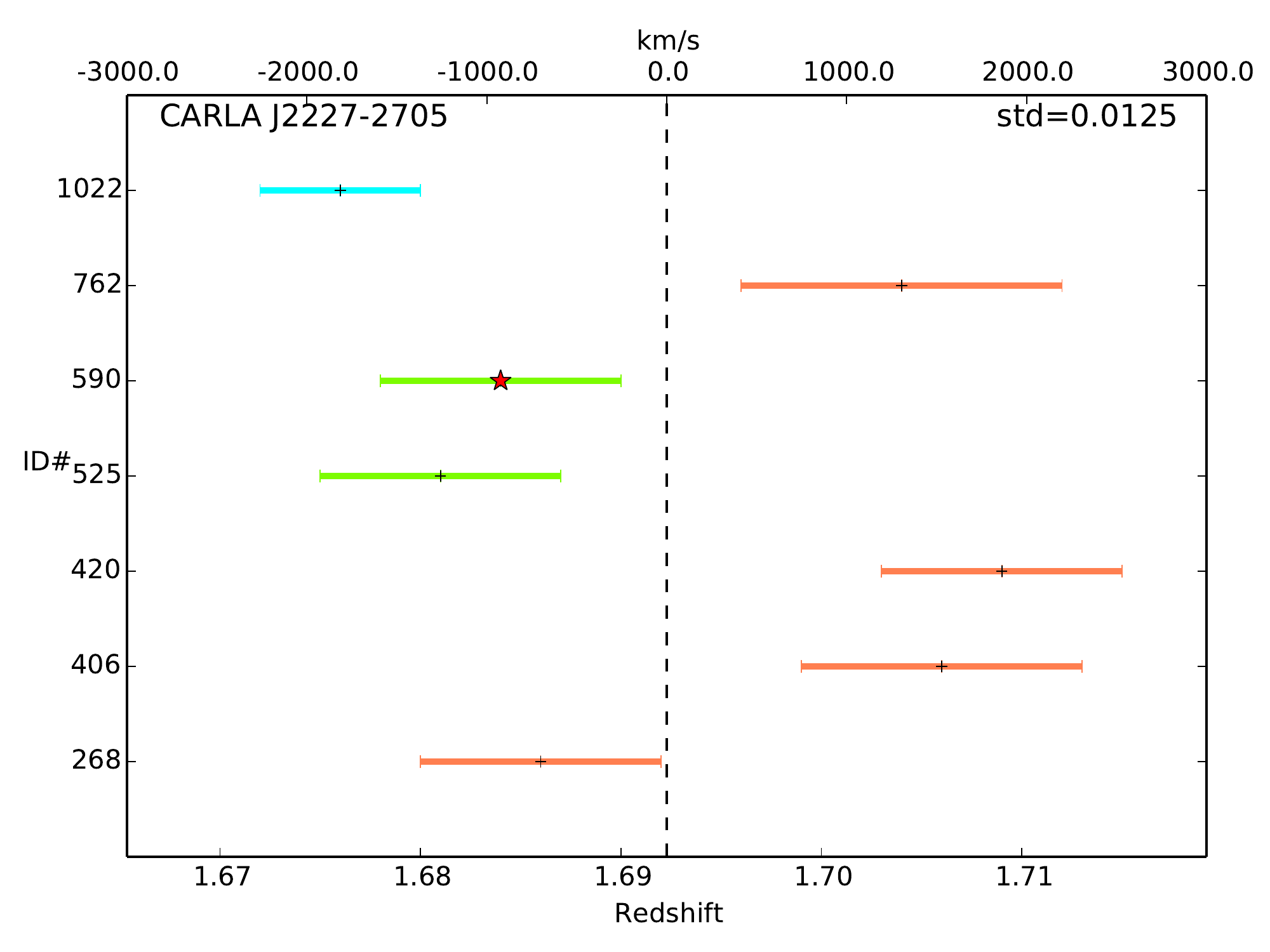} \hfill \includegraphics[scale=0.38]{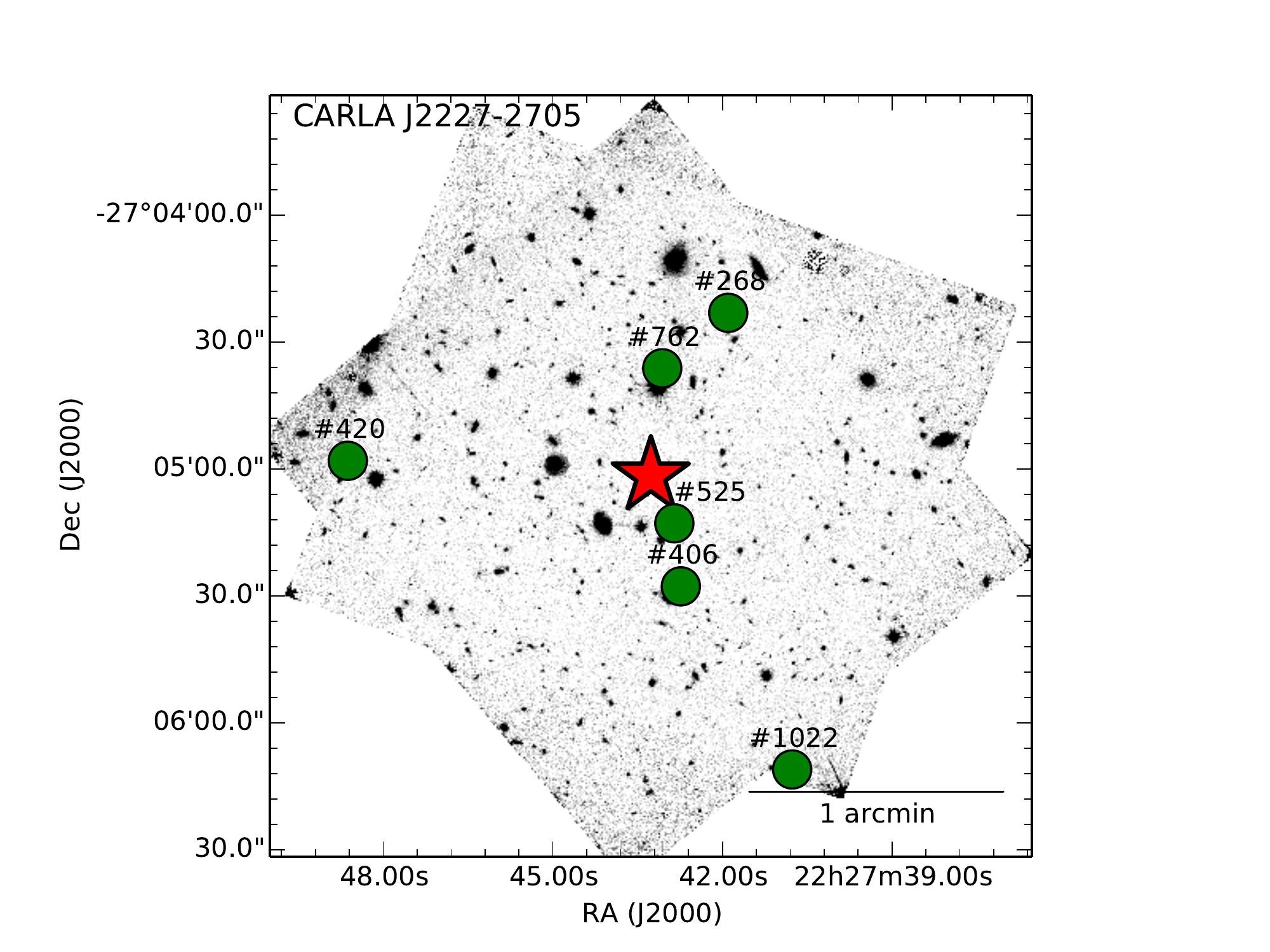} \mbox{(o)}}%
}\\%
{%
\setlength{\fboxsep}{0pt}%
\setlength{\fboxrule}{1pt}%
\fbox{\includegraphics[scale=0.38]{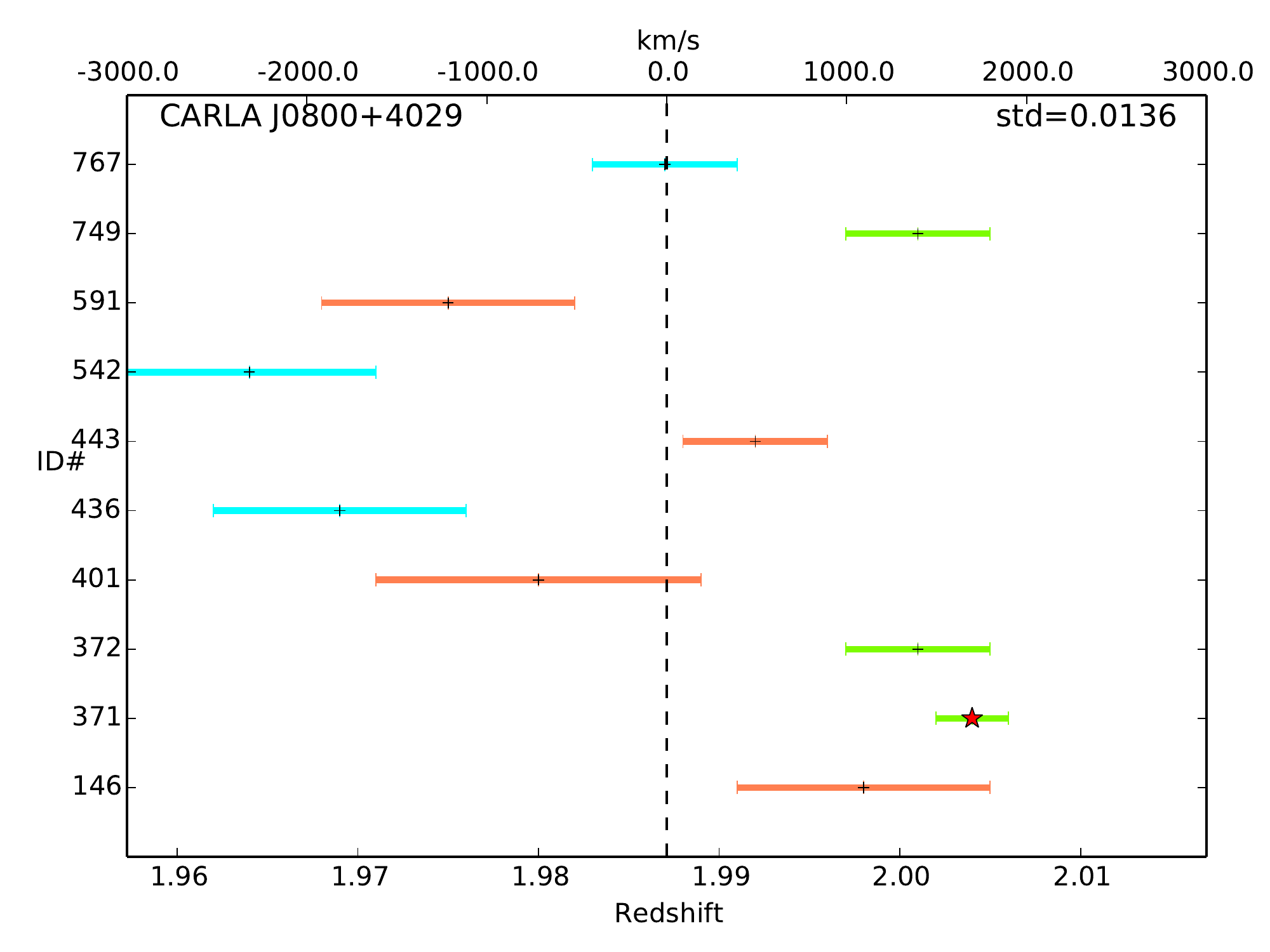} \hfill \includegraphics[scale=0.38]{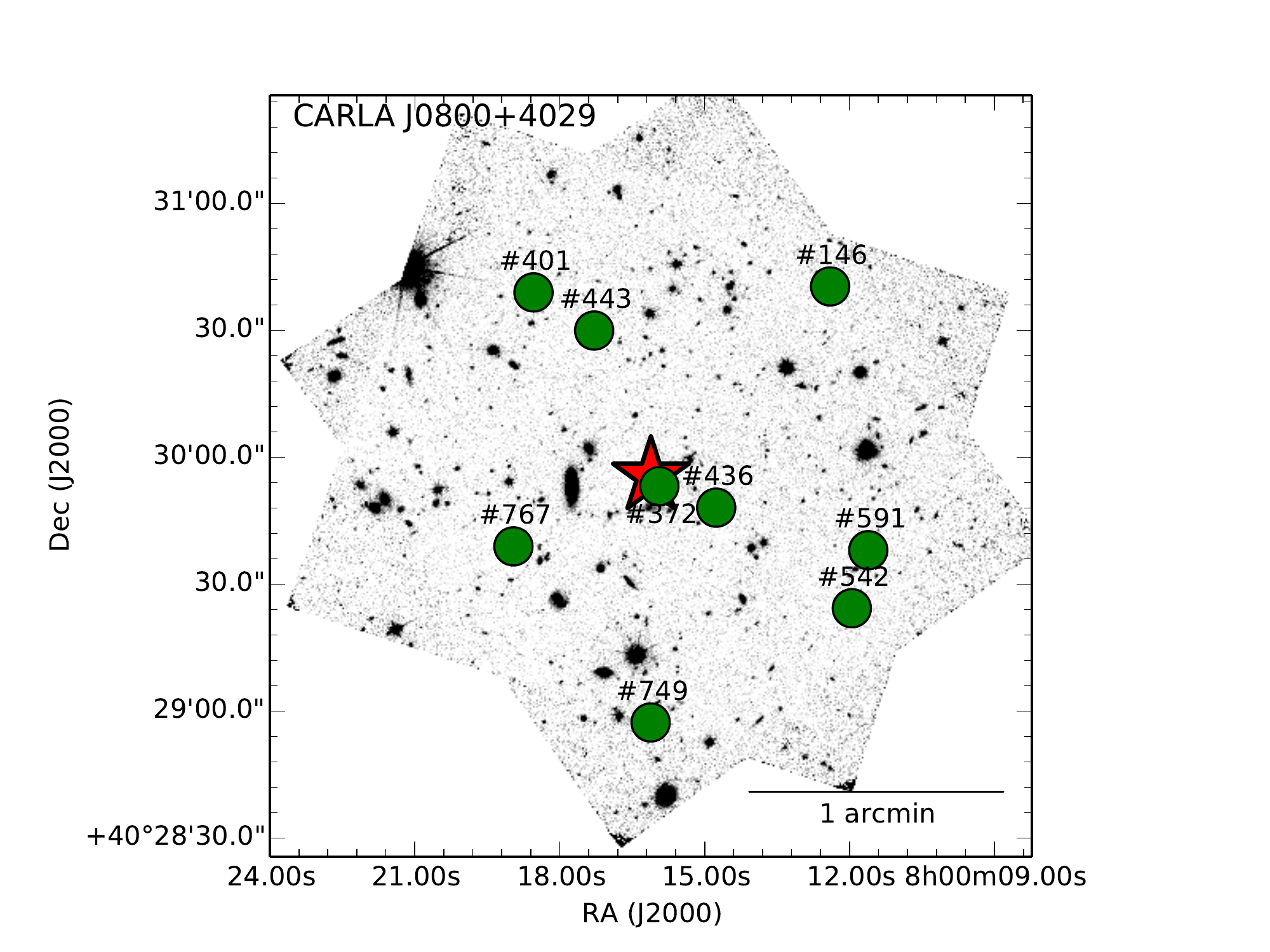} \mbox{(p)}}%
}\\%
\textbf{\mbox{}\\ Figure \ref{fig:velspadist}} --- Continued.
\end{figure*}

\section{Member Spectra}\label{app:spectra}
\begin{figure*}[!ht]
{%
\setlength{\fboxsep}{0pt}%
\setlength{\fboxrule}{1pt}%
\fbox{\includegraphics[page=2, scale=0.24]{CARLA_J0116-2052_45.pdf} \hfill \includegraphics[page=1, scale=0.20]{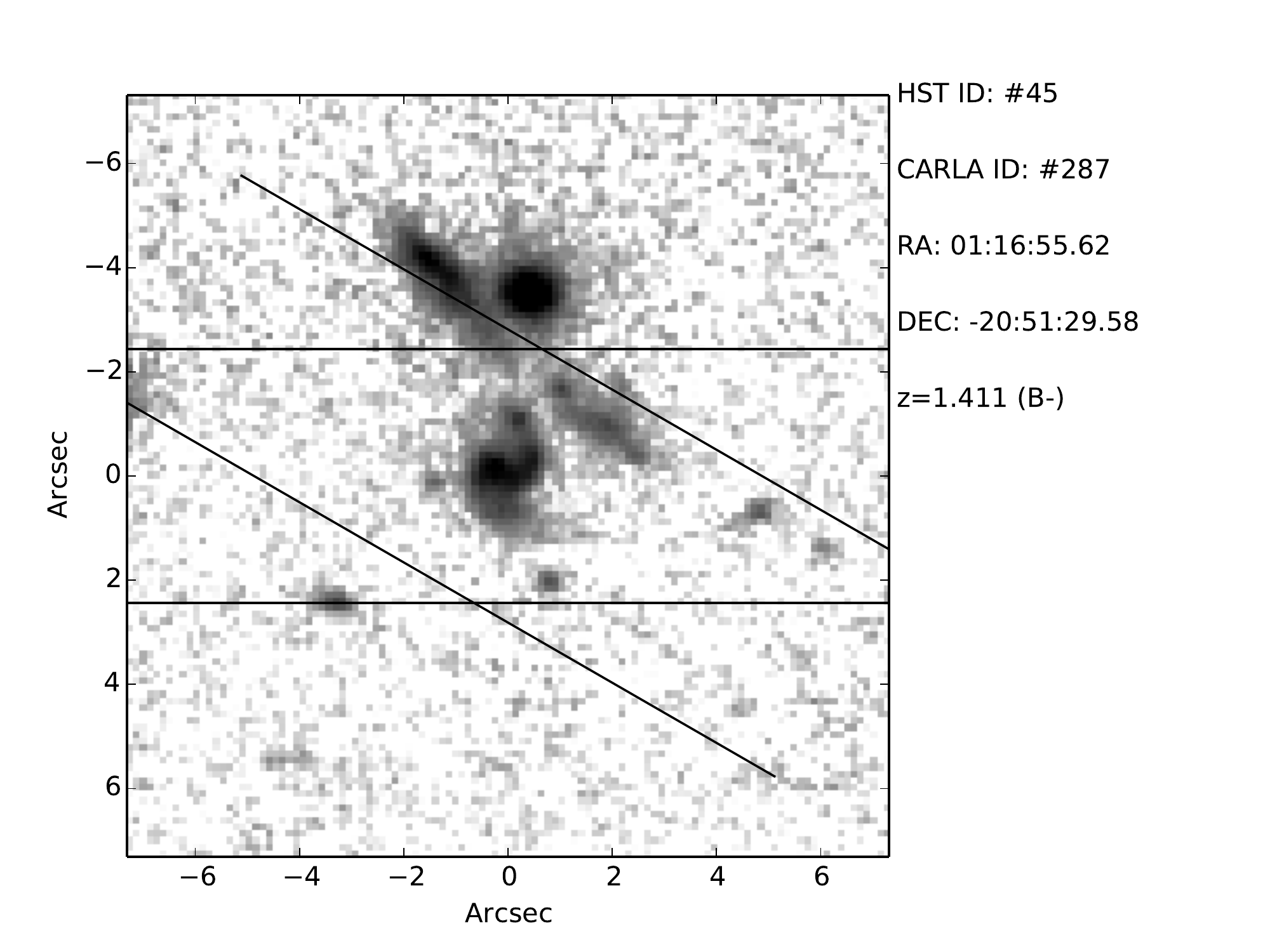} \mbox{(a)}}%
}%
{%
\setlength{\fboxsep}{0pt}%
\setlength{\fboxrule}{1pt}%
\fbox{\includegraphics[page=2, scale=0.24]{CARLA_J0116-2052_181.pdf} \hfill \includegraphics[page=1, scale=0.20]{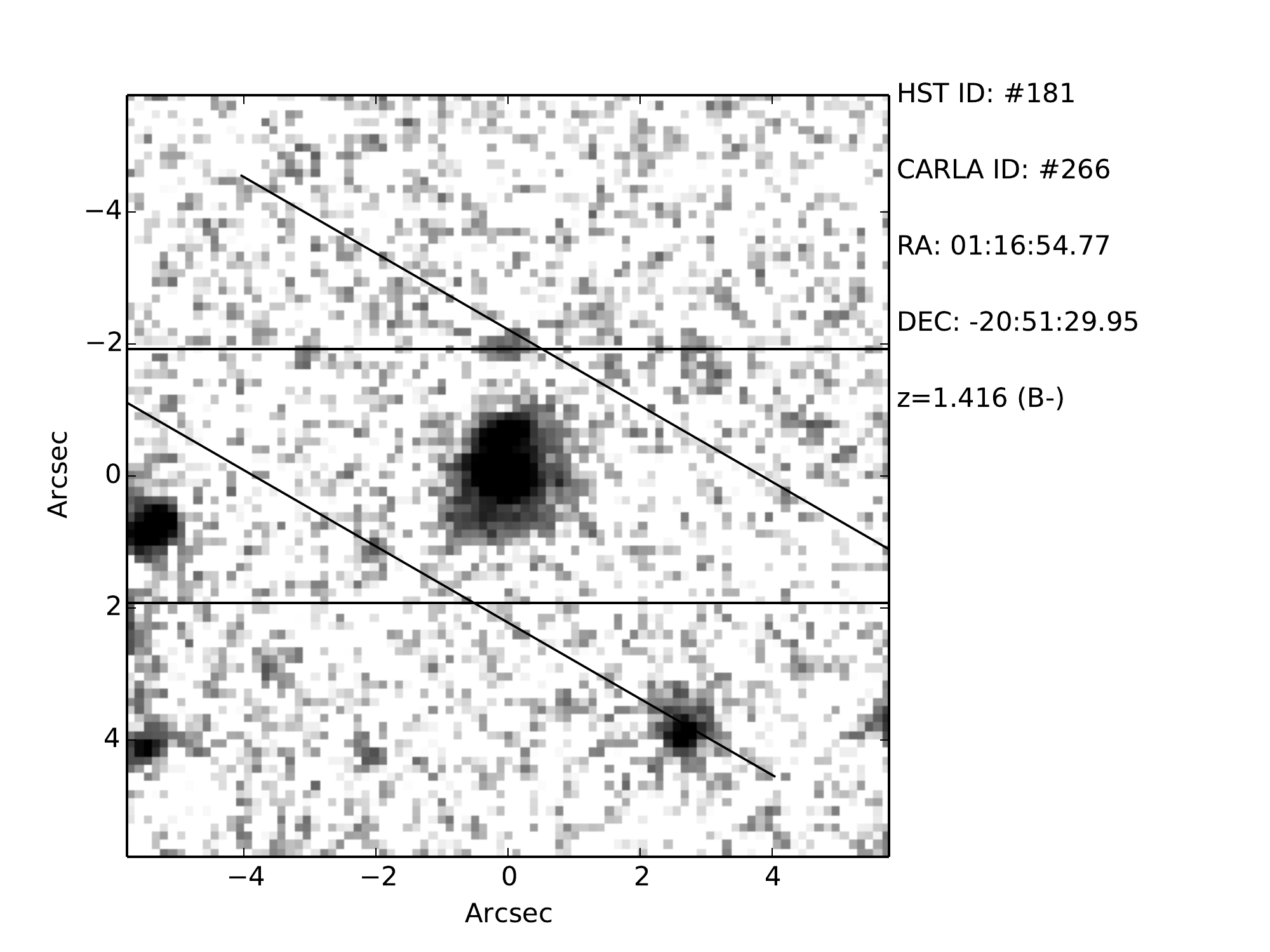} \mbox{(b)}}%
}\\%
{%
\setlength{\fboxsep}{0pt}%
\setlength{\fboxrule}{1pt}%
\fbox{\includegraphics[page=2, scale=0.24]{CARLA_J0116-2052_204.pdf} \hfill \includegraphics[page=1, scale=0.20]{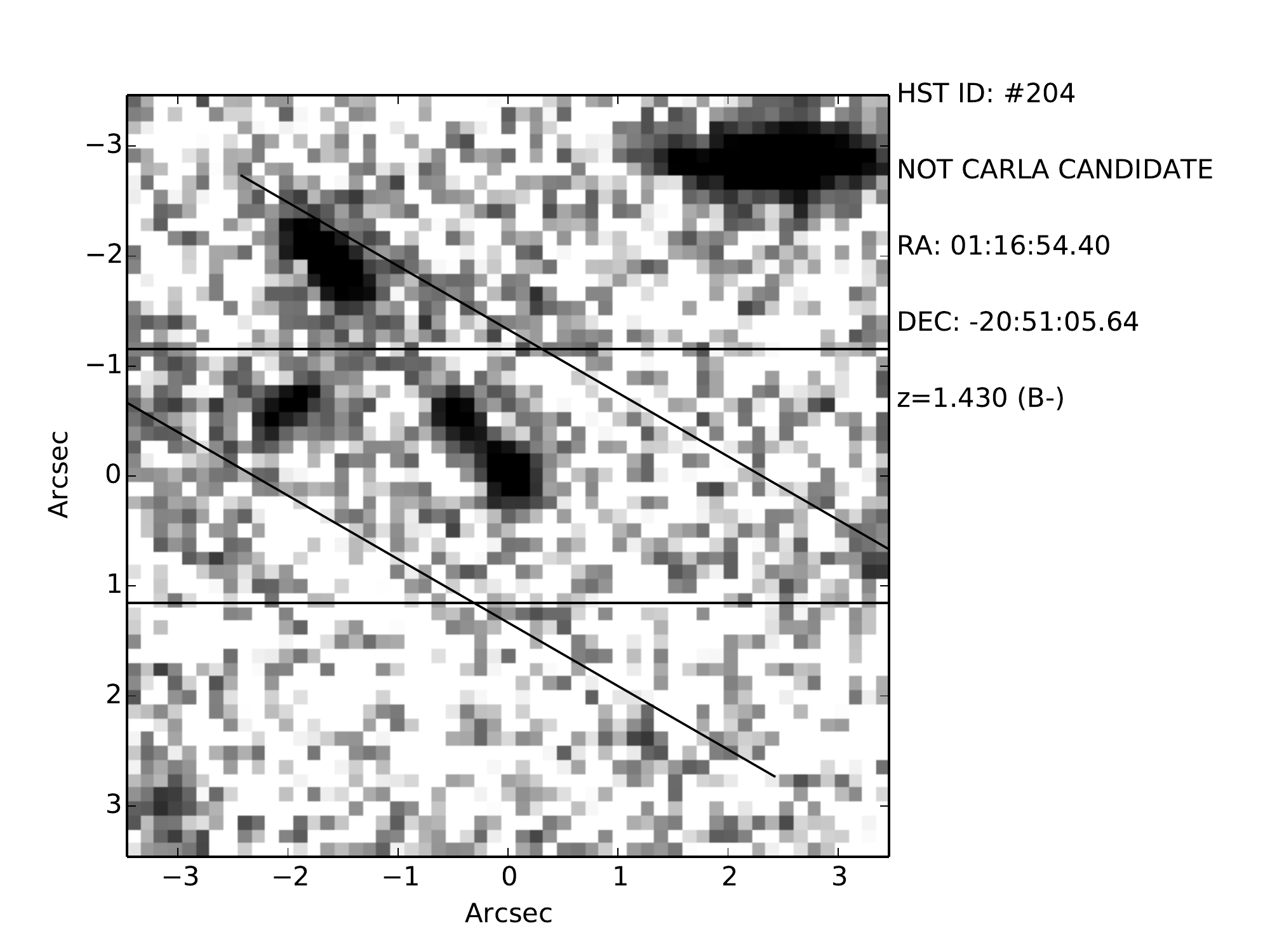} \mbox{(c)}}%
}%
{%
\setlength{\fboxsep}{0pt}%
\setlength{\fboxrule}{1pt}%
\fbox{\includegraphics[page=2, scale=0.24]{CARLA_J0116-2052_289.pdf} \hfill \includegraphics[page=1, scale=0.20]{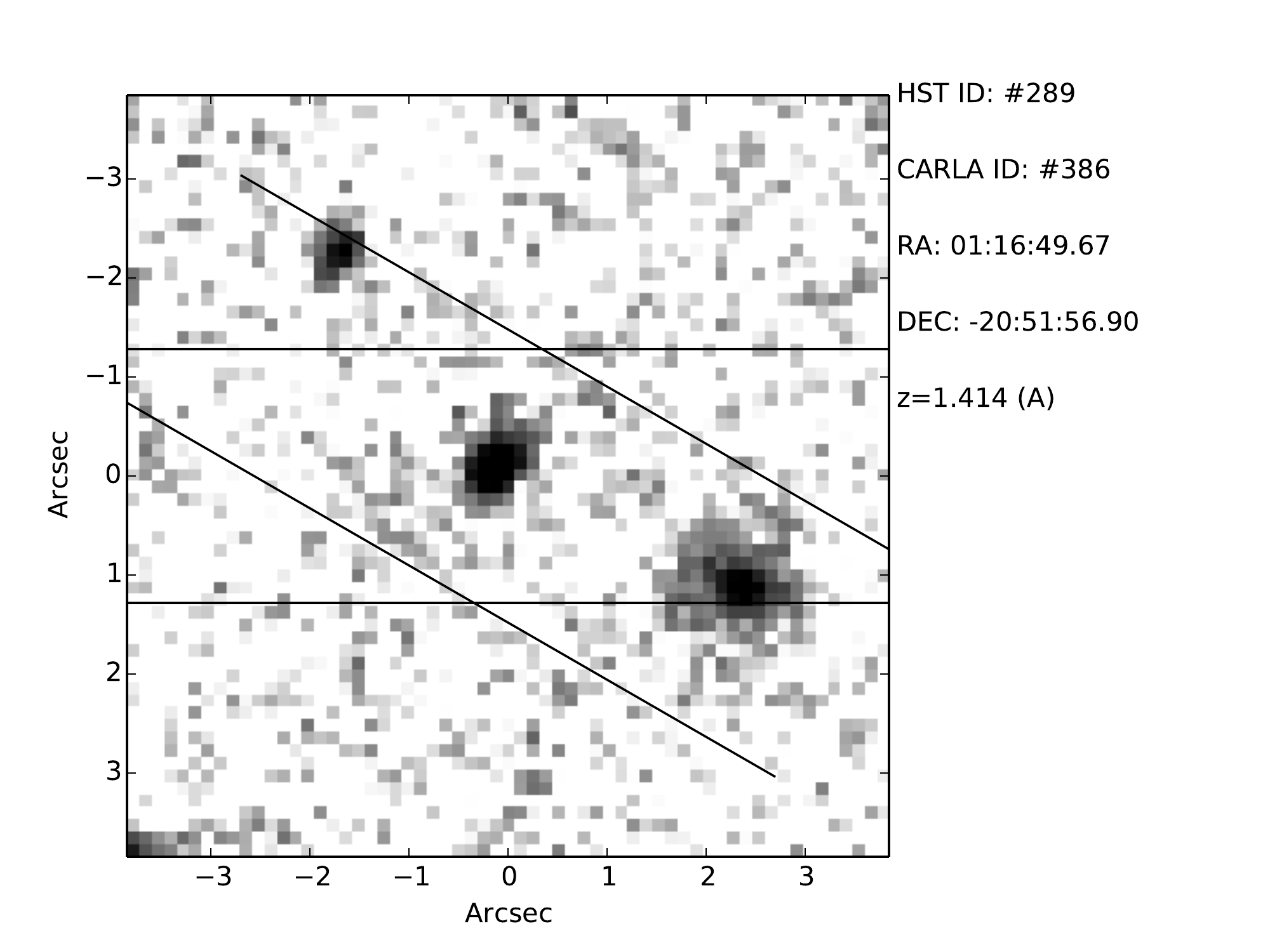} \mbox{(d)}}%
}\\%
{%
\setlength{\fboxsep}{0pt}%
\setlength{\fboxrule}{1pt}%
\fbox{\includegraphics[page=2, scale=0.24]{CARLA_J0116-2052_380.pdf} \hfill \includegraphics[page=1, scale=0.20]{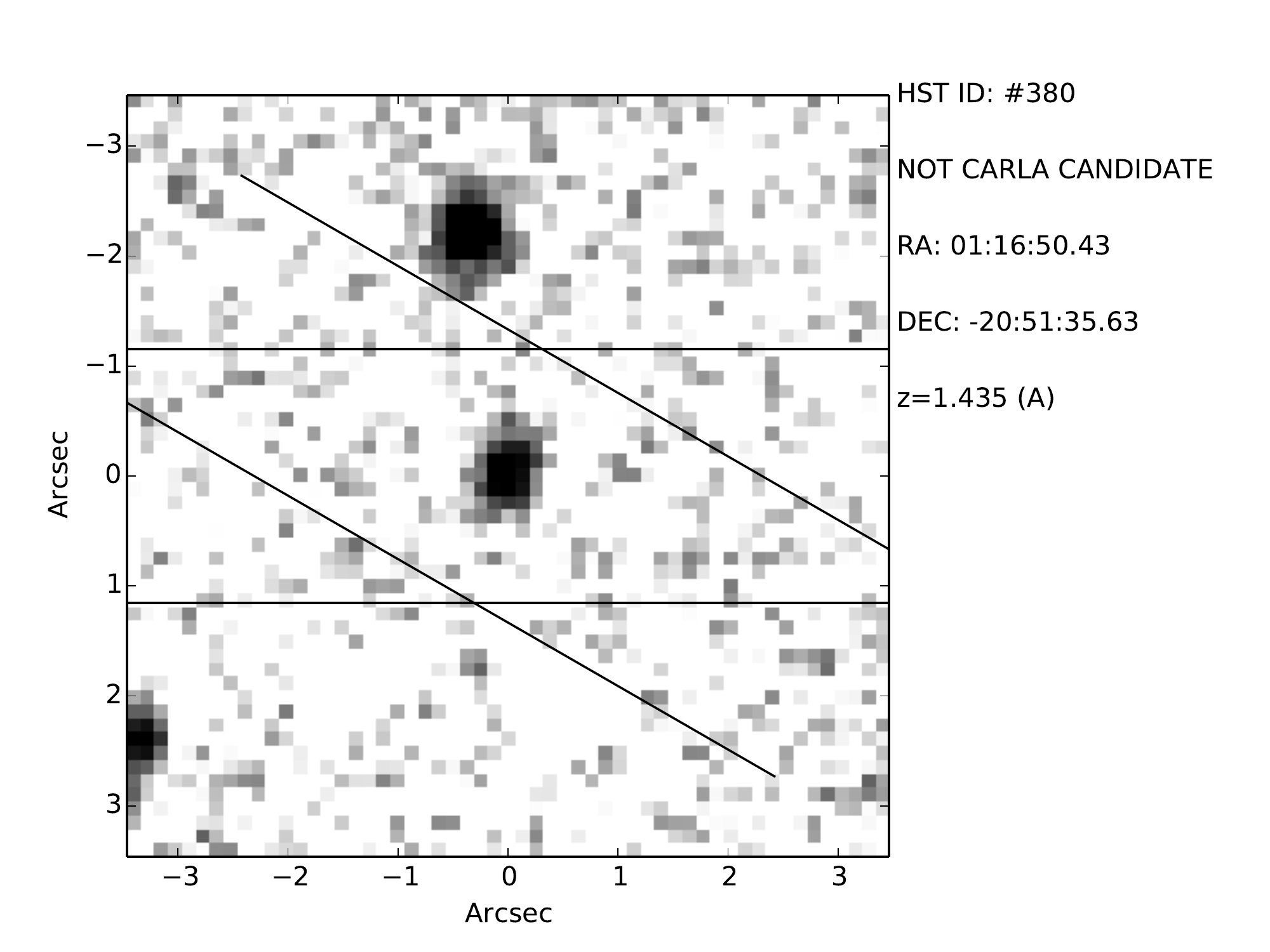} \mbox{(e)}}%
}%
{%
\setlength{\fboxsep}{0pt}%
\setlength{\fboxrule}{1pt}%
\fbox{\includegraphics[page=2, scale=0.24]{CARLA_J0116-2052_398.pdf} \hfill \includegraphics[page=1, scale=0.20]{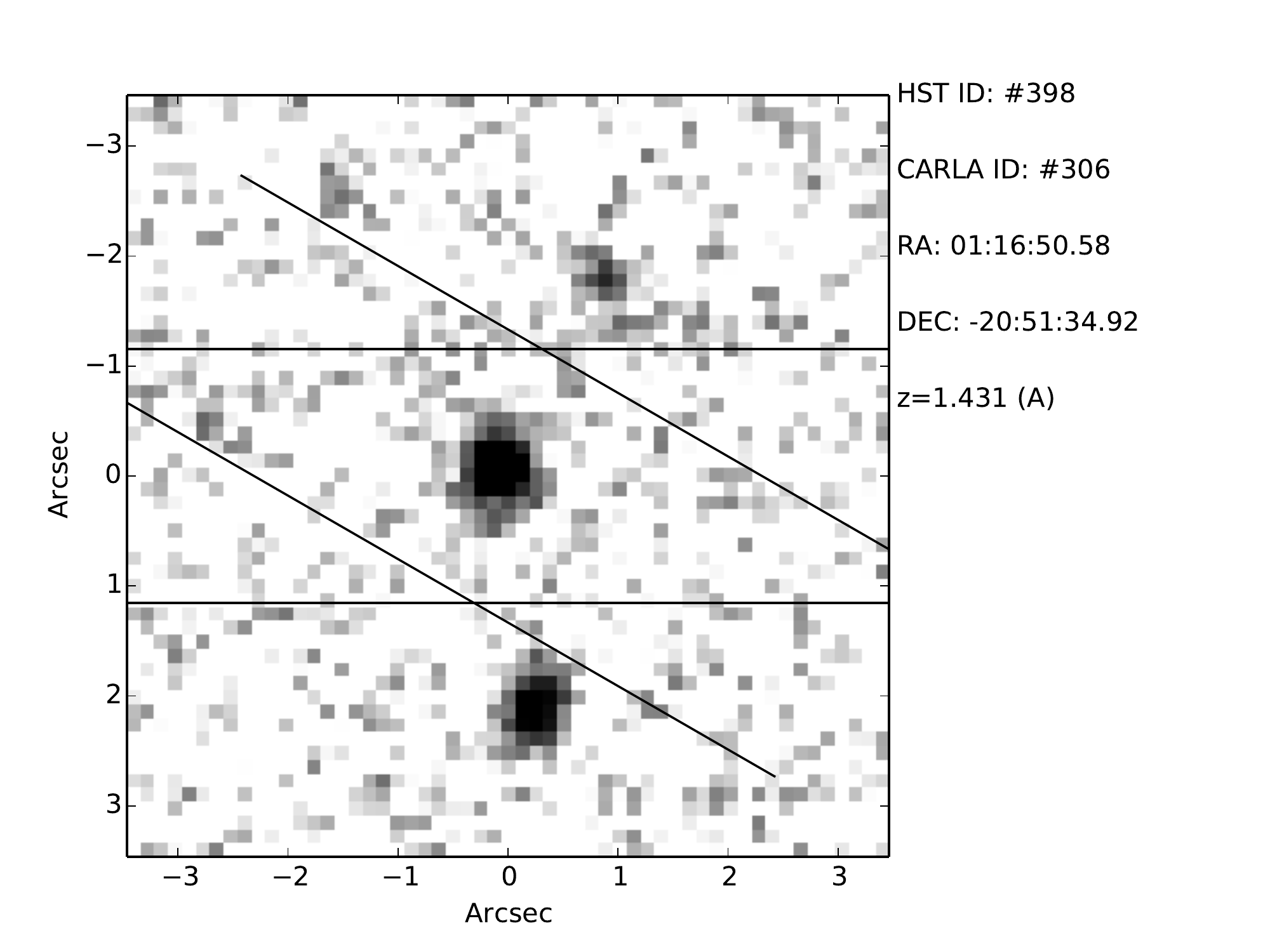} \mbox{(f)}}%
}\\%
{%
\setlength{\fboxsep}{0pt}%
\setlength{\fboxrule}{1pt}%
\fbox{\includegraphics[page=2, scale=0.24]{CARLA_J0116-2052_437.pdf} \hfill \includegraphics[page=1, scale=0.20]{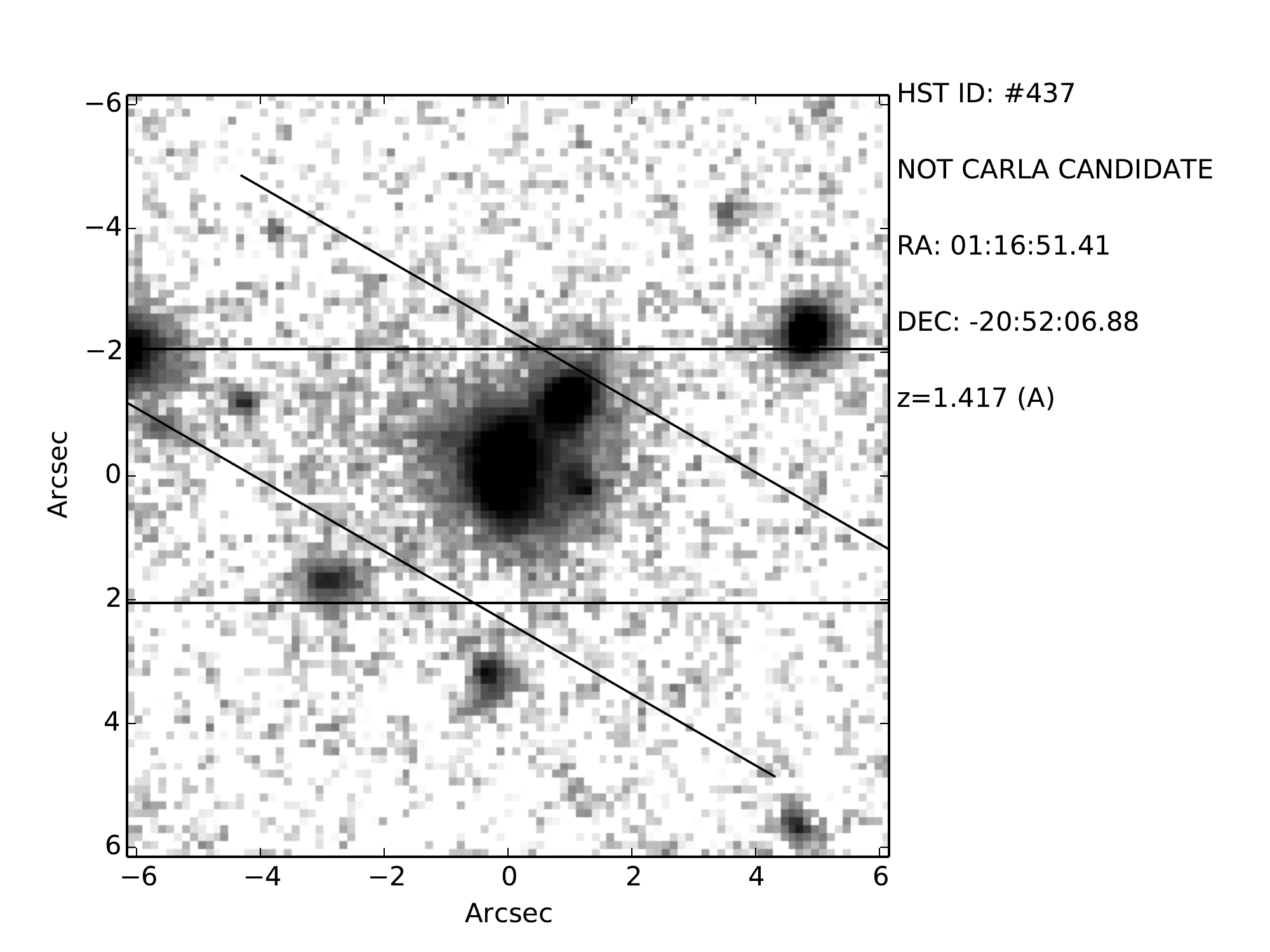} \mbox{(g)}}%
}%
{%
\setlength{\fboxsep}{0pt}%
\setlength{\fboxrule}{1pt}%
\fbox{\includegraphics[page=2, scale=0.24]{CARLA_J0116-2052_618.pdf} \hfill \includegraphics[page=1, scale=0.20]{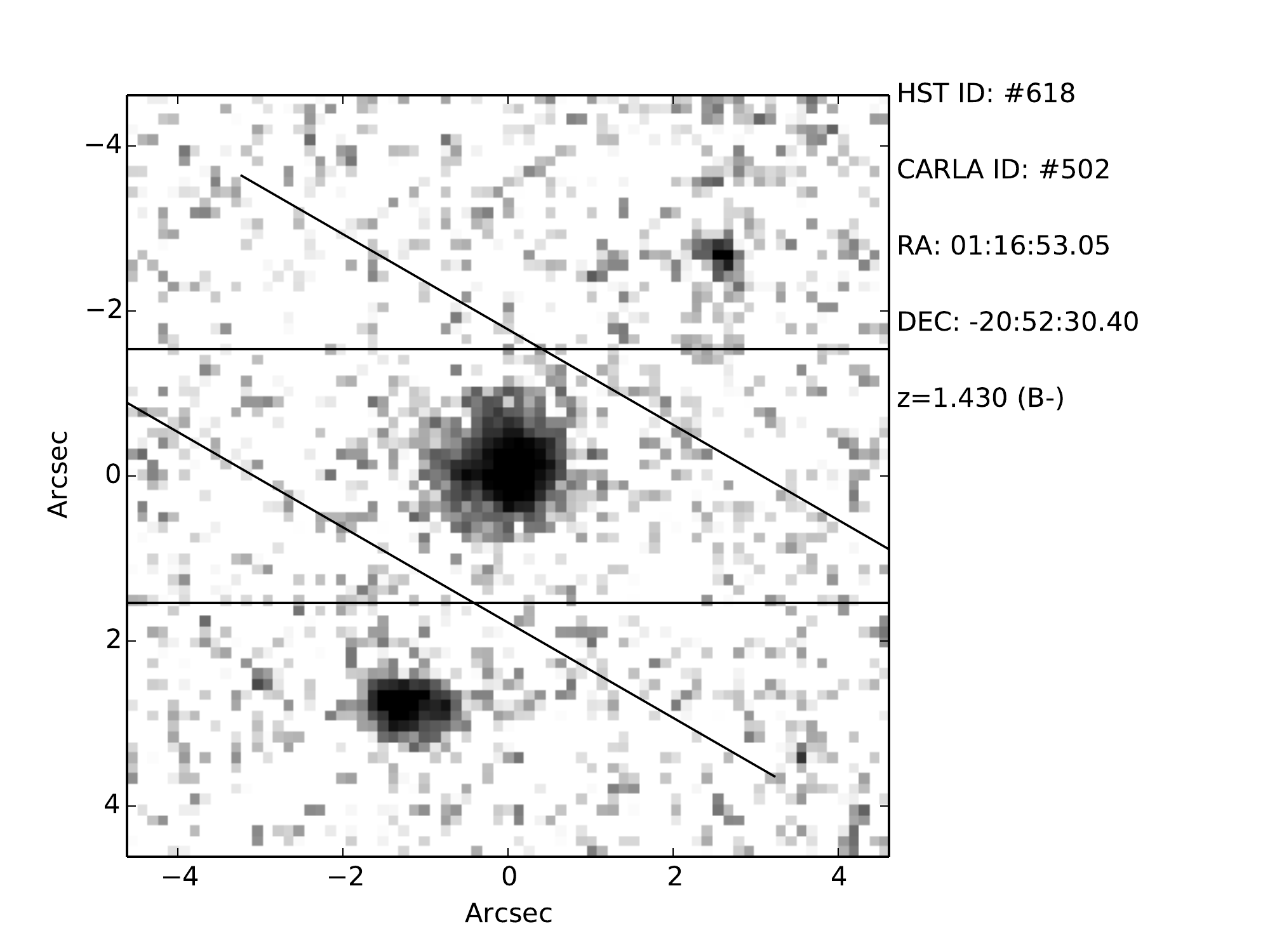} \mbox{(h)}}%
}\\%
{%
\setlength{\fboxsep}{0pt}%
\setlength{\fboxrule}{1pt}%
\fbox{\includegraphics[page=2, scale=0.24]{CARLA_J0116-2052_705.pdf} \hfill \includegraphics[page=1, scale=0.20]{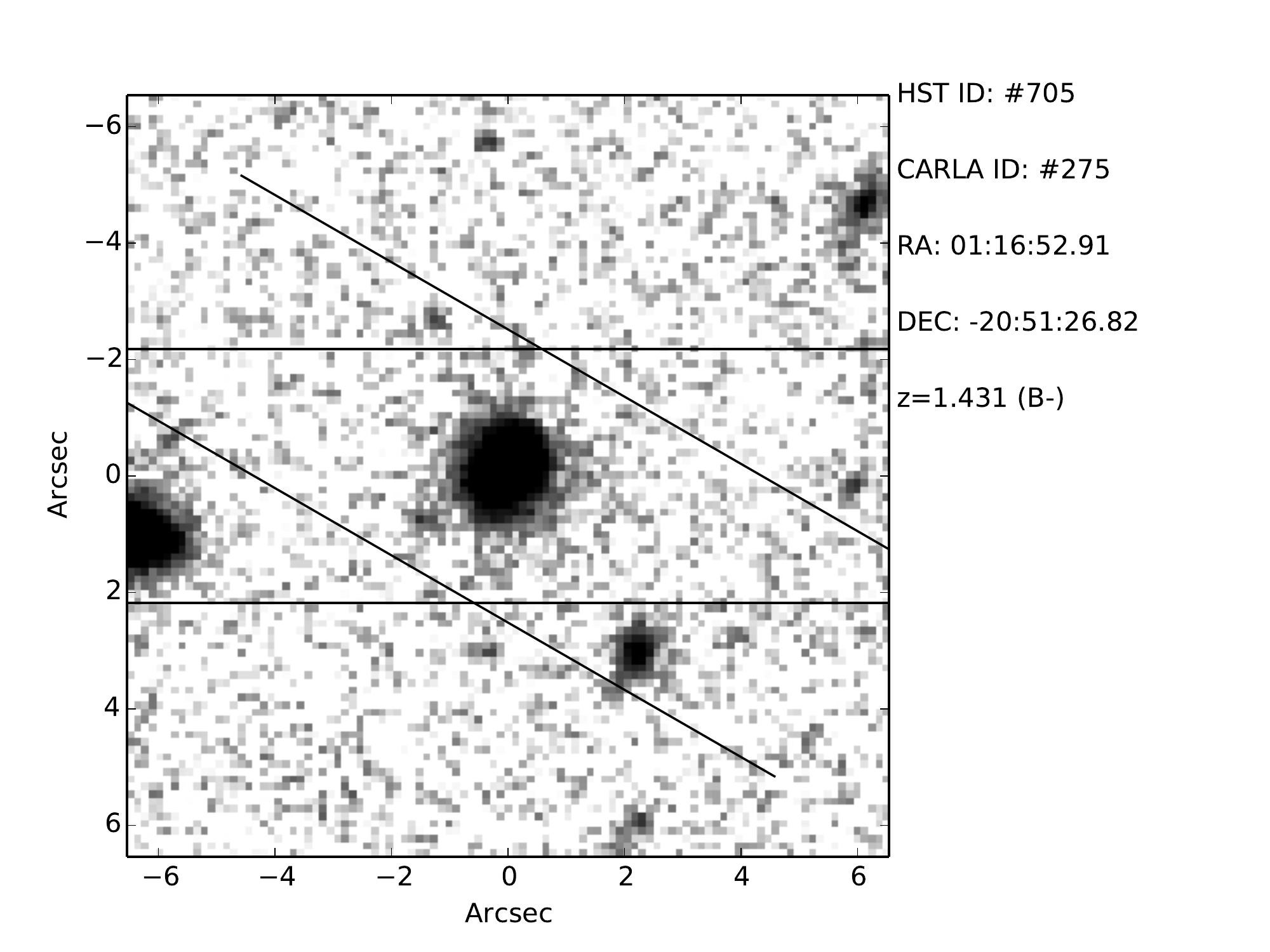} \mbox{(i)}}%
}%
{%
\setlength{\fboxsep}{0pt}%
\setlength{\fboxrule}{1pt}%
\fbox{\includegraphics[page=2, scale=0.24]{CARLA_J0116-2052_865.pdf} \hfill \includegraphics[page=1, scale=0.20]{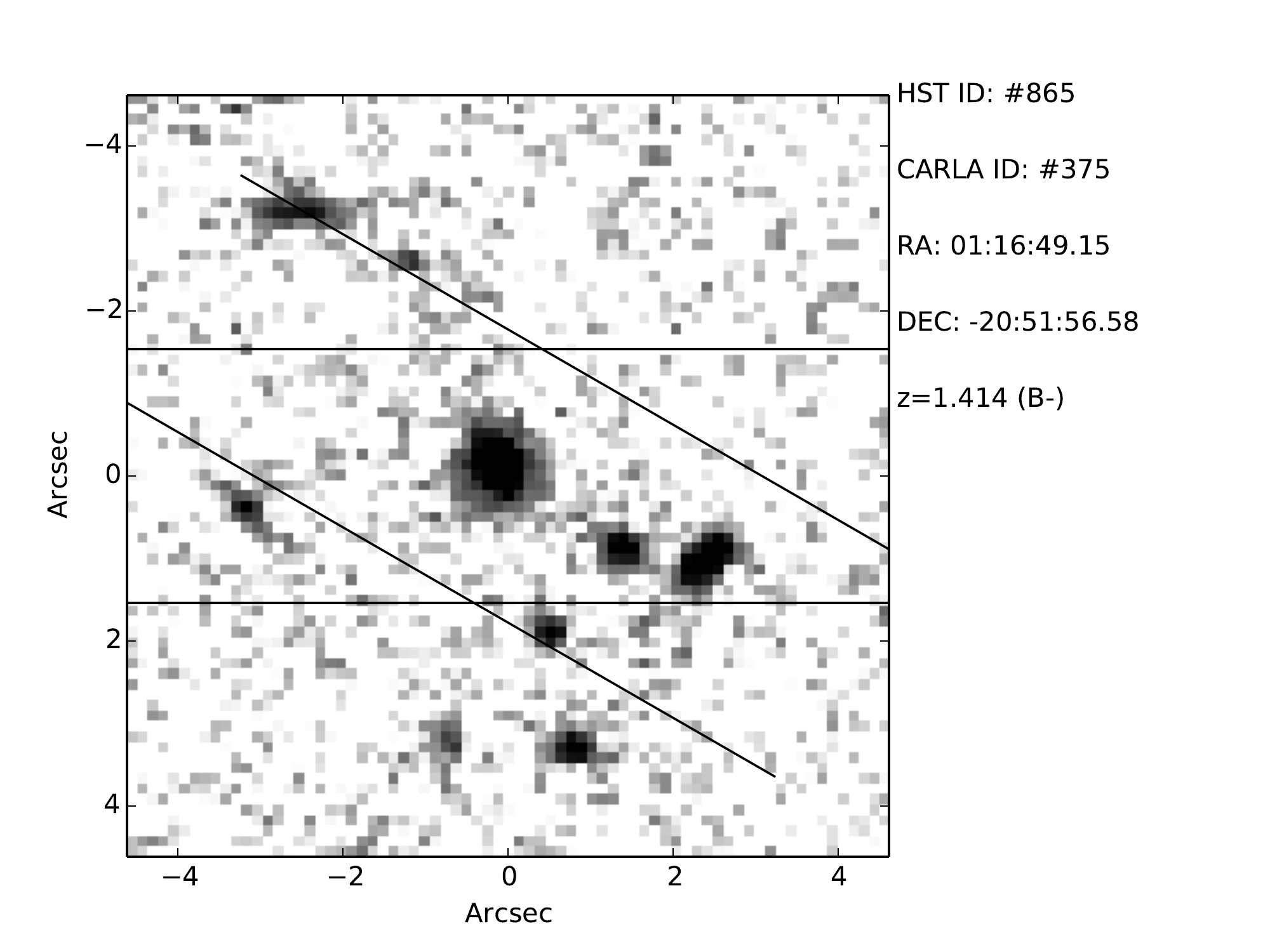} \mbox{(j)}}%
}\\%
{%
\setlength{\fboxsep}{0pt}%
\setlength{\fboxrule}{1pt}%
\fbox{\includegraphics[page=2, scale=0.24]{CARLA_J0116-2052_922.pdf} \hfill \includegraphics[page=1, scale=0.20]{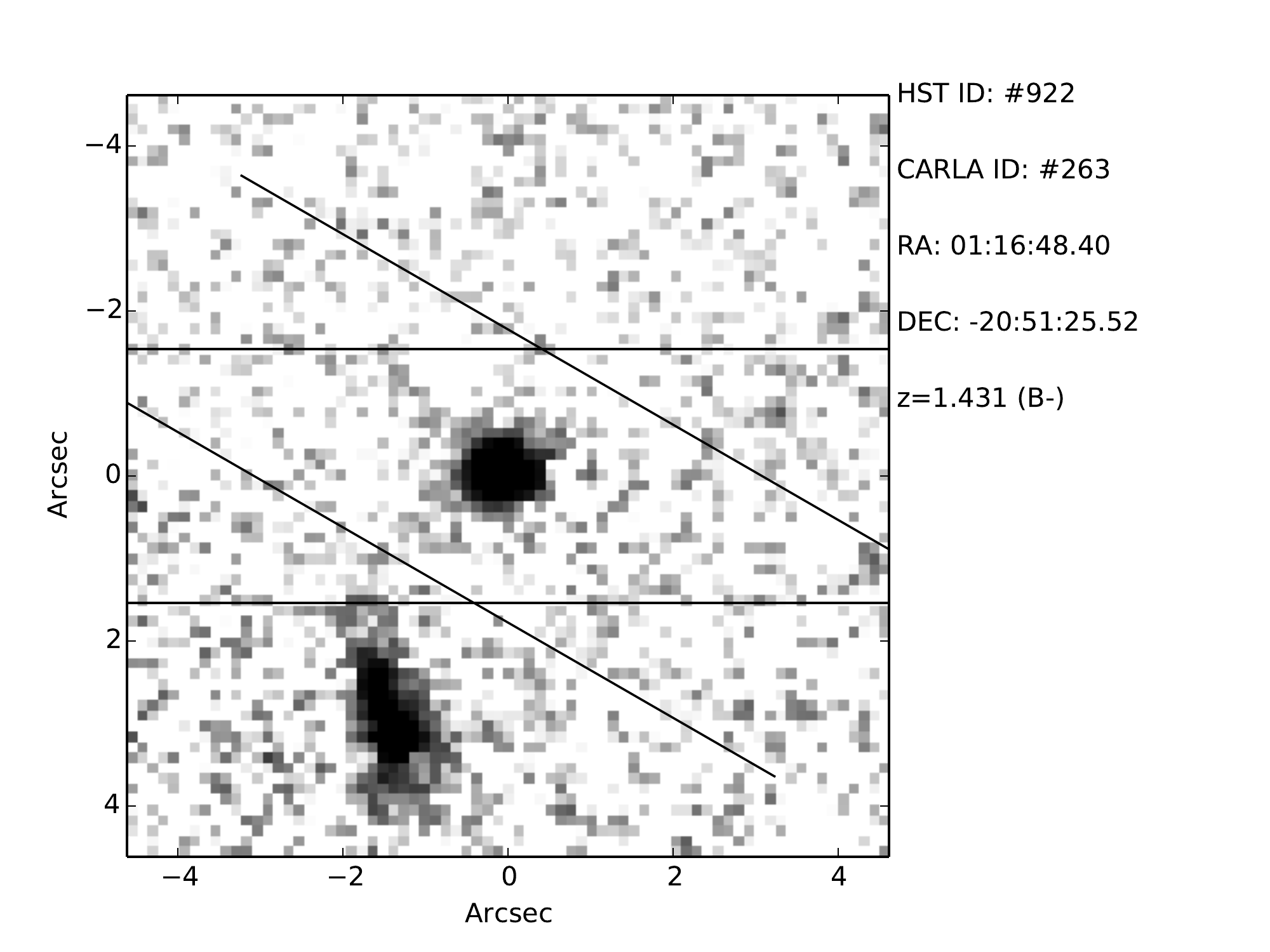} \mbox{(k)}}%
}%
{%
\setlength{\fboxsep}{0pt}%
\setlength{\fboxrule}{1pt}%
\fbox{\includegraphics[page=2, scale=0.24]{CARLA_J0116-2052_974.pdf} \hfill \includegraphics[page=1, scale=0.20]{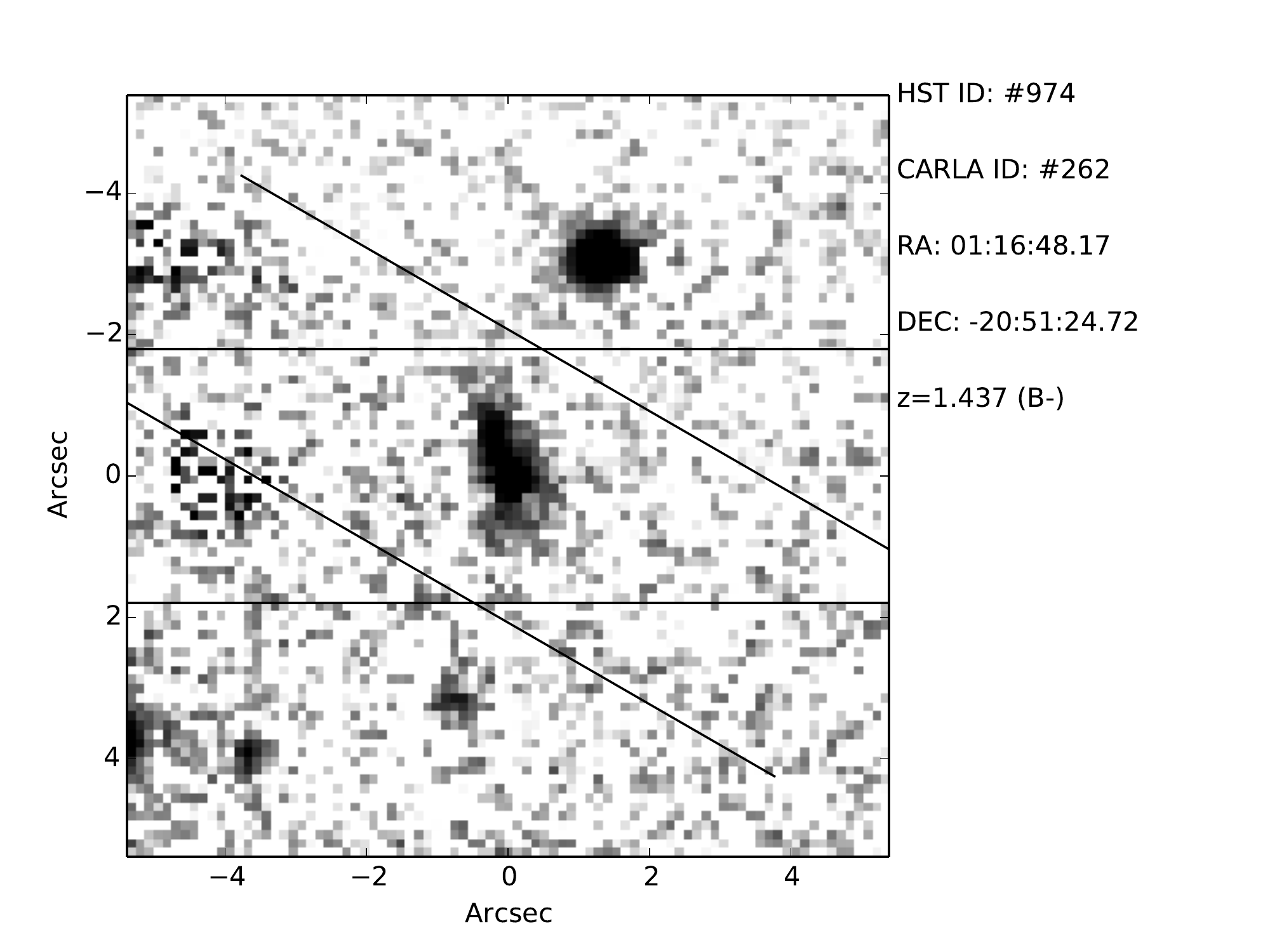} \mbox{(l)}}%
}\\%
\caption[CARLA~J0116-2052 member spectra]{CARLA~J0116$-$2052 member spectra. Each panel shows the G141 1D and 2D member spectra for both grism orientations, as well as the member F140W direct image stamp. Black slits on top of the F140W stamps represent the spectral dispersion directions of the two different grism orientations. Green contours overlaid on the 2D-spectra represent potential contamination from neighboring objects. Red slits on top of the 2D-spectra represent the regions from which 1D-spectra are extracted.
}
\label{fig:J0116-2052spectra}
\mbox{}\\
\end{figure*}

\begin{figure*}[!ht]
{%
\setlength{\fboxsep}{0pt}%
\setlength{\fboxrule}{1pt}%
\fbox{\includegraphics[page=2, scale=0.24]{CARLA_J0958-2904_255.pdf} \hfill \includegraphics[page=1, scale=0.20]{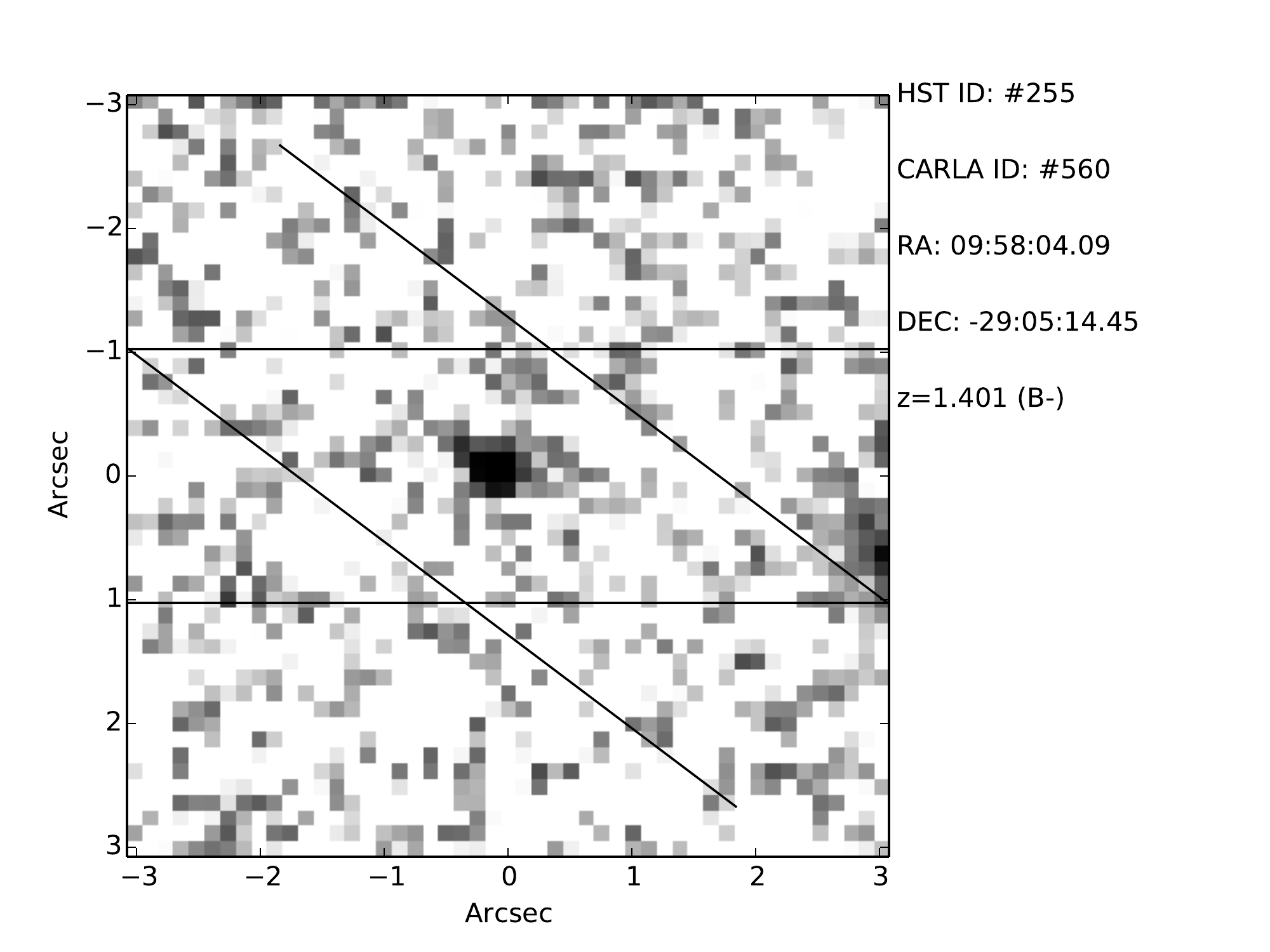} \mbox{(a)}}%
}%
{%
\setlength{\fboxsep}{0pt}%
\setlength{\fboxrule}{1pt}%
\fbox{\includegraphics[page=2, scale=0.24]{CARLA_J0958-2904_260.pdf} \hfill \includegraphics[page=1, scale=0.20]{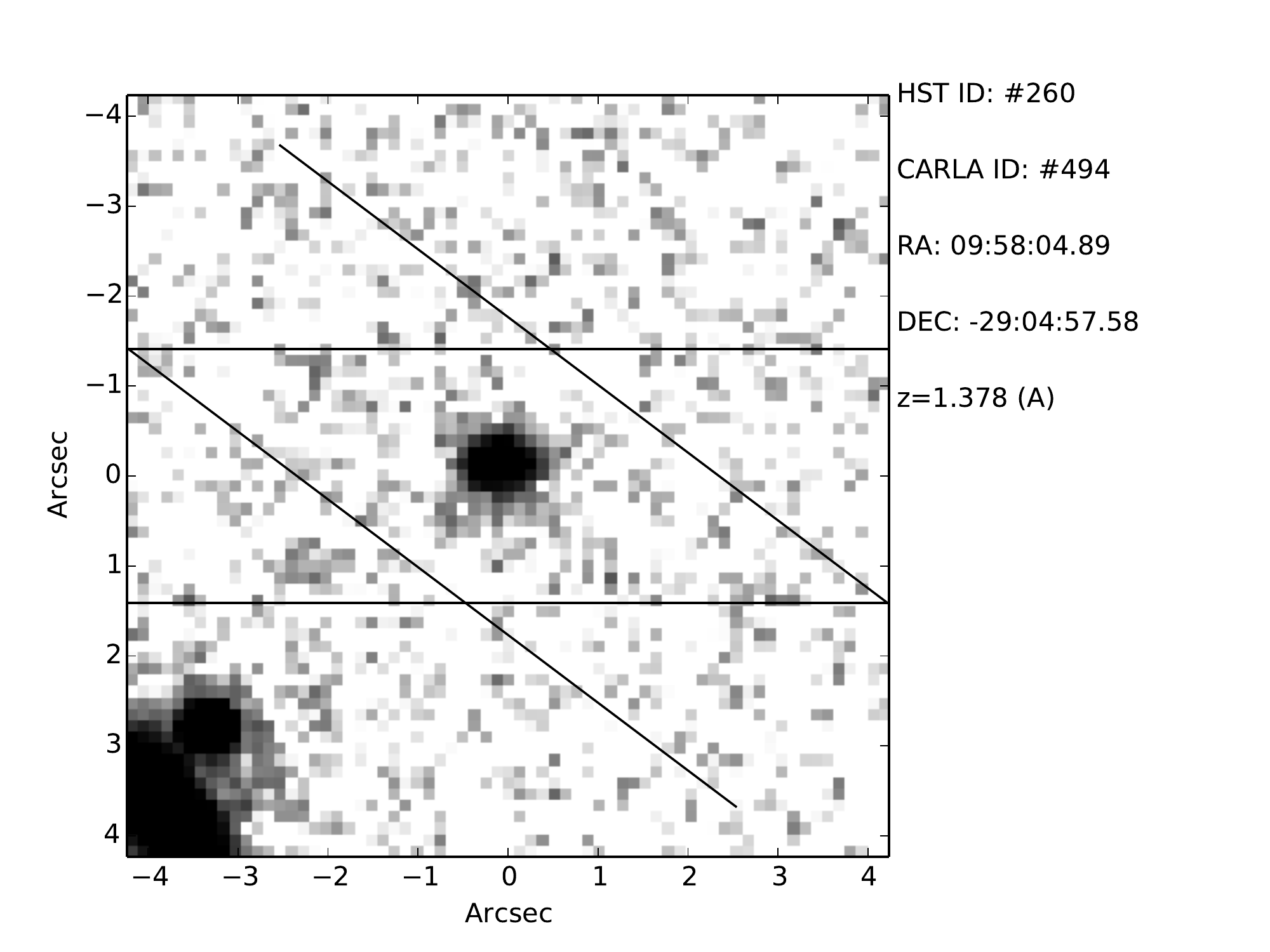} \mbox{(b)}}%
}\\%
{%
\setlength{\fboxsep}{0pt}%
\setlength{\fboxrule}{1pt}%
\fbox{\includegraphics[page=2, scale=0.24]{CARLA_J0958-2904_287.pdf} \hfill \includegraphics[page=1, scale=0.20]{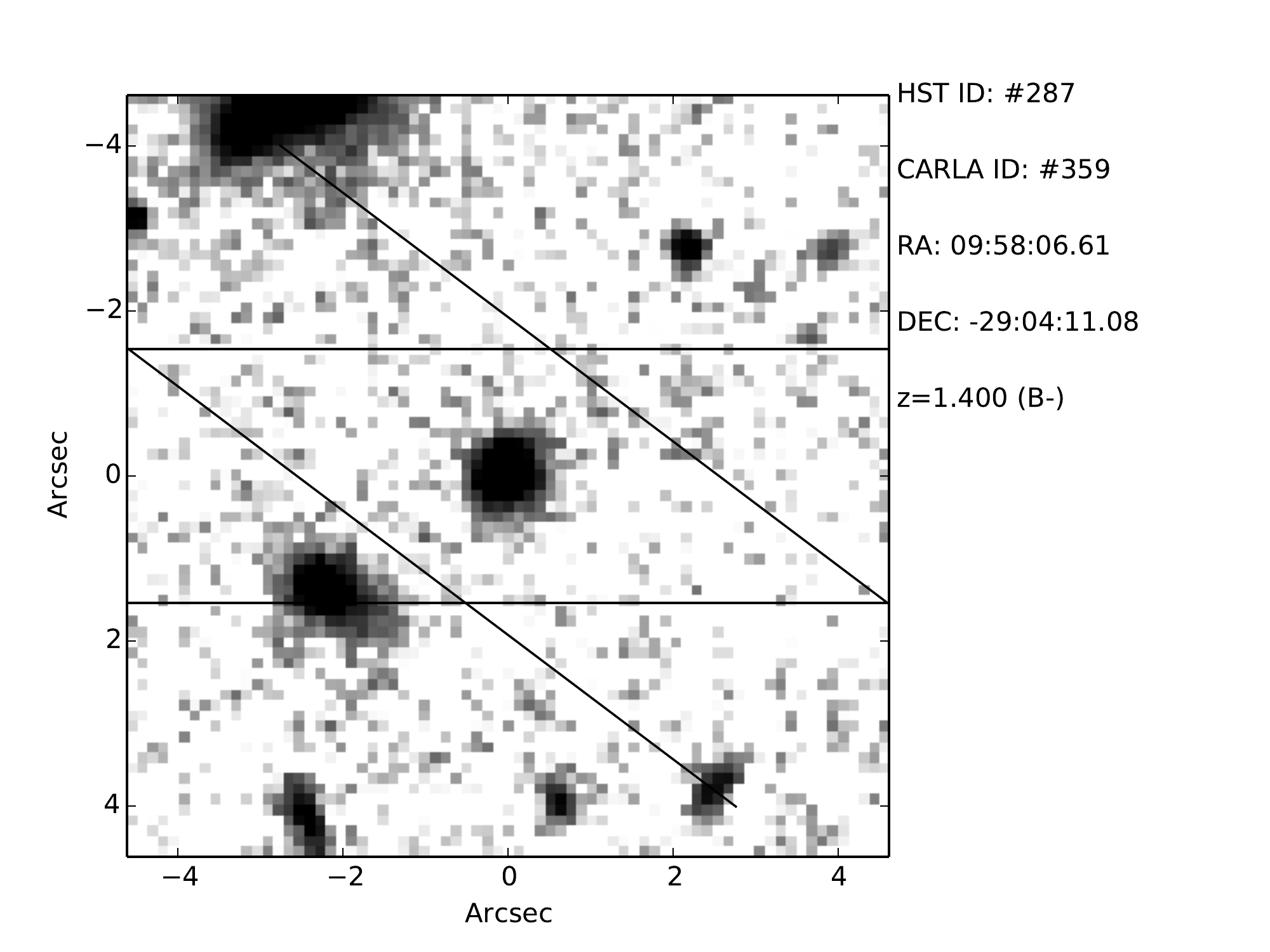} \mbox{(c)}}%
}%
{%
\setlength{\fboxsep}{0pt}%
\setlength{\fboxrule}{1pt}%
\fbox{\includegraphics[page=2, scale=0.24]{CARLA_J0958-2904_356.pdf} \hfill \includegraphics[page=1, scale=0.20]{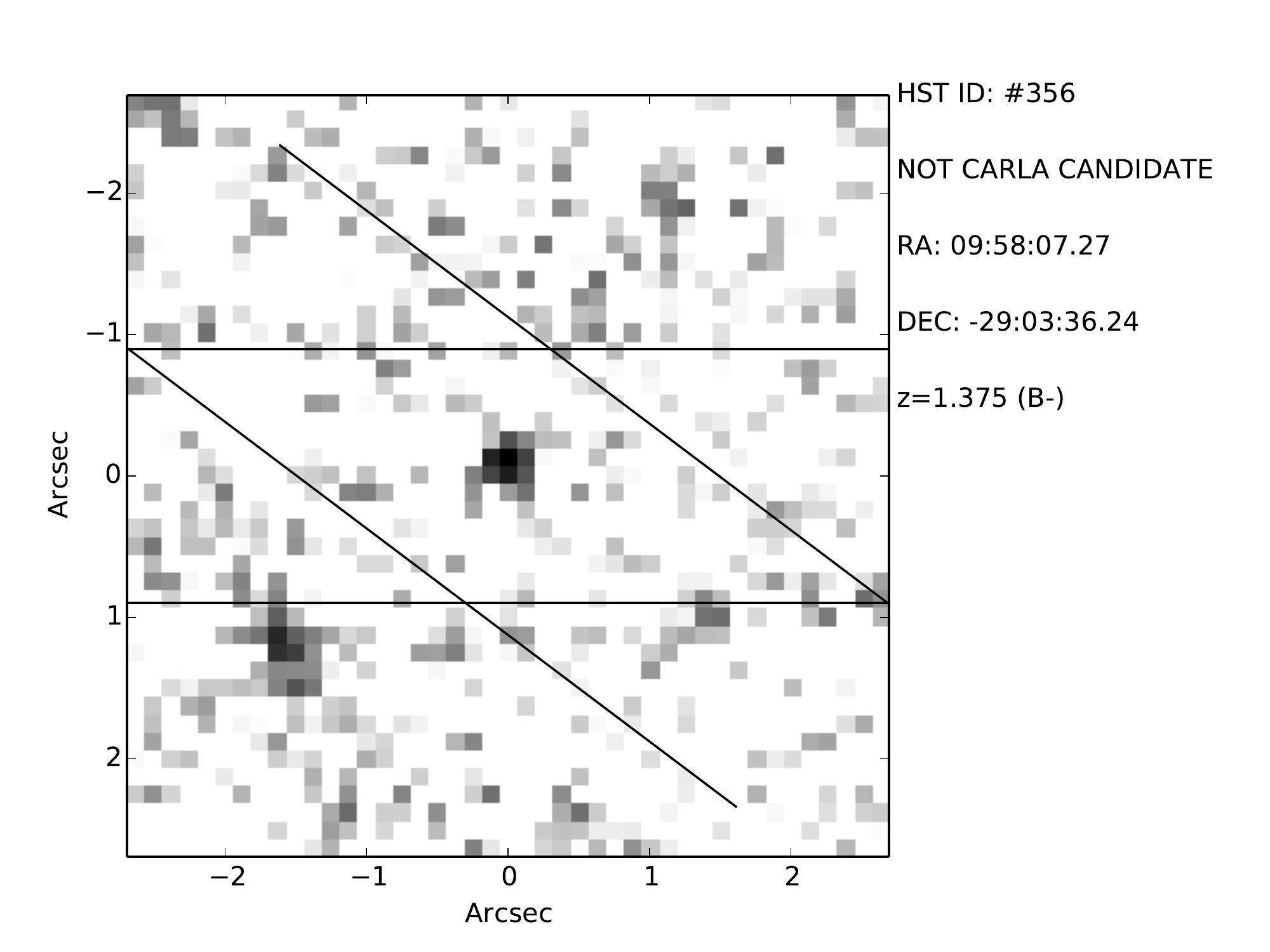} \mbox{(d)}}%
}\\%
{%
\setlength{\fboxsep}{0pt}%
\setlength{\fboxrule}{1pt}%
\fbox{\includegraphics[page=2, scale=0.24]{CARLA_J0958-2904_405.pdf} \hfill \includegraphics[page=1, scale=0.20]{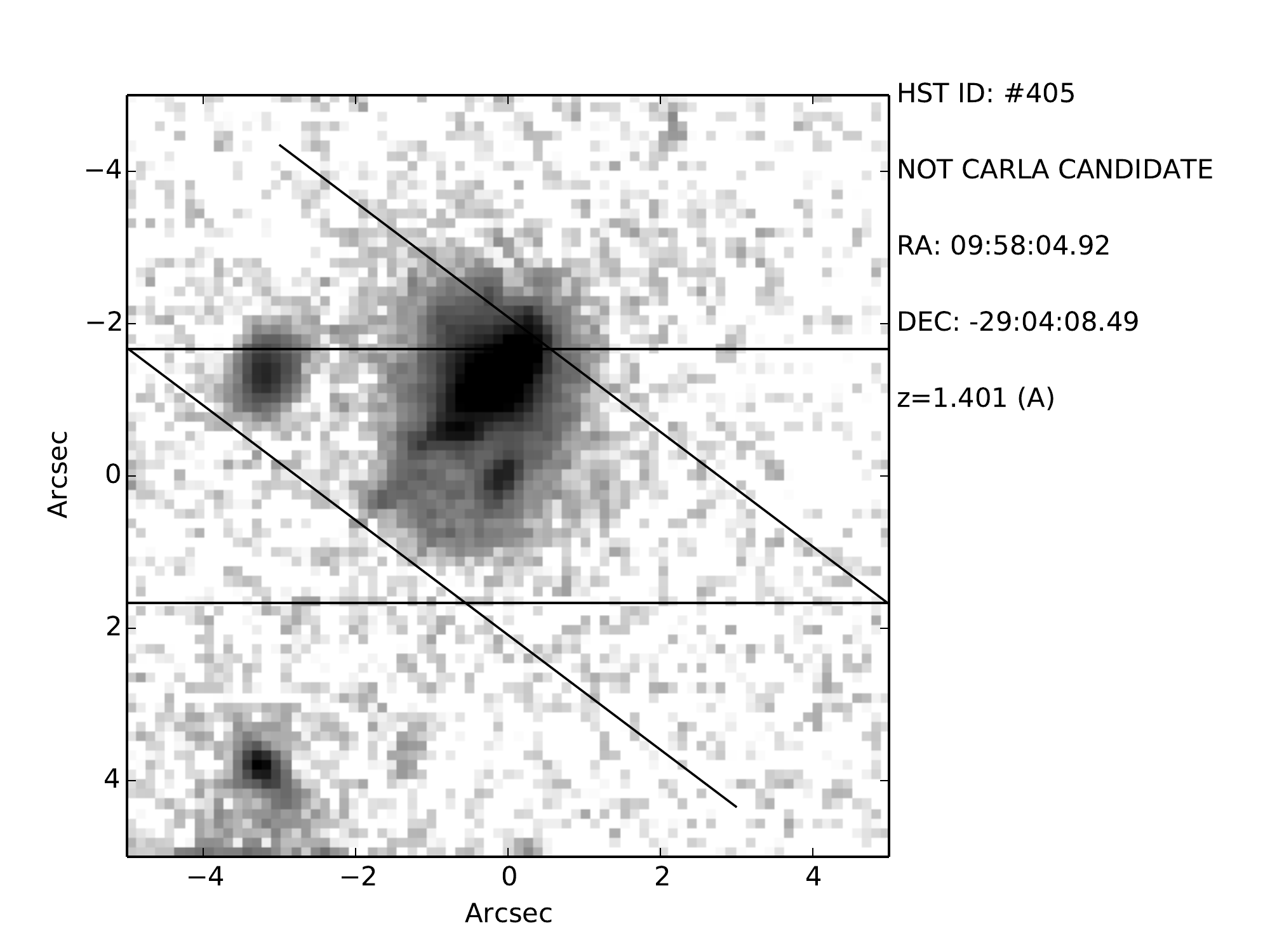} \mbox{(e)}}%
}%
{%
\setlength{\fboxsep}{0pt}%
\setlength{\fboxrule}{1pt}%
\fbox{\includegraphics[page=2, scale=0.24]{CARLA_J0958-2904_406.pdf} \hfill \includegraphics[page=1, scale=0.20]{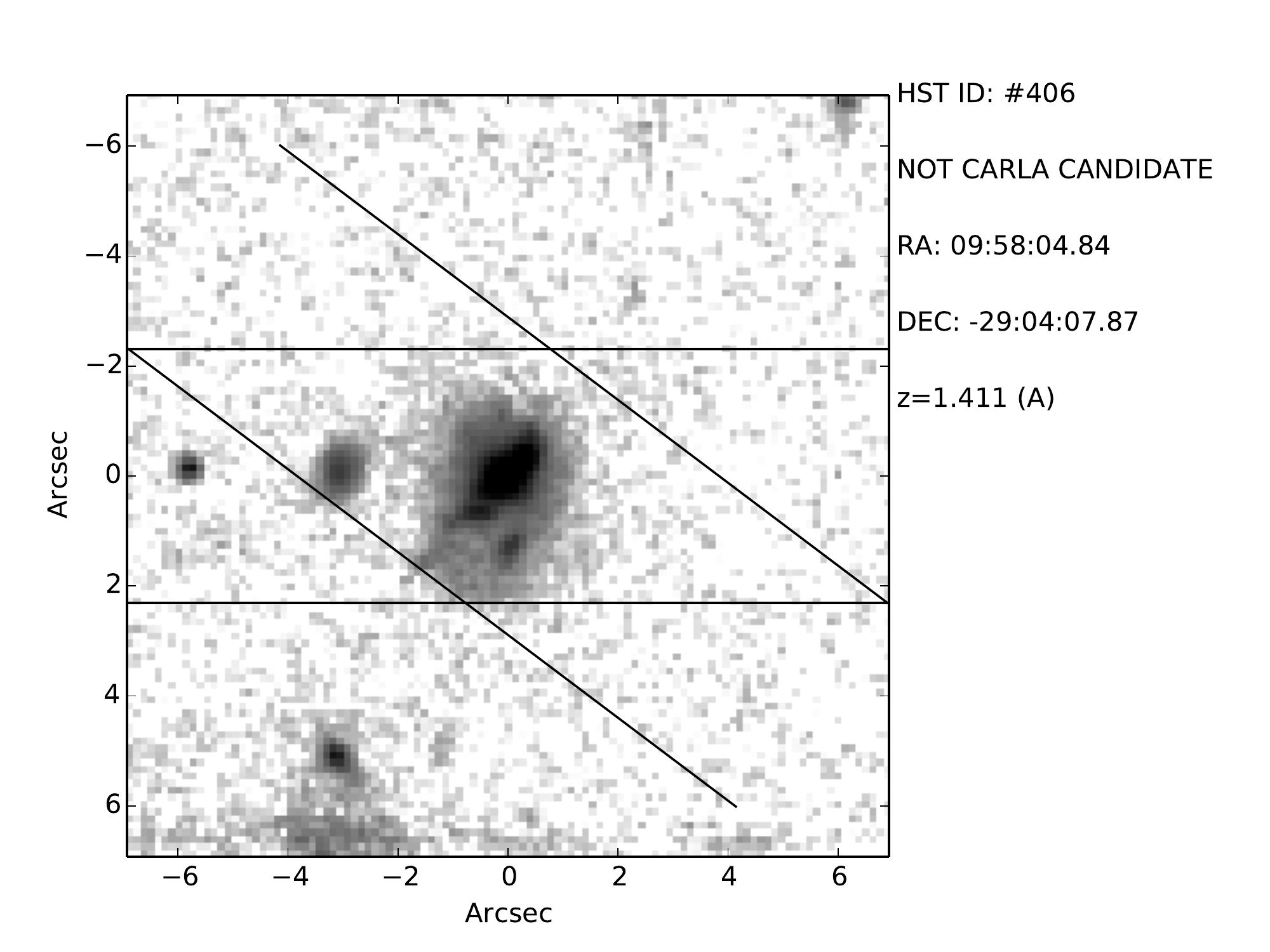} \mbox{(f)}}%
}\\%
{%
\setlength{\fboxsep}{0pt}%
\setlength{\fboxrule}{1pt}%
\fbox{\includegraphics[page=2, scale=0.24]{CARLA_J0958-2904_578.pdf} \hfill \includegraphics[page=1, scale=0.20]{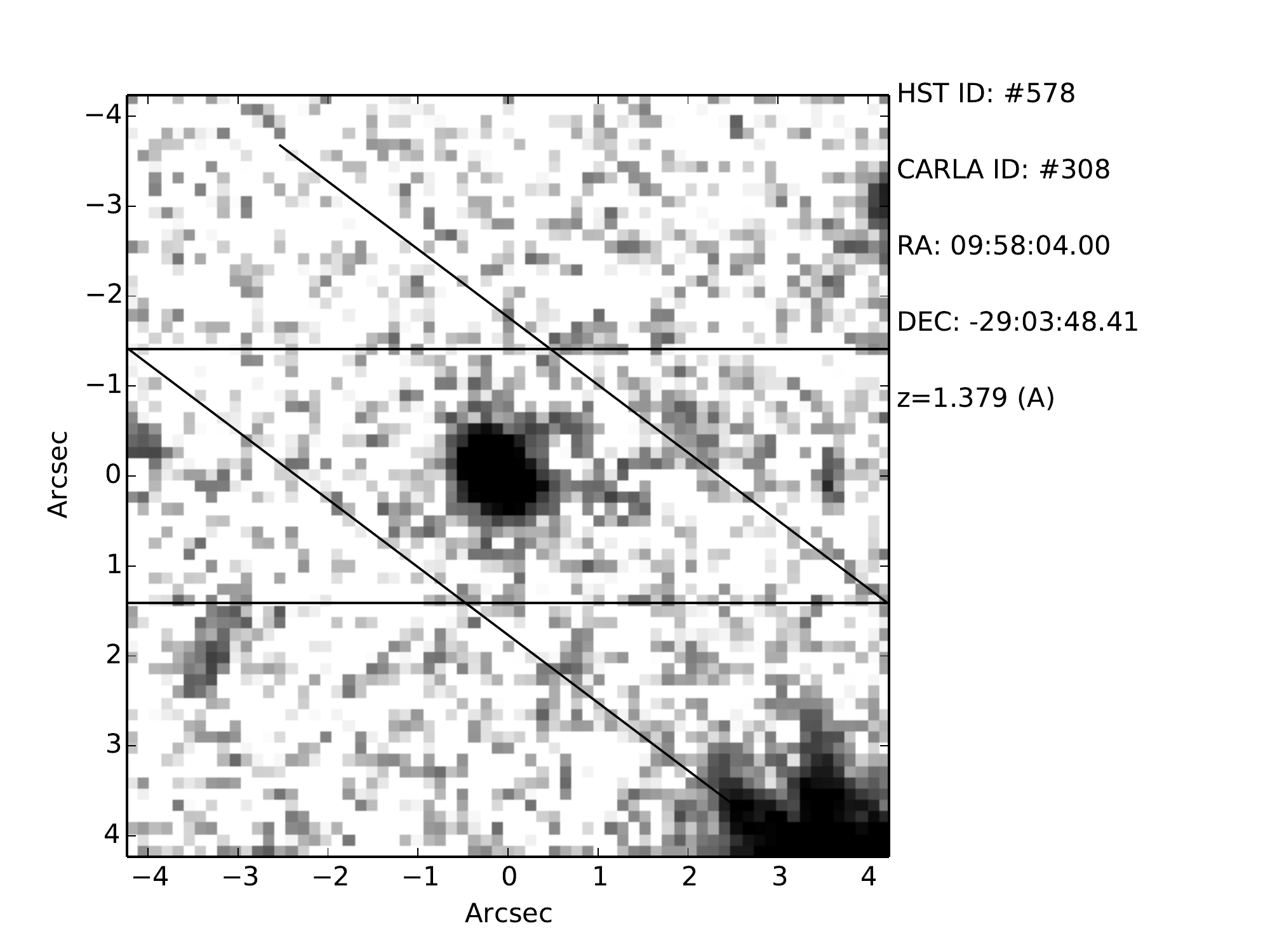} \mbox{(g)}}%
}%
{%
\setlength{\fboxsep}{0pt}%
\setlength{\fboxrule}{1pt}%
\fbox{\includegraphics[page=2, scale=0.24]{CARLA_J0958-2904_691.pdf} \hfill \includegraphics[page=1, scale=0.20]{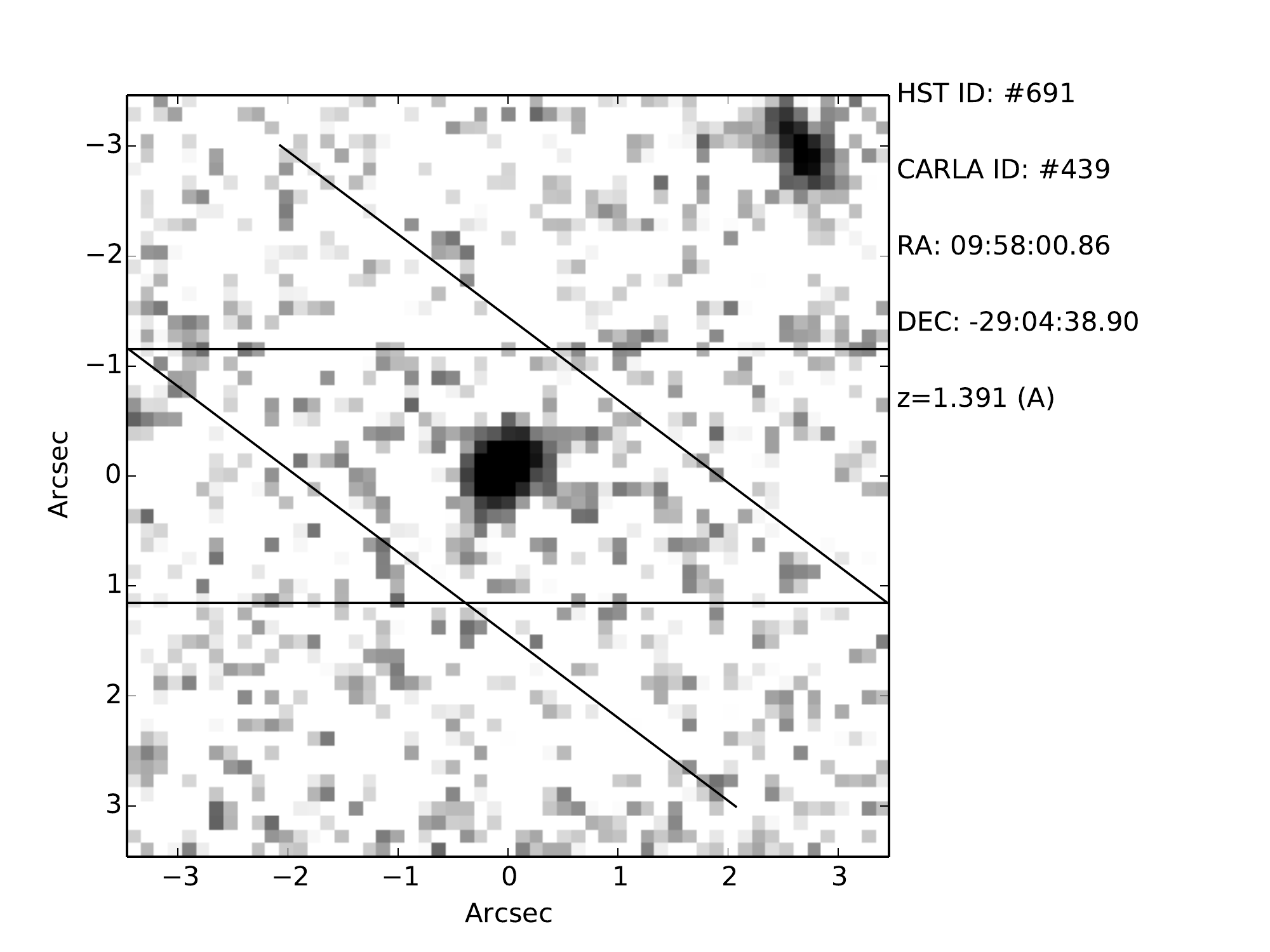} \mbox{(h)}}%
}\\%
\caption[CARLA~J0958$-$2904 member spectra]{CARLA~J0958$-$2904 member spectra.}
\label{fig:J0958-2904spectra}
\end{figure*}

\begin{figure*}[!ht]
{%
\setlength{\fboxsep}{0pt}%
\setlength{\fboxrule}{1pt}%
\fbox{\includegraphics[page=2, scale=0.24]{CARLA_J1017+6116_124.pdf} \hfill \includegraphics[page=1, scale=0.20]{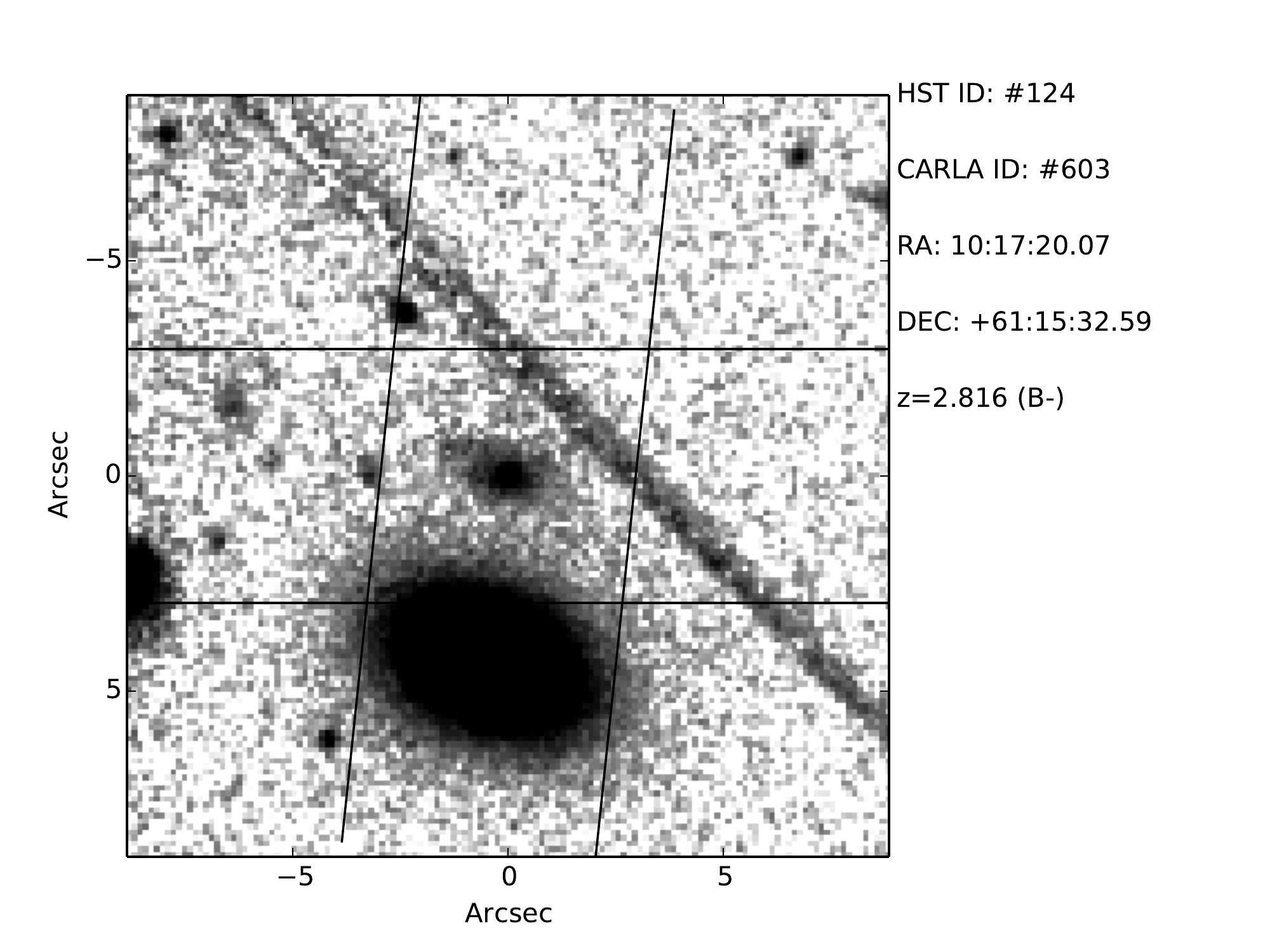} \mbox{(a)}}%
}%
{%
\setlength{\fboxsep}{0pt}%
\setlength{\fboxrule}{1pt}%
\fbox{\includegraphics[page=2, scale=0.24]{CARLA_J1017+6116_203.pdf} \hfill \includegraphics[page=1, scale=0.20]{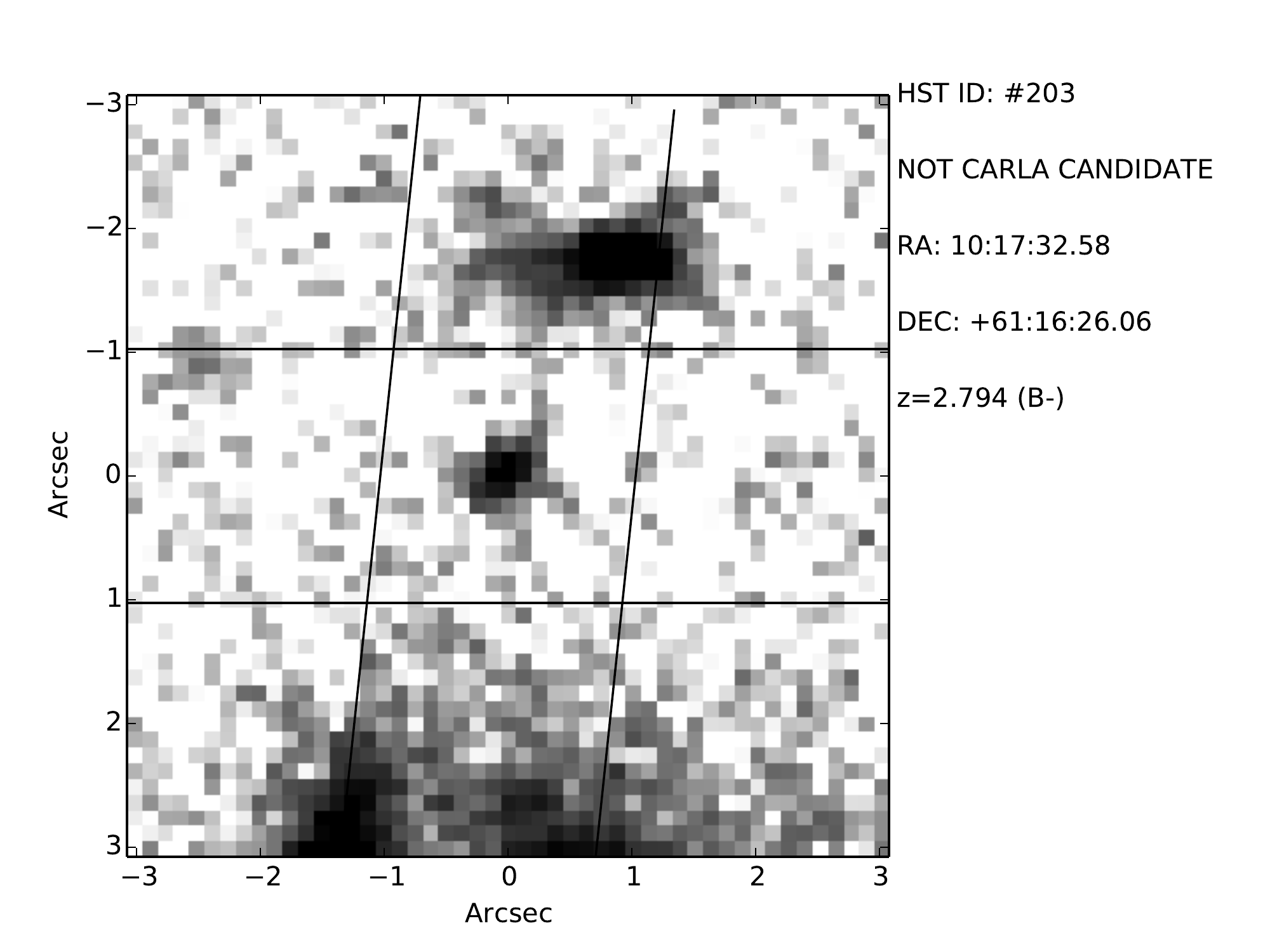} \mbox{(b)}}%
}\\%
{%
\setlength{\fboxsep}{0pt}%
\setlength{\fboxrule}{1pt}%
\fbox{\includegraphics[page=2, scale=0.24]{CARLA_J1017+6116_237.pdf} \hfill \includegraphics[page=1, scale=0.20]{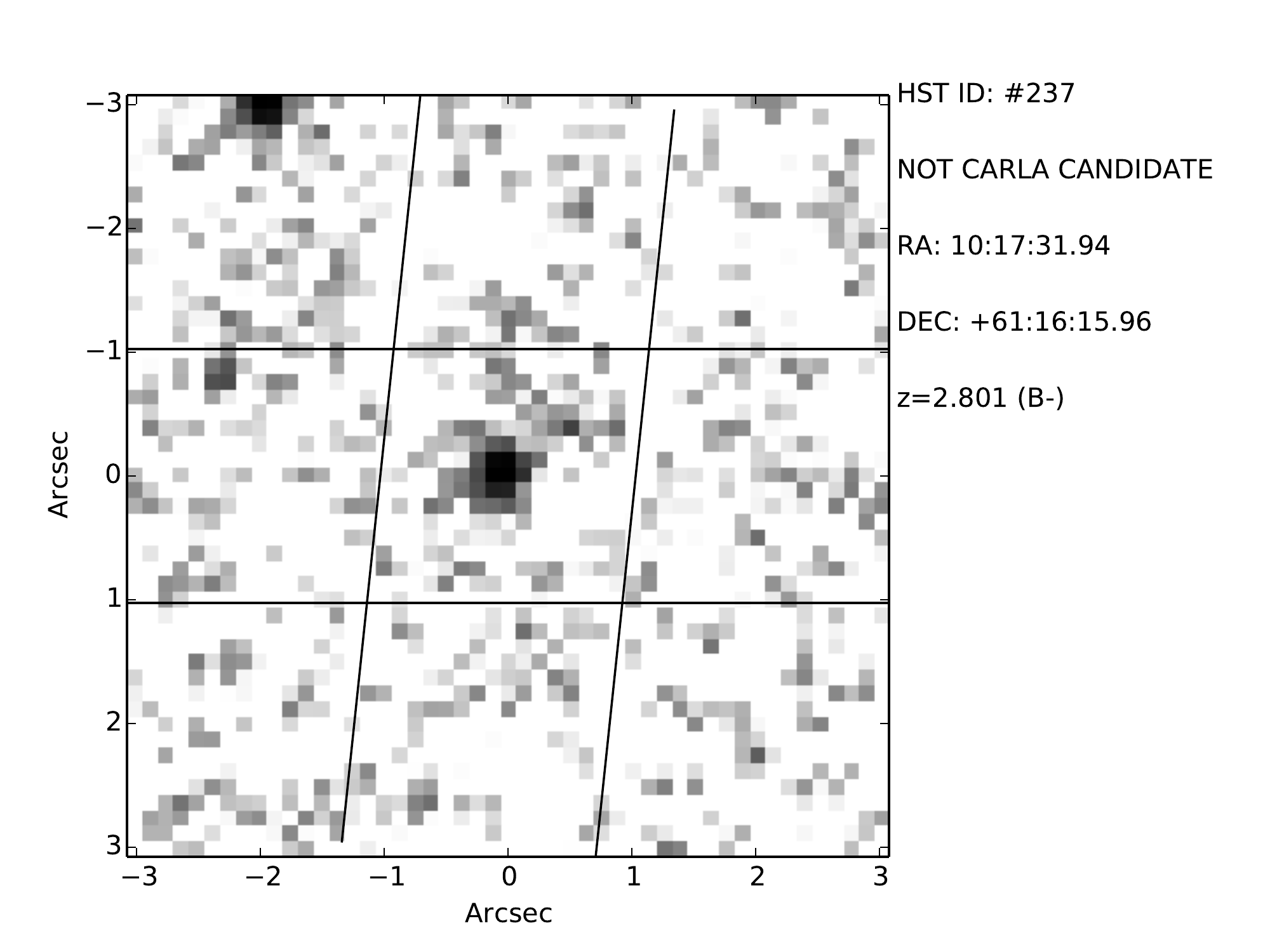} \mbox{(c)}}%
}%
{%
\setlength{\fboxsep}{0pt}%
\setlength{\fboxrule}{1pt}%
\fbox{\includegraphics[page=2, scale=0.24]{CARLA_J1017+6116_337.pdf} \hfill \includegraphics[page=1, scale=0.20]{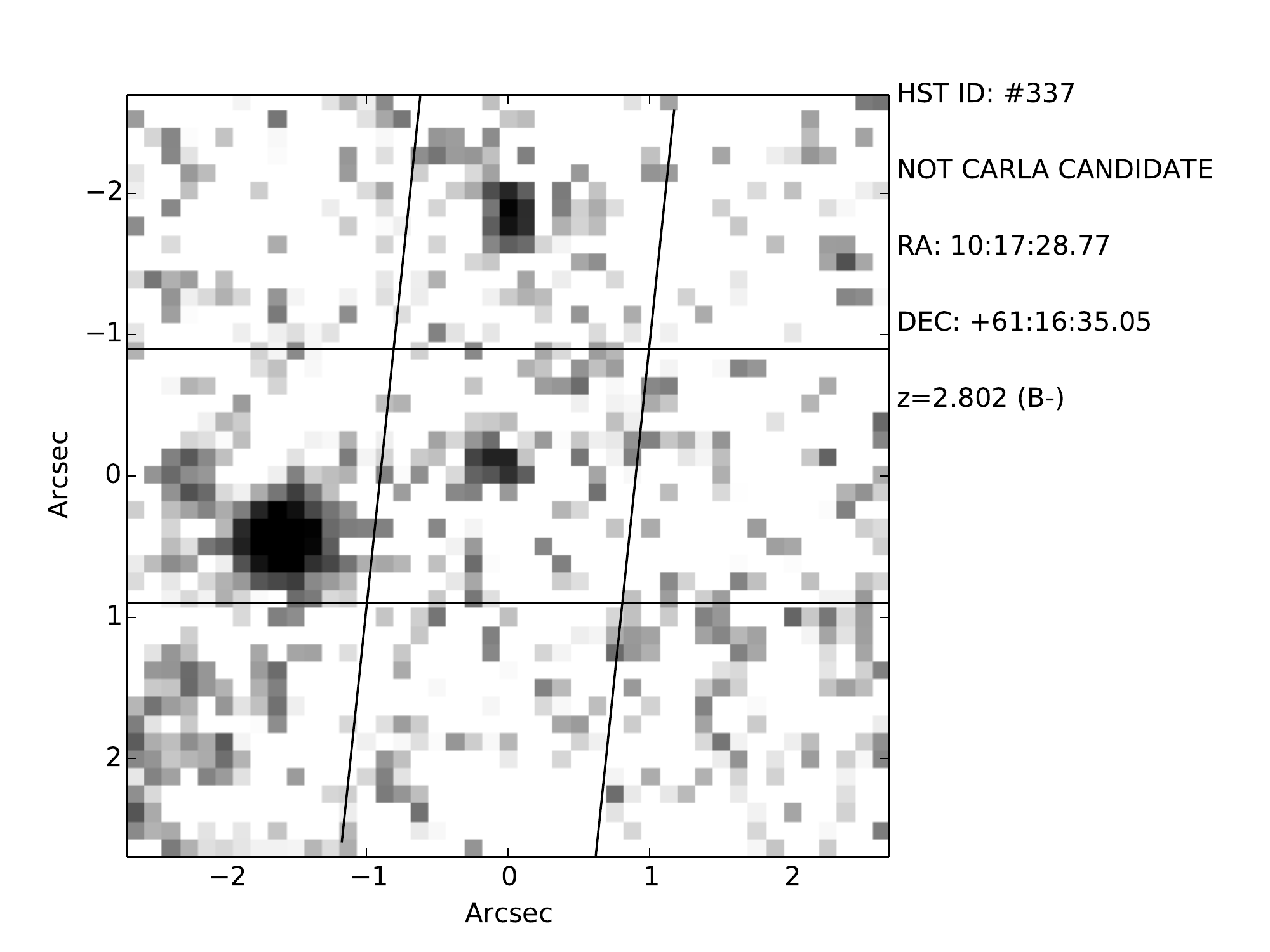} \mbox{(d)}}%
}\\%
{%
\setlength{\fboxsep}{0pt}%
\setlength{\fboxrule}{1pt}%
\fbox{\includegraphics[page=2, scale=0.24]{CARLA_J1017+6116_382.pdf} \hfill \includegraphics[page=1, scale=0.20]{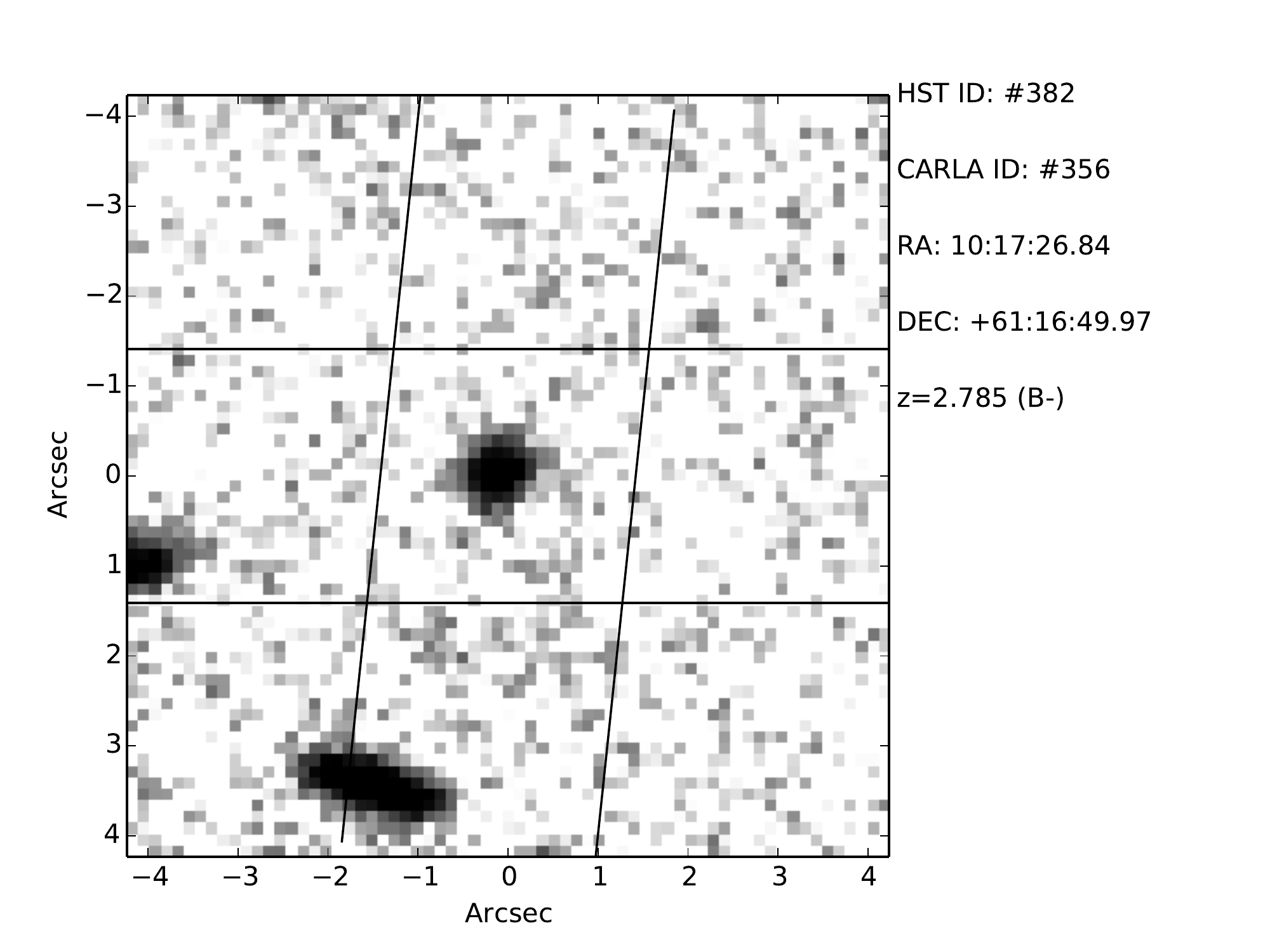} \mbox{(e)}}%
}%
{%
\setlength{\fboxsep}{0pt}%
\setlength{\fboxrule}{1pt}%
\fbox{\includegraphics[page=2, scale=0.24]{CARLA_J1017+6116_457.pdf} \hfill \includegraphics[page=1, scale=0.20]{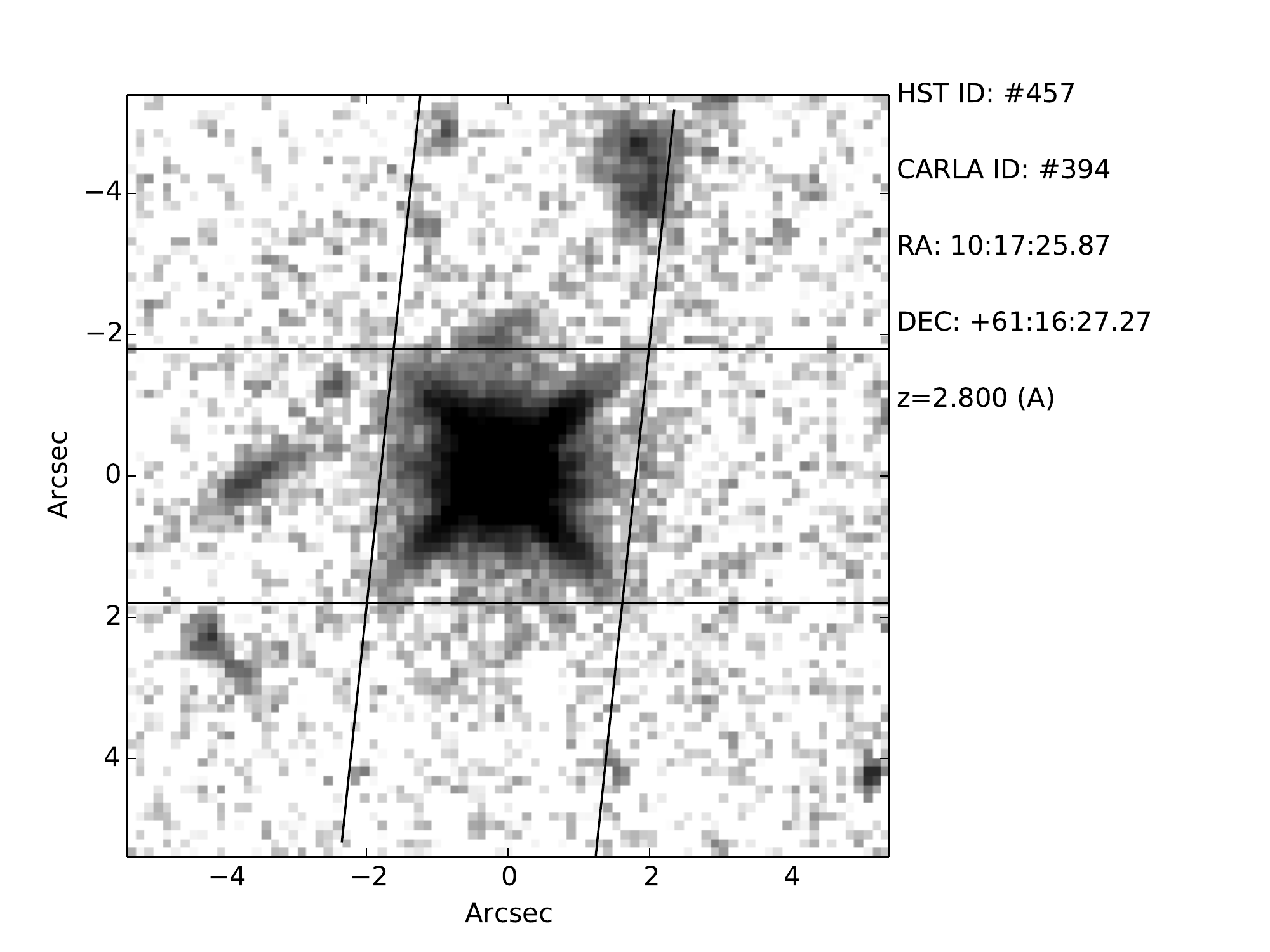} \mbox{(f)}}%
}\\%
{%
\setlength{\fboxsep}{0pt}%
\setlength{\fboxrule}{1pt}%
\fbox{\includegraphics[page=2, scale=0.24]{CARLA_J1017+6116_625.pdf} \hfill \includegraphics[page=1, scale=0.20]{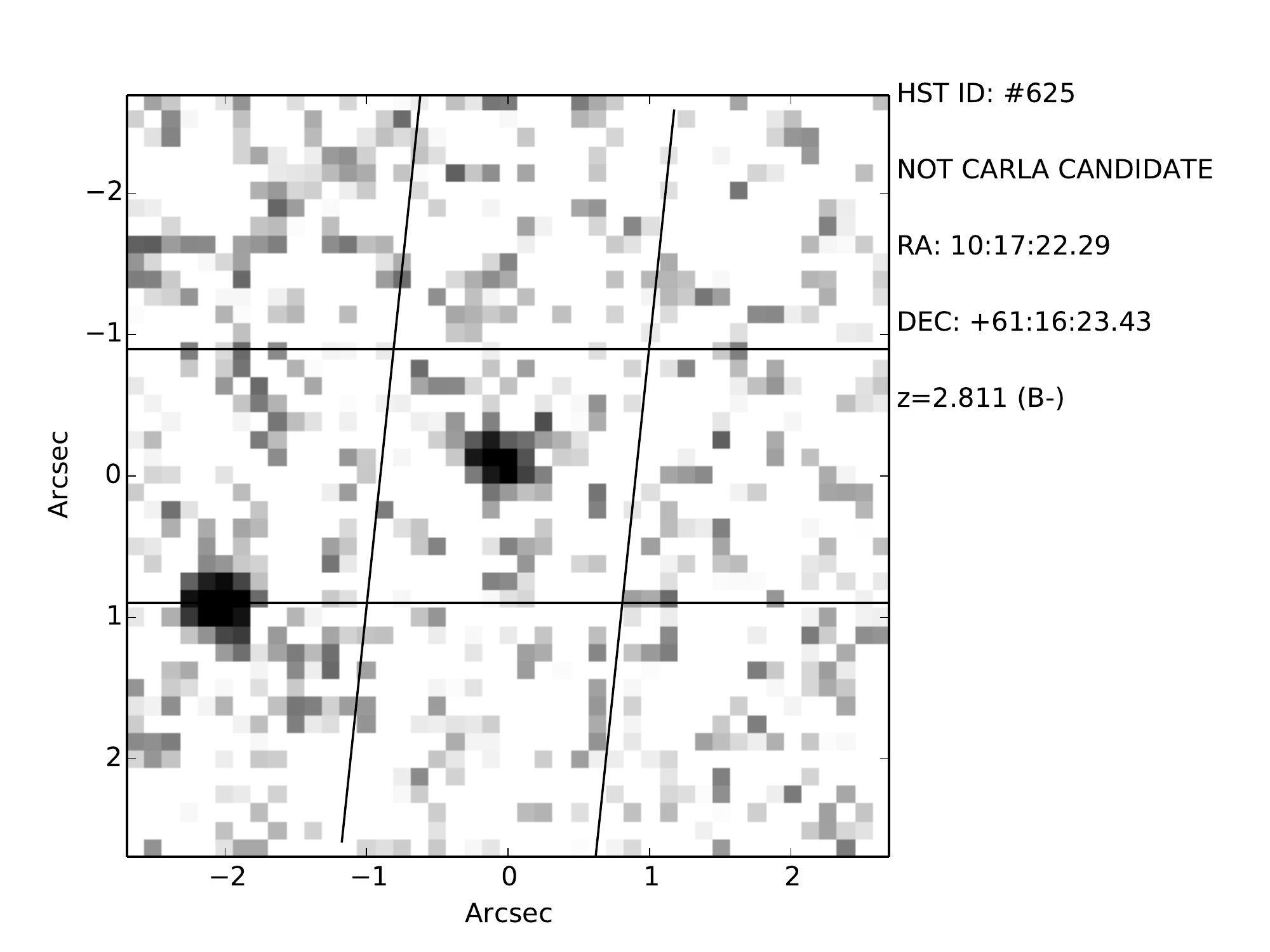} \mbox{(g)}}%
}%
\caption[CARLA~J1017+6116 member spectra]{CARLA~J1017+6116 member spectra.}
\label{fig:J1017+6116spectra}
\mbox{}\\
\end{figure*}

\begin{figure*}[]
{%
\setlength{\fboxsep}{0pt}%
\setlength{\fboxrule}{1pt}%
\fbox{\includegraphics[page=2, scale=0.24]{CARLA_J1018+0530_127.pdf} \hfill \includegraphics[page=1, scale=0.20]{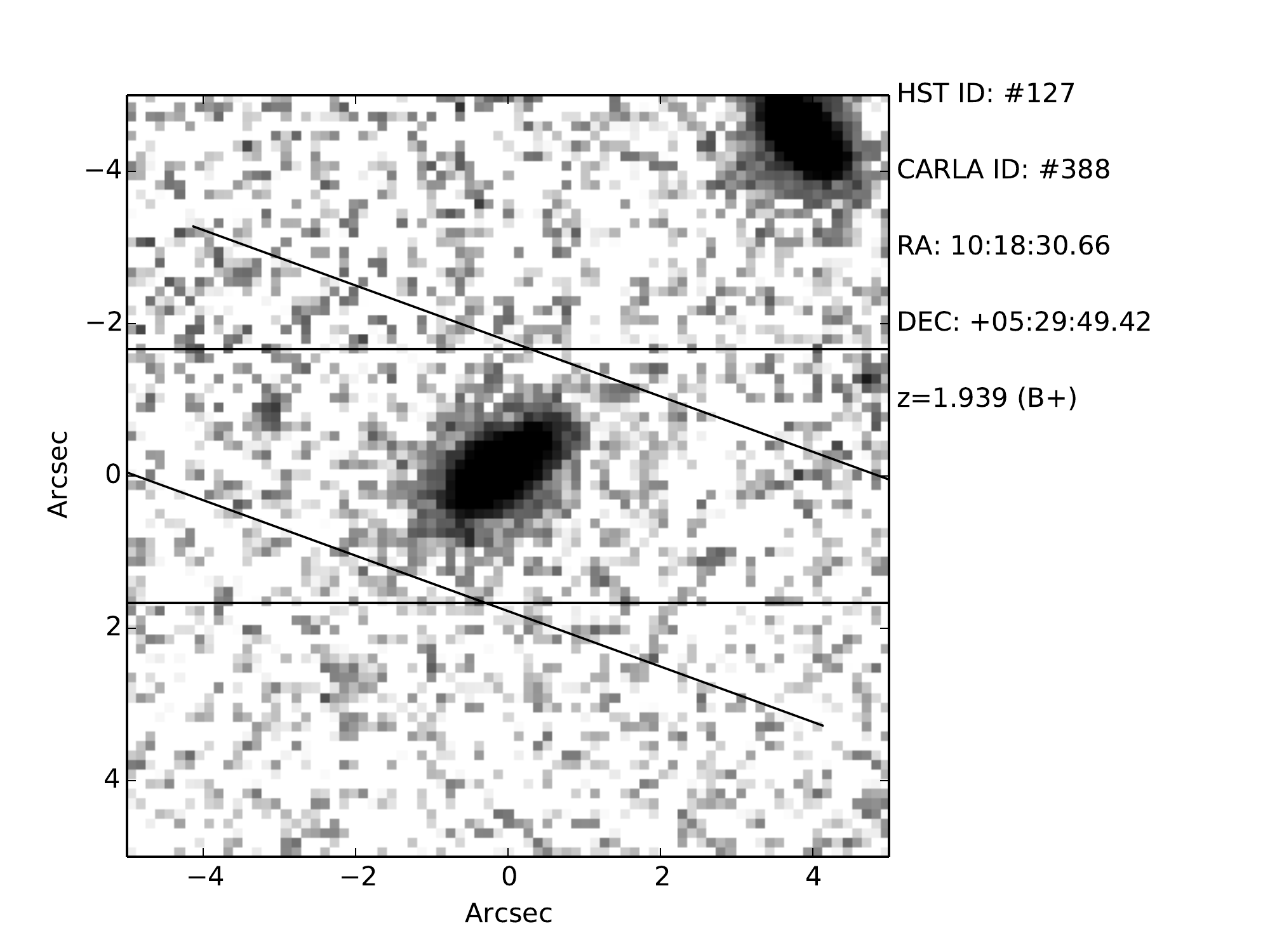} \mbox{(a)}}%
}%
{%
\setlength{\fboxsep}{0pt}%
\setlength{\fboxrule}{1pt}%
\fbox{\includegraphics[page=2, scale=0.24]{CARLA_J1018+0530_138.pdf} \hfill \includegraphics[page=1, scale=0.20]{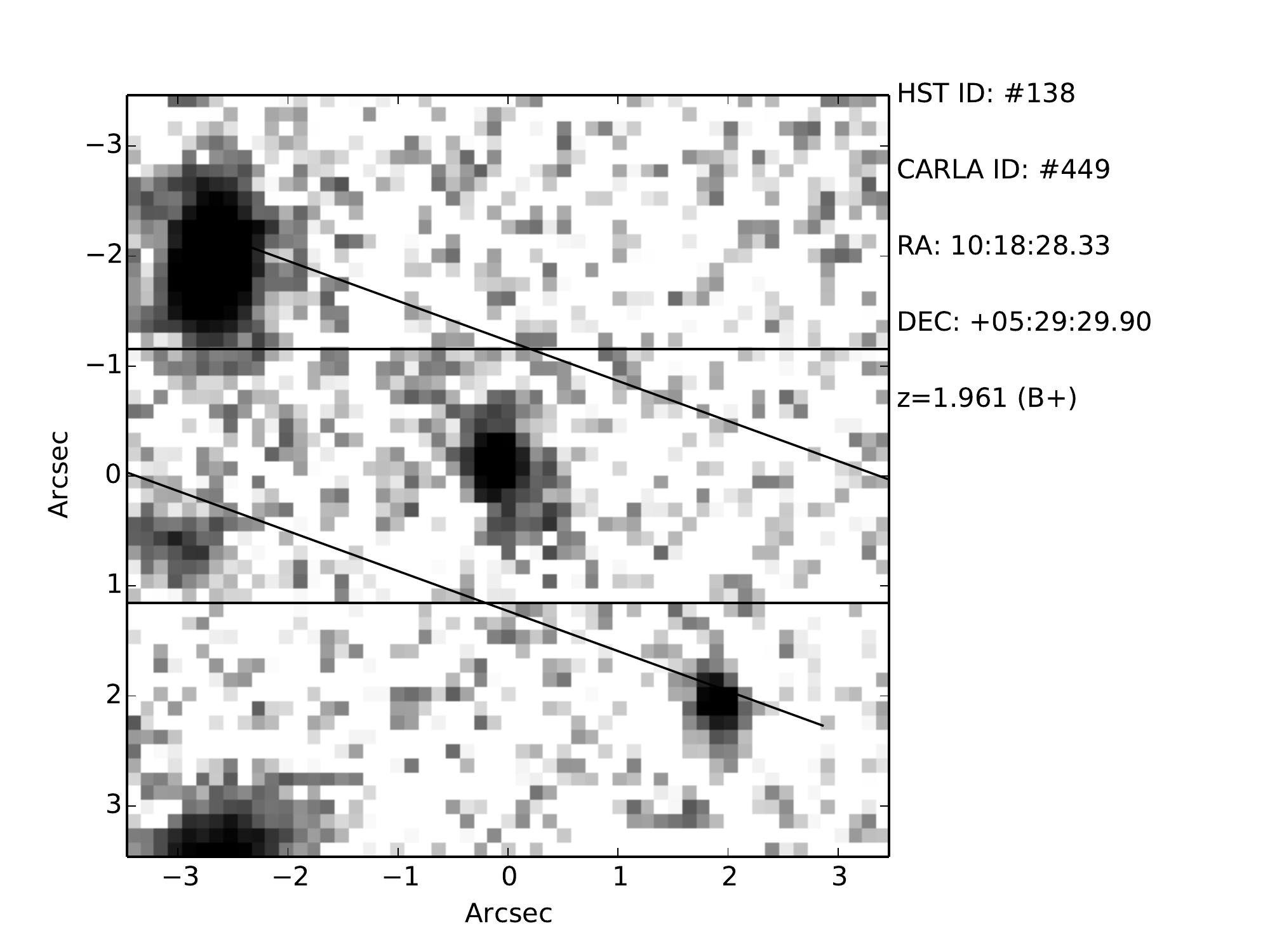} \mbox{(b)}}%
}\\%
{%
\setlength{\fboxsep}{0pt}%
\setlength{\fboxrule}{1pt}%
\fbox{\includegraphics[page=2, scale=0.24]{CARLA_J1018+0530_162.pdf} \hfill \includegraphics[page=1, scale=0.20]{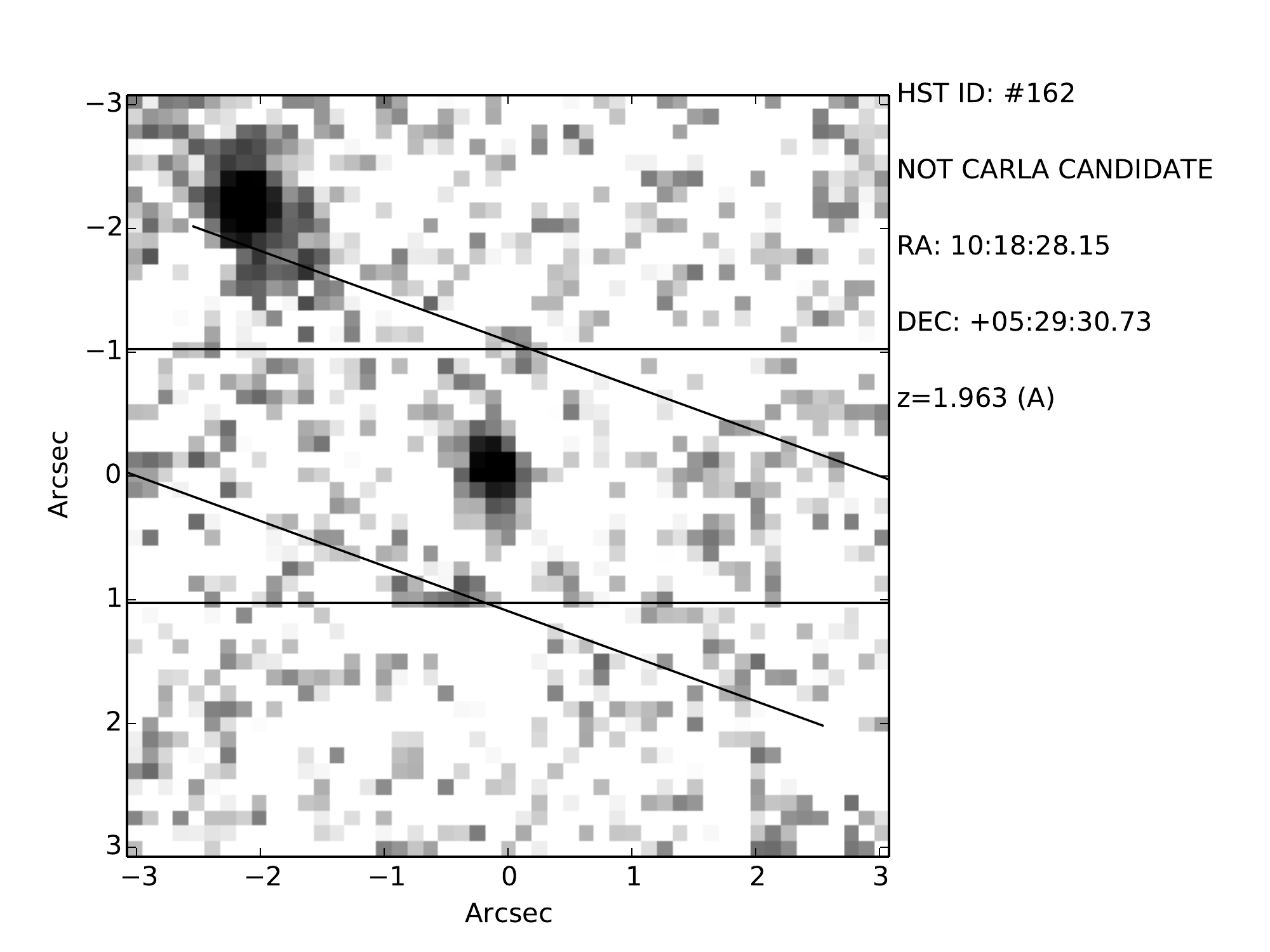} \mbox{(c)}}%
}%
{%
\setlength{\fboxsep}{0pt}%
\setlength{\fboxrule}{1pt}%
\fbox{\includegraphics[page=2, scale=0.24]{CARLA_J1018+0530_336.pdf} \hfill \includegraphics[page=1, scale=0.20]{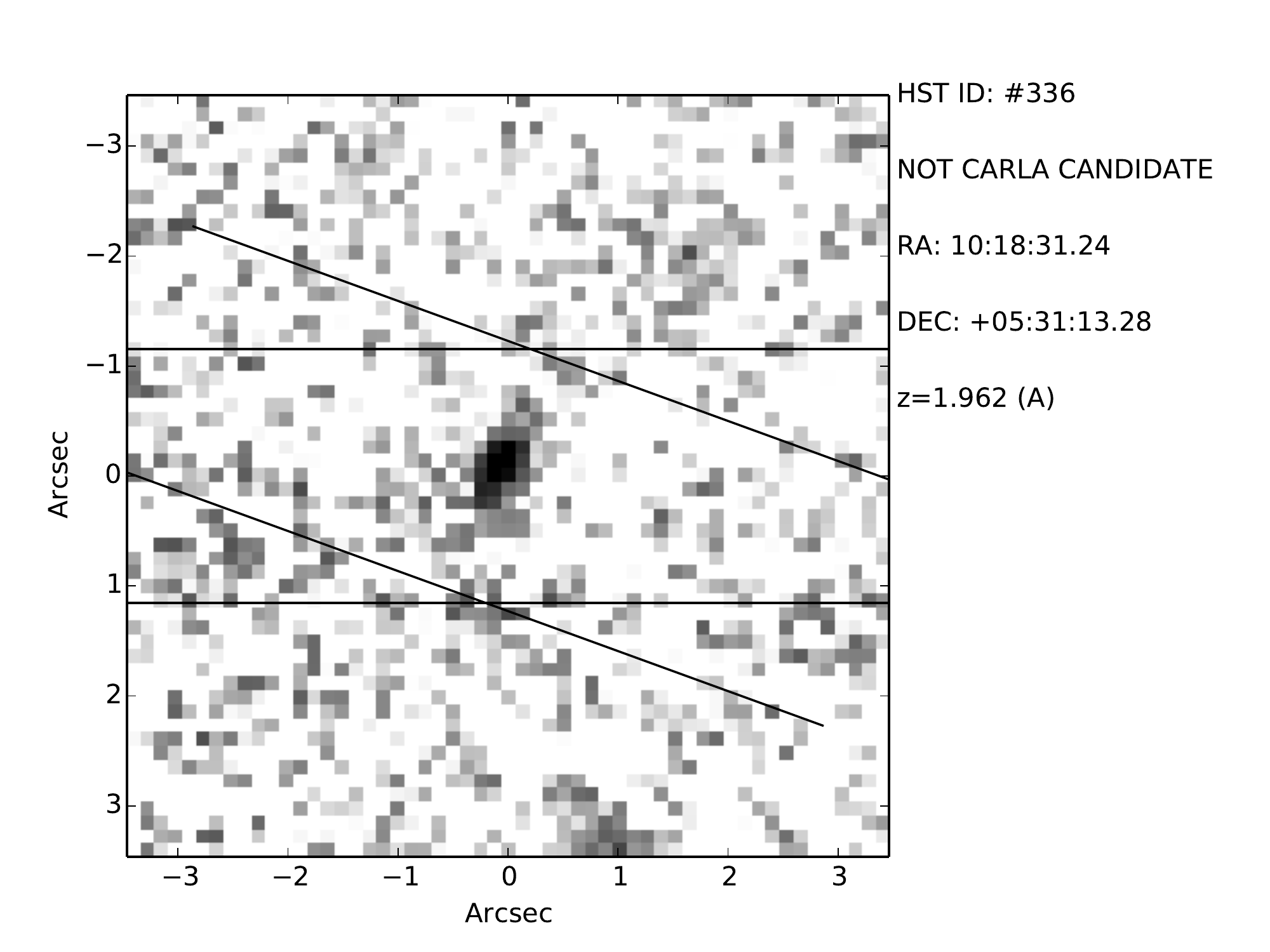} \mbox{(d)}}%
}\\%
{%
\setlength{\fboxsep}{0pt}%
\setlength{\fboxrule}{1pt}%
\fbox{\includegraphics[page=2, scale=0.24]{CARLA_J1018+0530_354.pdf} \hfill \includegraphics[page=1, scale=0.20]{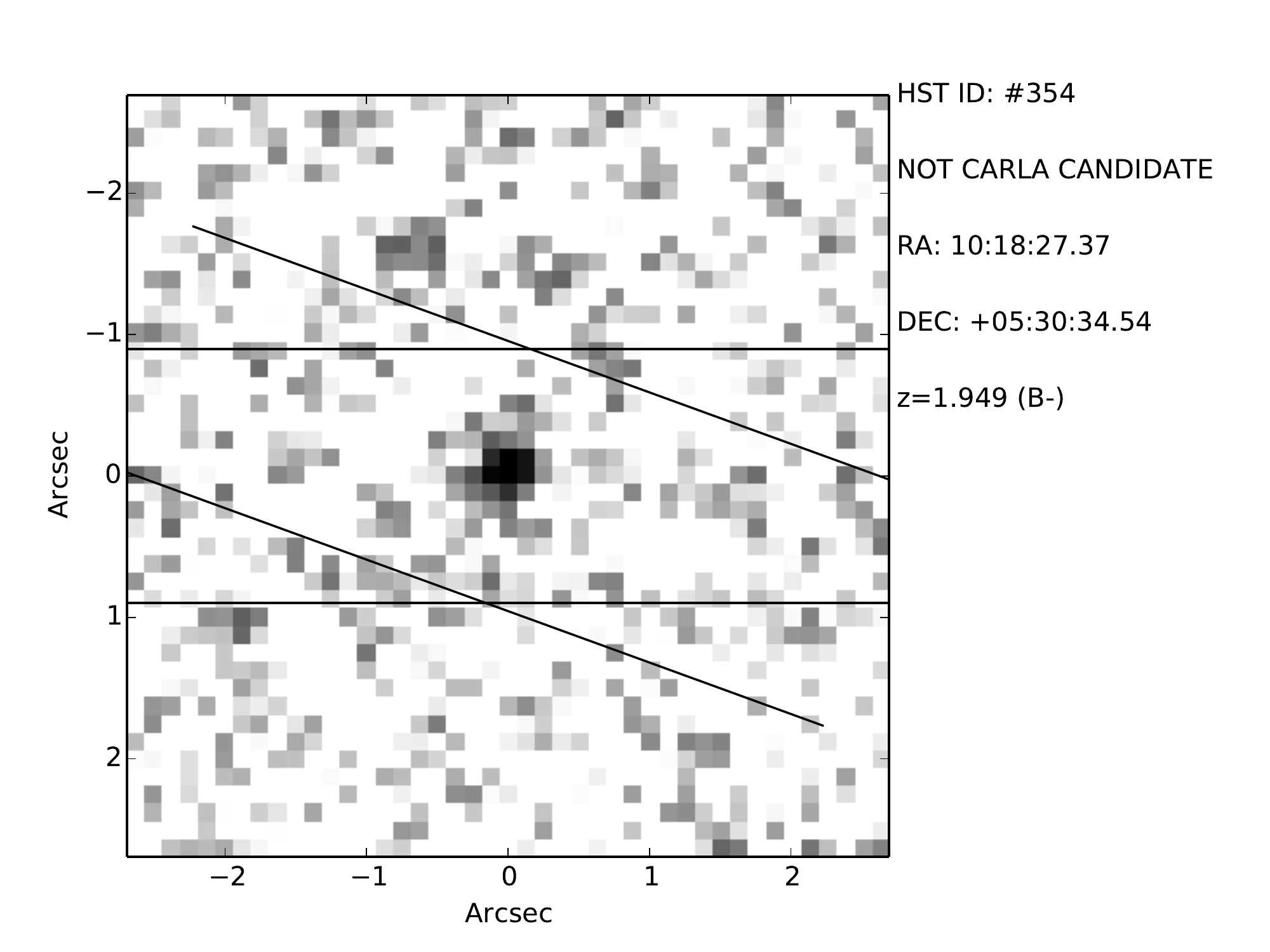} \mbox{(e)}}%
}%
{%
\setlength{\fboxsep}{0pt}%
\setlength{\fboxrule}{1pt}%
\fbox{\includegraphics[page=2, scale=0.24]{CARLA_J1018+0530_390.pdf} \hfill \includegraphics[page=1, scale=0.20]{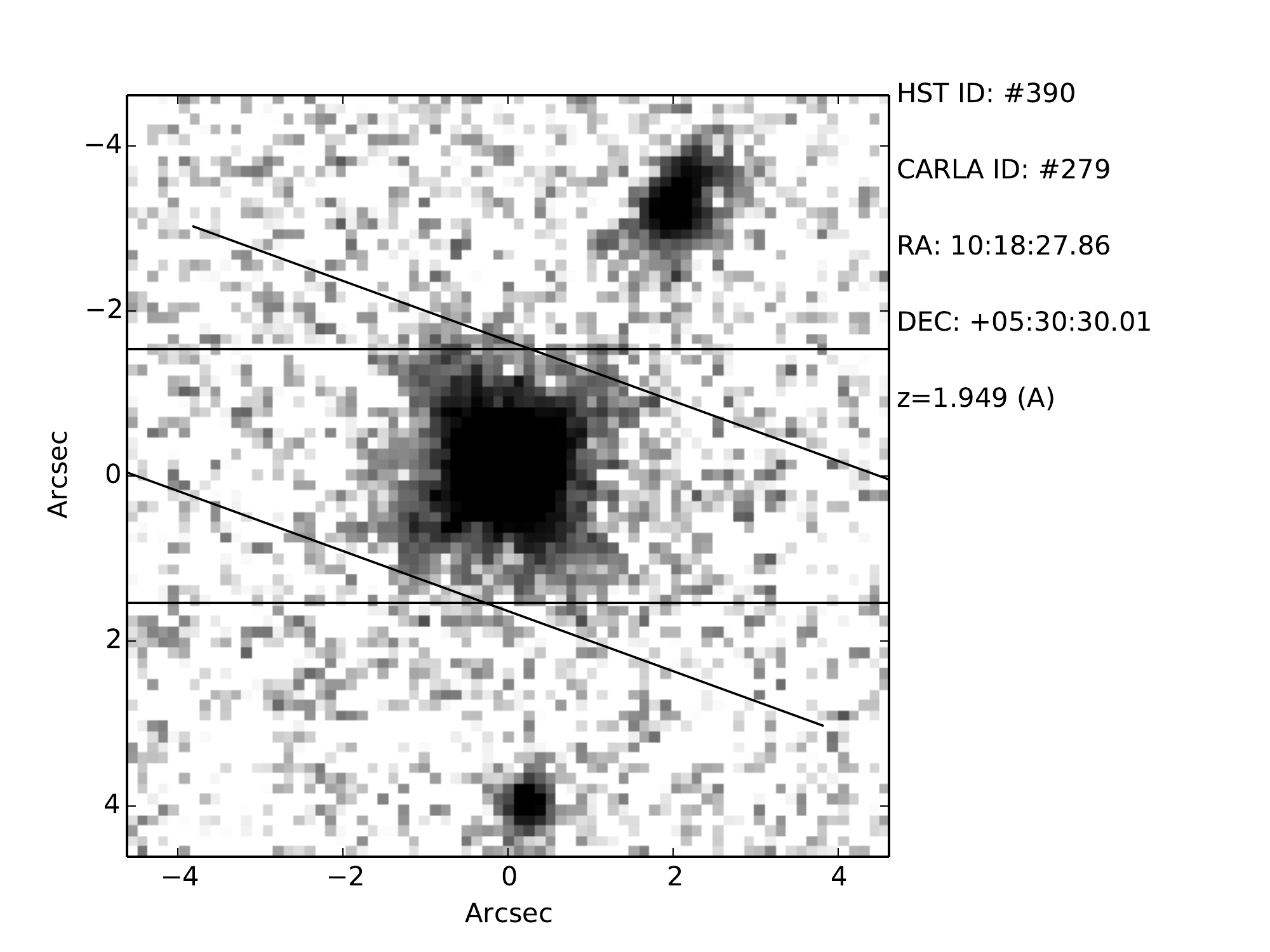} \mbox{(f)}}%
}\\%
{%
\setlength{\fboxsep}{0pt}%
\setlength{\fboxrule}{1pt}%
\fbox{\includegraphics[page=2, scale=0.24]{CARLA_J1018+0530_466.pdf} \hfill \includegraphics[page=1, scale=0.20]{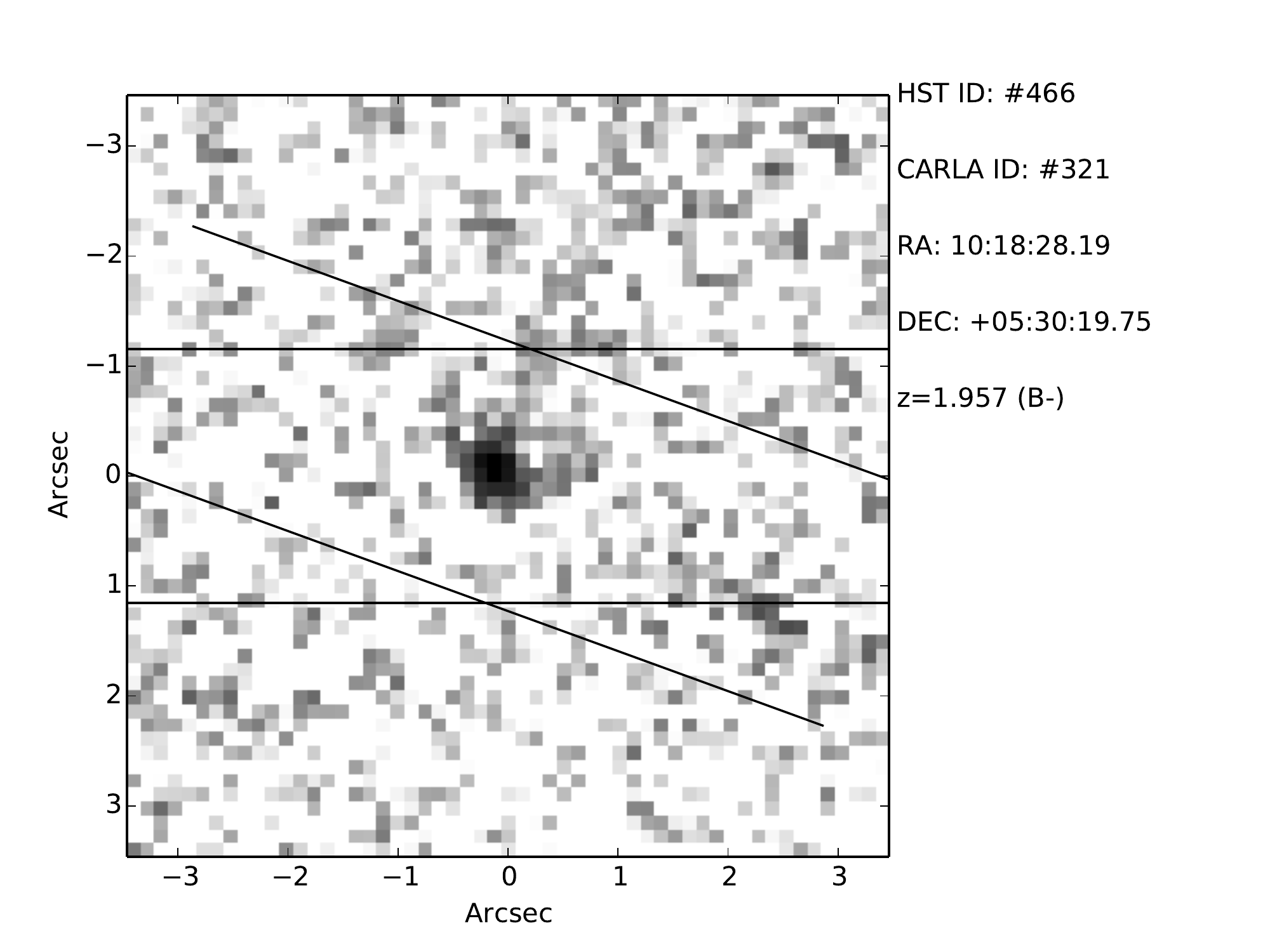} \mbox{(g)}}%
}%
{%
\setlength{\fboxsep}{0pt}%
\setlength{\fboxrule}{1pt}%
\fbox{\includegraphics[page=2, scale=0.24]{CARLA_J1018+0530_647.pdf} \hfill \includegraphics[page=1, scale=0.20]{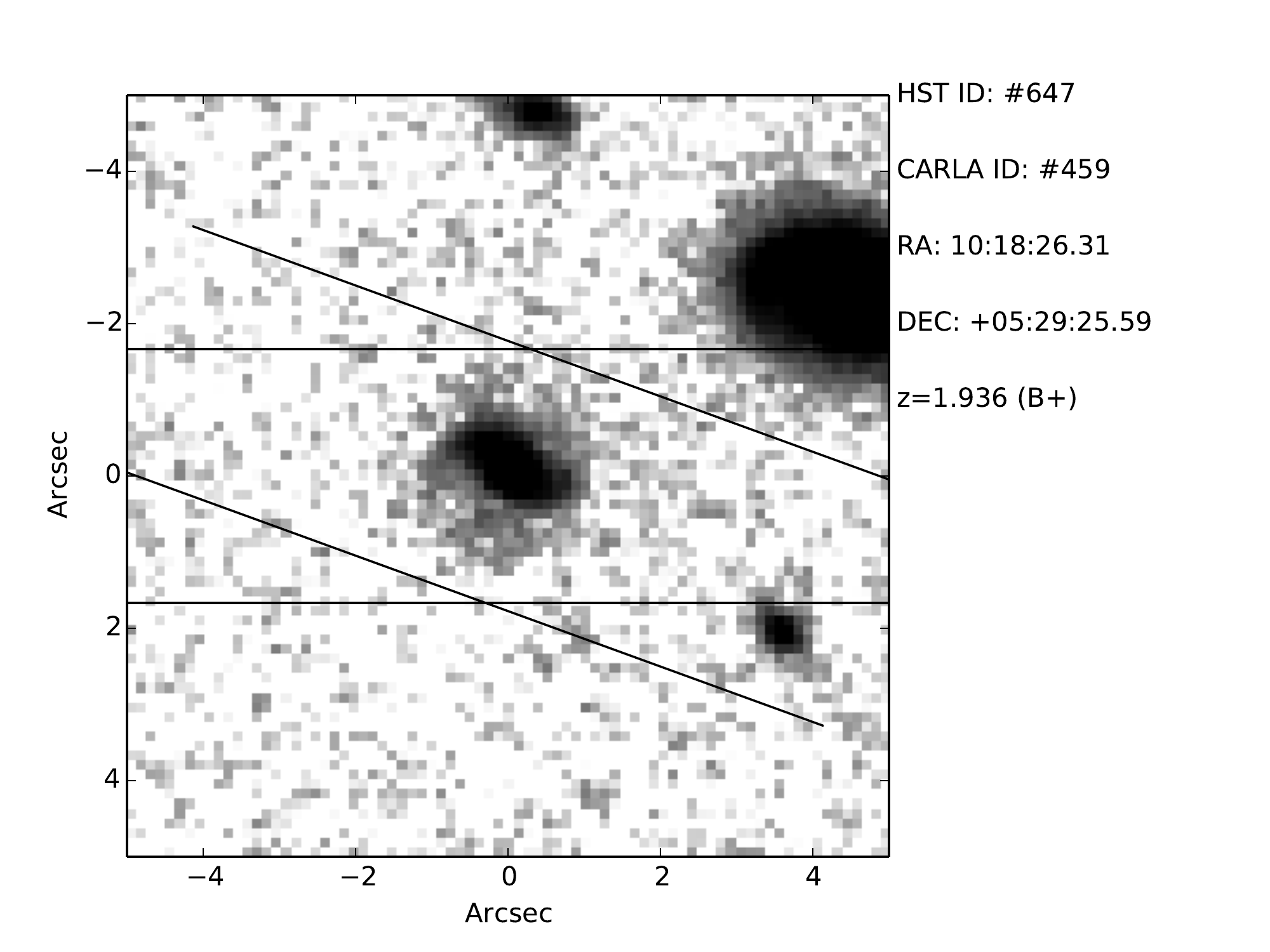} \mbox{(h)}}%
}\\%
\caption[CARLA~J1018+0530 member spectra]{CARLA~J1018+0530 member spectra.}
\label{fig:J1018+0530spectra}
\mbox{}\\
\end{figure*}

\begin{figure*}[]
{%
\setlength{\fboxsep}{0pt}%
\setlength{\fboxrule}{1pt}%
\fbox{\includegraphics[page=2, scale=0.24]{CARLA_J1052+0806_294.pdf} \hfill \includegraphics[page=1, scale=0.20]{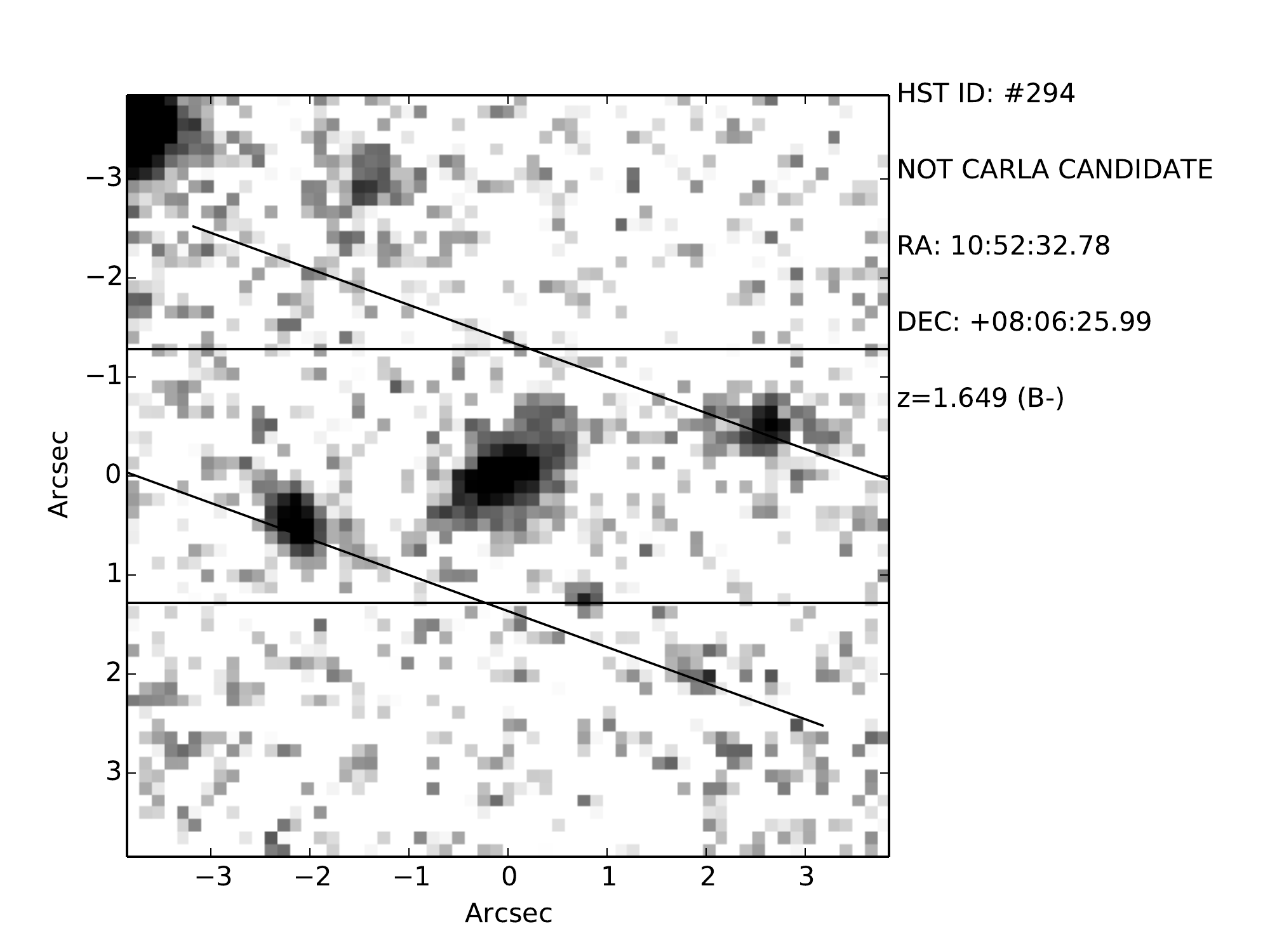} \mbox{(a)}}%
}%
{%
\setlength{\fboxsep}{0pt}%
\setlength{\fboxrule}{1pt}%
\fbox{\includegraphics[page=2, scale=0.24]{CARLA_J1052+0806_311.pdf} \hfill \includegraphics[page=1, scale=0.20]{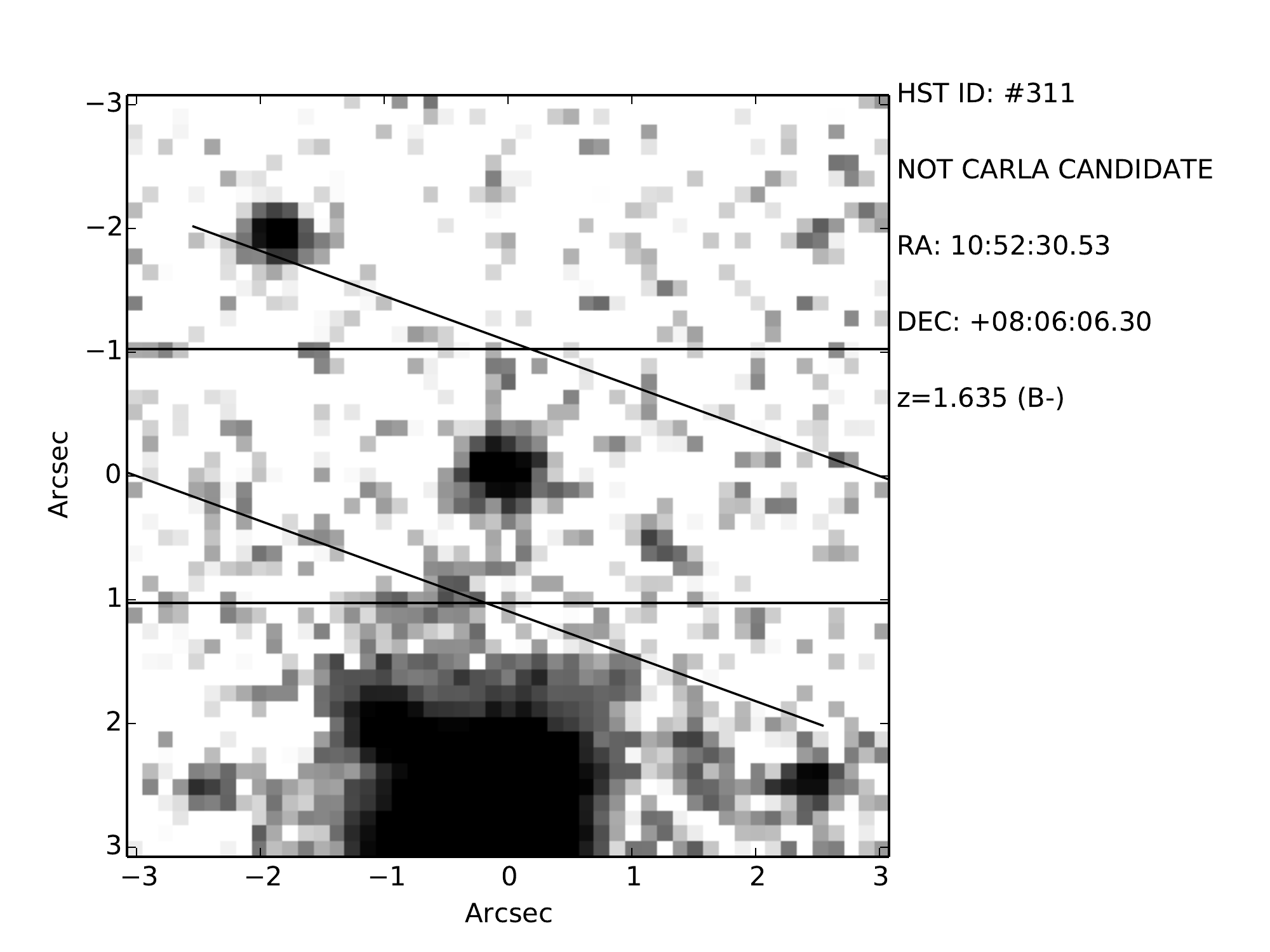} \mbox{(b)}}%
}\\%
{%
\setlength{\fboxsep}{0pt}%
\setlength{\fboxrule}{1pt}%
\fbox{\includegraphics[page=2, scale=0.24]{CARLA_J1052+0806_322.pdf} \hfill \includegraphics[page=1, scale=0.20]{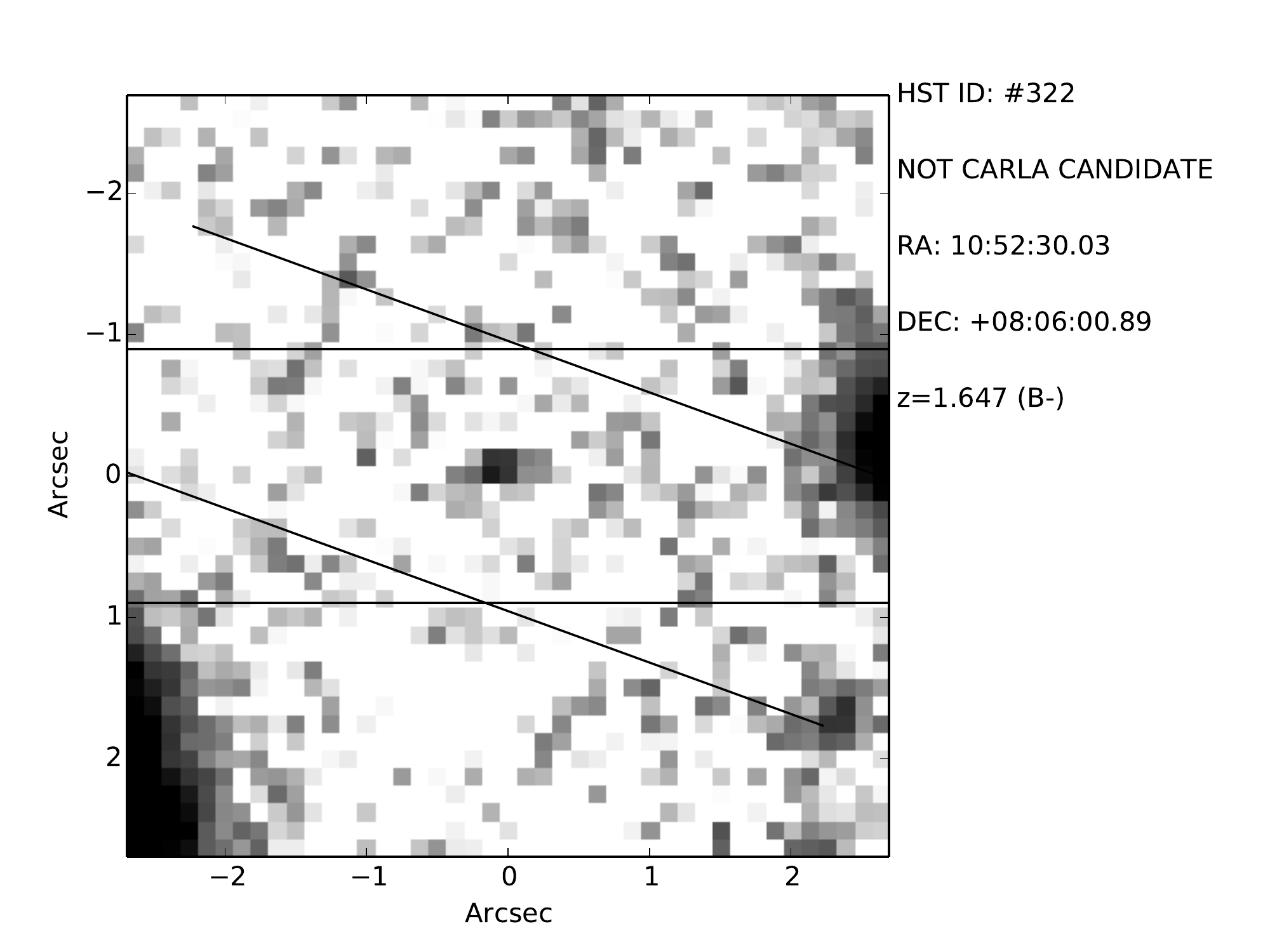} \mbox{(c)}}%
}%
{%
\setlength{\fboxsep}{0pt}%
\setlength{\fboxrule}{1pt}%
\fbox{\includegraphics[page=2, scale=0.24]{CARLA_J1052+0806_326.pdf} \hfill \includegraphics[page=1, scale=0.20]{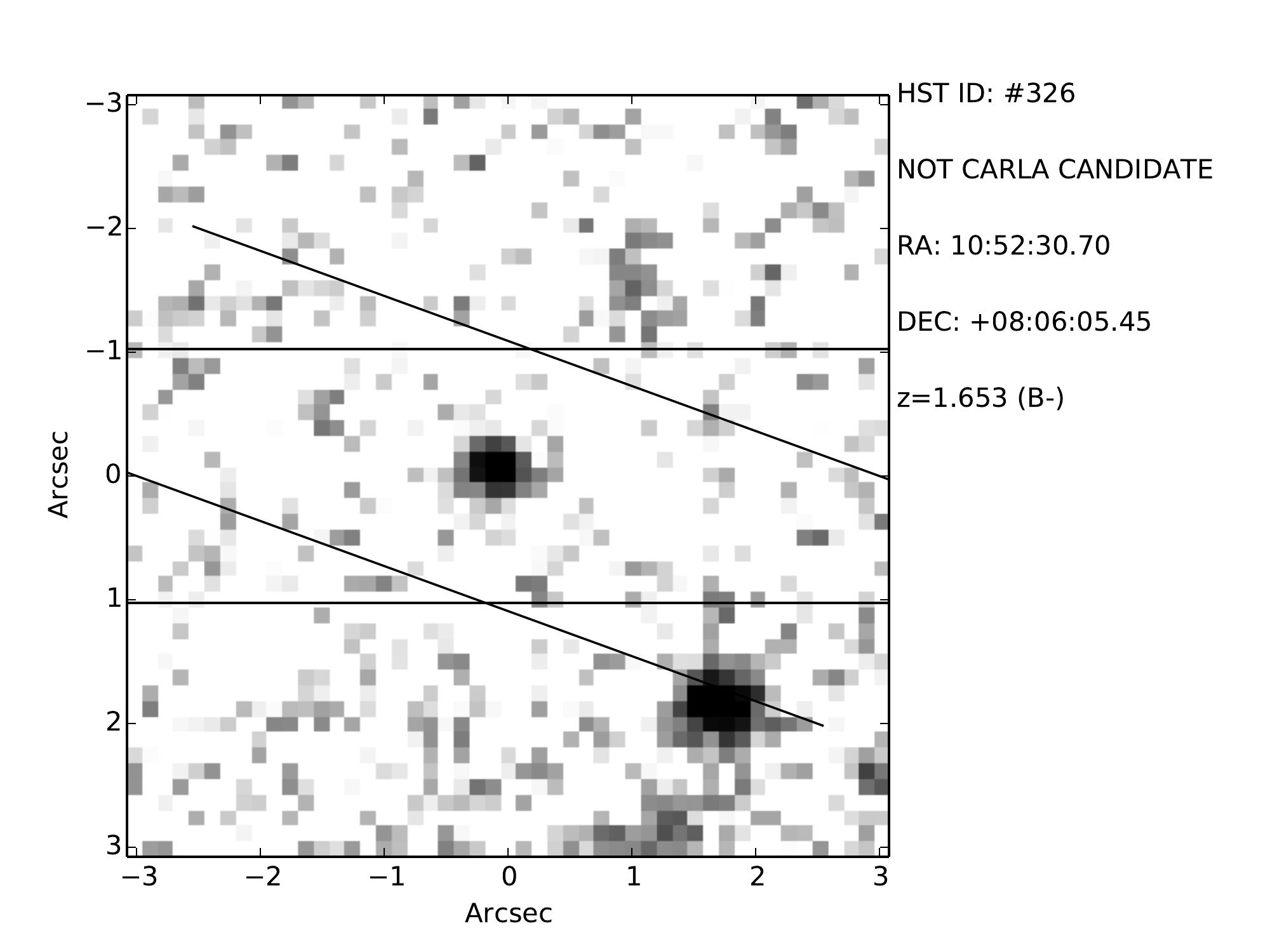} \mbox{(d)}}%
}\\%
{%
\setlength{\fboxsep}{0pt}%
\setlength{\fboxrule}{1pt}%
\fbox{\includegraphics[page=2, scale=0.24]{CARLA_J1052+0806_338.pdf} \hfill \includegraphics[page=1, scale=0.20]{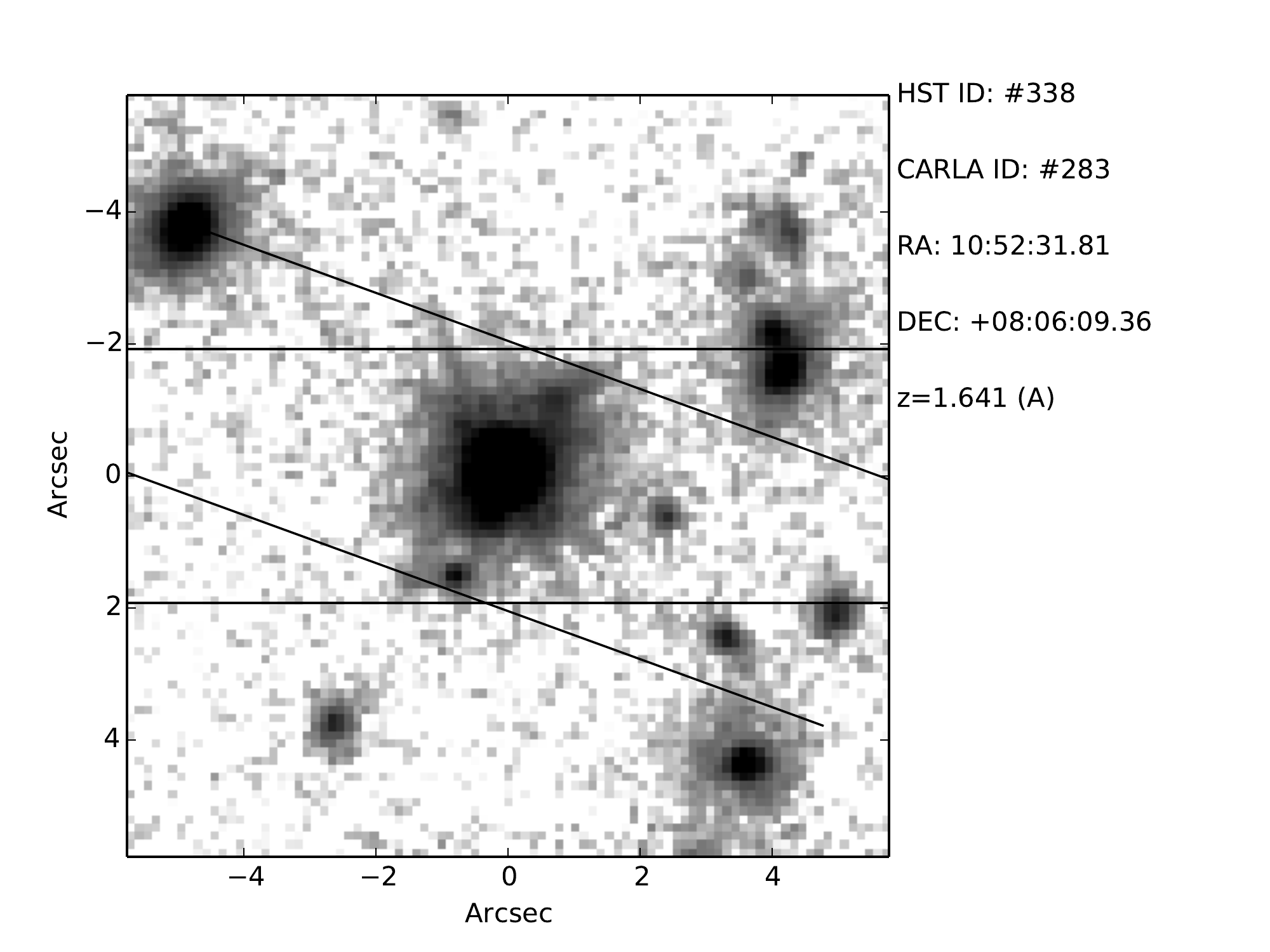} \mbox{(e)}}%
}%
{%
\setlength{\fboxsep}{0pt}%
\setlength{\fboxrule}{1pt}%
\fbox{\includegraphics[page=2, scale=0.24]{CARLA_J1052+0806_532.pdf} \hfill \includegraphics[page=1, scale=0.20]{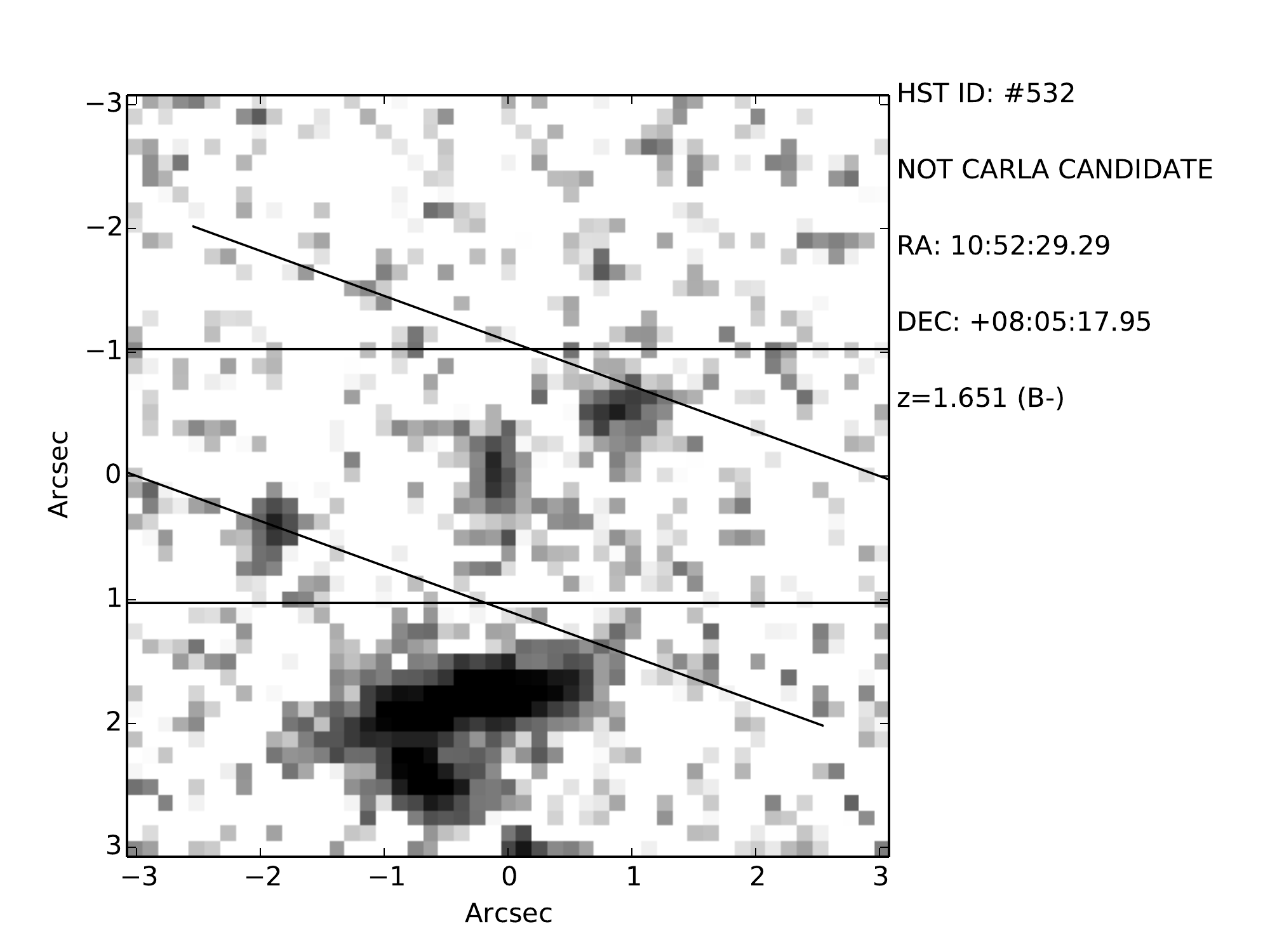} \mbox{(f)}}%
}\\%
\caption[CARLA~J1052+0806 member spectra]{CARLA~J1052+0806 member spectra.}
\label{fig:J1052+0806spectra}
\mbox{}\\
\end{figure*}

\begin{figure*}[!ht]
{%
\setlength{\fboxsep}{0pt}%
\setlength{\fboxrule}{1pt}%
\fbox{\includegraphics[page=2, scale=0.24]{CARLA_J1103+3449_93.pdf} \hfill \includegraphics[page=1, scale=0.20]{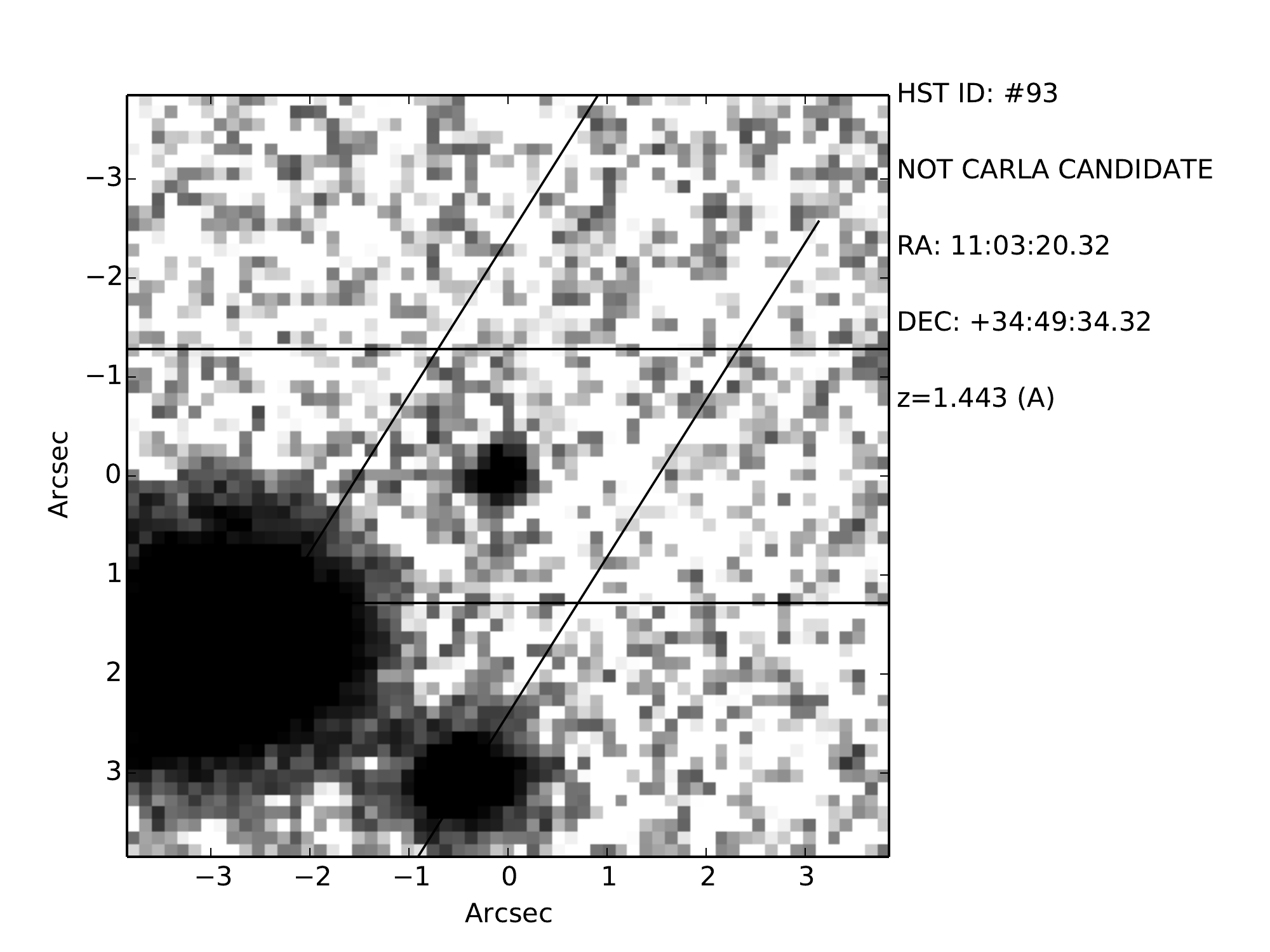} \mbox{(a)}}%
}%
{%
\setlength{\fboxsep}{0pt}%
\setlength{\fboxrule}{1pt}%
\fbox{\includegraphics[page=2, scale=0.24]{CARLA_J1103+3449_183.pdf} \hfill \includegraphics[page=1, scale=0.20]{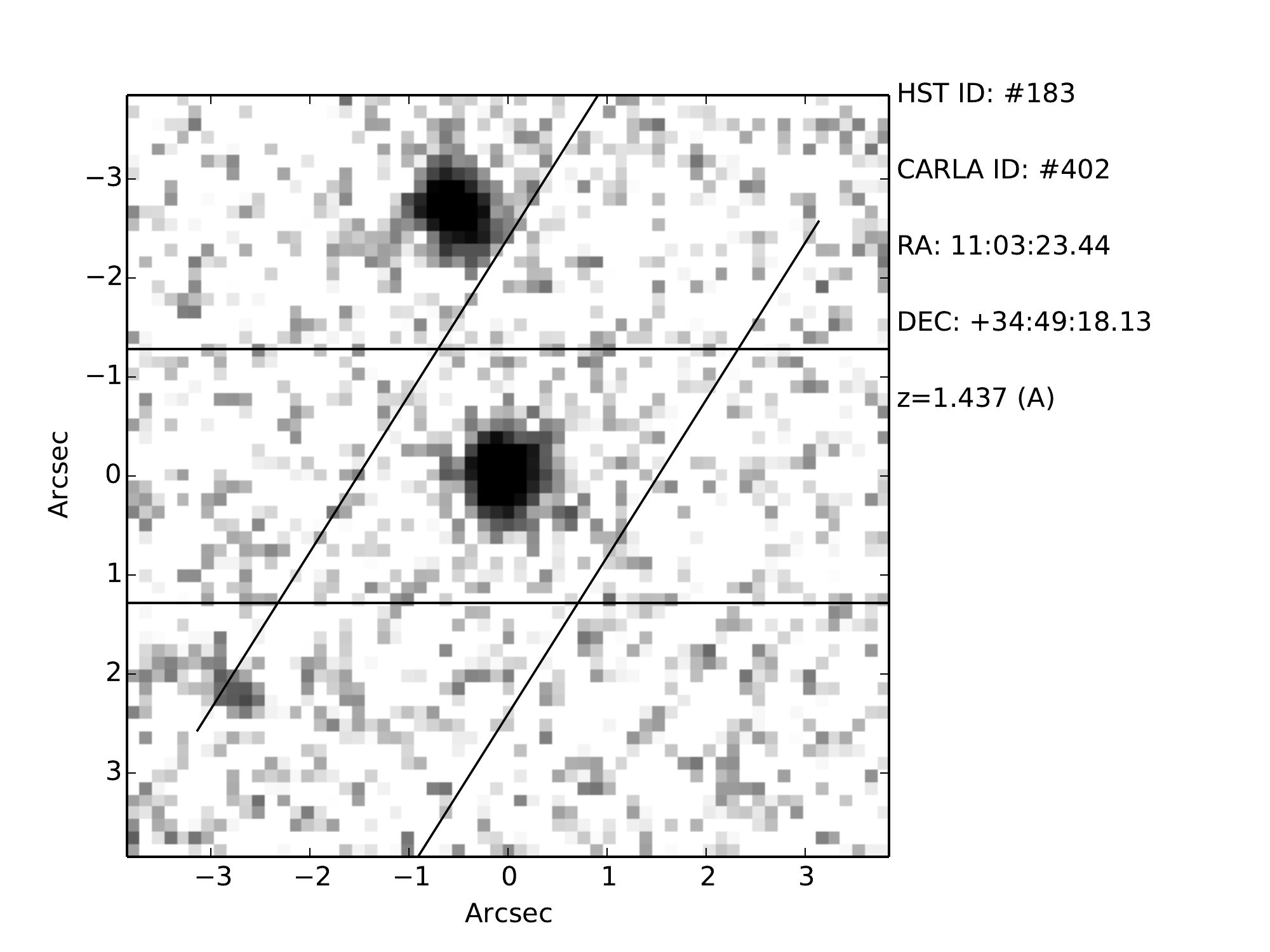} \mbox{(b)}}%
}\\%
{%
\setlength{\fboxsep}{0pt}%
\setlength{\fboxrule}{1pt}%
\fbox{\includegraphics[page=2, scale=0.24]{CARLA_J1103+3449_199.pdf} \hfill \includegraphics[page=1, scale=0.20]{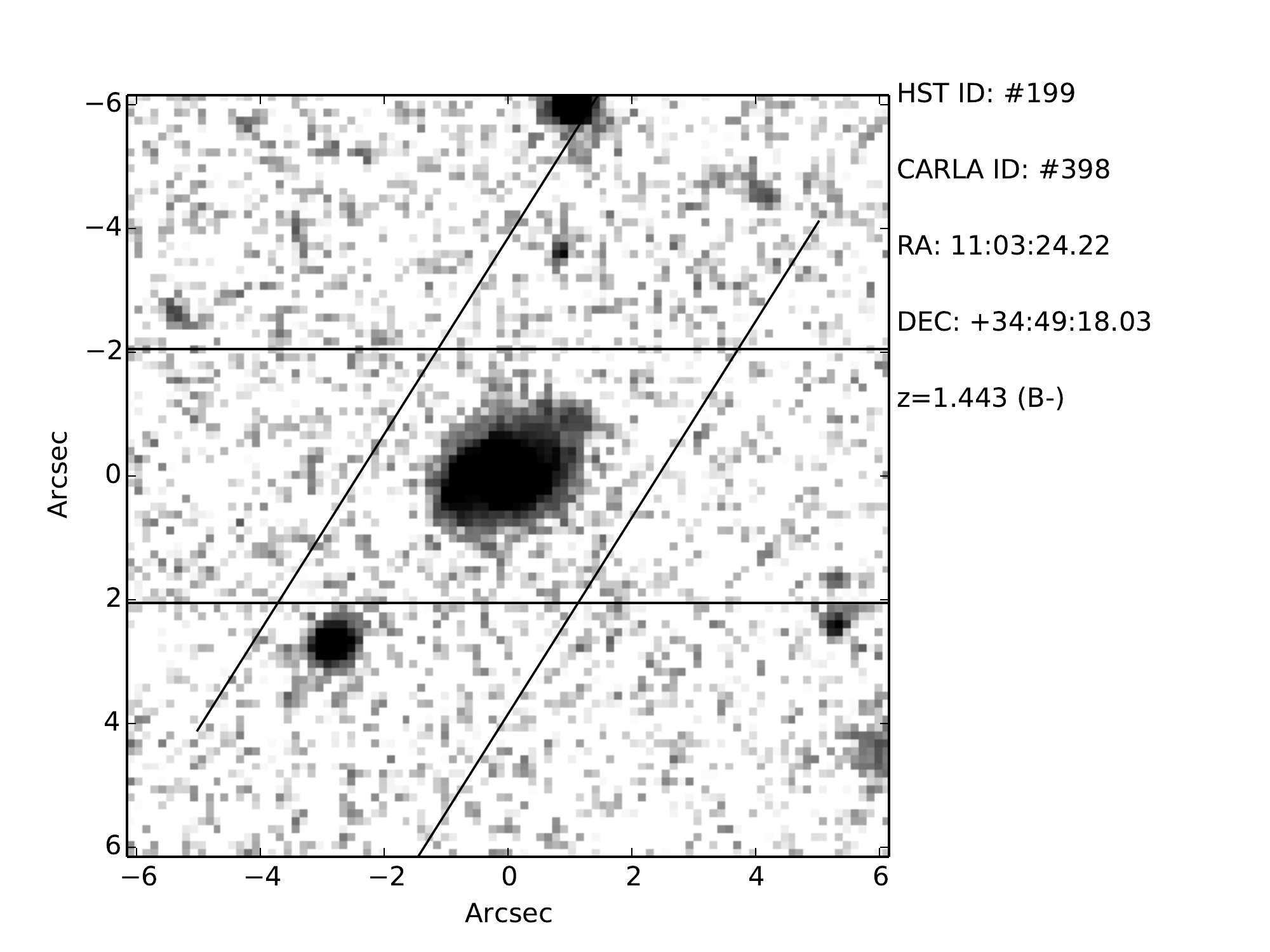} \mbox{(c)}}%
}%
{%
\setlength{\fboxsep}{0pt}%
\setlength{\fboxrule}{1pt}%
\fbox{\includegraphics[page=2, scale=0.24]{CARLA_J1103+3449_279.pdf} \hfill \includegraphics[page=1, scale=0.20]{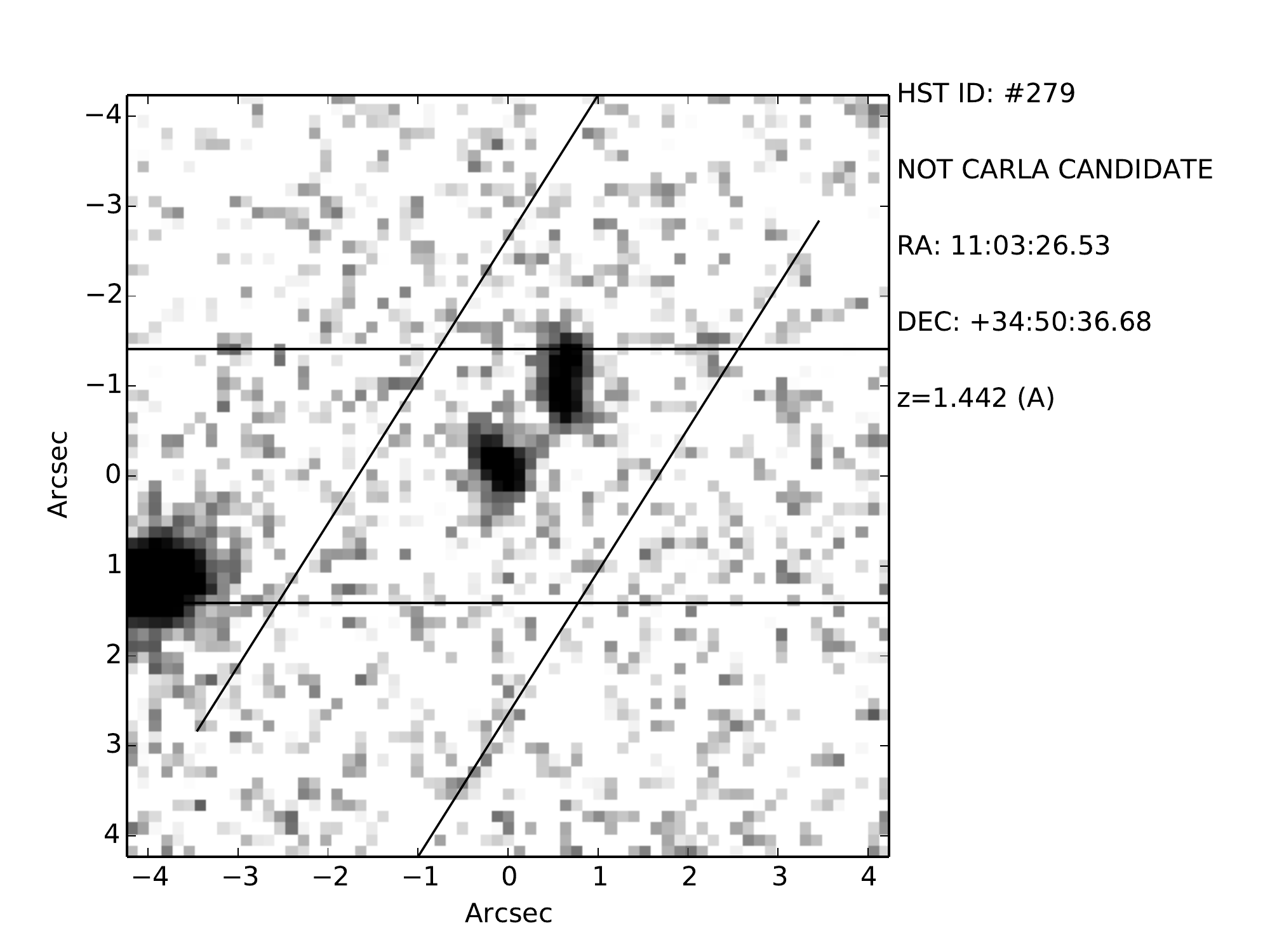} \mbox{(d)}}%
}\\%
{%
\setlength{\fboxsep}{0pt}%
\setlength{\fboxrule}{1pt}%
\fbox{\includegraphics[page=2, scale=0.24]{CARLA_J1103+3449_283.pdf} \hfill \includegraphics[page=1, scale=0.20]{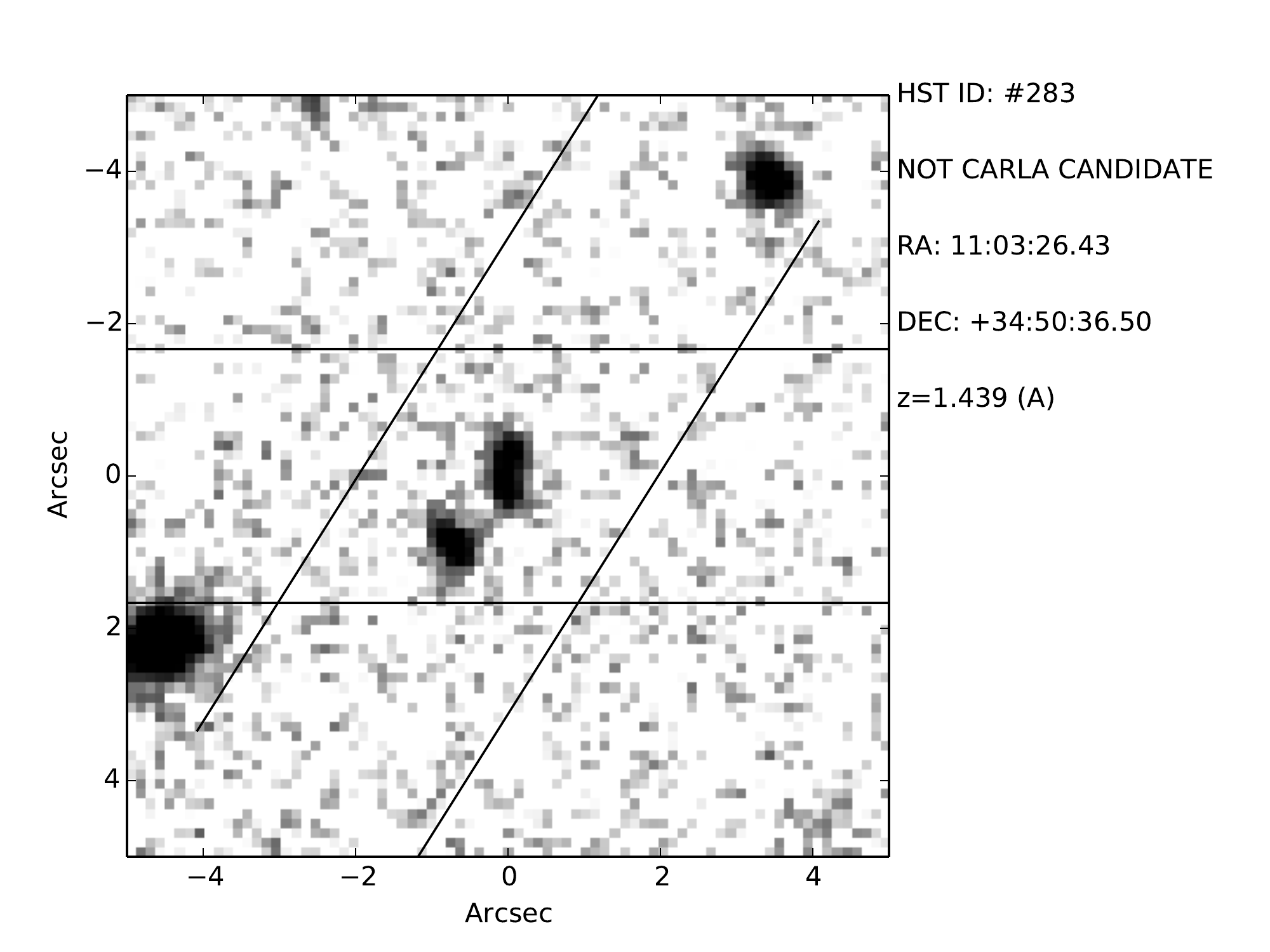} \mbox{(e)}}%
}%
{%
\setlength{\fboxsep}{0pt}%
\setlength{\fboxrule}{1pt}%
\fbox{\includegraphics[page=2, scale=0.24]{CARLA_J1103+3449_320.pdf} \hfill \includegraphics[page=1, scale=0.20]{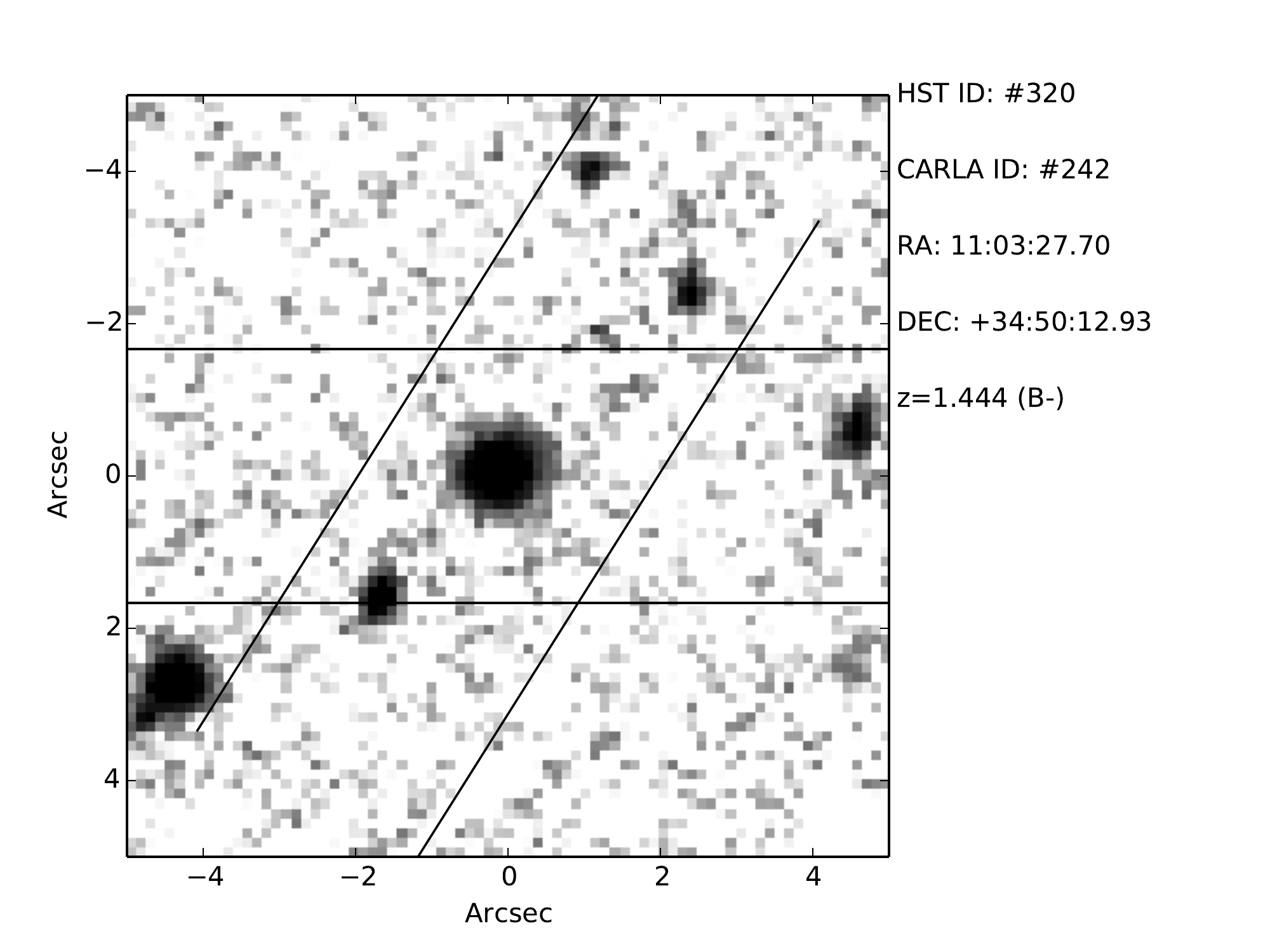} \mbox{(f)}}%
}\\%
{%
\setlength{\fboxsep}{0pt}%
\setlength{\fboxrule}{1pt}%
\fbox{\includegraphics[page=2, scale=0.24]{CARLA_J1103+3449_490.pdf} \hfill \includegraphics[page=1, scale=0.20]{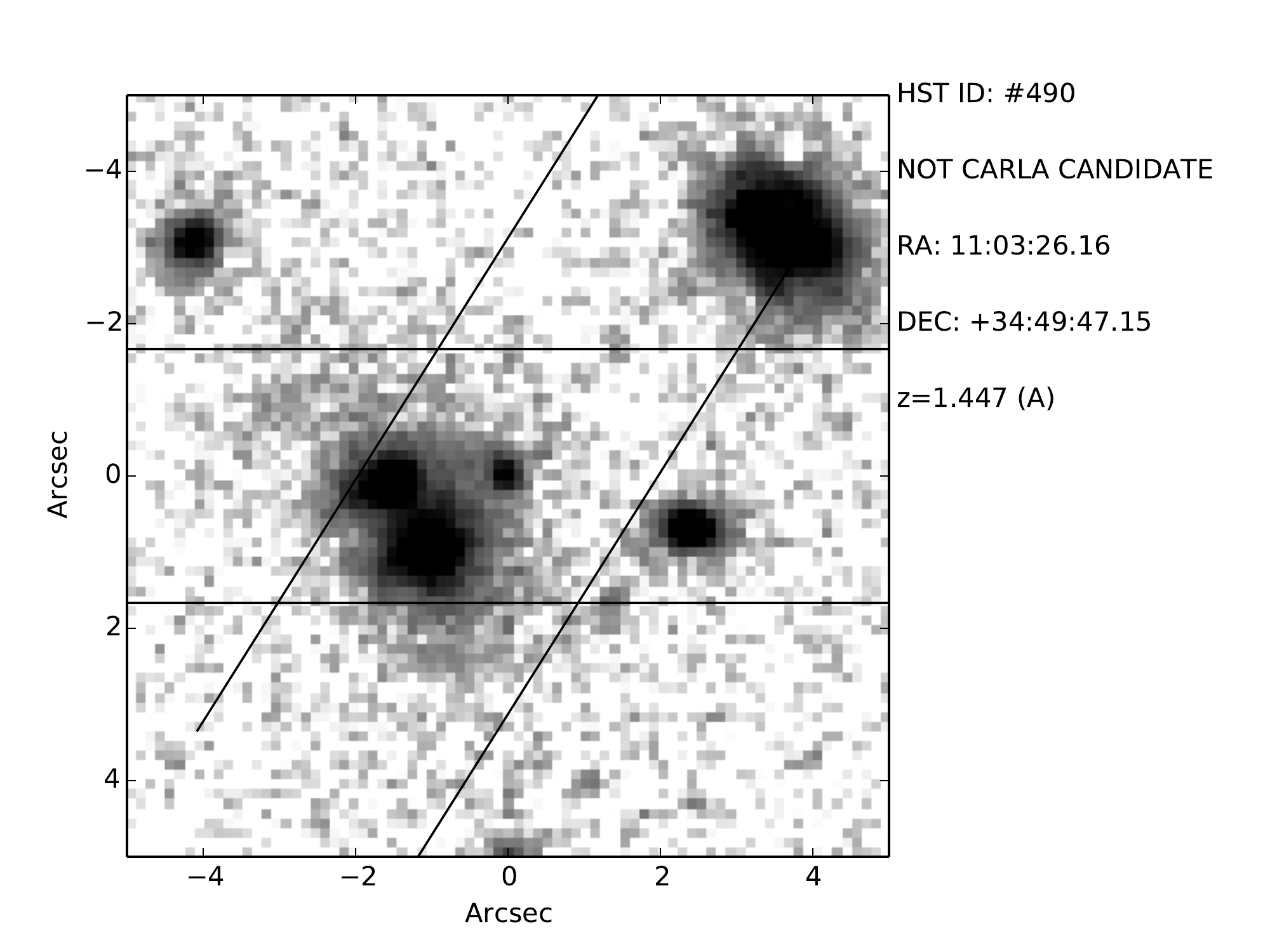} \mbox{(g)}}%
}%
{%
\setlength{\fboxsep}{0pt}%
\setlength{\fboxrule}{1pt}%
\fbox{\includegraphics[page=2, scale=0.24]{CARLA_J1103+3449_491.pdf} \hfill \includegraphics[page=1, scale=0.20]{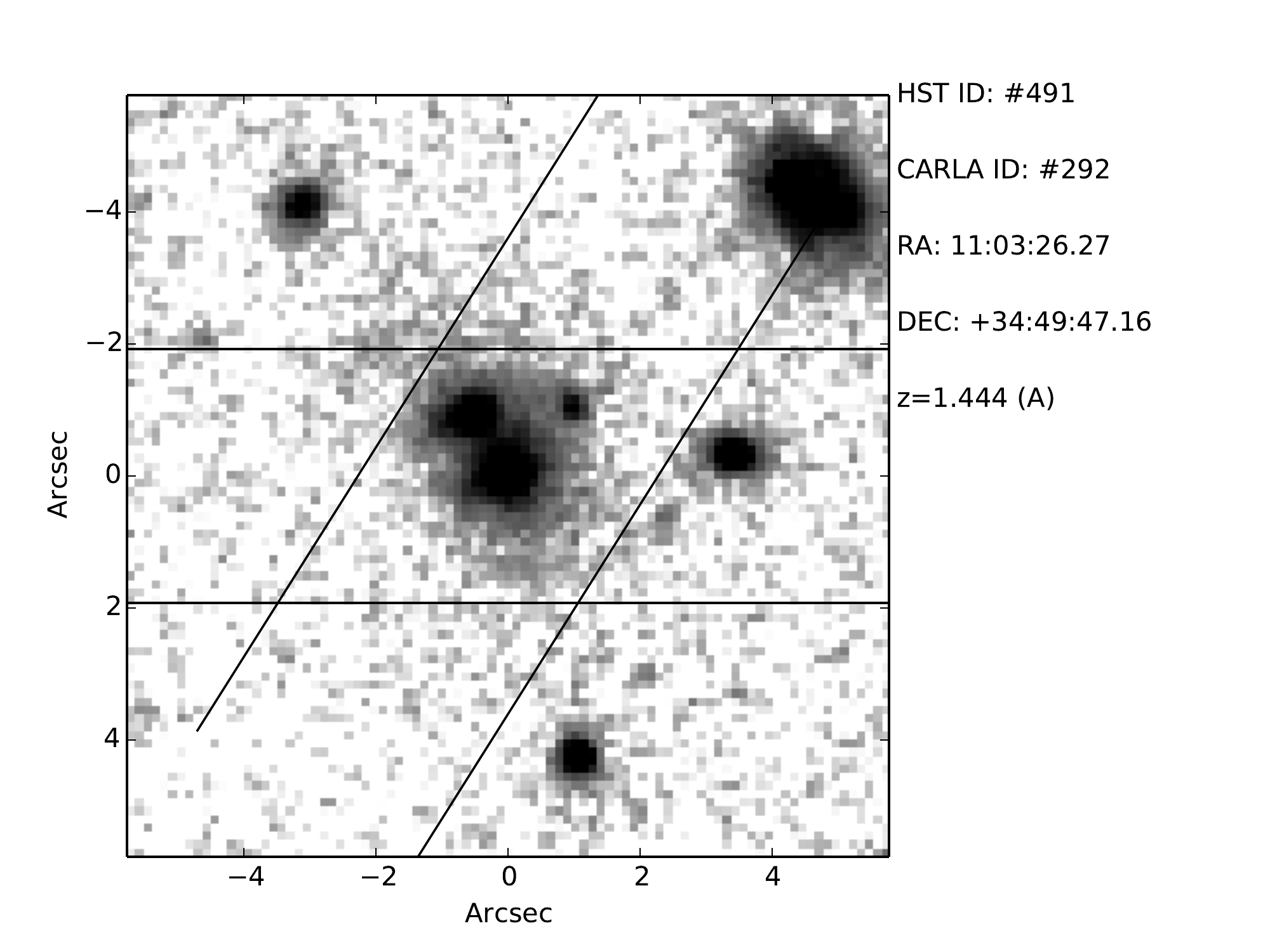} \mbox{(h)}}%
}\\%
\caption[CARLA~J1103+3449 member spectra]{CARLA~J1103+3449 member spectra.}
\label{fig:J1103+3449spectra}
\end{figure*}

\begin{figure*}[!ht]
{%
\setlength{\fboxsep}{0pt}%
\setlength{\fboxrule}{1pt}%
\fbox{\includegraphics[page=2, scale=0.24]{CARLA_J1129+0951_126.pdf} \hfill \includegraphics[page=1, scale=0.20]{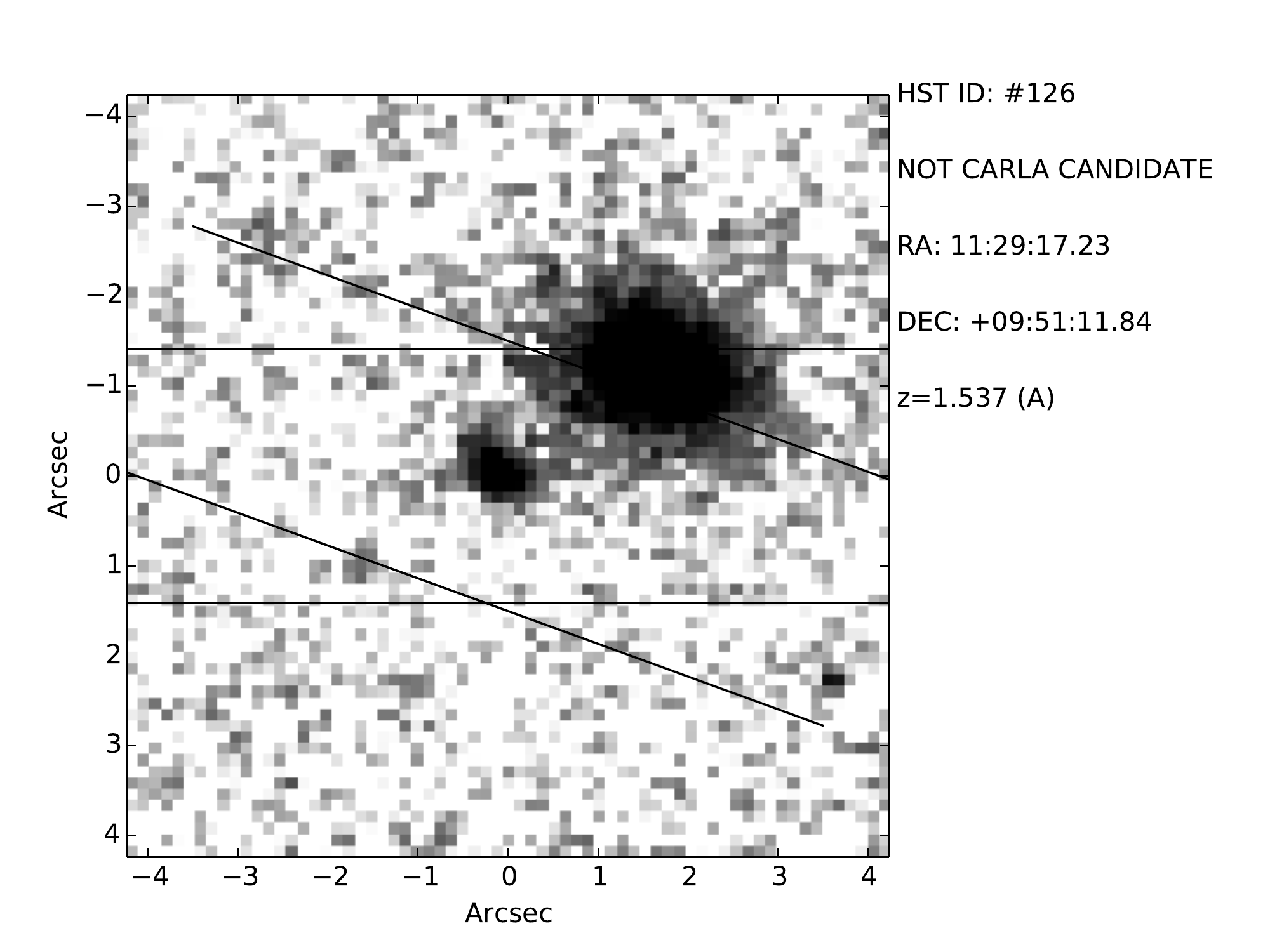} \mbox{(a)}}%
}%
{%
\setlength{\fboxsep}{0pt}%
\setlength{\fboxrule}{1pt}%
\fbox{\includegraphics[page=2, scale=0.24]{CARLA_J1129+0951_240.pdf} \hfill \includegraphics[page=1, scale=0.20]{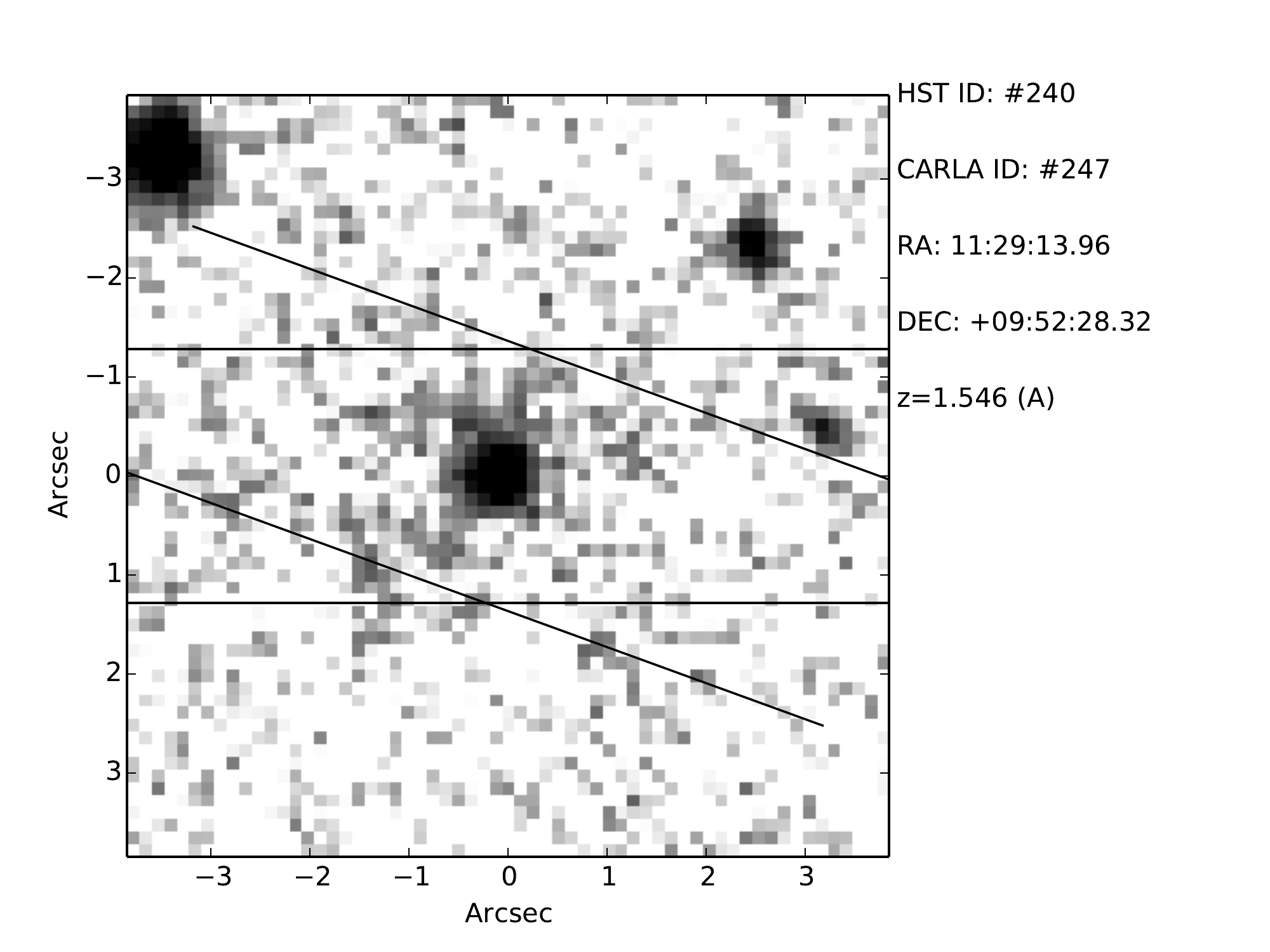} \mbox{(b)}}%
}\\%
{%
\setlength{\fboxsep}{0pt}%
\setlength{\fboxrule}{1pt}%
\fbox{\includegraphics[page=2, scale=0.24]{CARLA_J1129+0951_261.pdf} \hfill \includegraphics[page=1, scale=0.20]{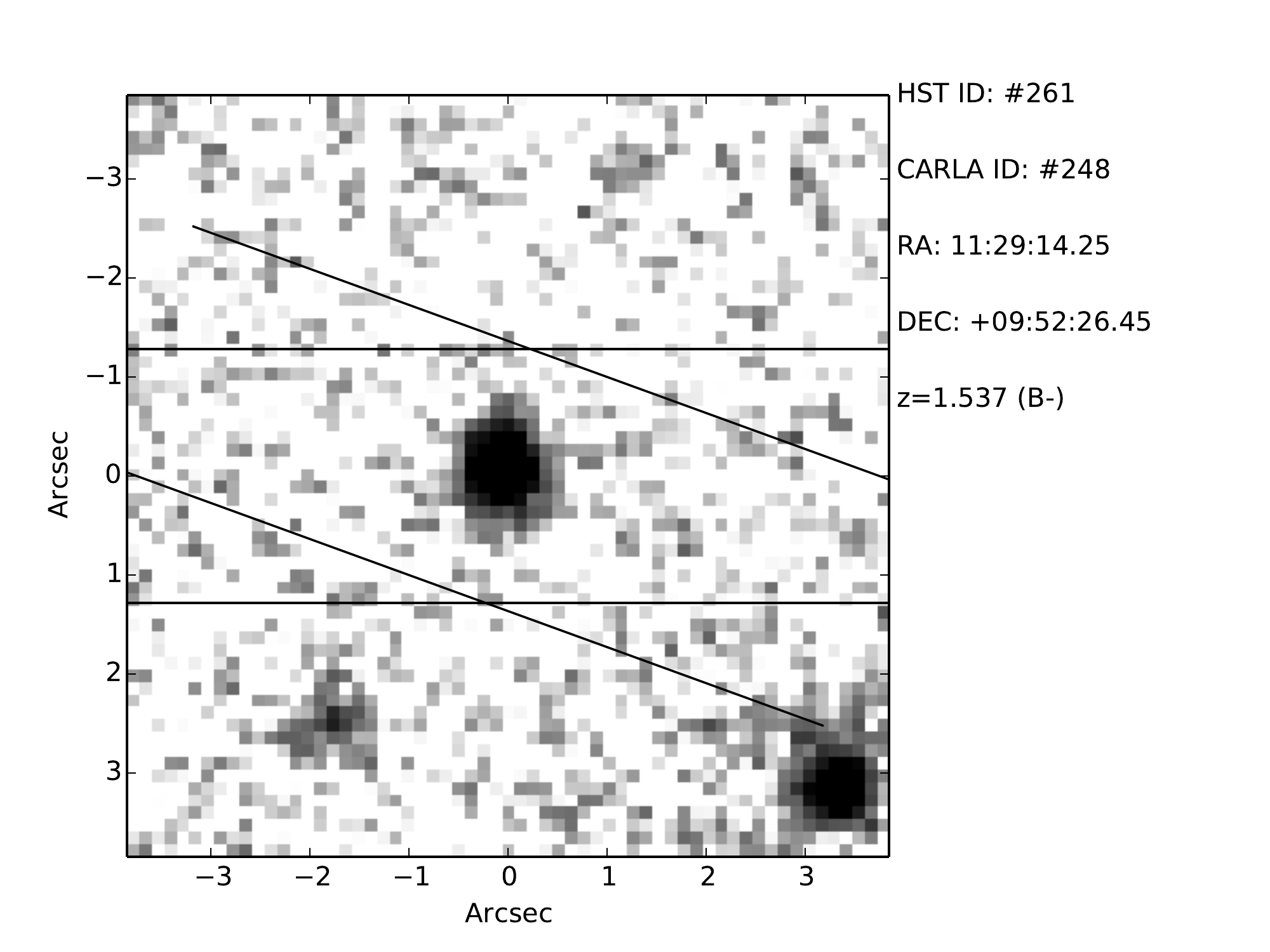} \mbox{(c)}}%
}%
{%
\setlength{\fboxsep}{0pt}%
\setlength{\fboxrule}{1pt}%
\fbox{\includegraphics[page=2, scale=0.24]{CARLA_J1129+0951_335.pdf} \hfill \includegraphics[page=1, scale=0.20]{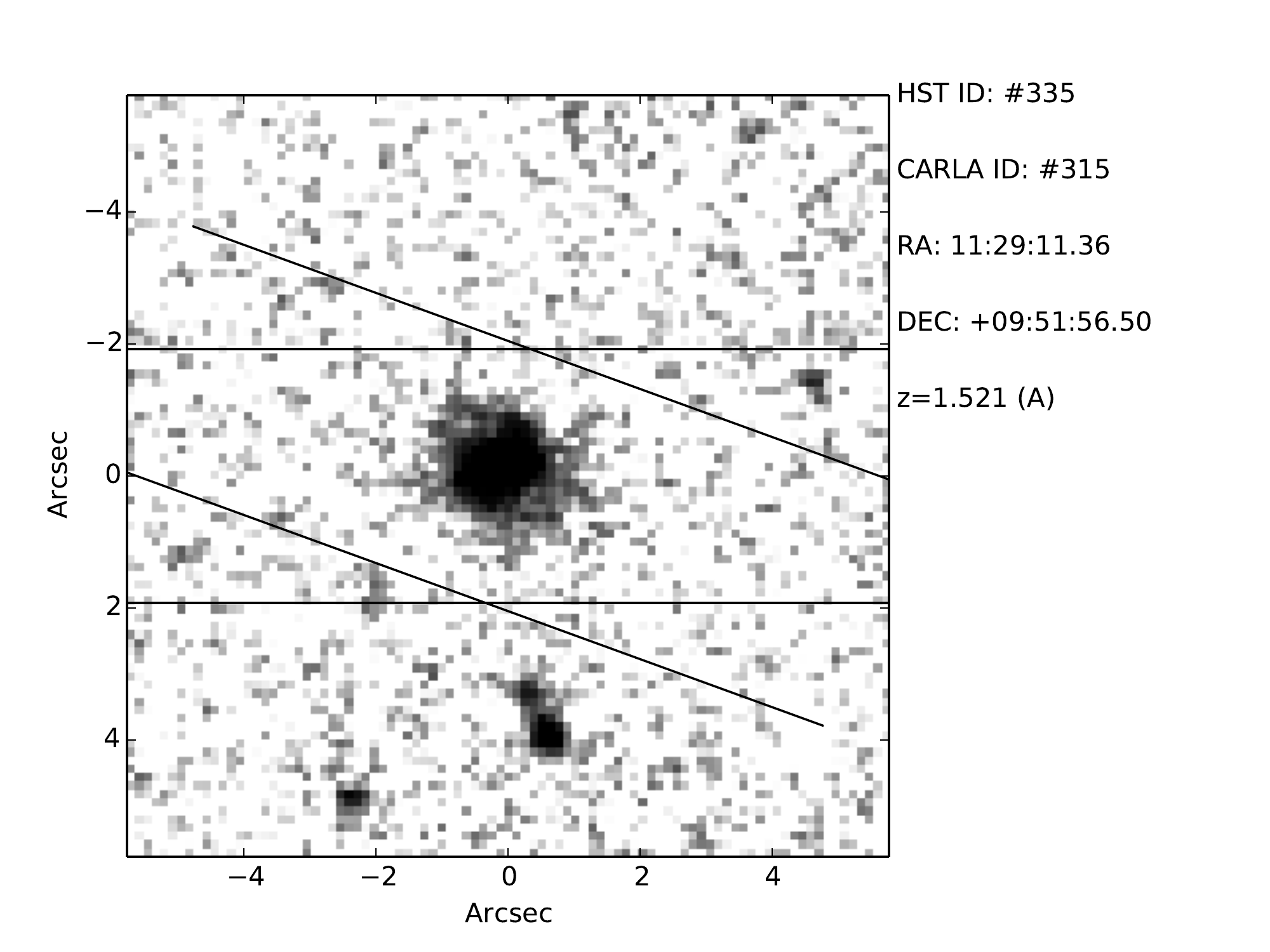} \mbox{(d)}}%
}\\%
{%
\setlength{\fboxsep}{0pt}%
\setlength{\fboxrule}{1pt}%
\fbox{\includegraphics[page=2, scale=0.24]{CARLA_J1129+0951_377.pdf} \hfill \includegraphics[page=1, scale=0.20]{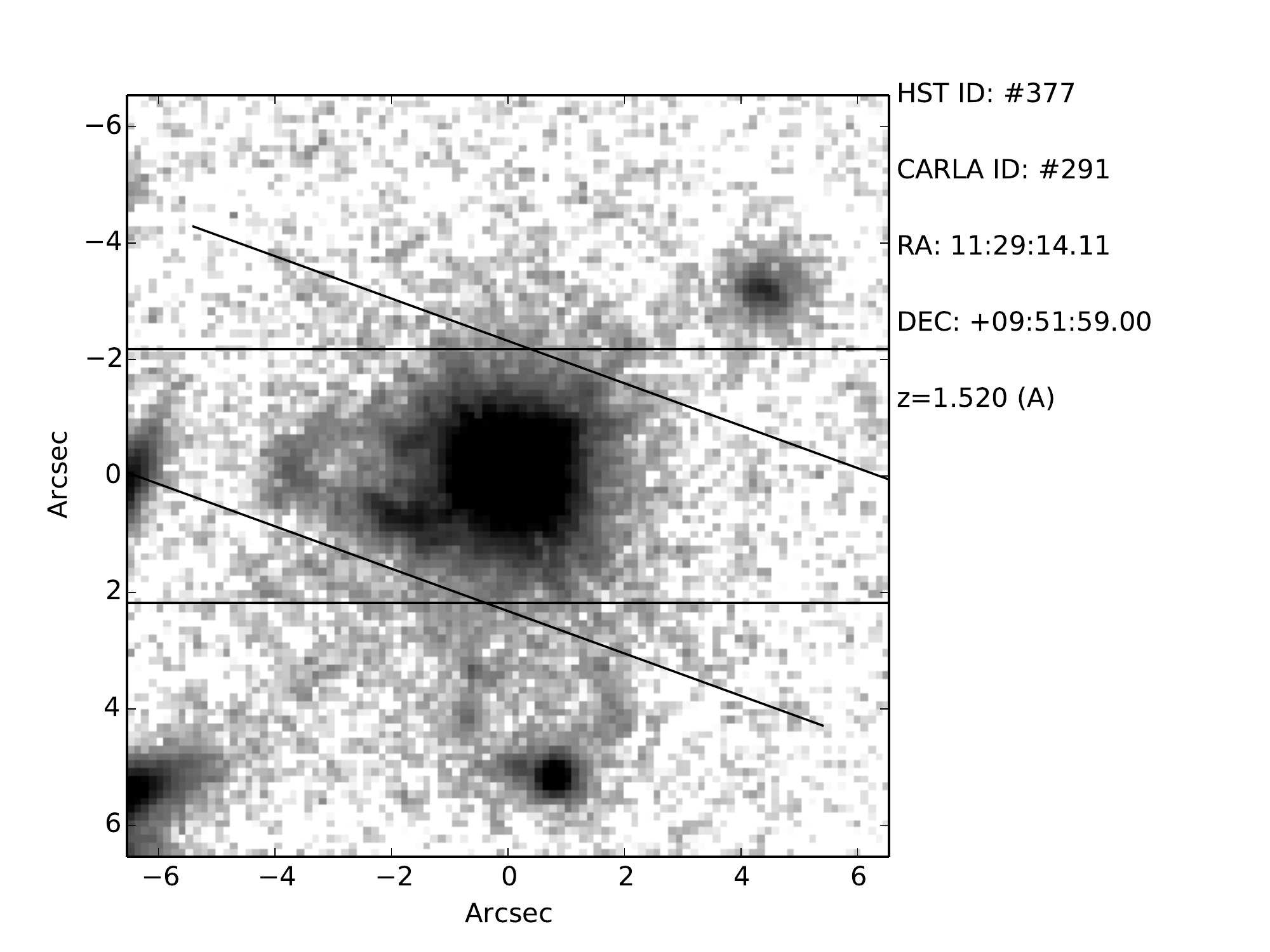} \mbox{(e)}}%
}%
{%
\setlength{\fboxsep}{0pt}%
\setlength{\fboxrule}{1pt}%
\fbox{\includegraphics[page=2, scale=0.24]{CARLA_J1129+0951_527.pdf} \hfill \includegraphics[page=1, scale=0.20]{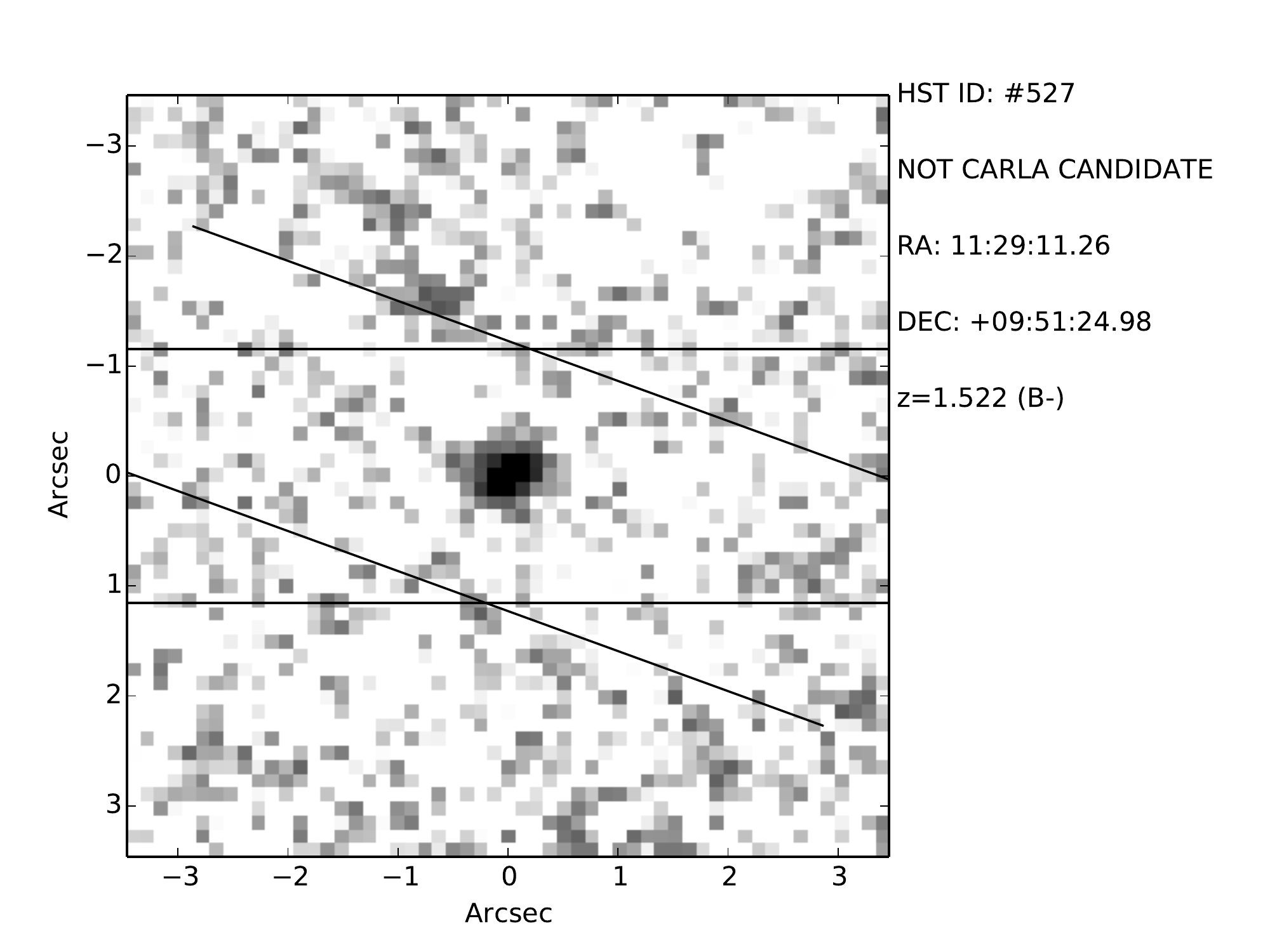} \mbox{(f)}}%
}\\%
{%
\setlength{\fboxsep}{0pt}%
\setlength{\fboxrule}{1pt}%
\fbox{\includegraphics[page=2, scale=0.24]{CARLA_J1129+0951_565.pdf} \hfill \includegraphics[page=1, scale=0.20]{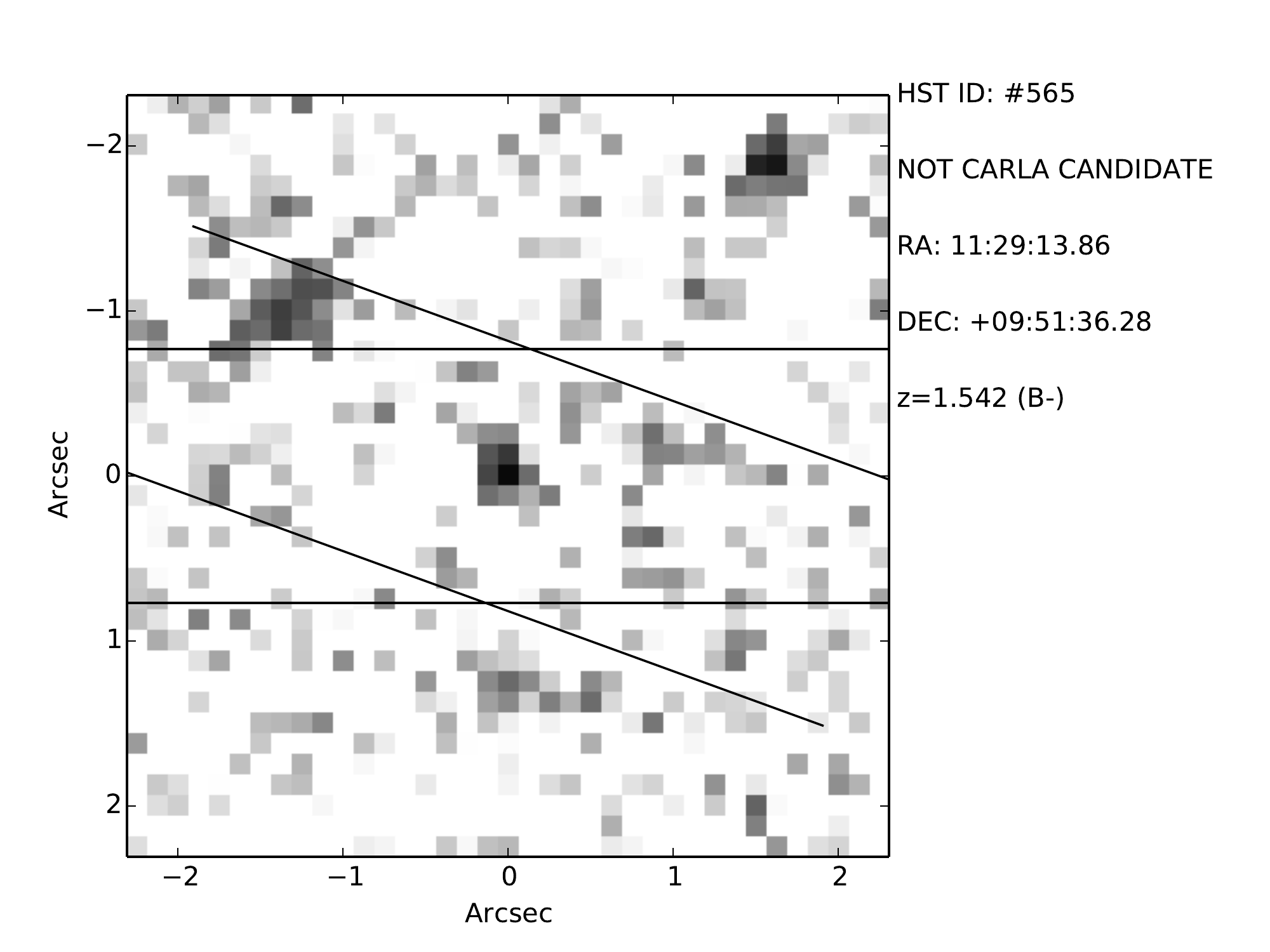} \mbox{(g)}}%
}%
{%
\setlength{\fboxsep}{0pt}%
\setlength{\fboxrule}{1pt}%
\fbox{\includegraphics[page=2, scale=0.24]{CARLA_J1129+0951_607.pdf} \hfill \includegraphics[page=1, scale=0.20]{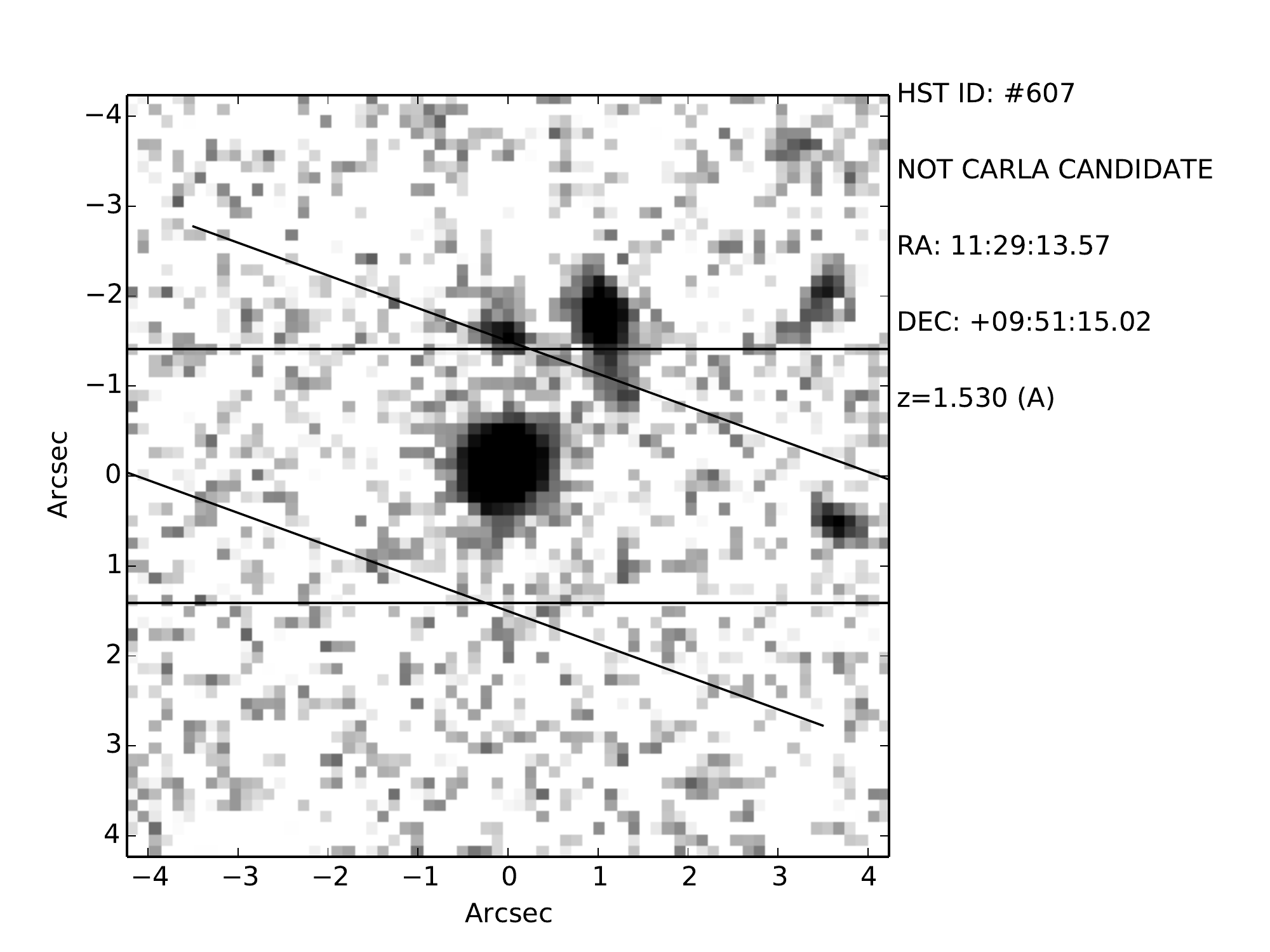} \mbox{(h)}}%
}\\%
{%
\setlength{\fboxsep}{0pt}%
\setlength{\fboxrule}{1pt}%
\fbox{\includegraphics[page=2, scale=0.24]{CARLA_J1129+0951_634.pdf} \hfill \includegraphics[page=1, scale=0.20]{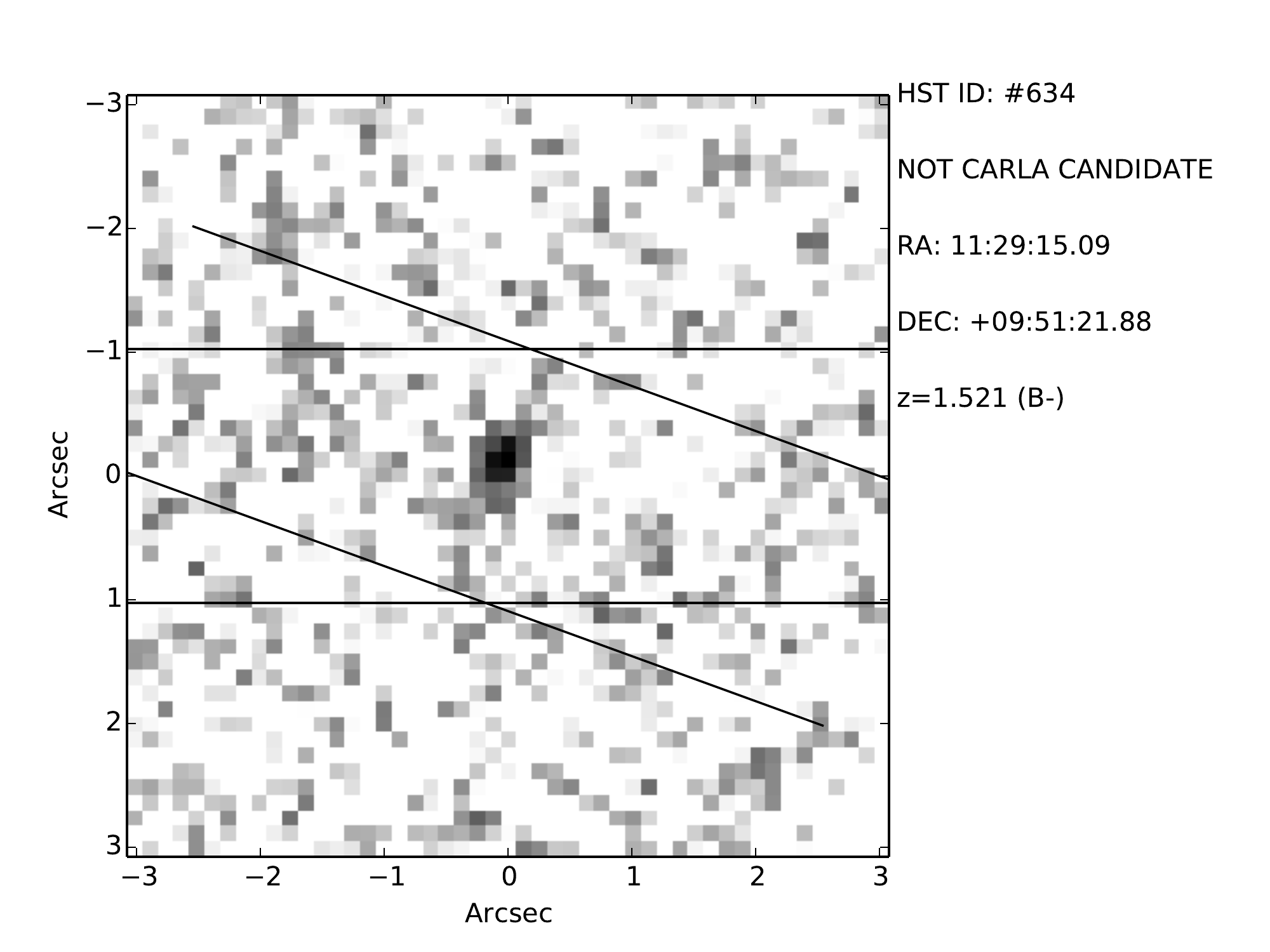} \mbox{(i)}}%
}%
{%
\setlength{\fboxsep}{0pt}%
\setlength{\fboxrule}{1pt}%
\fbox{\includegraphics[page=2, scale=0.24]{CARLA_J1129+0951_637.pdf} \hfill \includegraphics[page=1, scale=0.20]{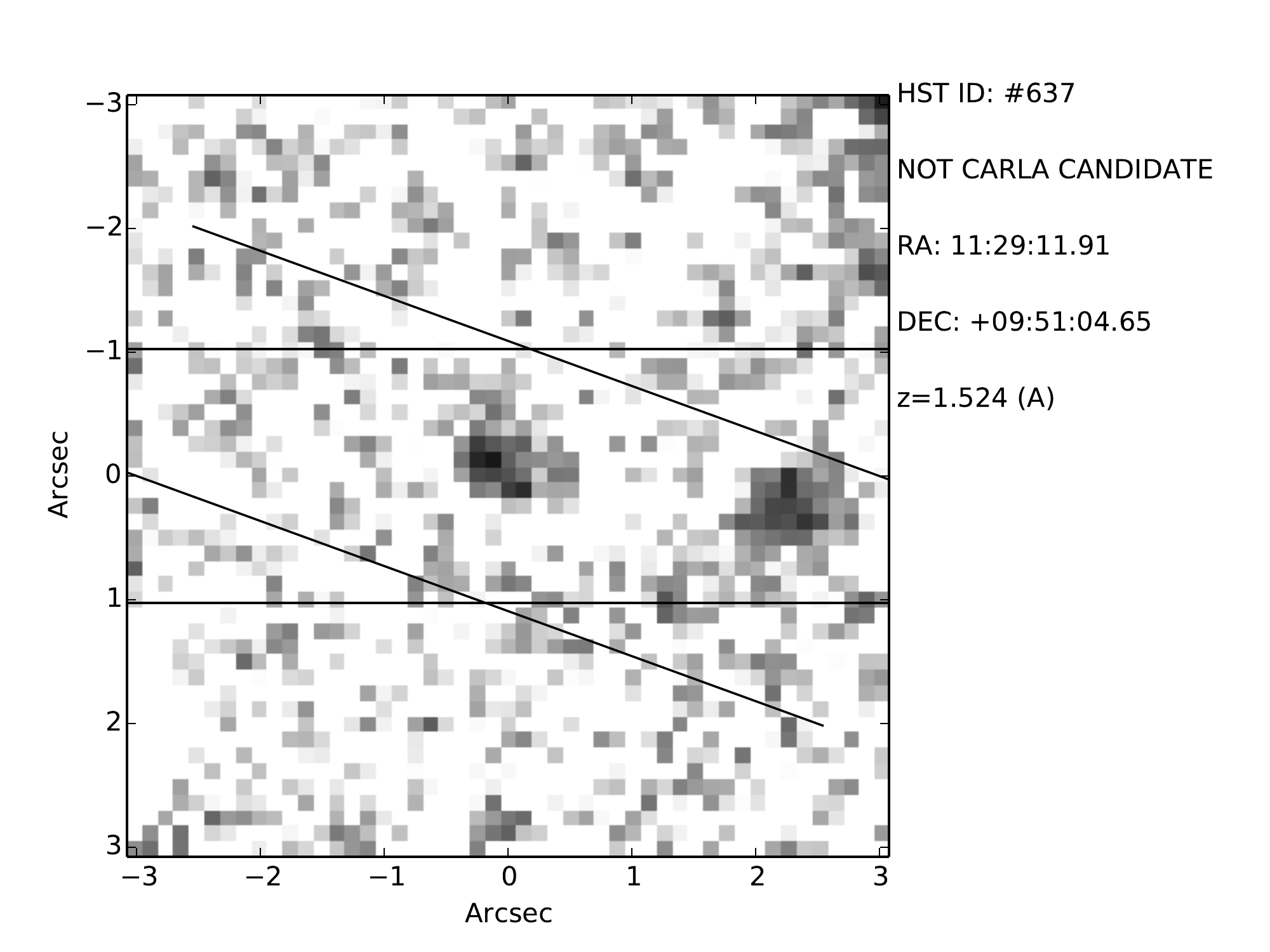} \mbox{(j)}}%
}\\%
{%
\setlength{\fboxsep}{0pt}%
\setlength{\fboxrule}{1pt}%
\fbox{\includegraphics[page=2, scale=0.24]{CARLA_J1129+0951_674.pdf} \hfill \includegraphics[page=1, scale=0.20]{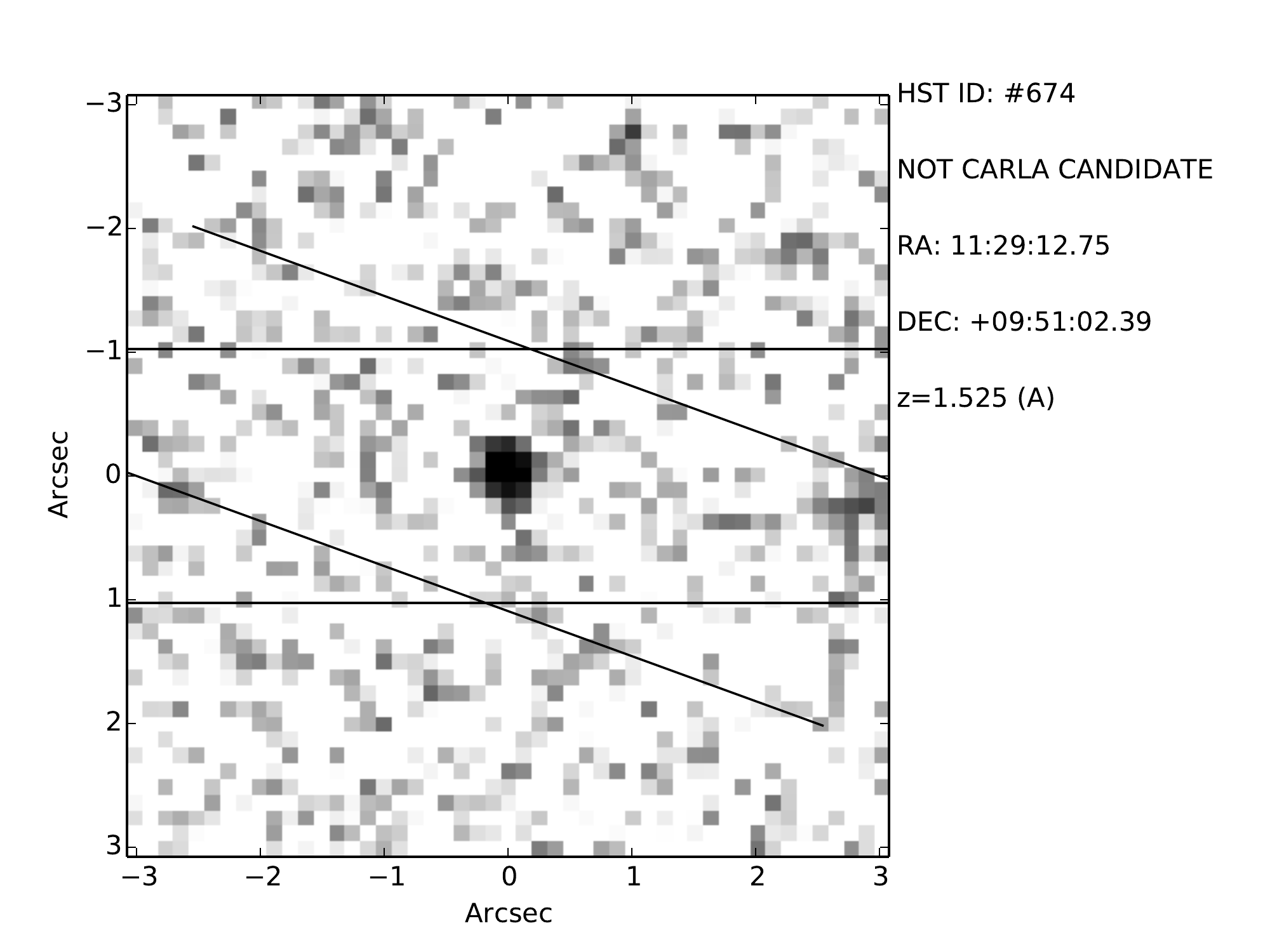} \mbox{(k)}}%
}%
{%
\setlength{\fboxsep}{0pt}%
\setlength{\fboxrule}{1pt}%
\fbox{\includegraphics[page=2, scale=0.24]{CARLA_J1129+0951_805.pdf} \hfill \includegraphics[page=1, scale=0.20]{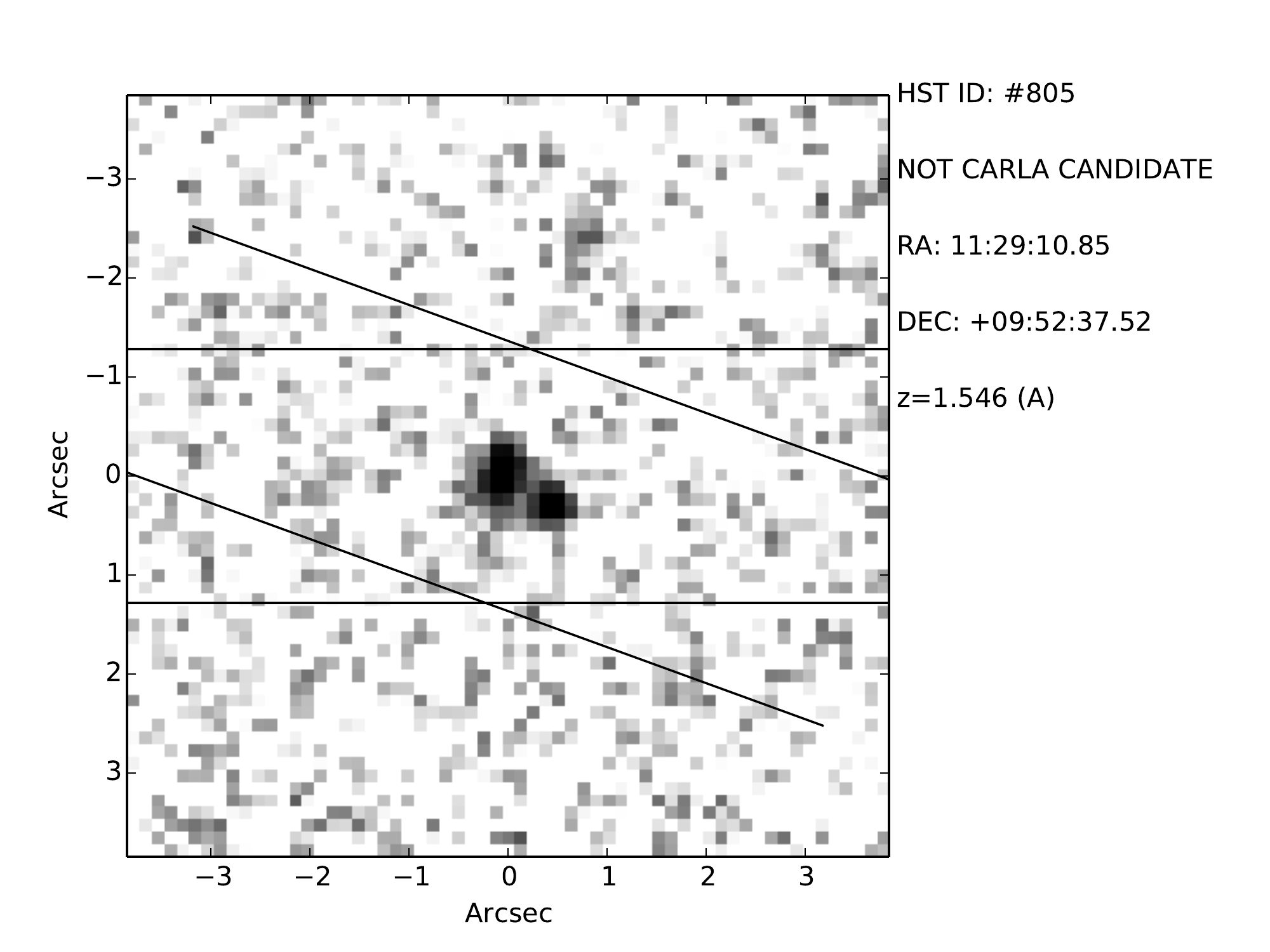} \mbox{(l)}}%
}\\%
\caption[CARLA~J1129+0951 member spectra]{CARLA~J1129+0951 member spectra.}
\label{fig:J1129+0951spectra}
\mbox{}\\
\end{figure*}

\begin{figure*}[!ht]
{%
\setlength{\fboxsep}{0pt}%
\setlength{\fboxrule}{1pt}%
\fbox{\includegraphics[page=2, scale=0.24]{CARLA_J1131-2705_115.pdf} \hfill \includegraphics[page=1, scale=0.20]{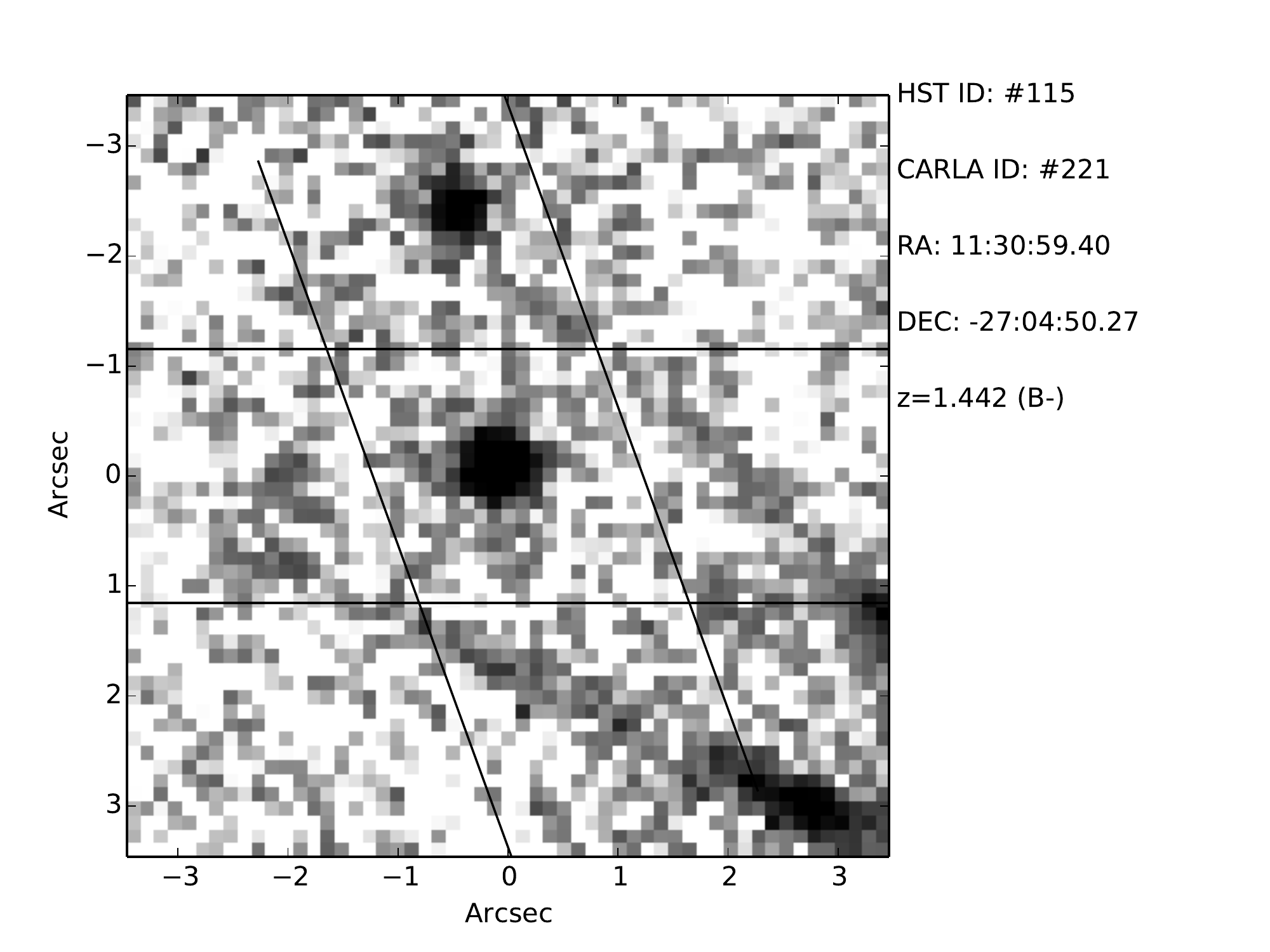} \mbox{(a)}}%
}%
{%
\setlength{\fboxsep}{0pt}%
\setlength{\fboxrule}{1pt}%
\fbox{\includegraphics[page=2, scale=0.24]{CARLA_J1131-2705_227.pdf} \hfill \includegraphics[page=1, scale=0.20]{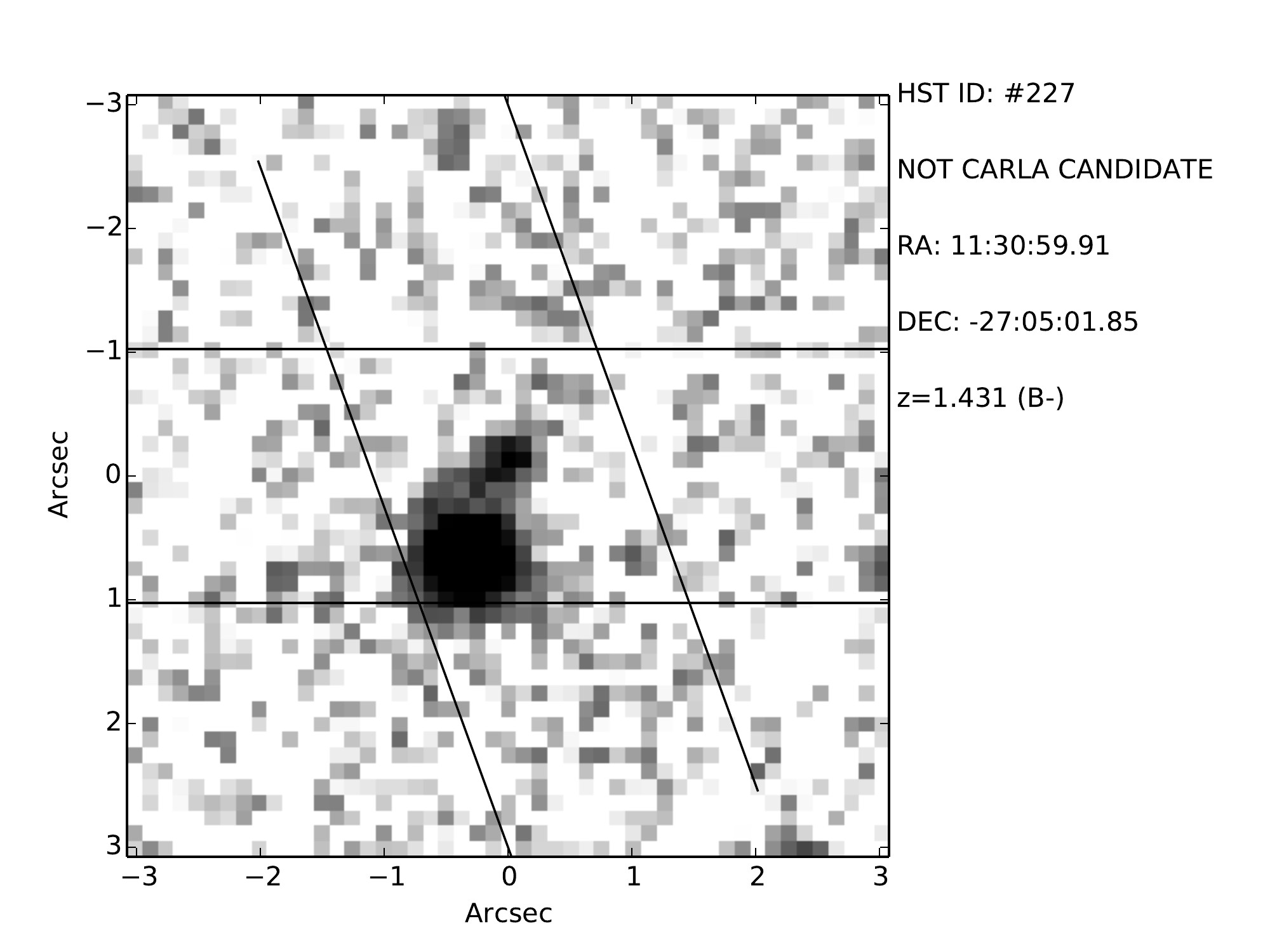} \mbox{(b)}}%
}\\%
{%
\setlength{\fboxsep}{0pt}%
\setlength{\fboxrule}{1pt}%
\fbox{\includegraphics[page=2, scale=0.24]{CARLA_J1131-2705_276.pdf} \hfill \includegraphics[page=1, scale=0.20]{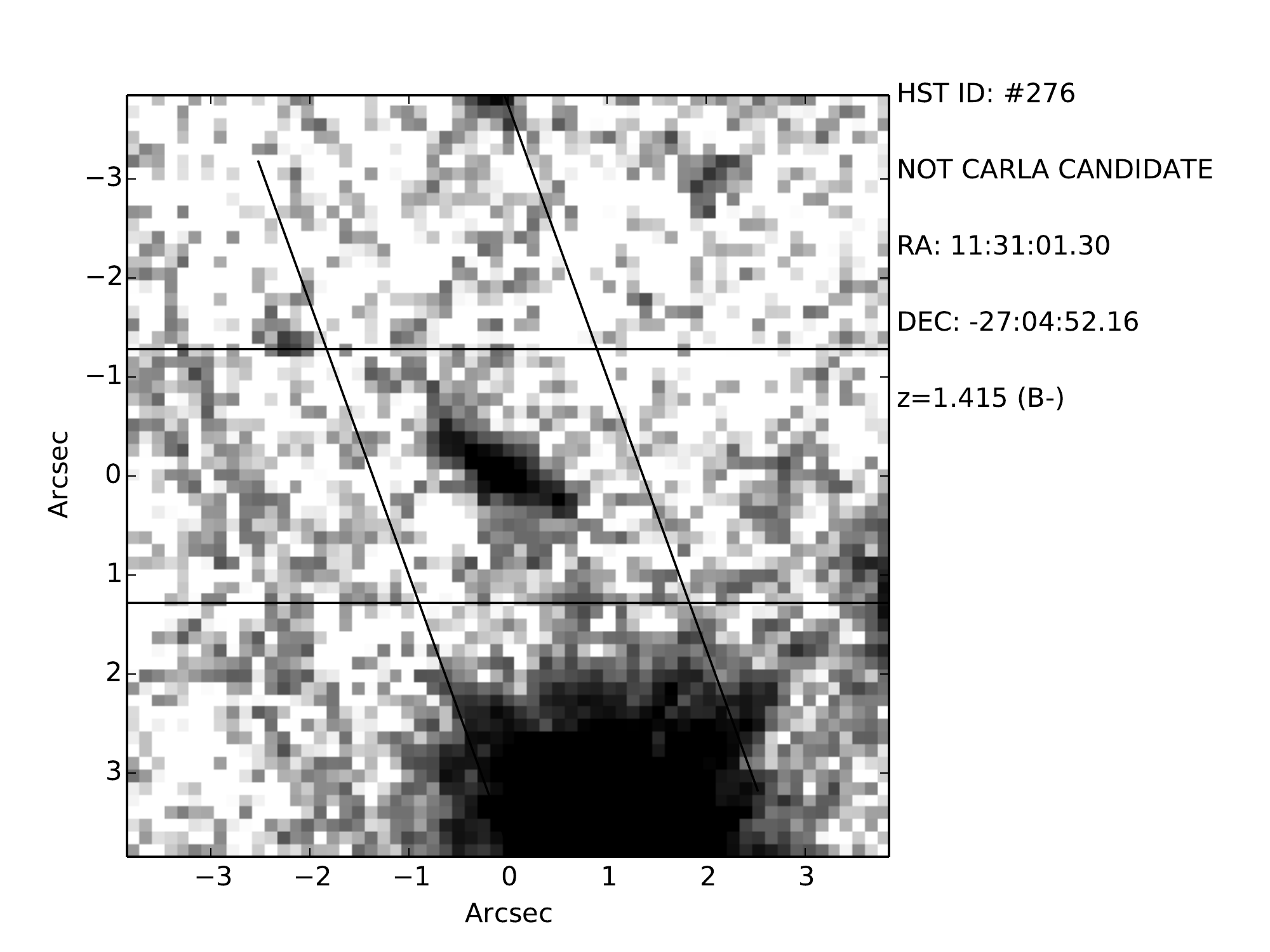} \mbox{(c)}}%
}%
{%
\setlength{\fboxsep}{0pt}%
\setlength{\fboxrule}{1pt}%
\fbox{\includegraphics[page=2, scale=0.24]{CARLA_J1131-2705_389.pdf} \hfill \includegraphics[page=1, scale=0.20]{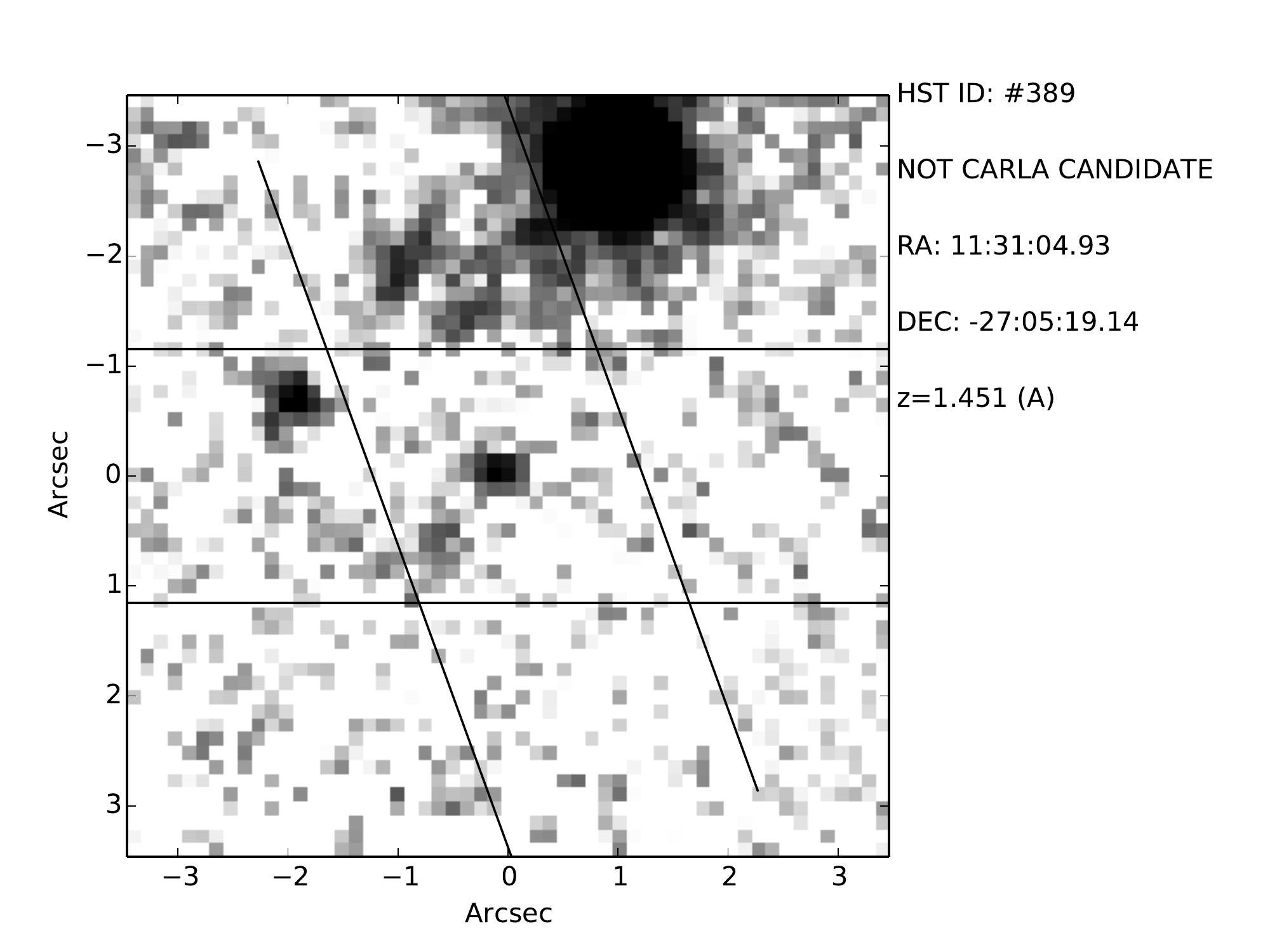} \mbox{(d)}}%
}\\%
{%
\setlength{\fboxsep}{0pt}%
\setlength{\fboxrule}{1pt}%
\fbox{\includegraphics[page=2, scale=0.24]{CARLA_J1131-2705_442.pdf} \hfill \includegraphics[page=1, scale=0.20]{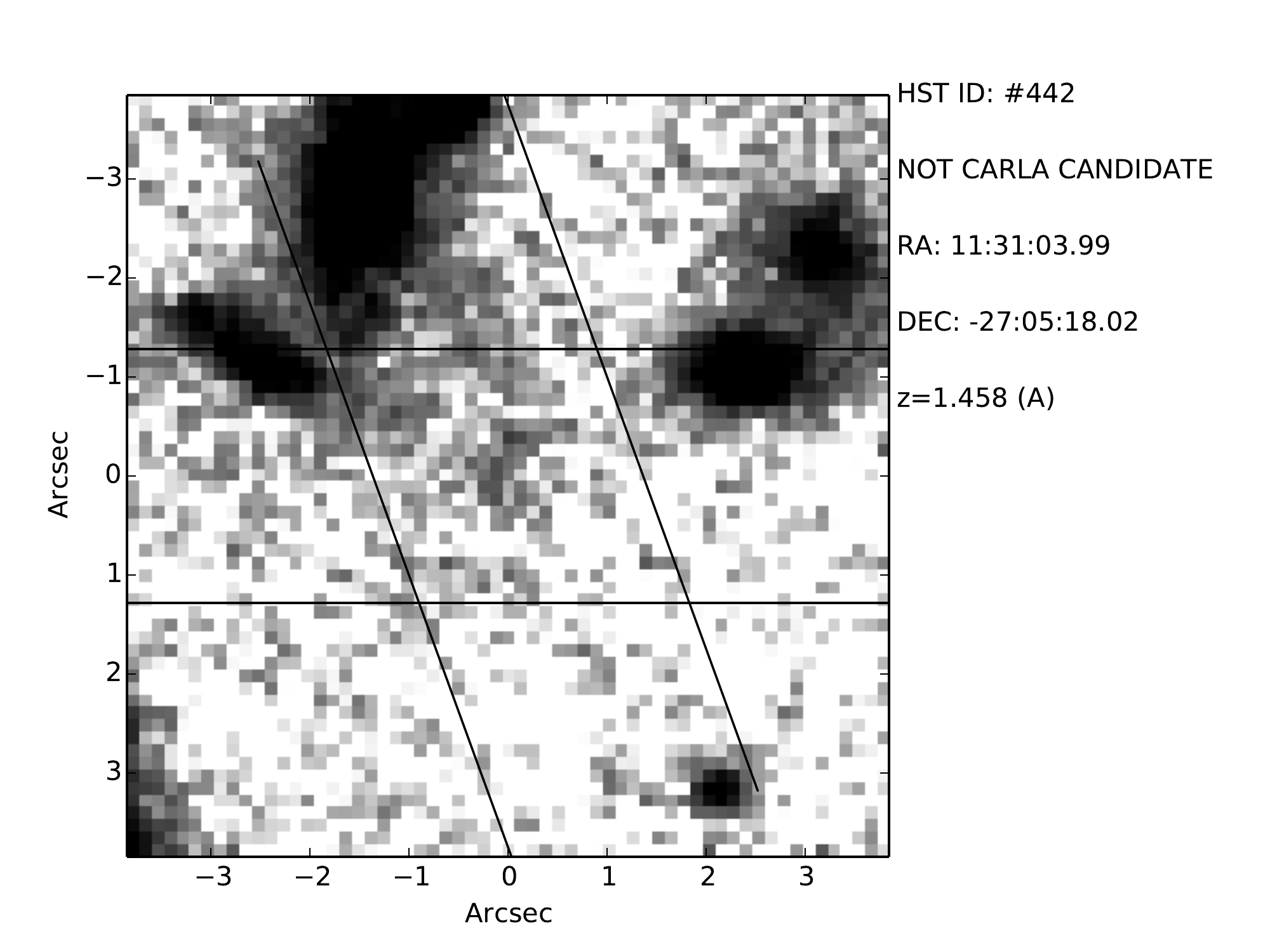} \mbox{(e)}}%
}%
{%
\setlength{\fboxsep}{0pt}%
\setlength{\fboxrule}{1pt}%
\fbox{\includegraphics[page=2, scale=0.24]{CARLA_J1131-2705_447.pdf} \hfill \includegraphics[page=1, scale=0.20]{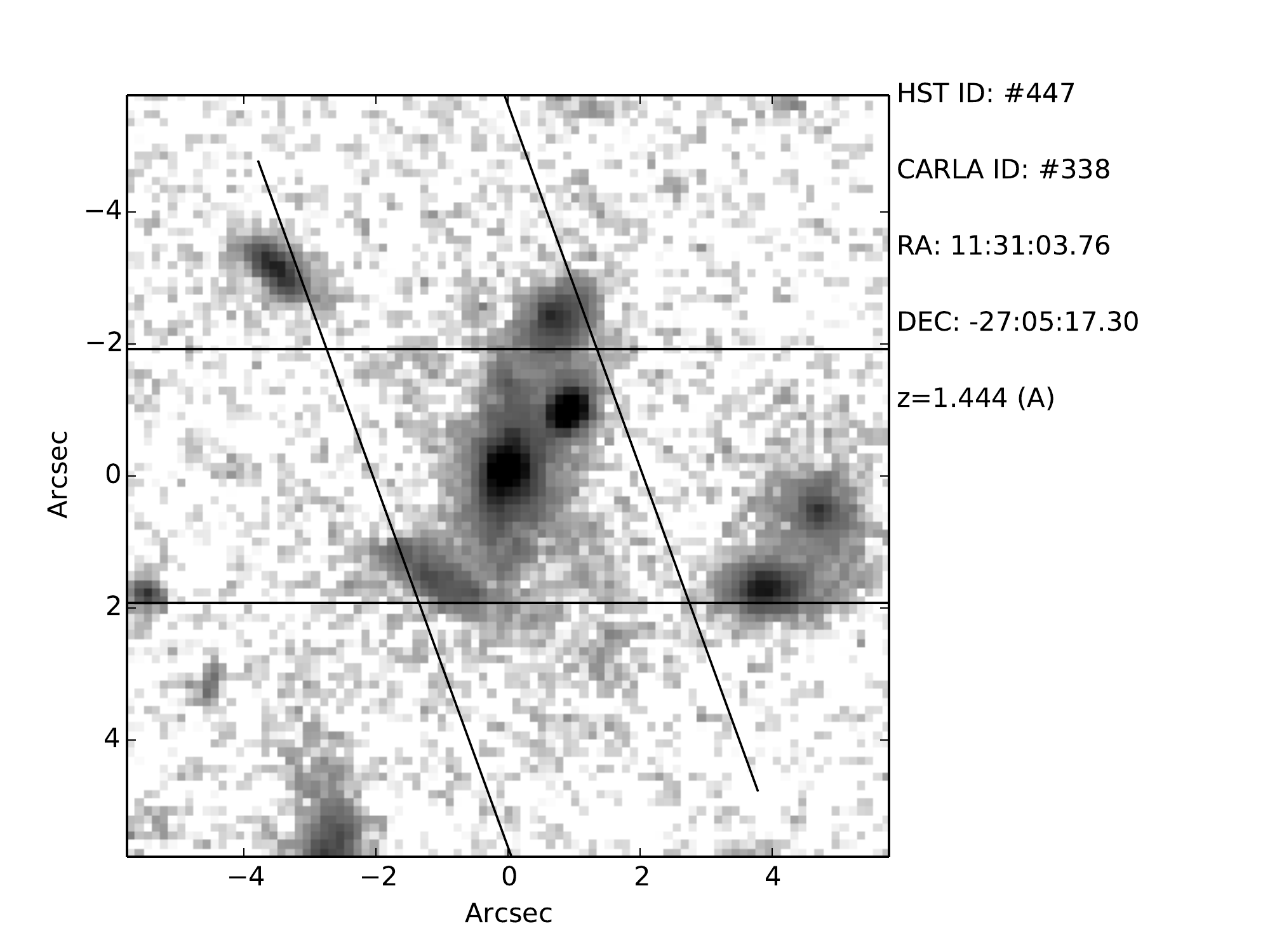} \mbox{(f)}}%
}\\%
{%
\setlength{\fboxsep}{0pt}%
\setlength{\fboxrule}{1pt}%
\fbox{\includegraphics[page=2, scale=0.24]{CARLA_J1131-2705_504.pdf} \hfill \includegraphics[page=1, scale=0.20]{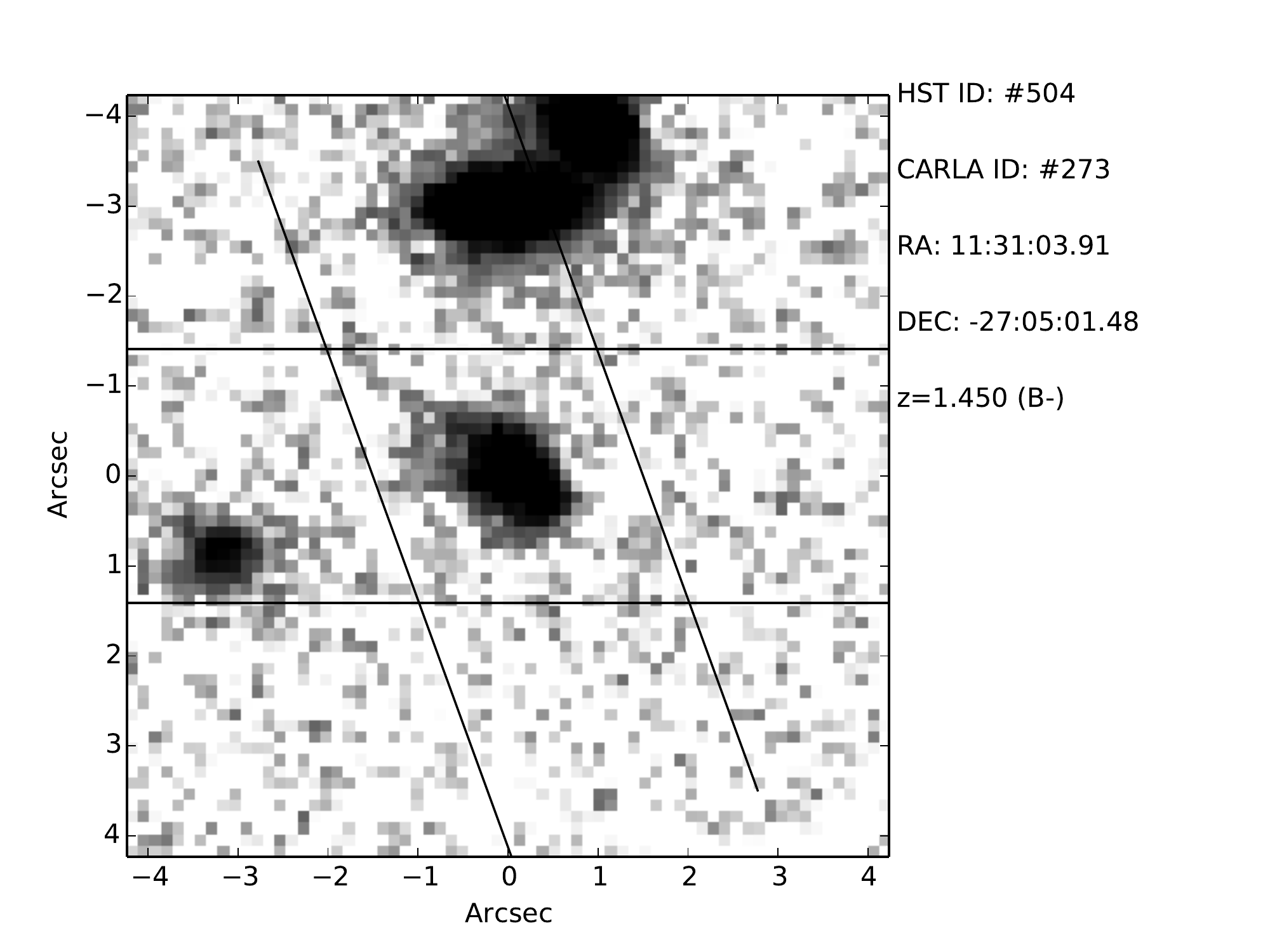} \mbox{(g)}}%
}%
{%
\setlength{\fboxsep}{0pt}%
\setlength{\fboxrule}{1pt}%
\fbox{\includegraphics[page=2, scale=0.24]{CARLA_J1131-2705_701.pdf} \hfill \includegraphics[page=1, scale=0.20]{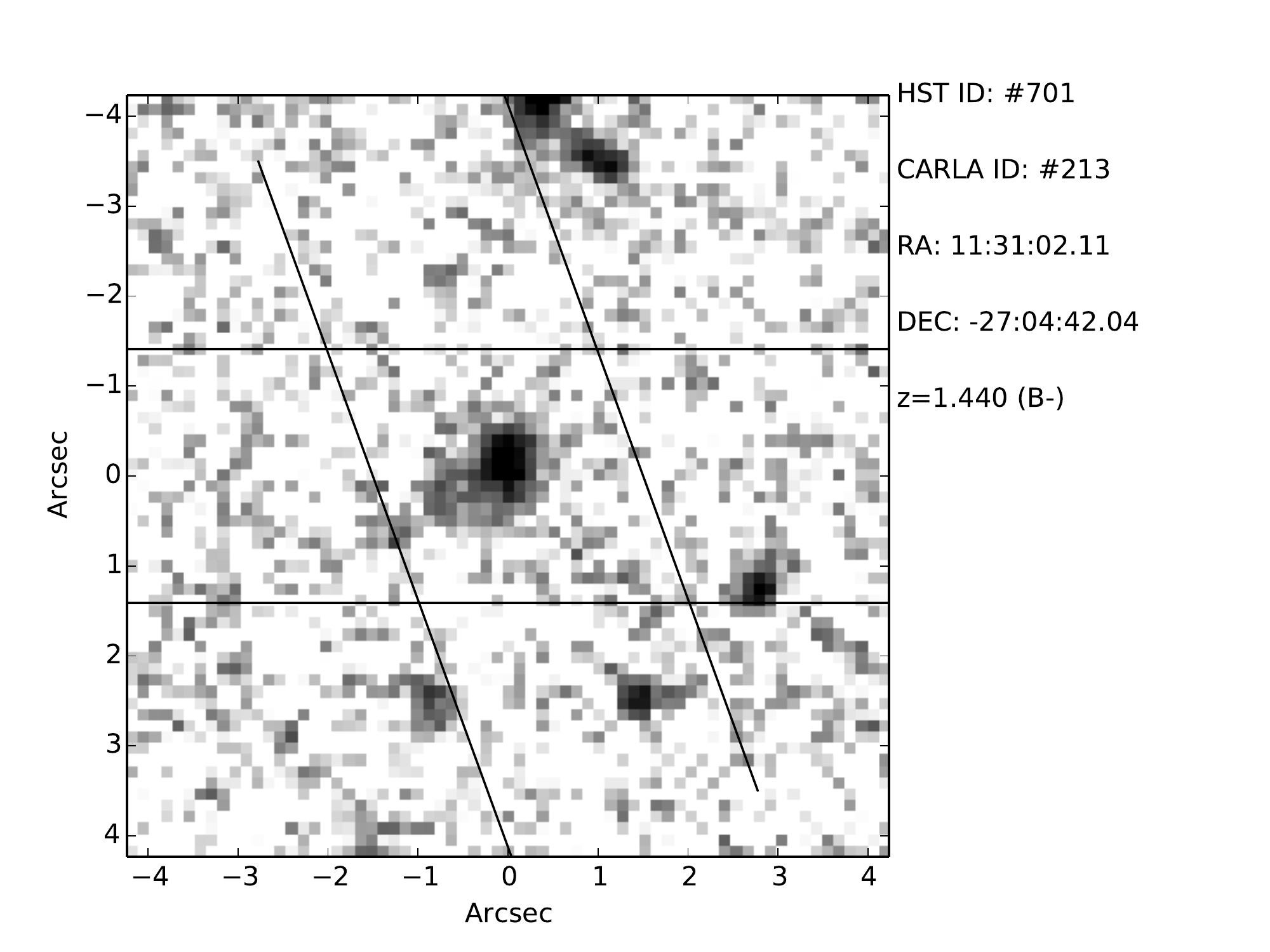} \mbox{(h)}}%
}\\%
{%
\setlength{\fboxsep}{0pt}%
\setlength{\fboxrule}{1pt}%
\fbox{\includegraphics[page=2, scale=0.24]{CARLA_J1131-2705_712.pdf} \hfill \includegraphics[page=1, scale=0.20]{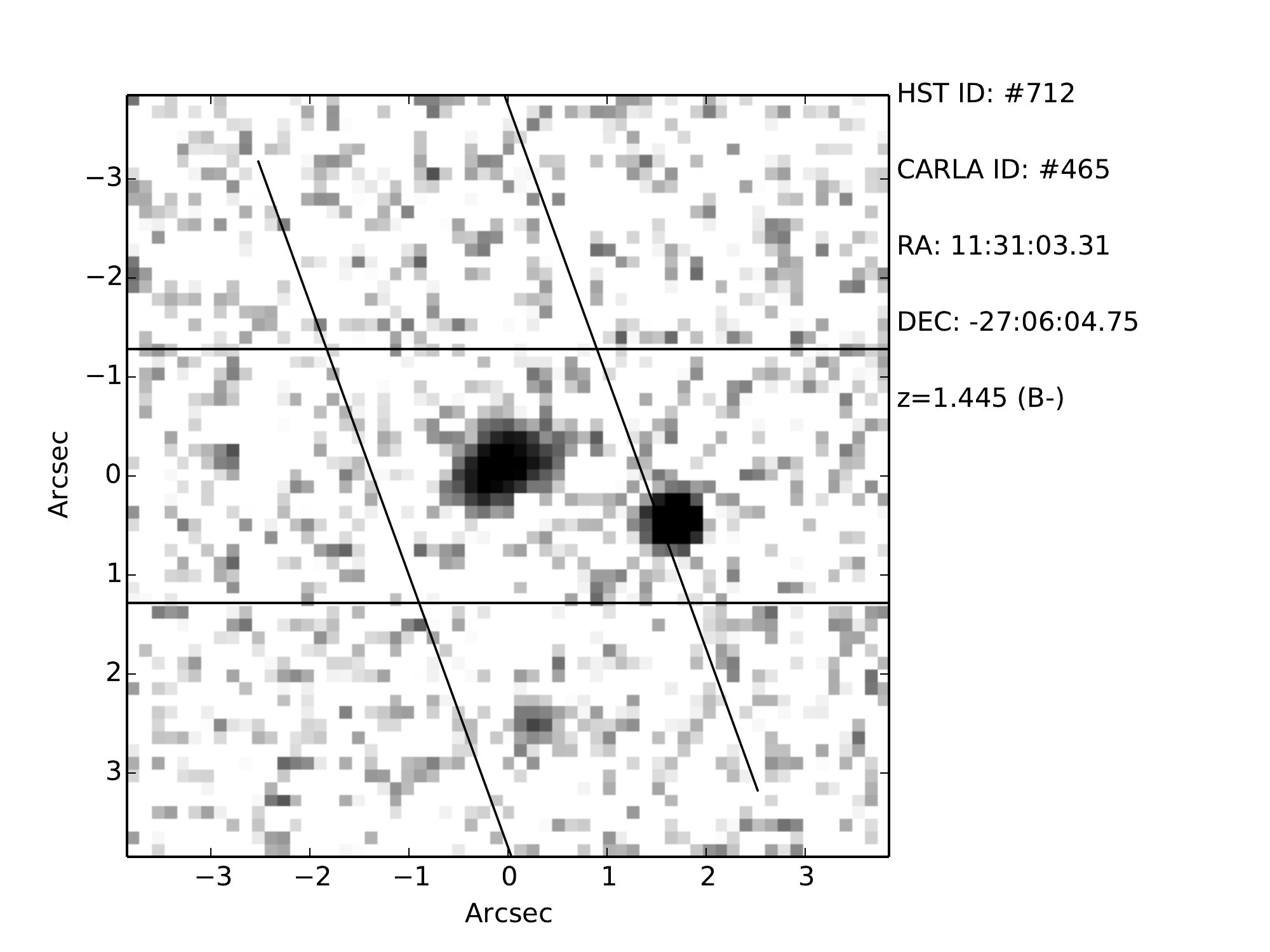} \mbox{(i)}}%
}%
{%
\setlength{\fboxsep}{0pt}%
\setlength{\fboxrule}{1pt}%
\fbox{\includegraphics[page=2, scale=0.24]{CARLA_J1131-2705_747.pdf} \hfill \includegraphics[page=1, scale=0.20]{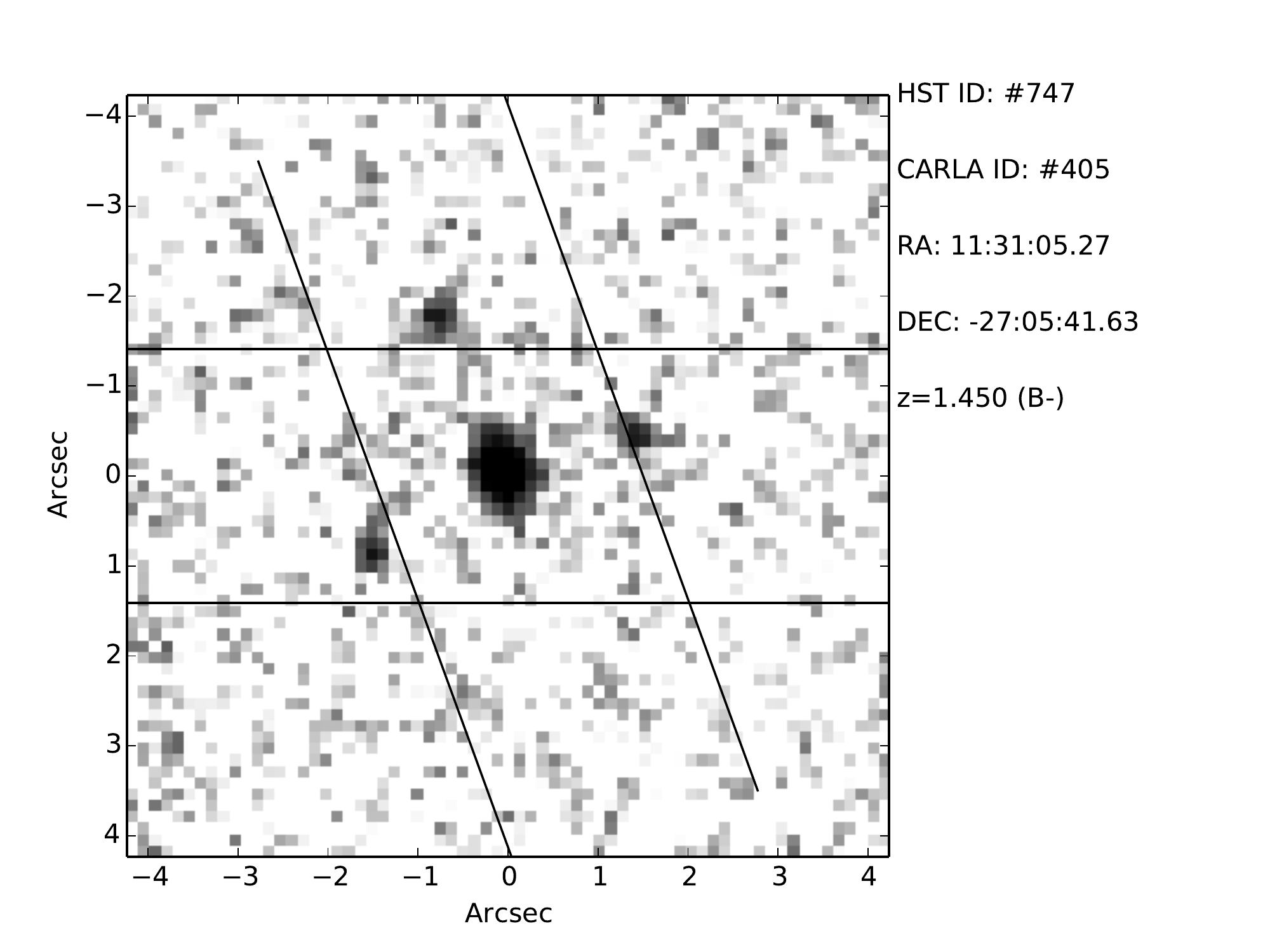} \mbox{(j)}}%
}\\%
\caption[CARLA~J1131-2705 member spectra]{CARLA~J1131$-$2705 member spectra.}
\label{fig:J1131-2705spectra}
\mbox{}\\
\end{figure*}

\begin{figure*}[!ht]
{%
\setlength{\fboxsep}{0pt}%
\setlength{\fboxrule}{1pt}%
\fbox{\includegraphics[page=2, scale=0.24]{CARLA_J1300+4009_391.pdf} \hfill \includegraphics[page=1, scale=0.20]{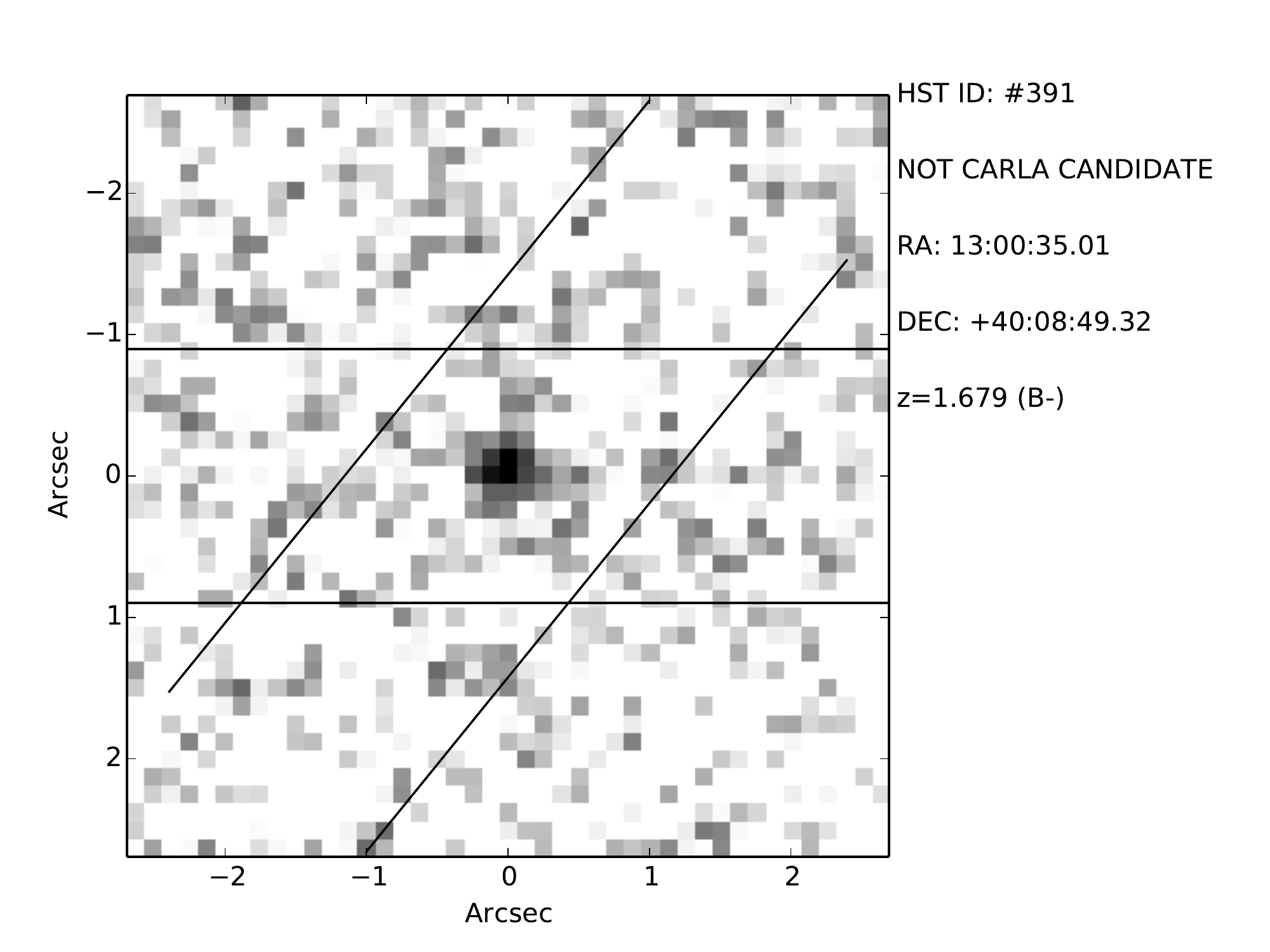} \mbox{(a)}}%
}%
{%
\setlength{\fboxsep}{0pt}%
\setlength{\fboxrule}{1pt}%
\fbox{\includegraphics[page=2, scale=0.24]{CARLA_J1300+4009_445.pdf} \hfill \includegraphics[page=1, scale=0.20]{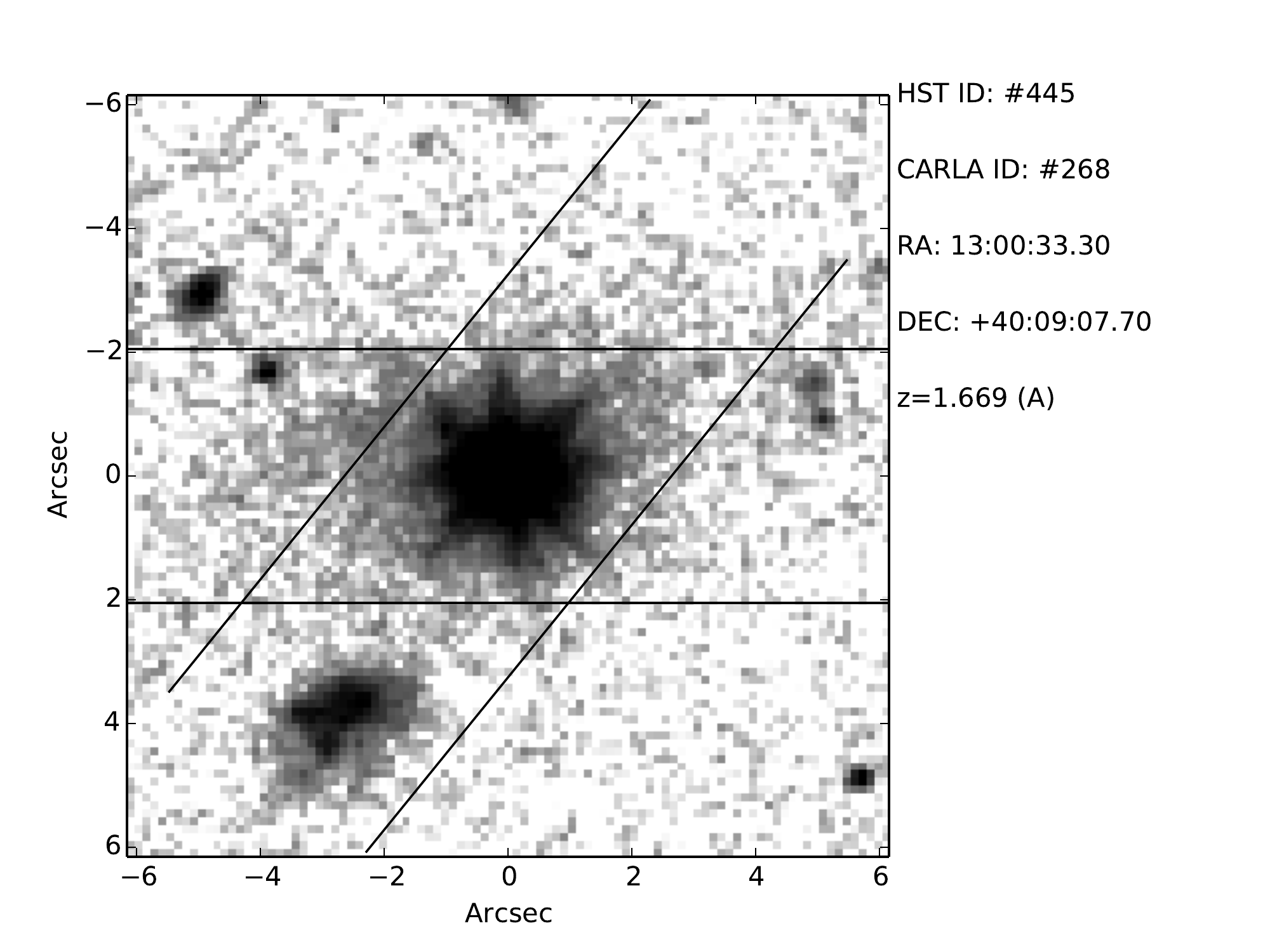} \mbox{(b)}}%
}\\%
{%
\setlength{\fboxsep}{0pt}%
\setlength{\fboxrule}{1pt}%
\fbox{\includegraphics[page=2, scale=0.24]{CARLA_J1300+4009_479.pdf} \hfill \includegraphics[page=1, scale=0.20]{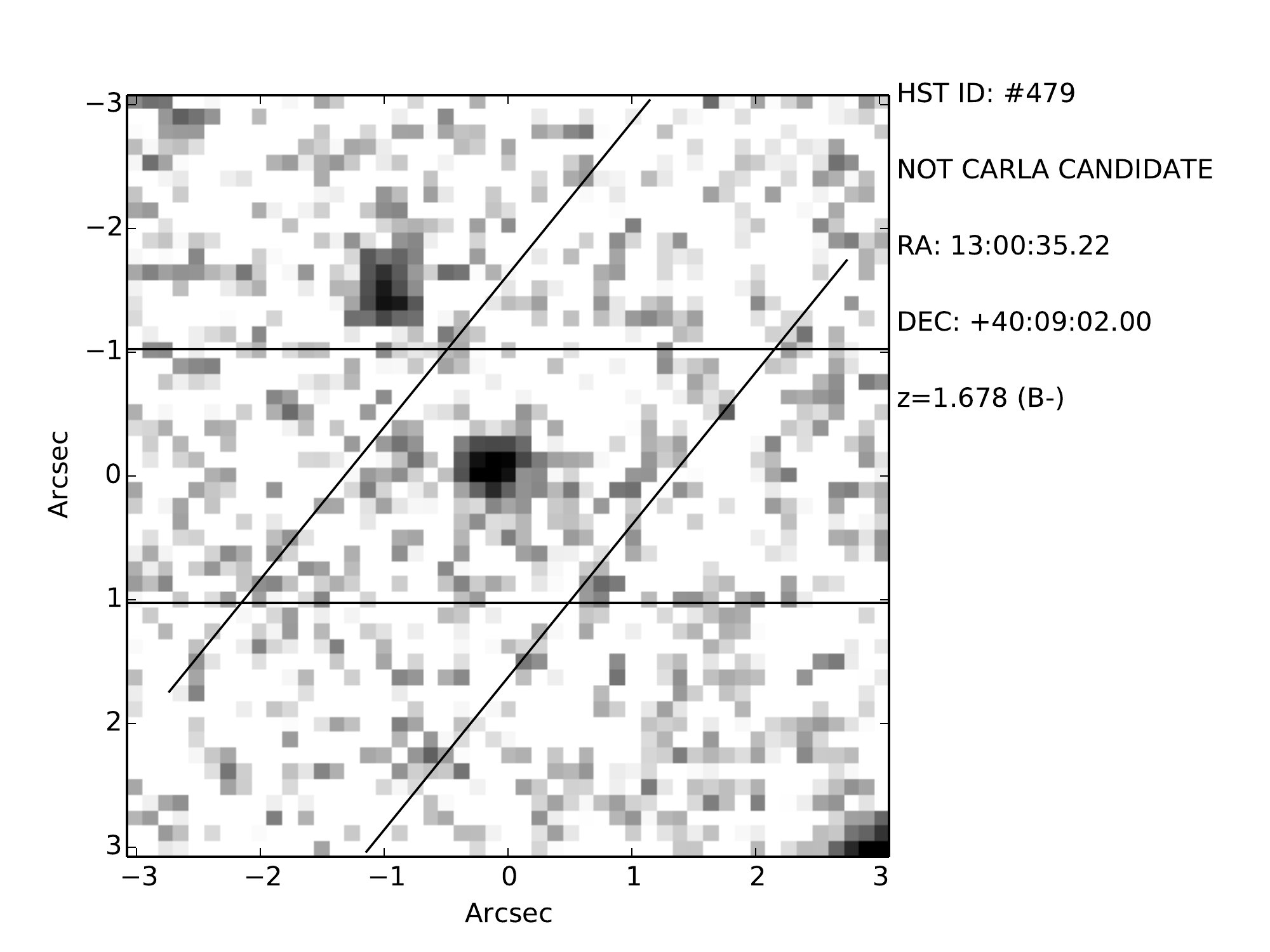} \mbox{(c)}}%
}%
{%
\setlength{\fboxsep}{0pt}%
\setlength{\fboxrule}{1pt}%
\fbox{\includegraphics[page=2, scale=0.24]{CARLA_J1300+4009_527.pdf} \hfill \includegraphics[page=1, scale=0.20]{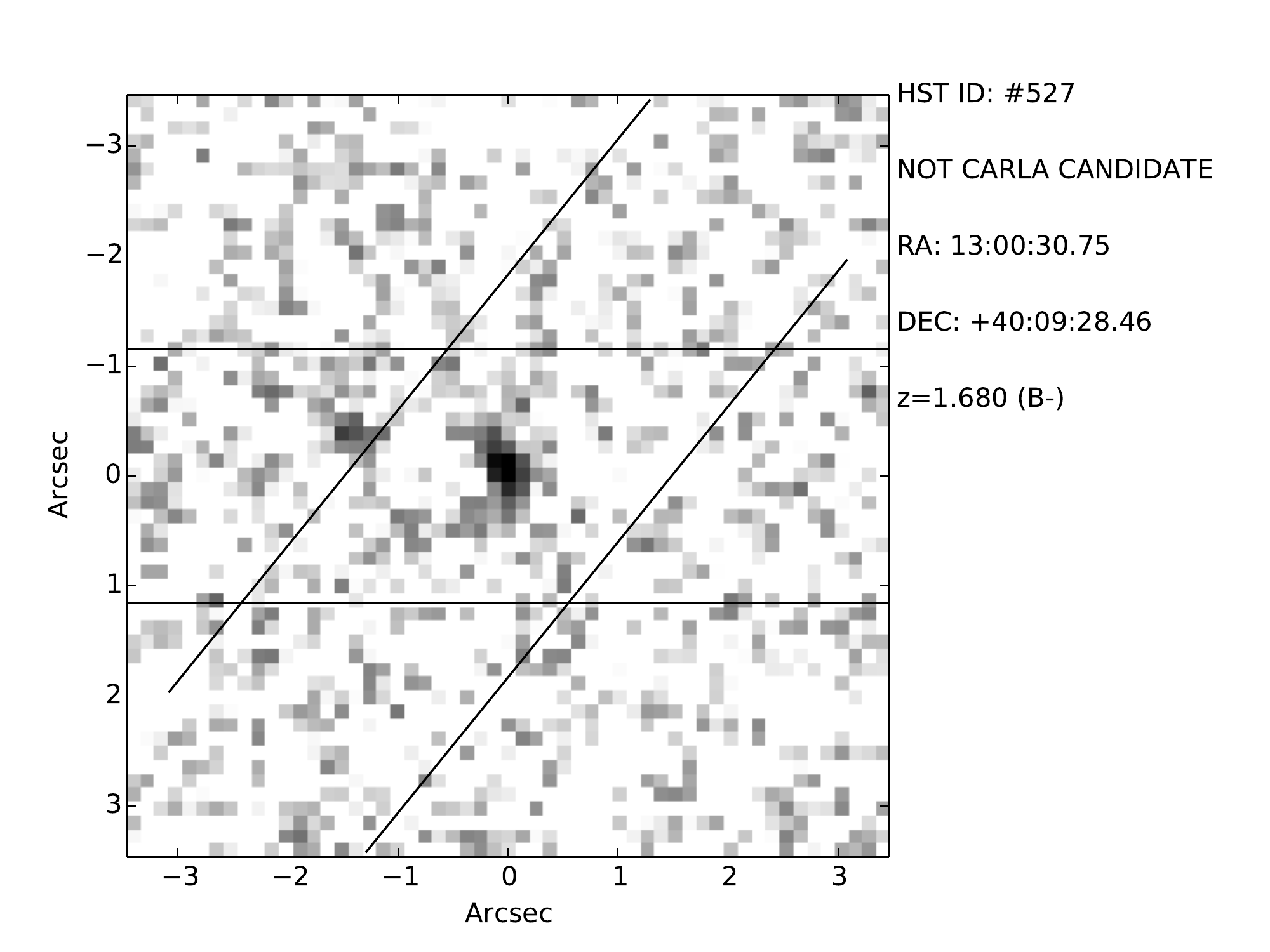} \mbox{(d)}}%
}\\%
{%
\setlength{\fboxsep}{0pt}%
\setlength{\fboxrule}{1pt}%
\fbox{\includegraphics[page=2, scale=0.24]{CARLA_J1300+4009_589.pdf} \hfill \includegraphics[page=1, scale=0.20]{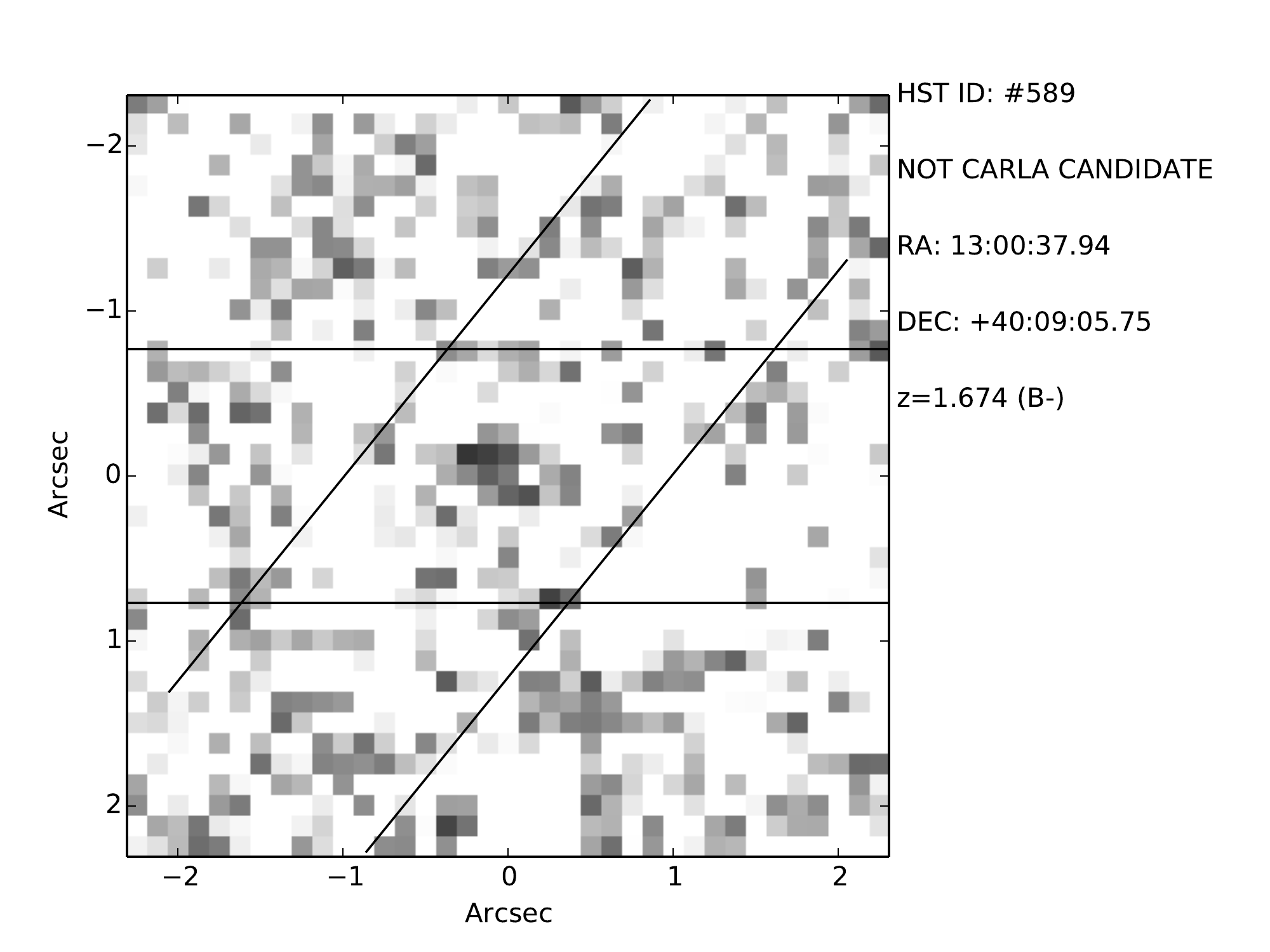} \mbox{(e)}}%
}%
{%
\setlength{\fboxsep}{0pt}%
\setlength{\fboxrule}{1pt}%
\fbox{\includegraphics[page=2, scale=0.24]{CARLA_J1300+4009_658.pdf} \hfill \includegraphics[page=1, scale=0.20]{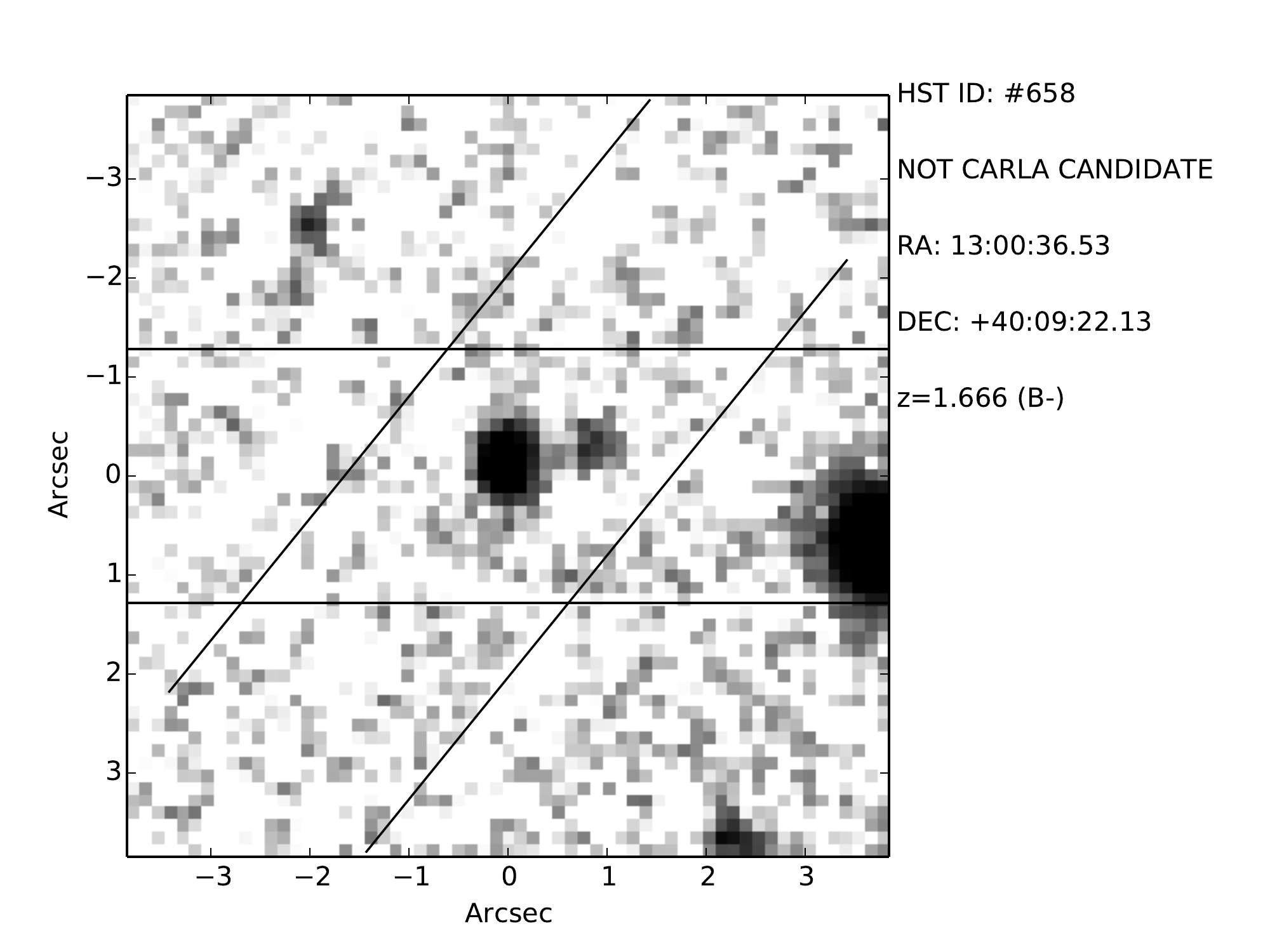} \mbox{(f)}}%
}\\%
{%
\setlength{\fboxsep}{0pt}%
\setlength{\fboxrule}{1pt}%
\fbox{\includegraphics[page=2, scale=0.24]{CARLA_J1300+4009_676.pdf} \hfill \includegraphics[page=1, scale=0.20]{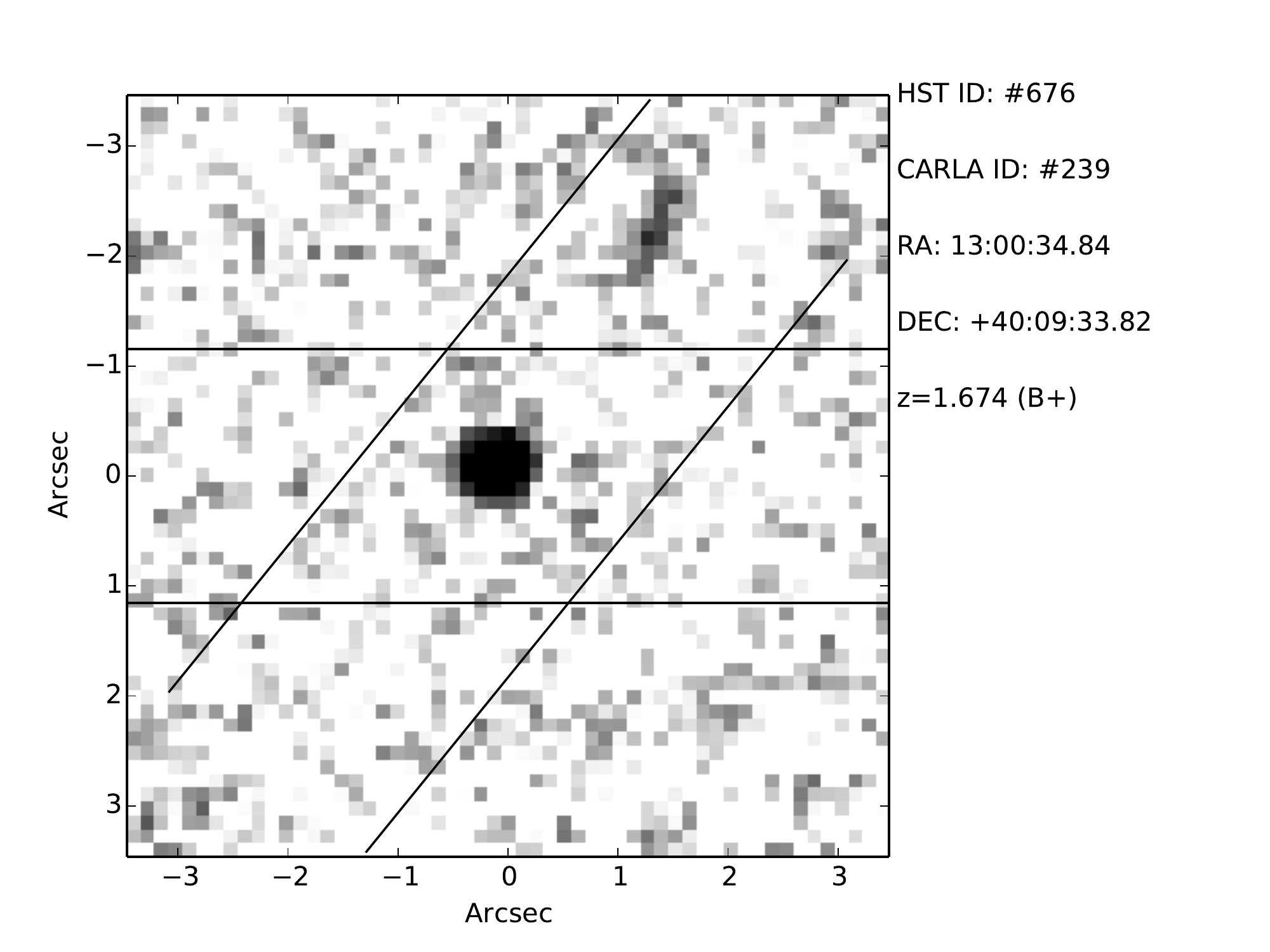} \mbox{(g)}}%
}%
{%
\setlength{\fboxsep}{0pt}%
\setlength{\fboxrule}{1pt}%
\fbox{\includegraphics[page=2, scale=0.24]{CARLA_J1300+4009_836.pdf} \hfill \includegraphics[page=1, scale=0.20]{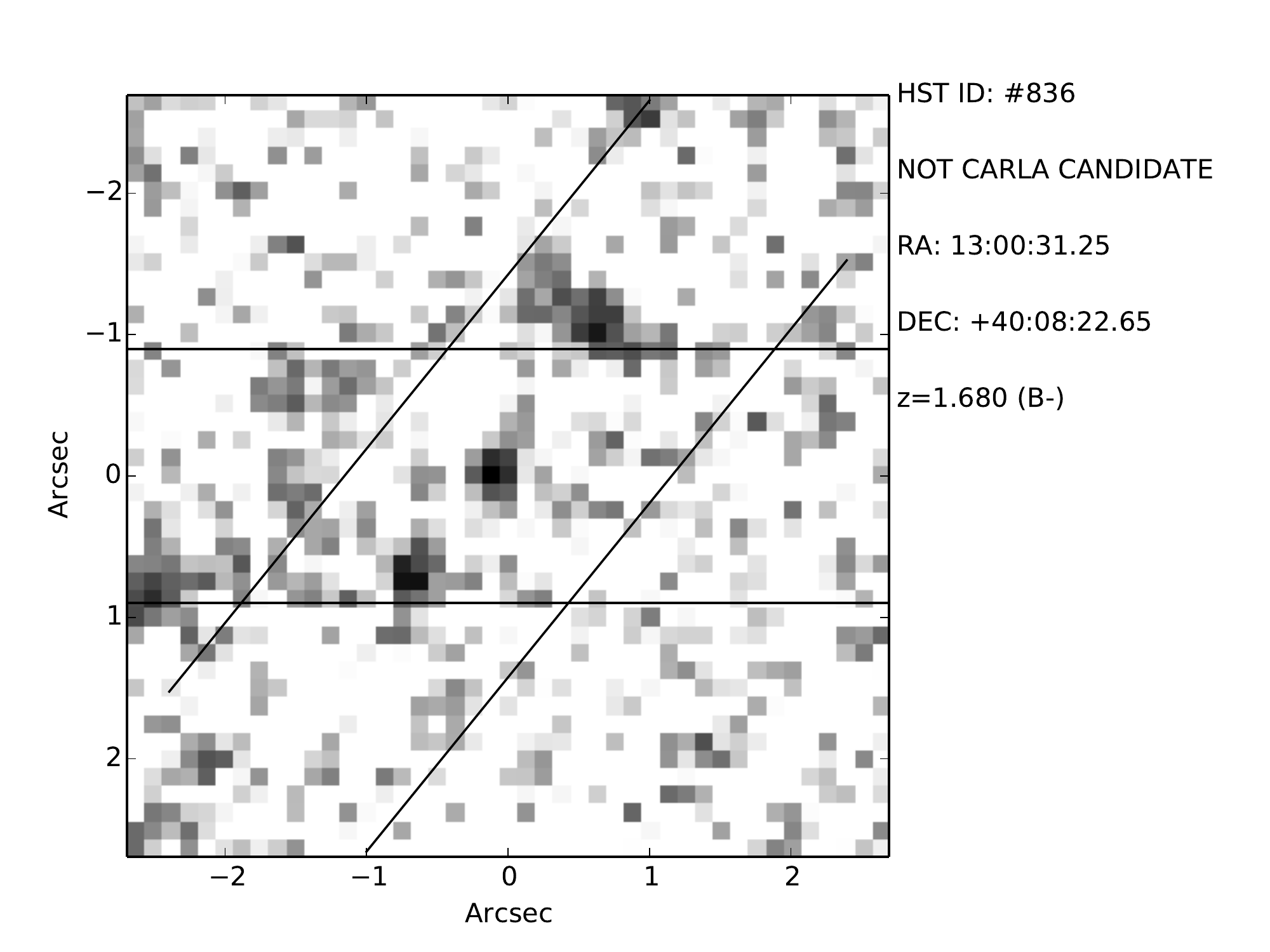} \mbox{(h)}}%
}\\%
\caption[CARLA~J1300+4009 member spectra]{CARLA~J1300+4009 member spectra.}
\label{fig:J1300+4009spectra}
\mbox{}\\
\end{figure*}

\begin{figure*}[]
{%
\setlength{\fboxsep}{0pt}%
\setlength{\fboxrule}{1pt}%
\fbox{\includegraphics[page=2, scale=0.24]{CARLA_J1358+5752_44.pdf} \hfill \includegraphics[page=1, scale=0.20]{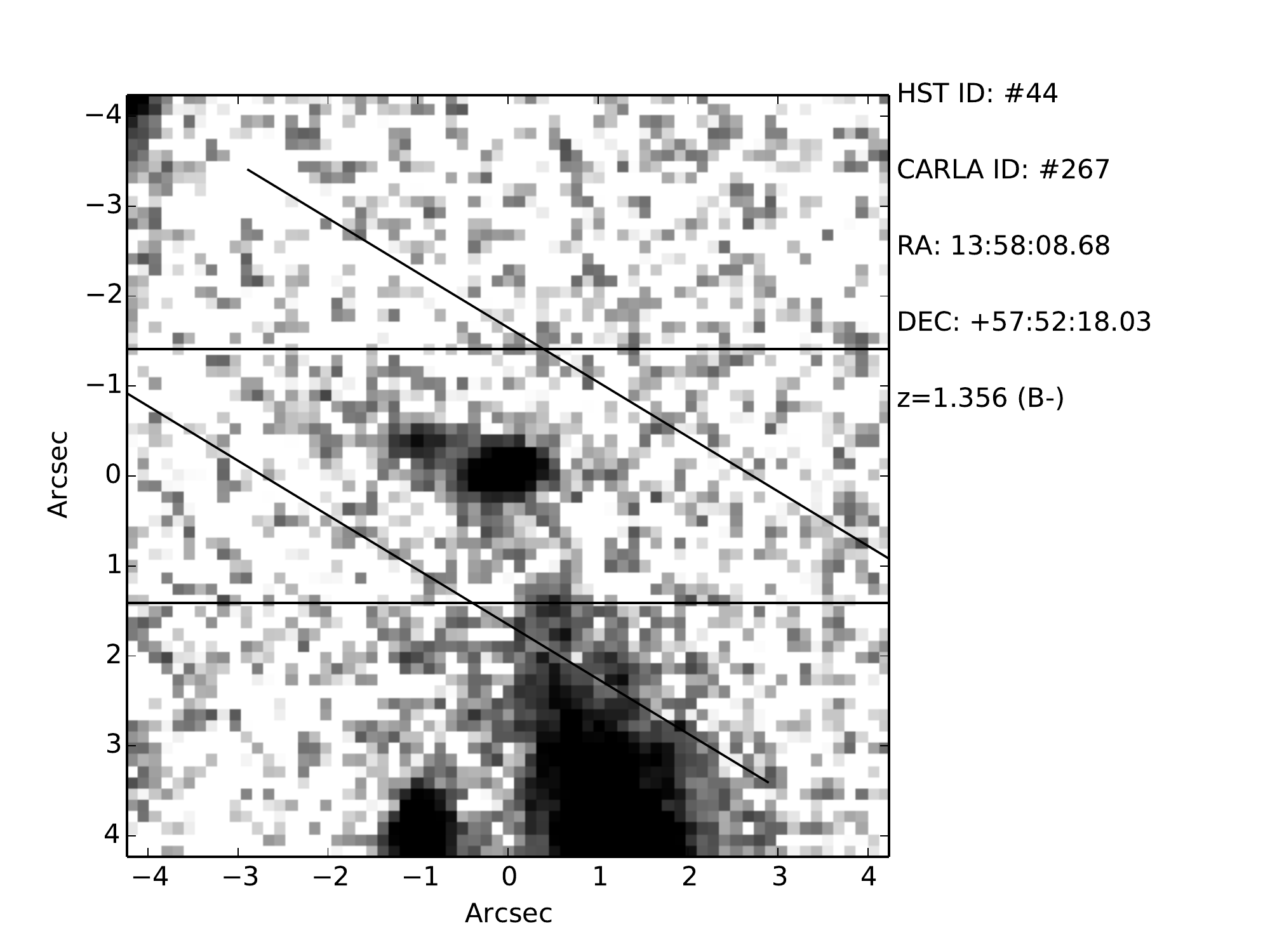} \mbox{(a)}}%
}%
{%
\setlength{\fboxsep}{0pt}%
\setlength{\fboxrule}{1pt}%
\fbox{\includegraphics[page=2, scale=0.24]{CARLA_J1358+5752_50.pdf} \hfill \includegraphics[page=1, scale=0.20]{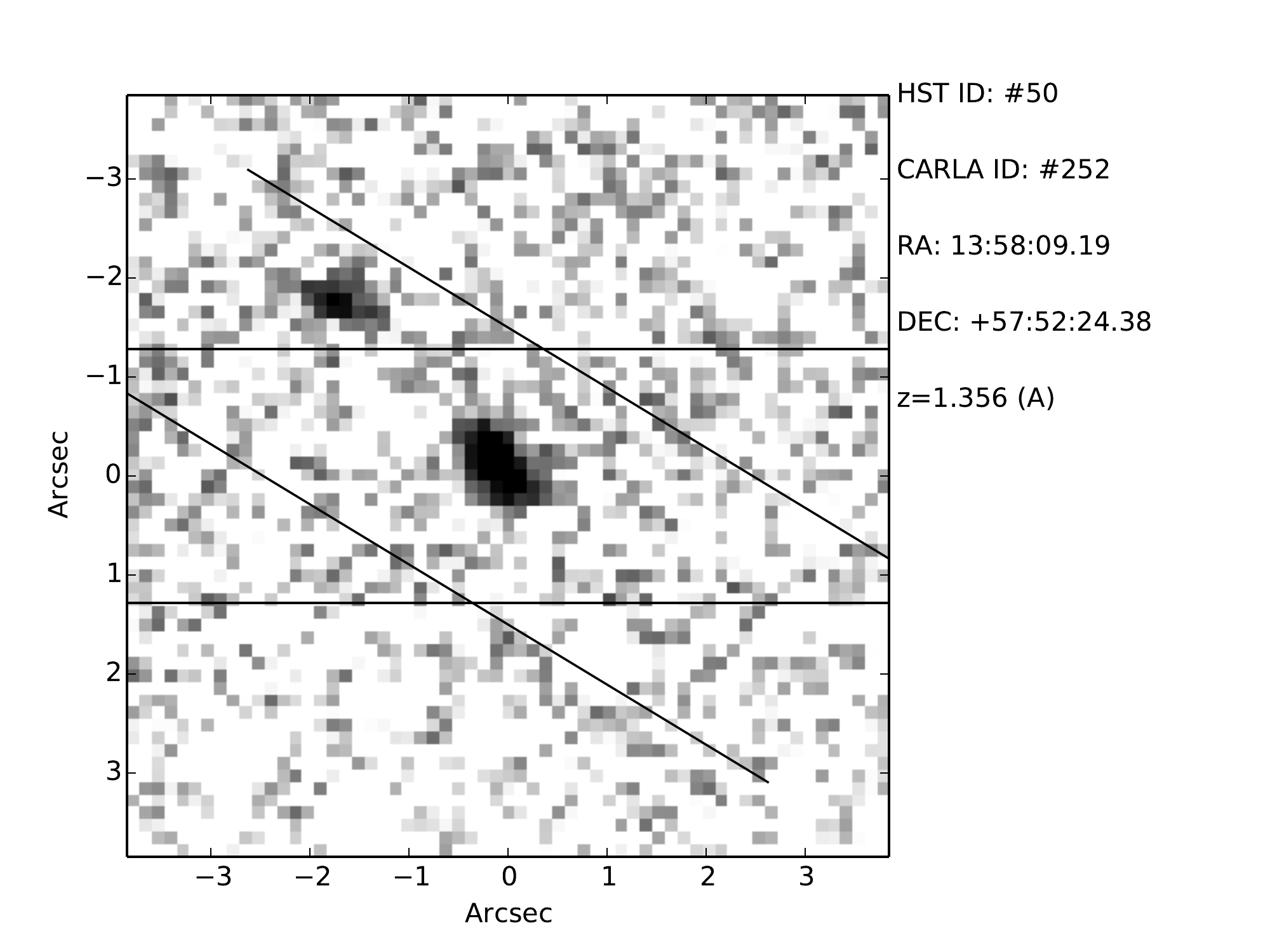} \mbox{(b)}}
}\\%
{%
\setlength{\fboxsep}{0pt}%
\setlength{\fboxrule}{1pt}%
\fbox{\includegraphics[page=2, scale=0.24]{CARLA_J1358+5752_52.pdf} \hfill \includegraphics[page=1, scale=0.20]{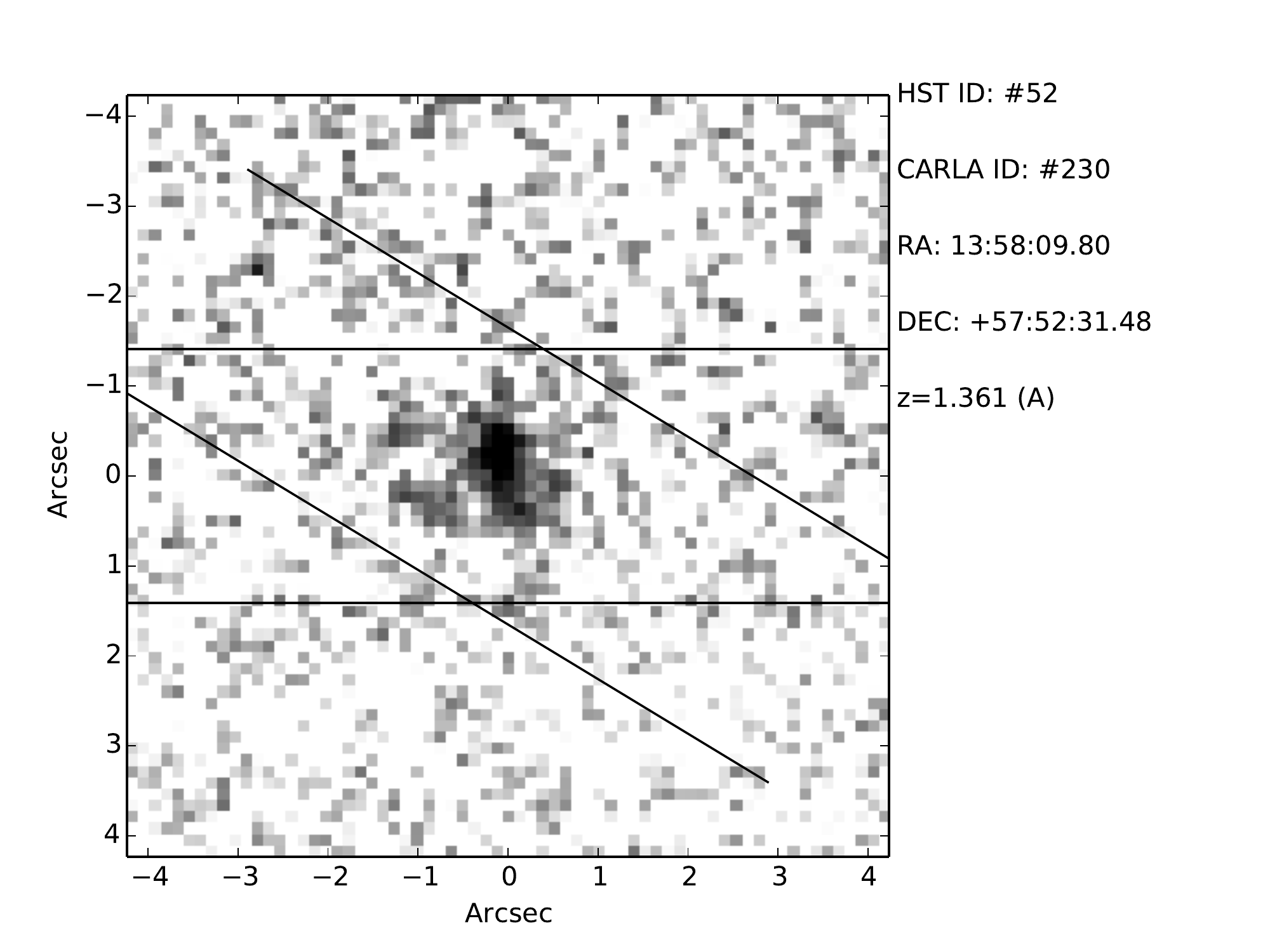} \mbox{(c)}}%
}%
{%
\setlength{\fboxsep}{0pt}%
\setlength{\fboxrule}{1pt}%
\fbox{\includegraphics[page=2, scale=0.24]{CARLA_J1358+5752_223.pdf} \hfill \includegraphics[page=1, scale=0.20]{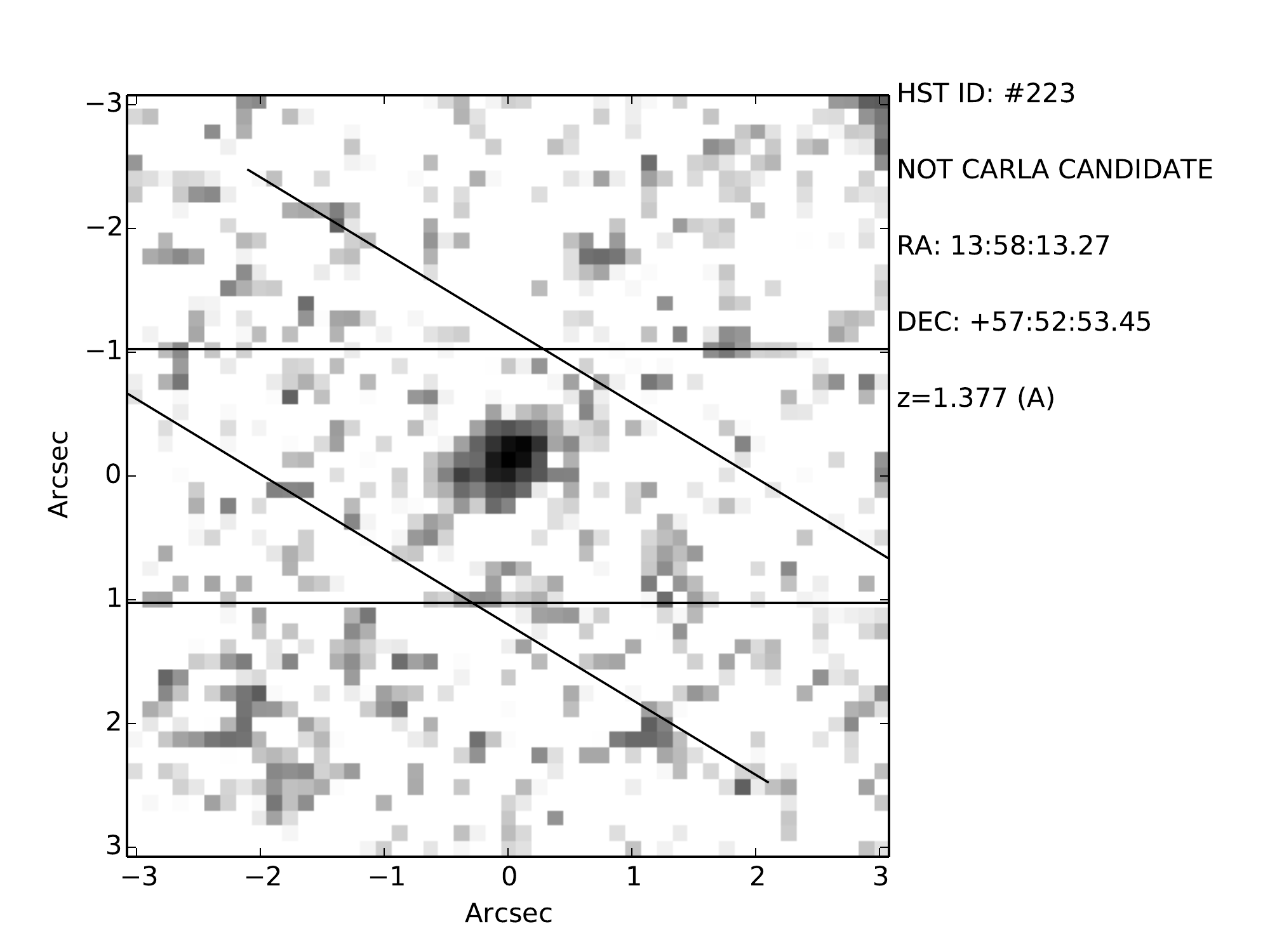} \mbox{(d)}}%
}\\%
{%
\setlength{\fboxsep}{0pt}%
\setlength{\fboxrule}{1pt}%
\fbox{\includegraphics[page=2, scale=0.24]{CARLA_J1358+5752_250.pdf} \hfill \includegraphics[page=1, scale=0.20]{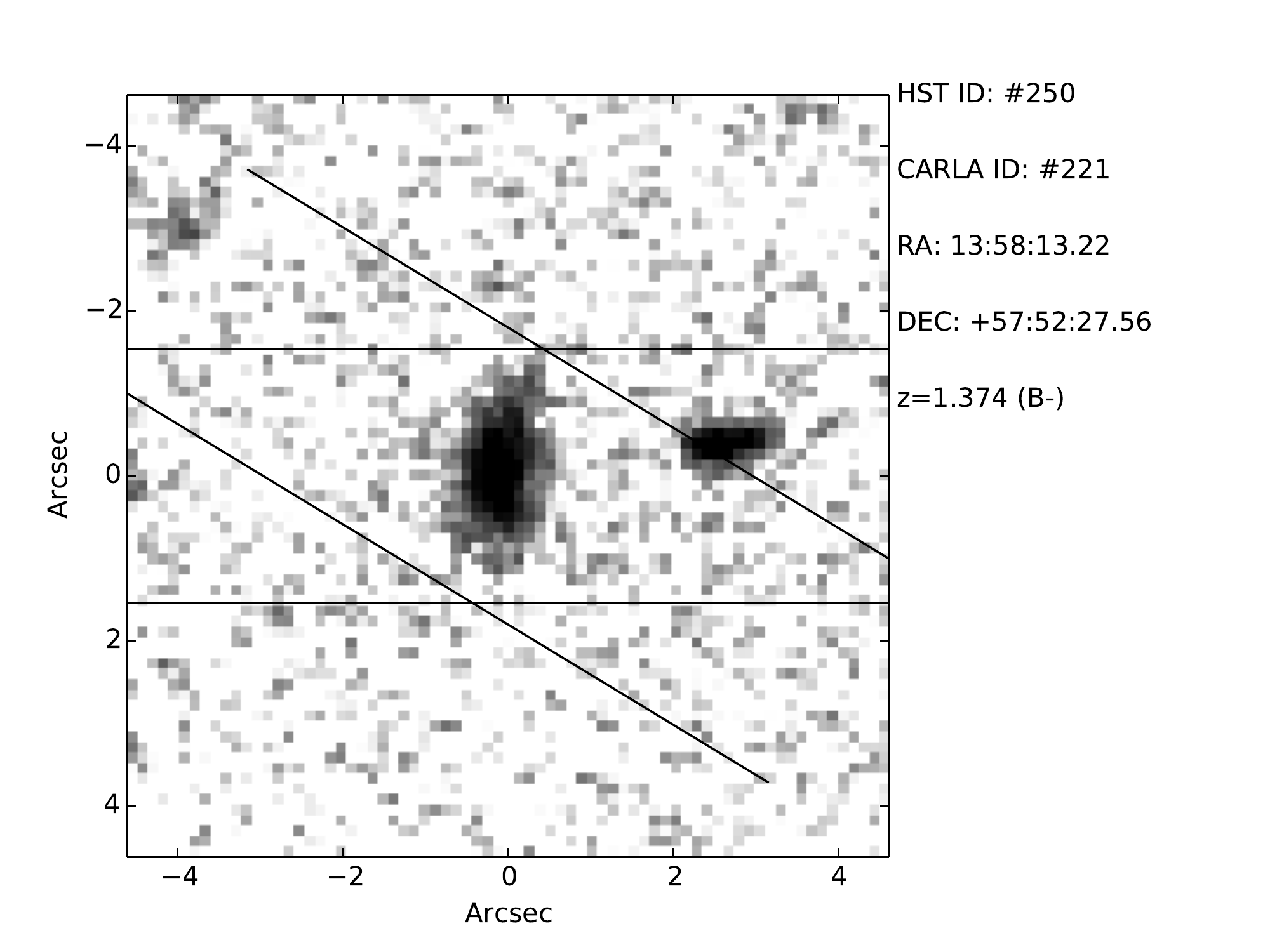} \mbox{(e)}}%
}%
{%
\setlength{\fboxsep}{0pt}%
\setlength{\fboxrule}{1pt}%
\fbox{\includegraphics[page=2, scale=0.24]{CARLA_J1358+5752_414.pdf} \hfill \includegraphics[page=1, scale=0.20]{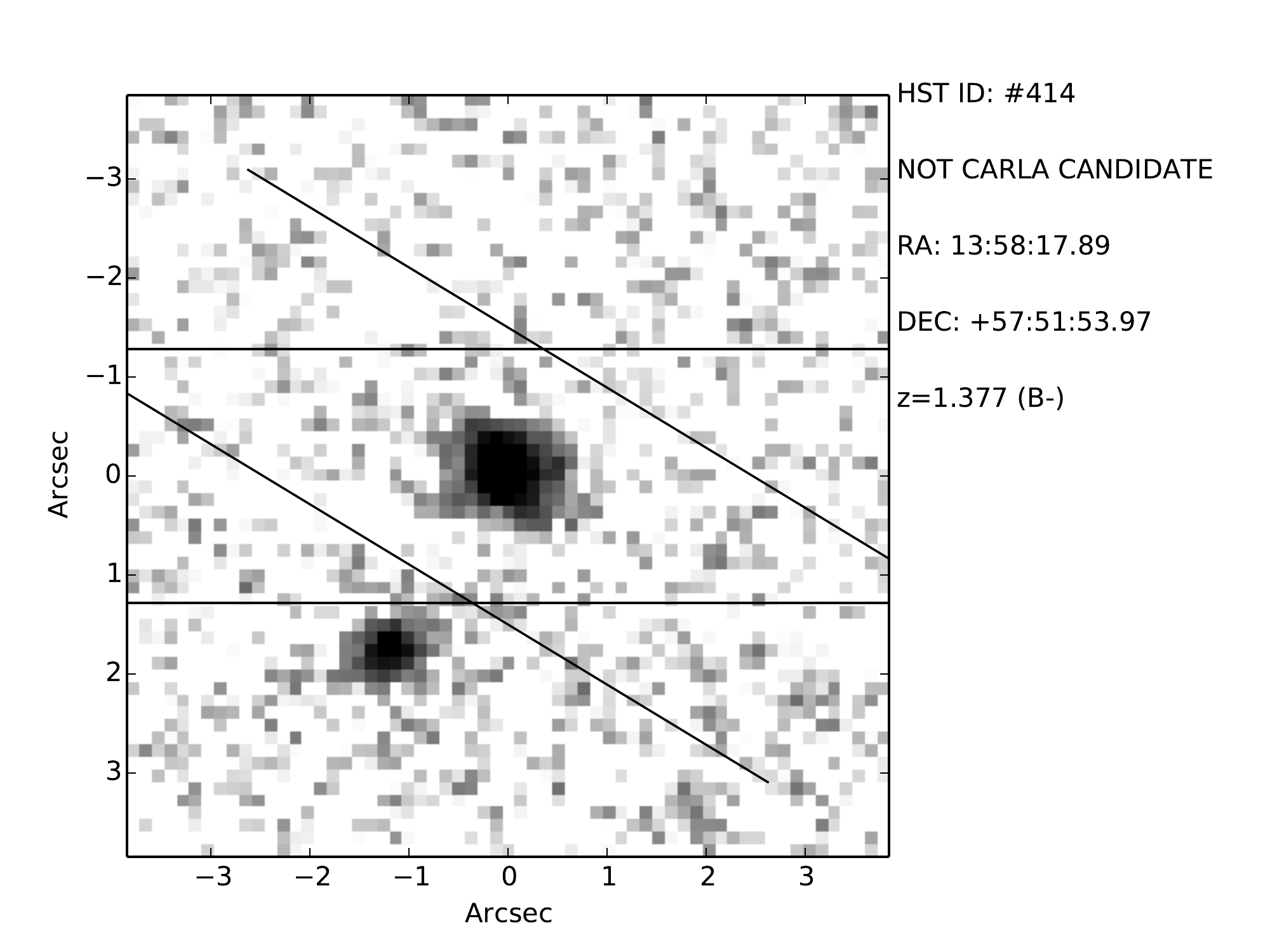} \mbox{(f)}}%
}\\%
{%
\setlength{\fboxsep}{0pt}%
\setlength{\fboxrule}{1pt}%
\fbox{\includegraphics[page=2, scale=0.24]{CARLA_J1358+5752_445.pdf} \hfill \includegraphics[page=1, scale=0.20]{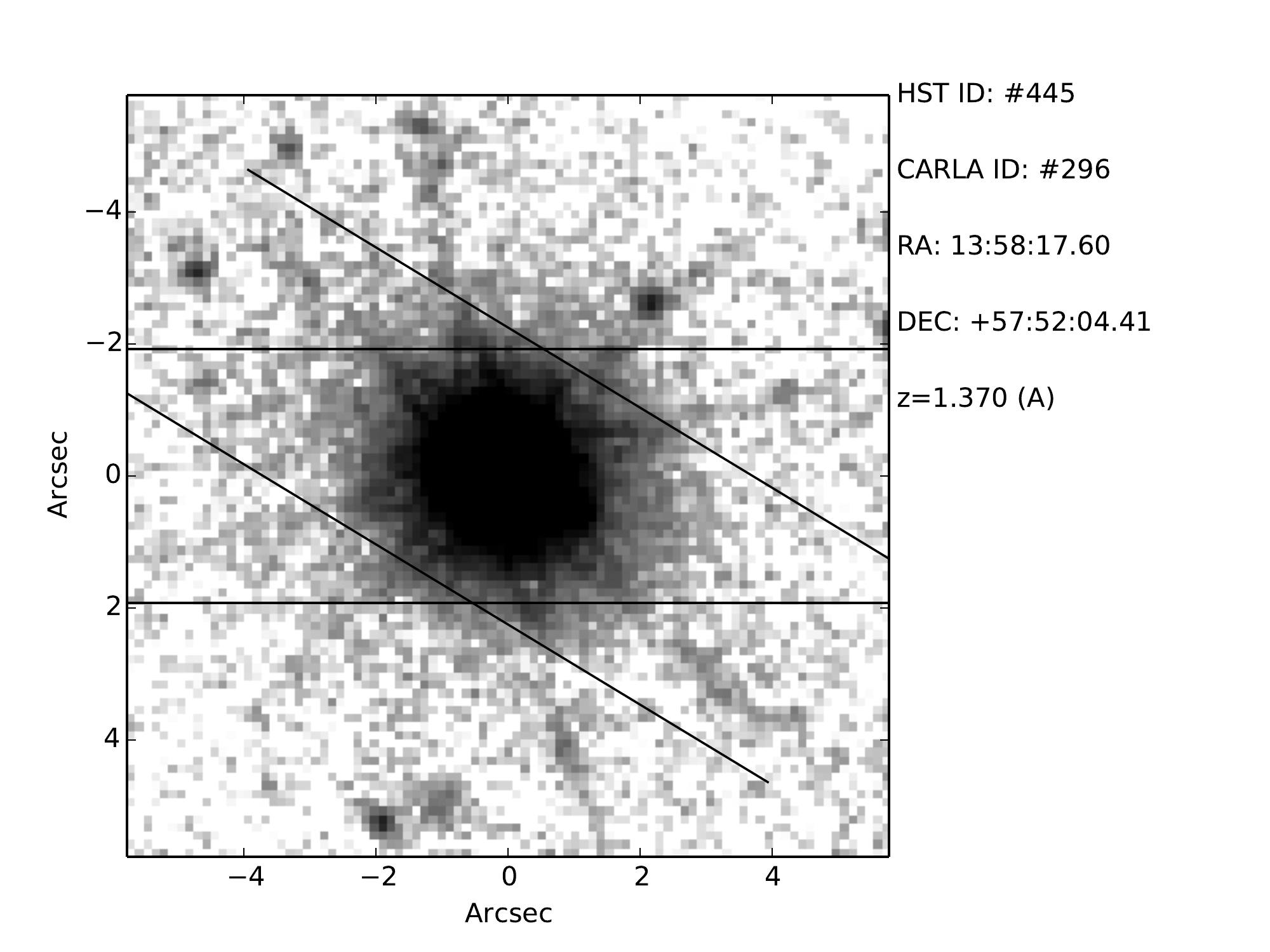} \mbox{(g)}}%
}%
{%
\setlength{\fboxsep}{0pt}%
\setlength{\fboxrule}{1pt}%
\fbox{\includegraphics[page=2, scale=0.24]{CARLA_J1358+5752_661.pdf} \hfill \includegraphics[page=1, scale=0.20]{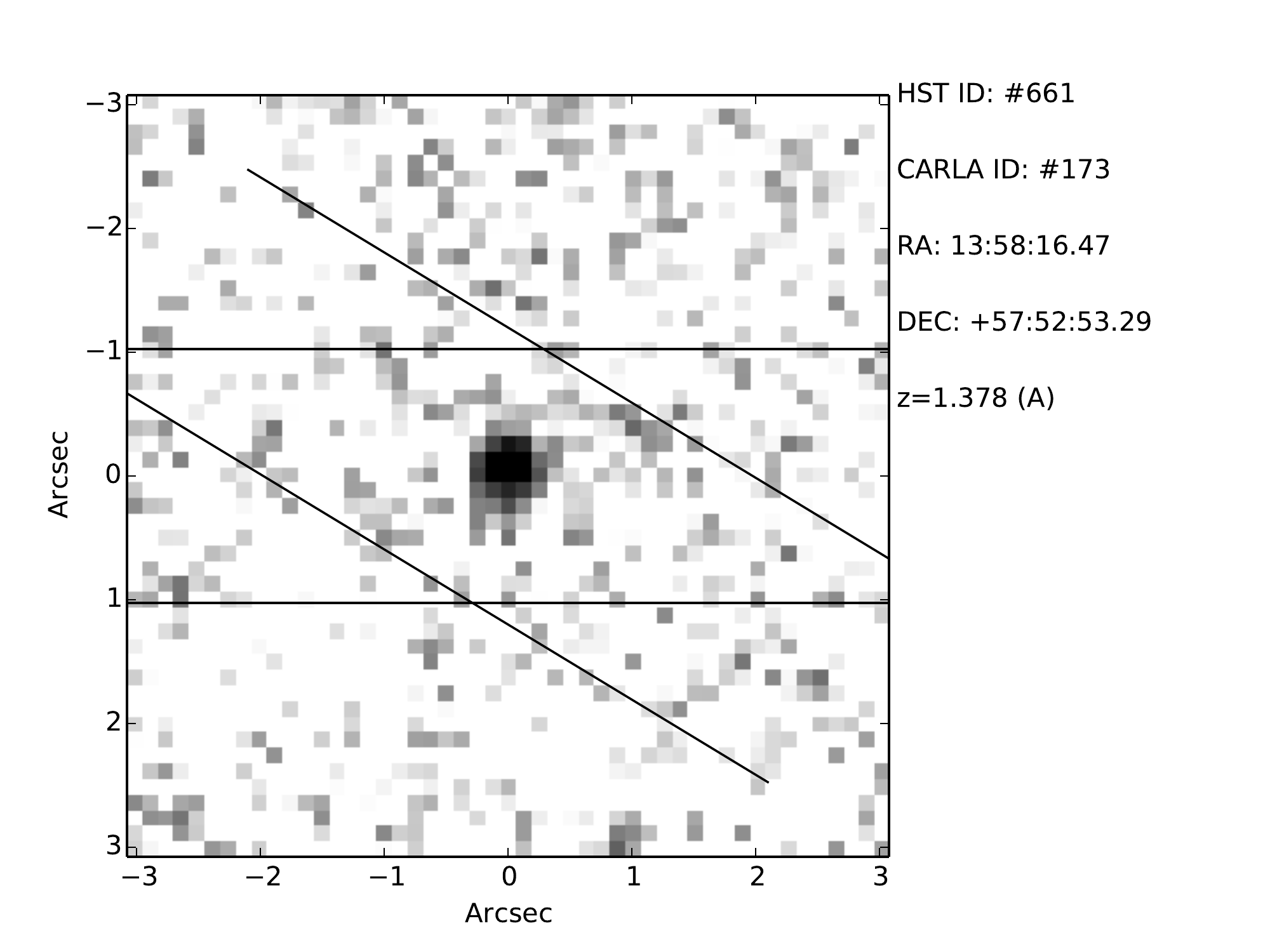} \mbox{(h)}}%
}\\%
{%
\setlength{\fboxsep}{0pt}%
\setlength{\fboxrule}{1pt}%
\fbox{\includegraphics[page=2, scale=0.24]{CARLA_J1358+5752_681.pdf} \hfill \includegraphics[page=1, scale=0.20]{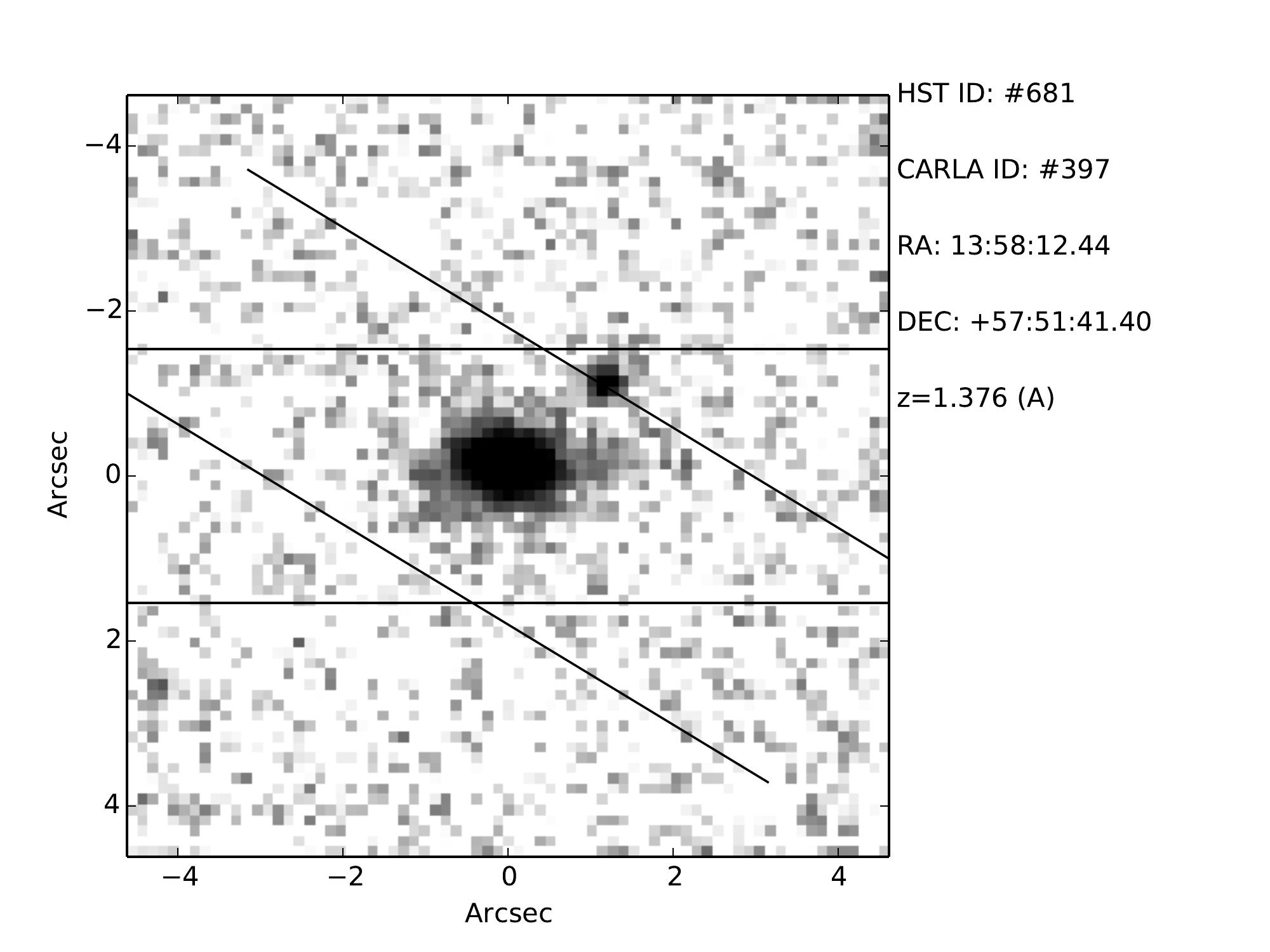} \mbox{(i)}}%
}%
{%
\setlength{\fboxsep}{0pt}%
\setlength{\fboxrule}{1pt}%
\fbox{\includegraphics[page=2, scale=0.24]{CARLA_J1358+5752_685.pdf} \hfill \includegraphics[page=1, scale=0.20]{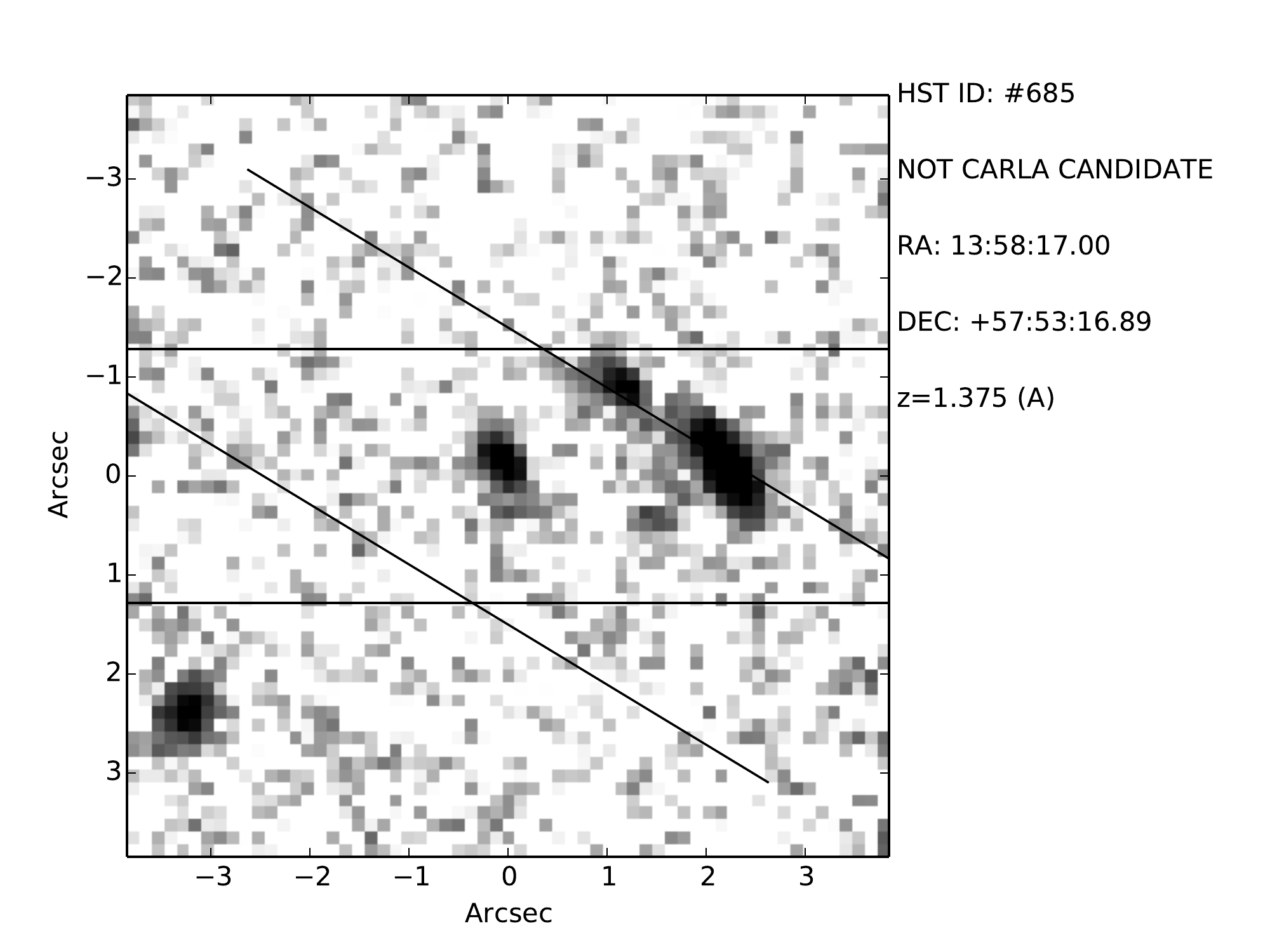} \mbox{(j)}}%
}\\%
{%
\setlength{\fboxsep}{0pt}%
\setlength{\fboxrule}{1pt}%
\fbox{\includegraphics[page=2, scale=0.24]{CARLA_J1358+5752_689.pdf} \hfill \includegraphics[page=1, scale=0.20]{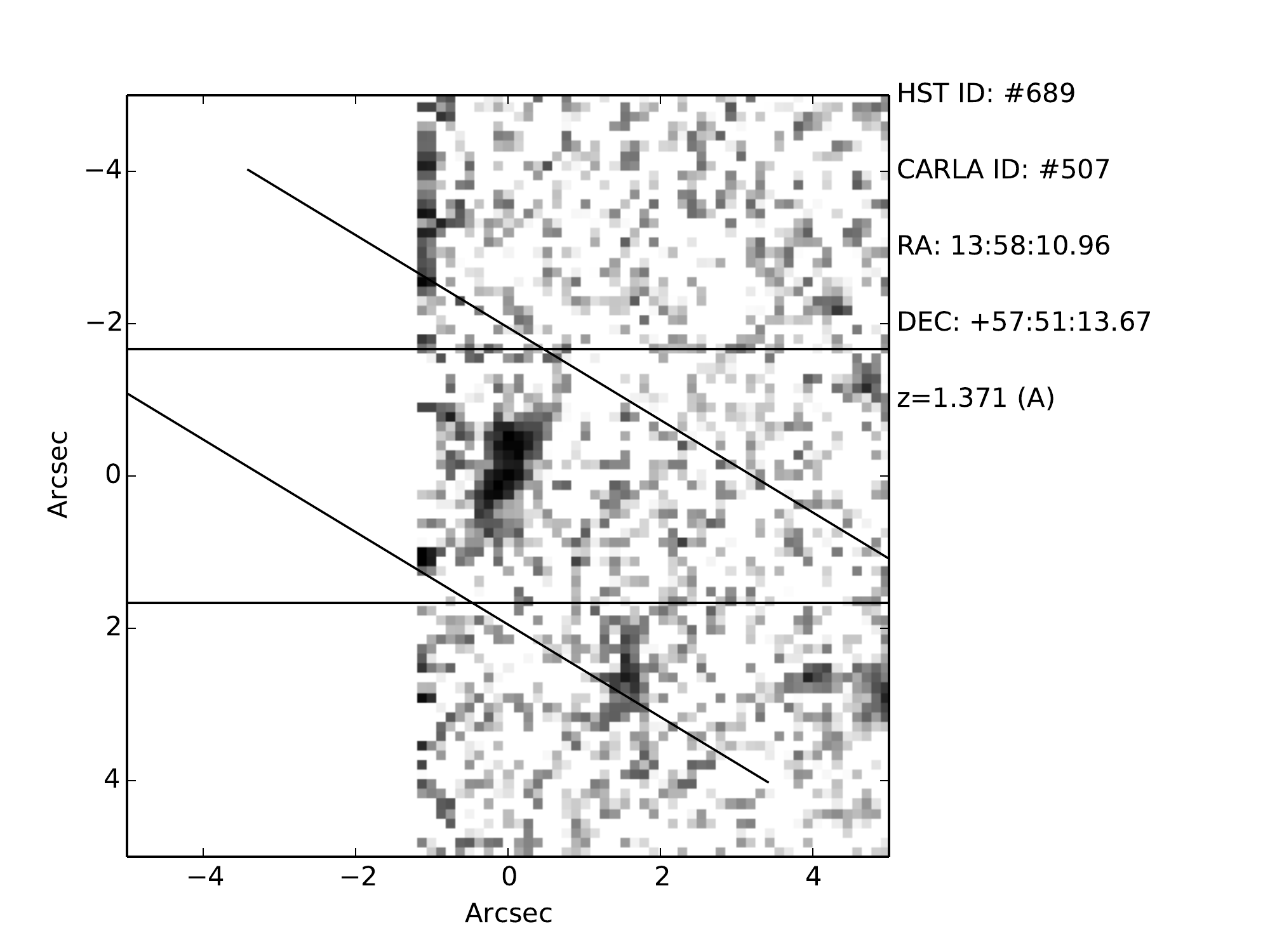} \mbox{(k)}}%
}%
{%
\setlength{\fboxsep}{0pt}%
\setlength{\fboxrule}{1pt}%
\fbox{\includegraphics[page=2, scale=0.24]{CARLA_J1358+5752_694.pdf} \hfill \includegraphics[page=1, scale=0.20]{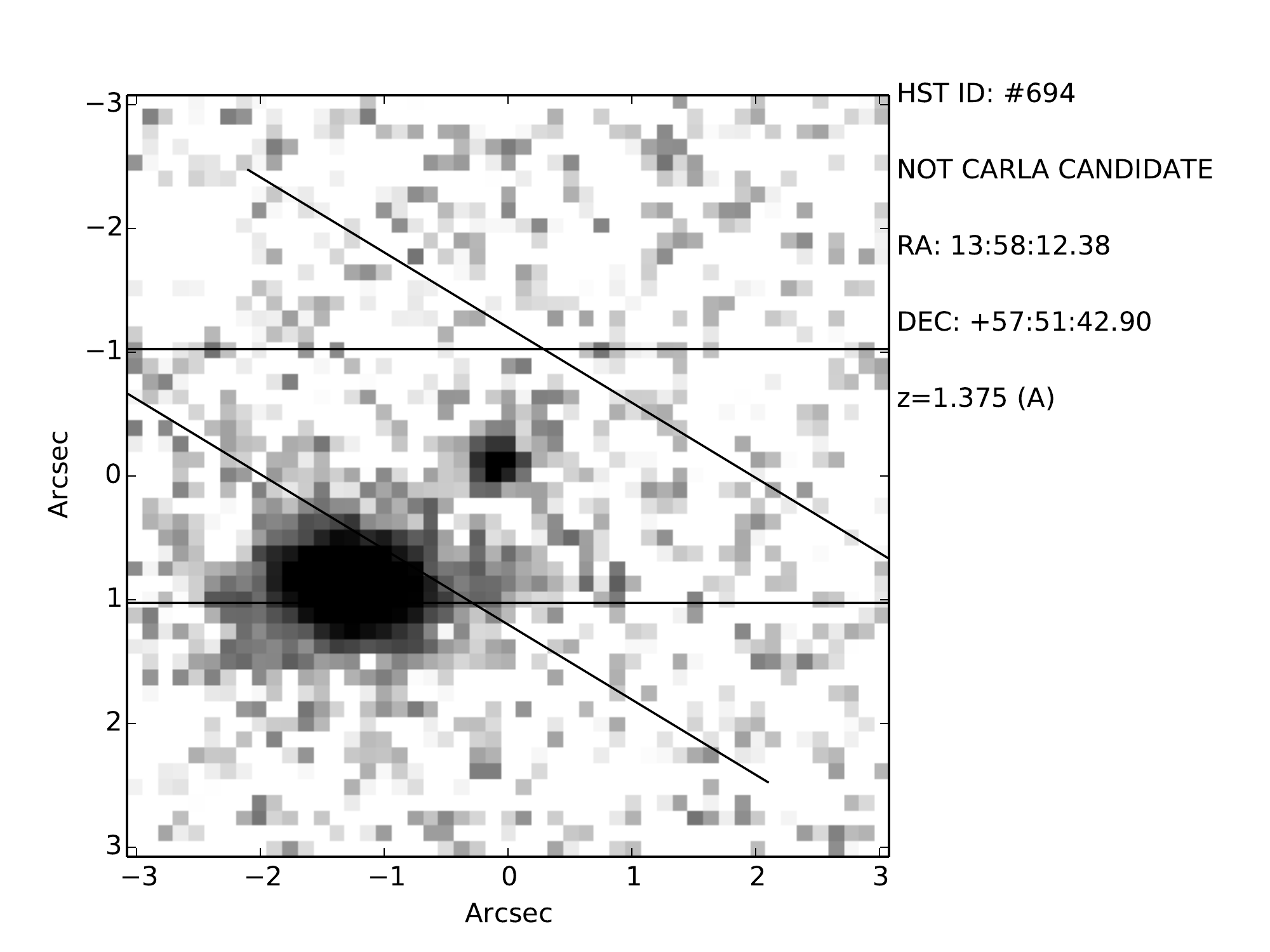} \mbox{(l)}}%
}\\%
\caption[CARLA~J1358+5752 member spectra]{CARLA~J1358+5752 member spectra.}
\label{fig:J1358+5752spectra}
\mbox{}\\
\end{figure*}

\begin{figure*}[!ht]
{%
\setlength{\fboxsep}{0pt}%
\setlength{\fboxrule}{1pt}%
\fbox{\includegraphics[page=2, scale=0.24]{CARLA_J1510+5958_49.pdf} \hfill \includegraphics[page=1, scale=0.20]{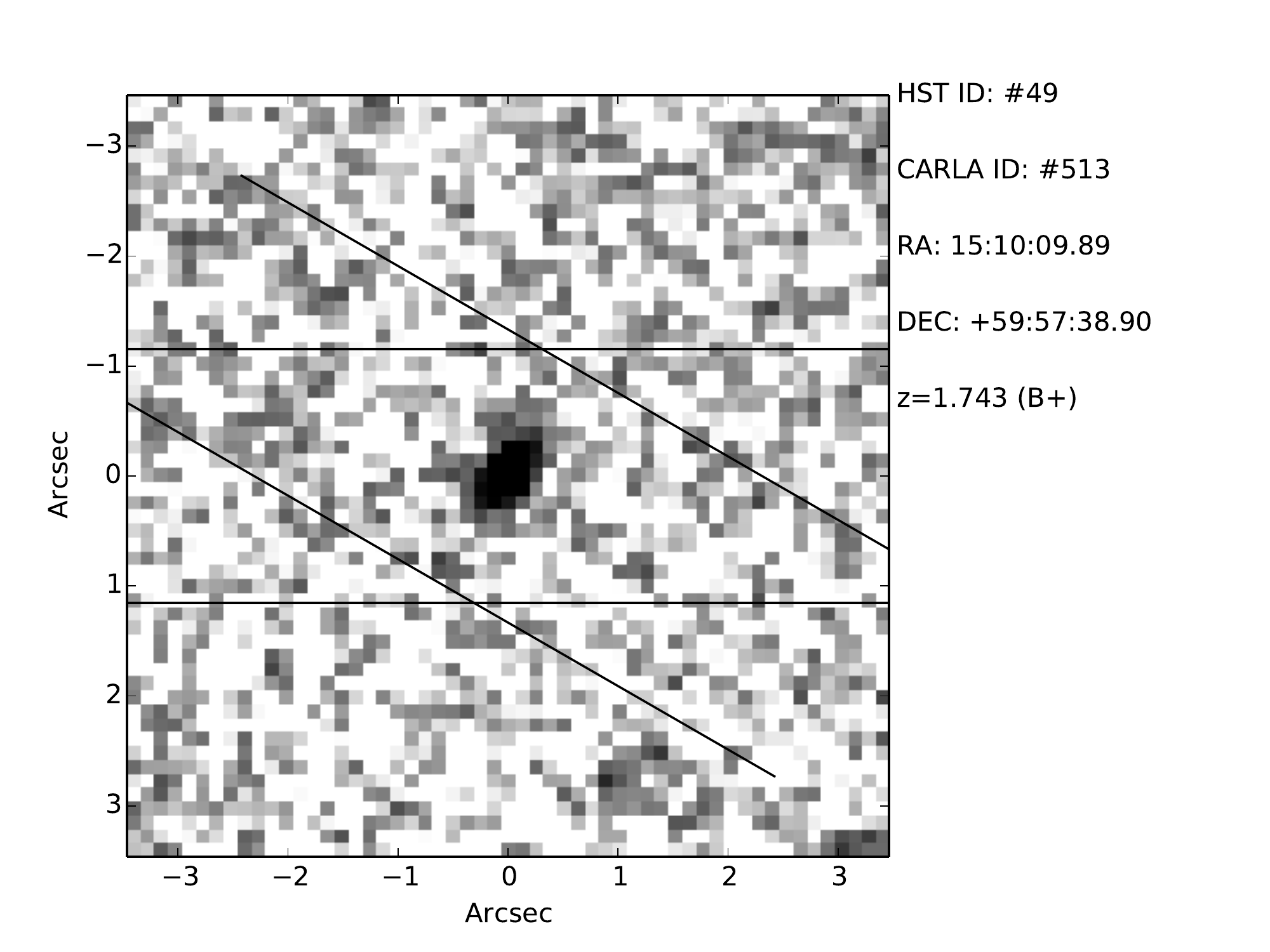} \mbox{(a)}}%
}%
{%
\setlength{\fboxsep}{0pt}%
\setlength{\fboxrule}{1pt}%
\fbox{\includegraphics[page=2, scale=0.24]{CARLA_J1510+5958_221.pdf} \hfill \includegraphics[page=1, scale=0.20]{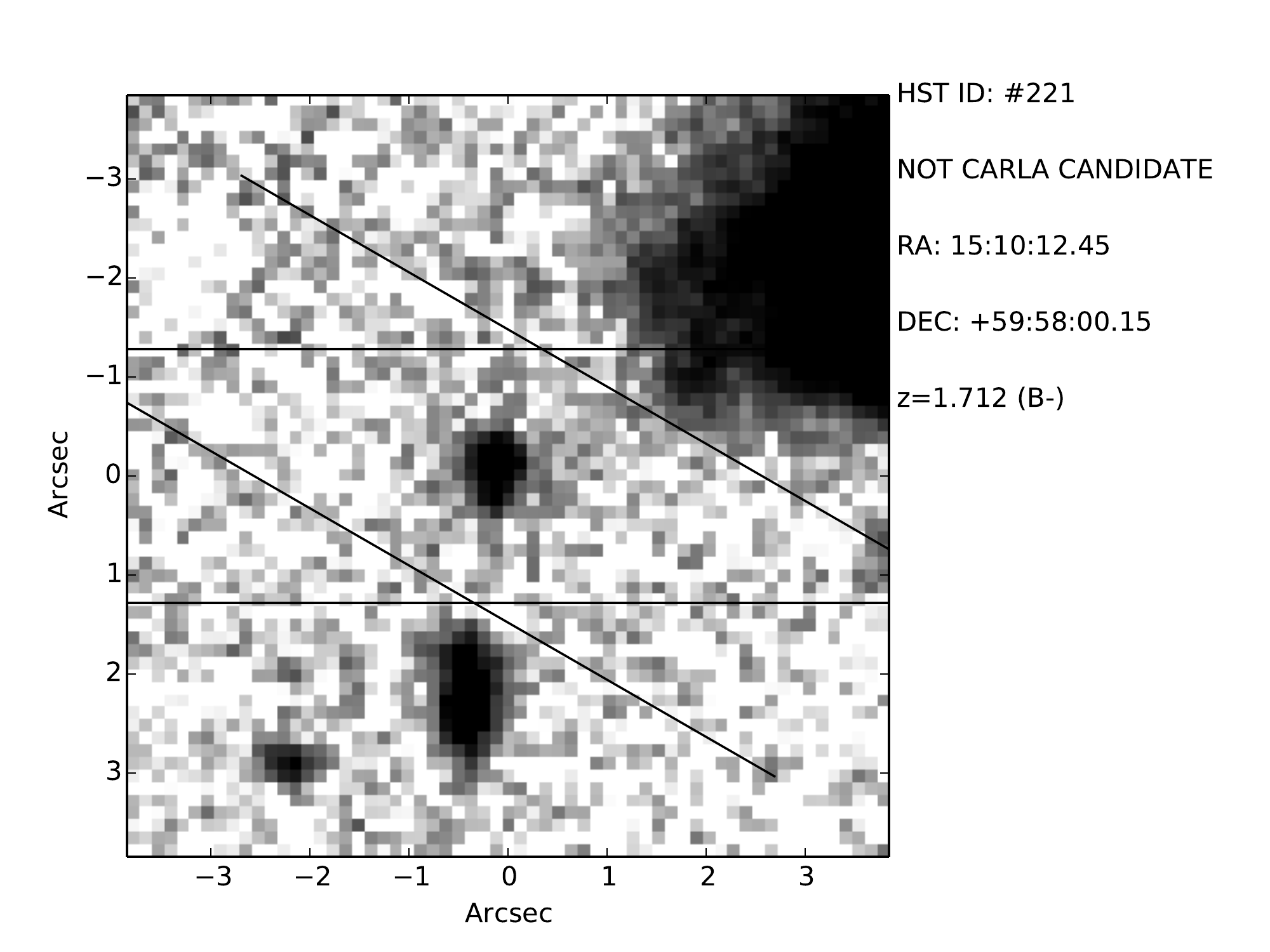} \mbox{(b)}}%
}\\%
{%
\setlength{\fboxsep}{0pt}%
\setlength{\fboxrule}{1pt}%
\fbox{\includegraphics[page=2, scale=0.24]{CARLA_J1510+5958_402.pdf} \hfill \includegraphics[page=1, scale=0.20]{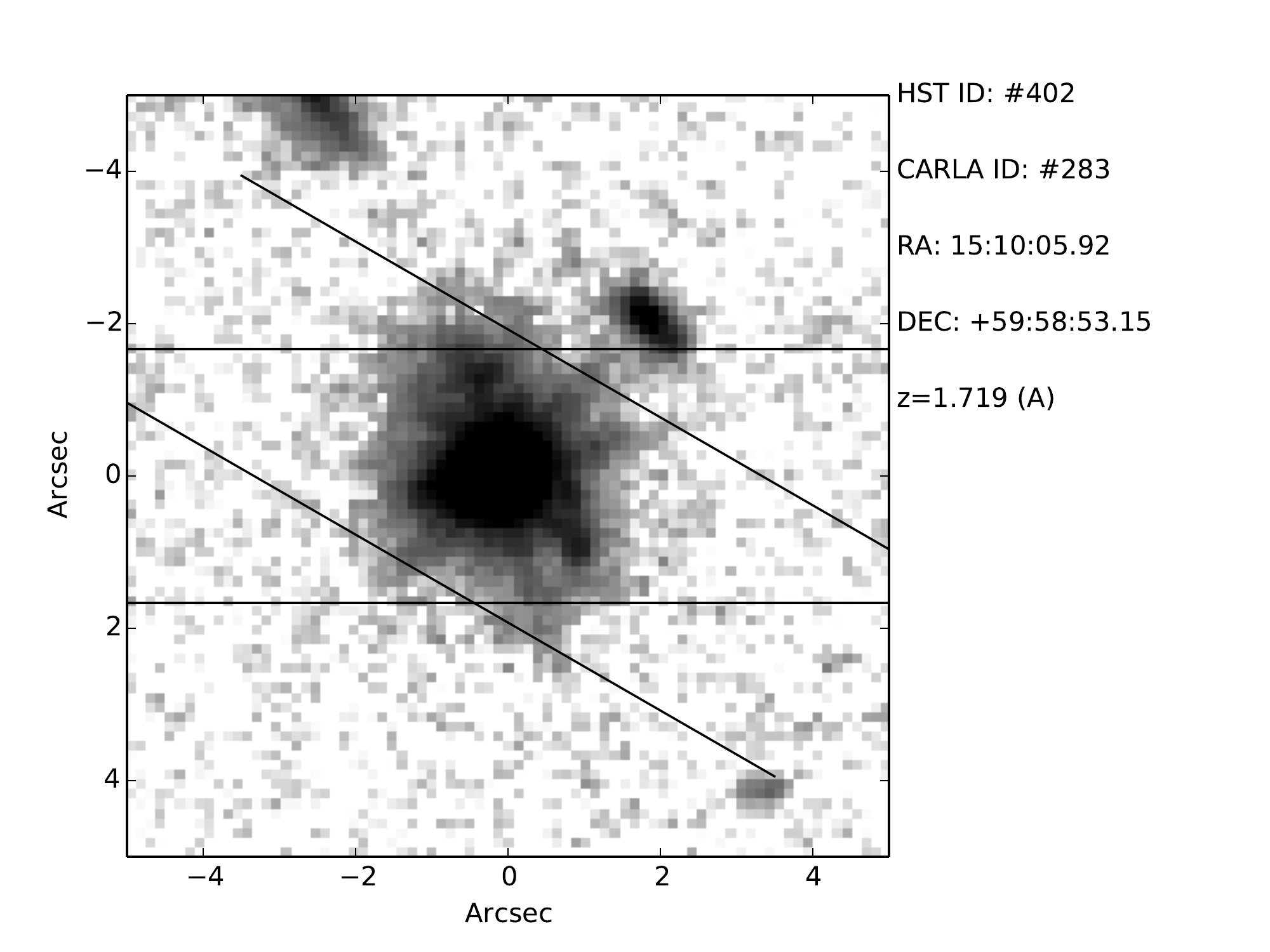} \mbox{(c)}}%
}%
{%
\setlength{\fboxsep}{0pt}%
\setlength{\fboxrule}{1pt}%
\fbox{\includegraphics[page=2, scale=0.24]{CARLA_J1510+5958_771.pdf} \hfill \includegraphics[page=1, scale=0.20]{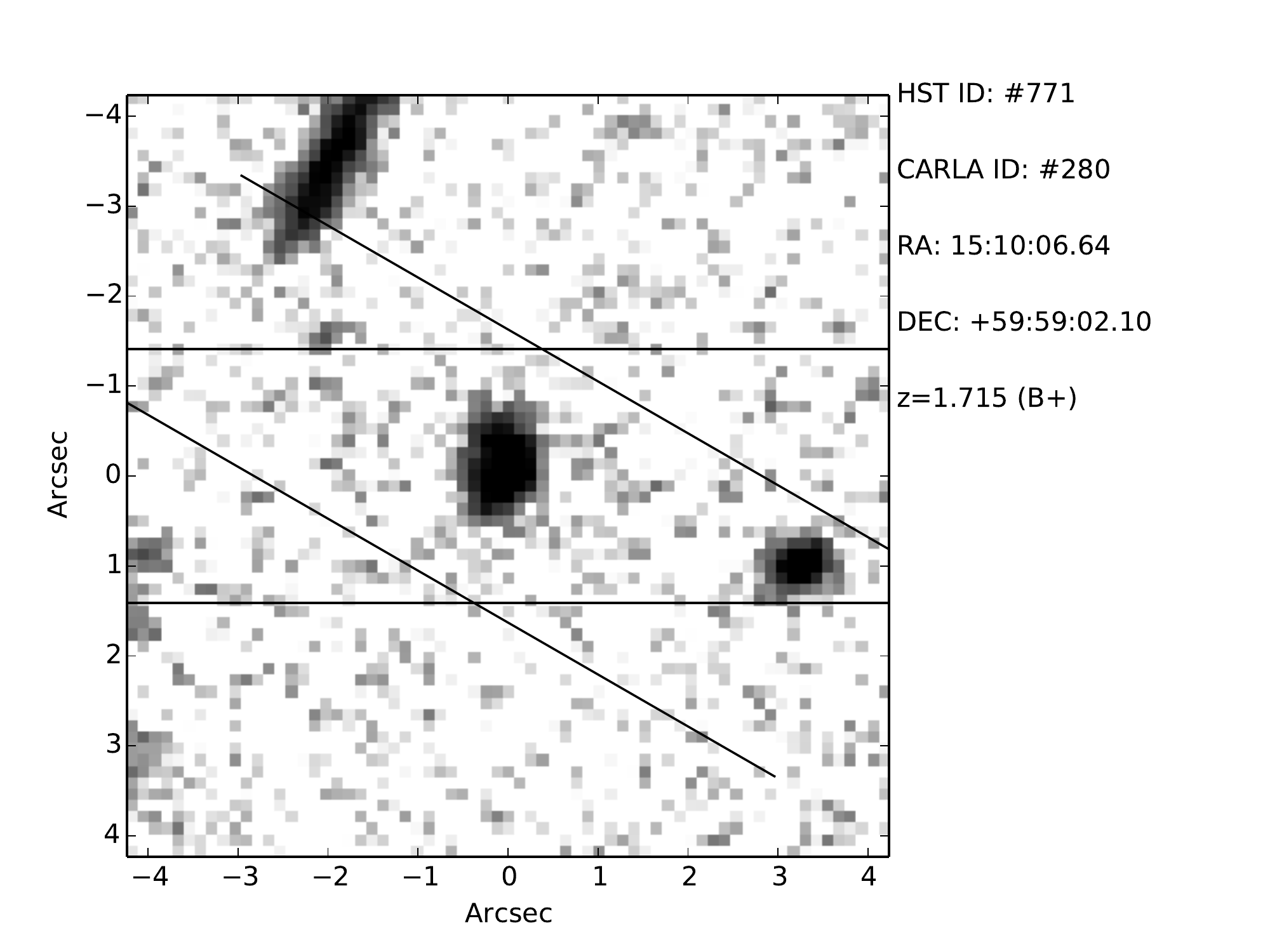} \mbox{(d)}}%
}\\%
{%
\setlength{\fboxsep}{0pt}%
\setlength{\fboxrule}{1pt}%
\fbox{\includegraphics[page=2, scale=0.24]{CARLA_J1510+5958_982.pdf} \hfill \includegraphics[page=1, scale=0.20]{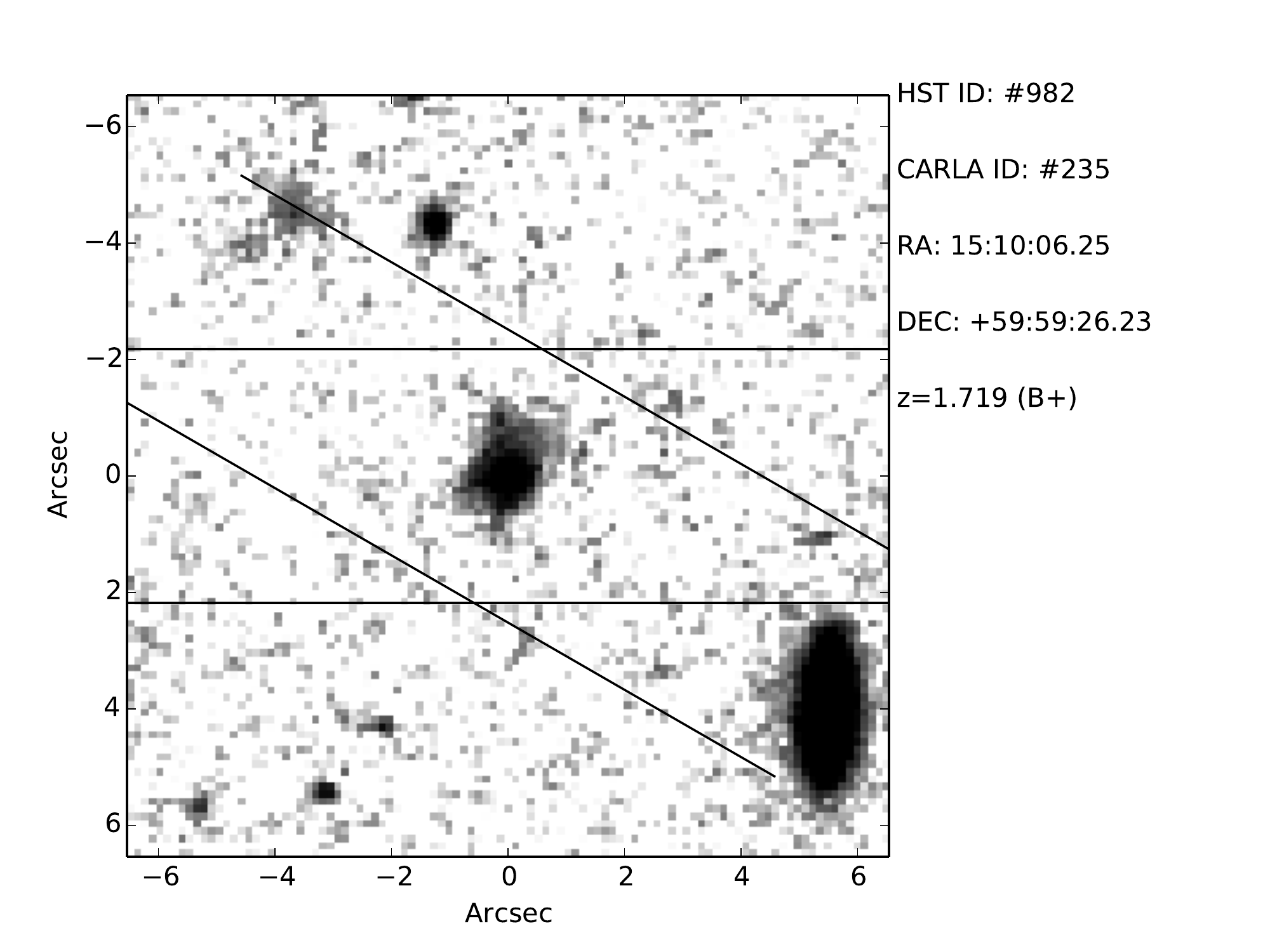} \mbox{(e)}}%
}%
{%
\setlength{\fboxsep}{0pt}%
\setlength{\fboxrule}{1pt}%
\fbox{\includegraphics[page=2, scale=0.24]{CARLA_J1510+5958_1122.pdf} \hfill \includegraphics[page=1, scale=0.20]{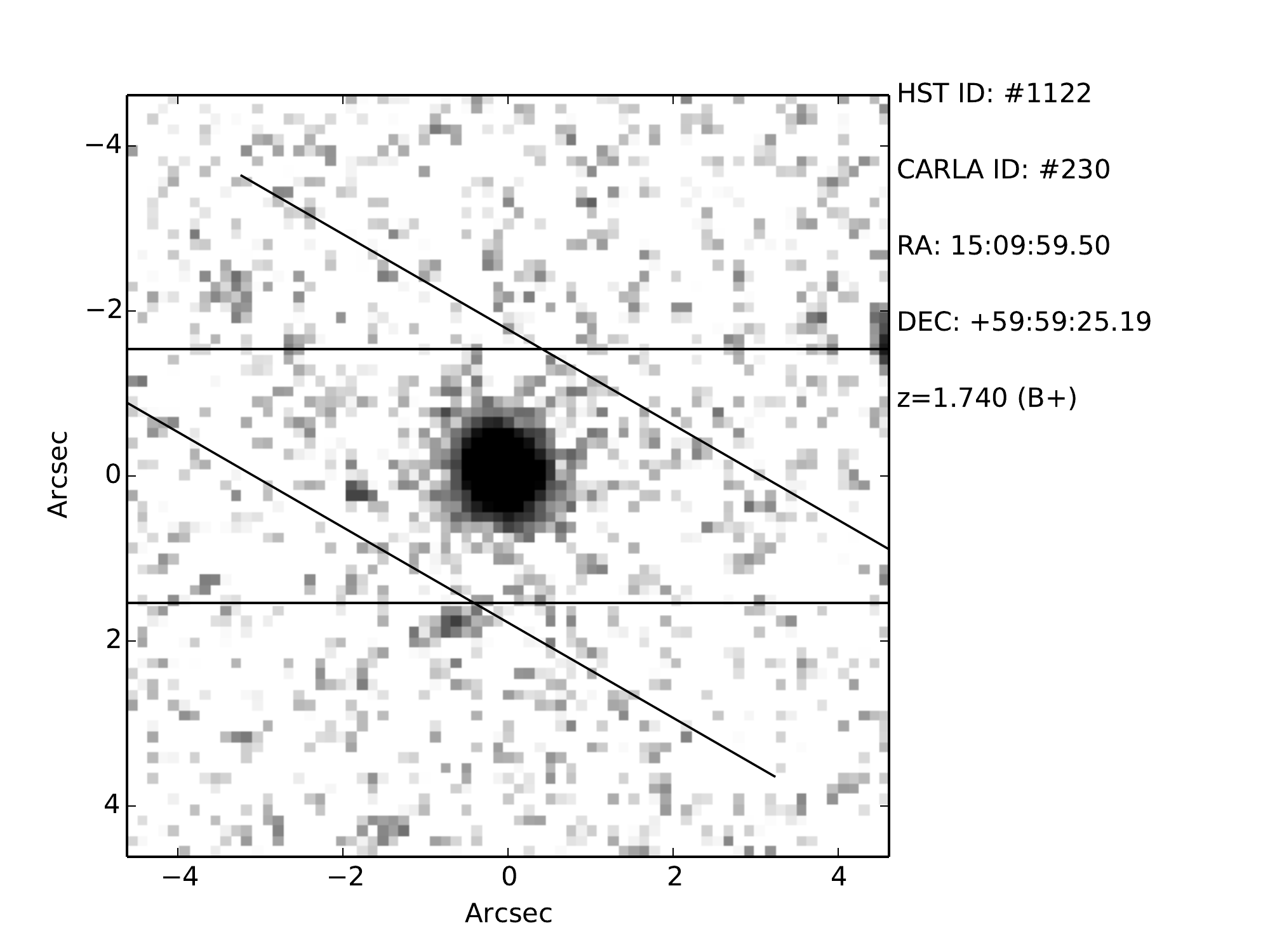} \mbox{(f)}}%
}\\%
\caption[CARLA~J1510+5958 member spectra]{CARLA~J1510+5958 member spectra.}
\label{fig:JJ1510+5958spectra}
\mbox{}\\
\end{figure*}

\begin{figure*}[!ht]
{%
\setlength{\fboxsep}{0pt}%
\setlength{\fboxrule}{1pt}%
\fbox{\includegraphics[page=2, scale=0.24]{CARLA_J1753+6310_157.pdf} \hfill \includegraphics[page=1, scale=0.20]{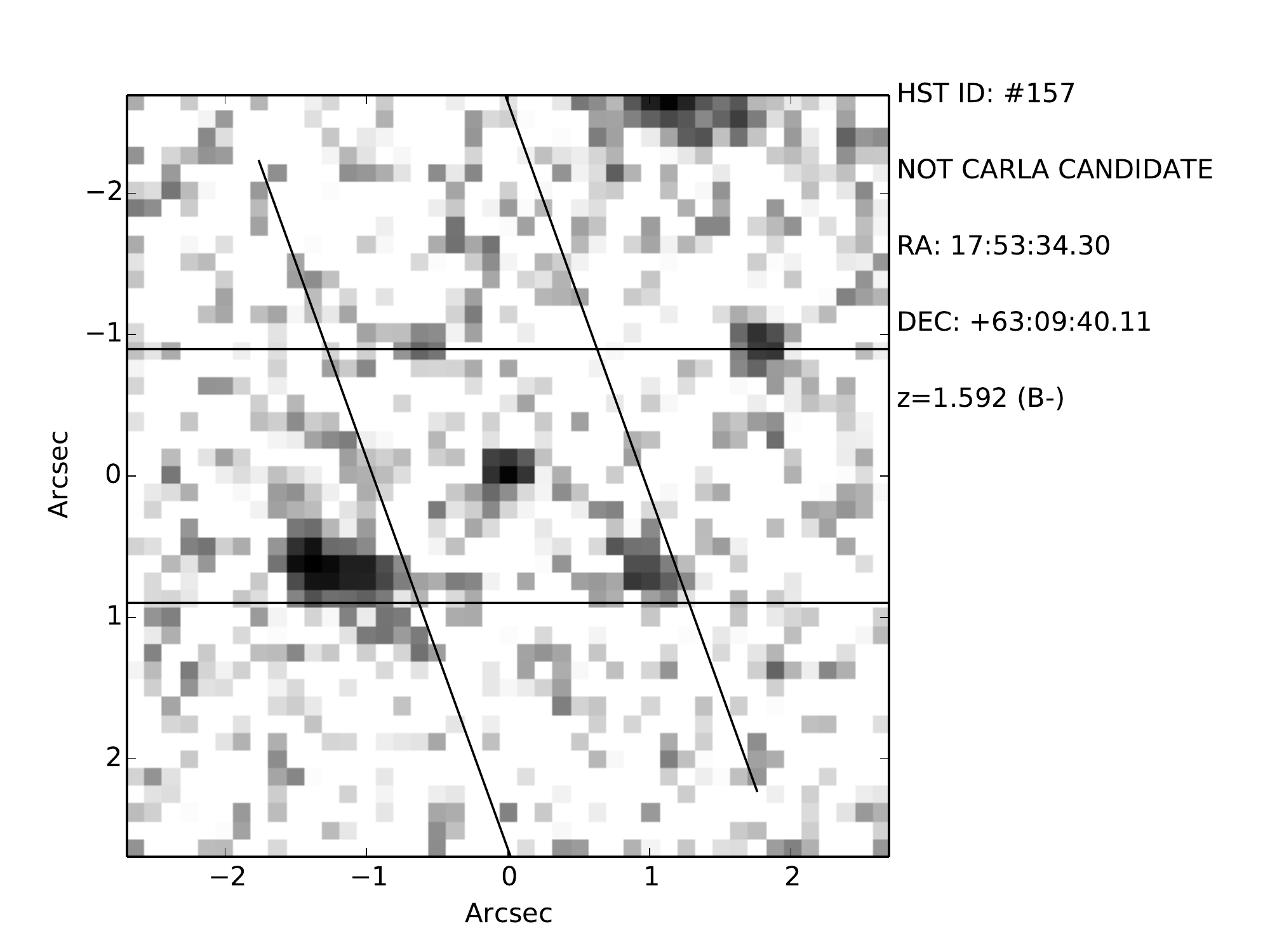} \mbox{(a)}}%
}%
{%
\setlength{\fboxsep}{0pt}%
\setlength{\fboxrule}{1pt}%
\fbox{\includegraphics[page=2, scale=0.24]{CARLA_J1753+6310_457.pdf} \hfill \includegraphics[page=1, scale=0.20]{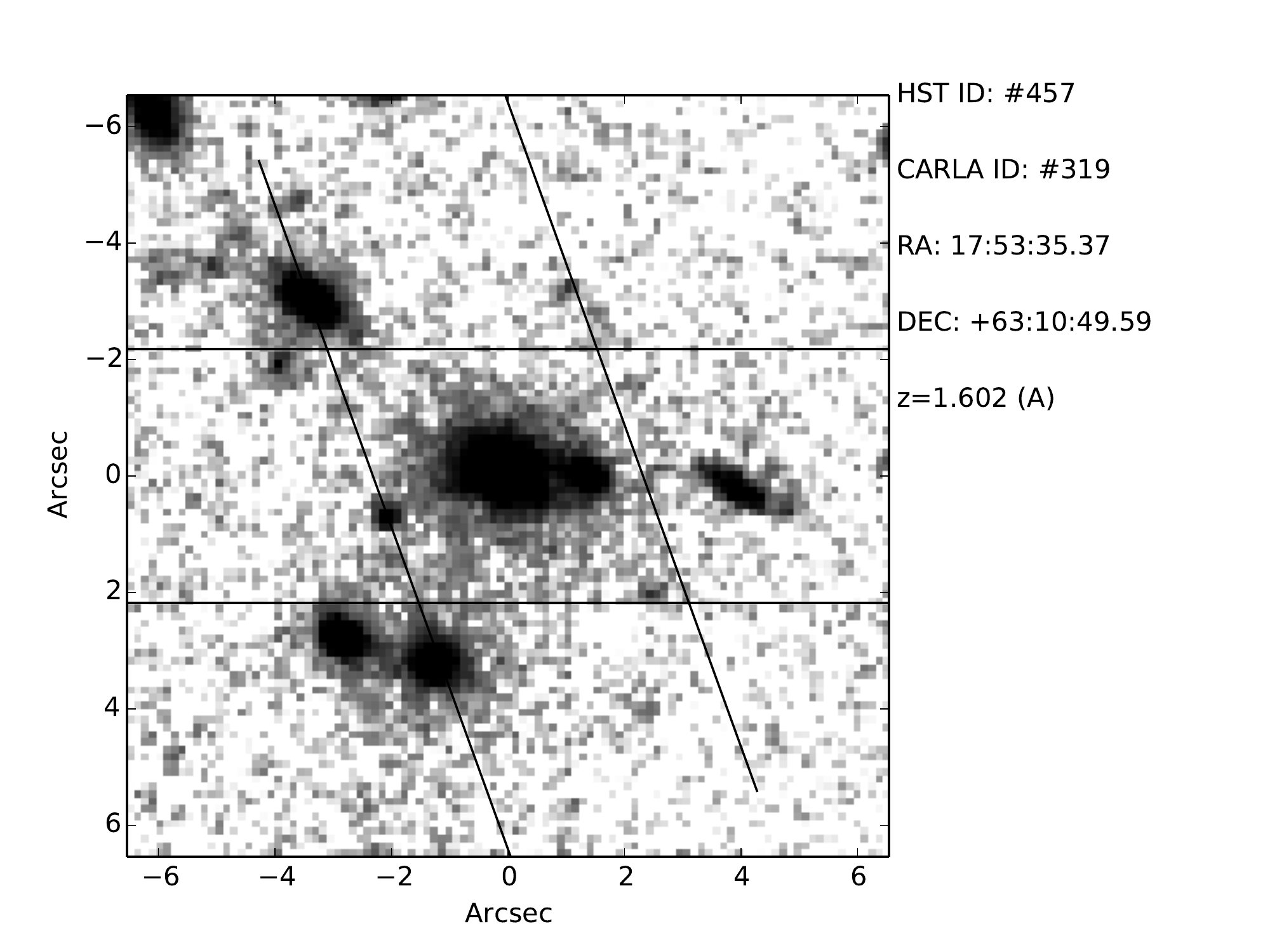} \mbox{(b)}}%
}\\%
{%
\setlength{\fboxsep}{0pt}%
\setlength{\fboxrule}{1pt}%
\fbox{\includegraphics[page=2, scale=0.24]{CARLA_J1753+6310_470.pdf} \hfill \includegraphics[page=1, scale=0.20]{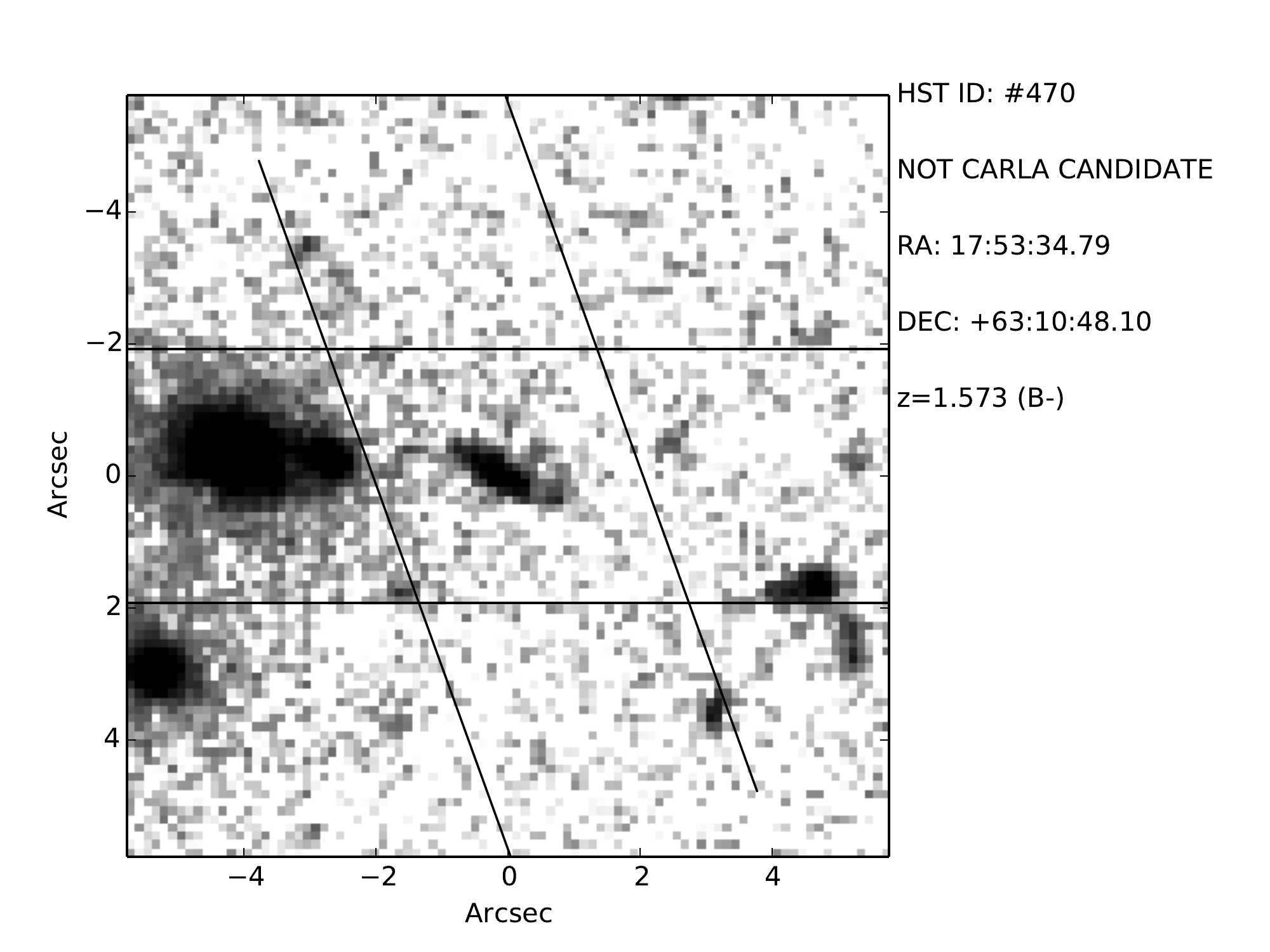} \mbox{(c)}}%
}%
{%
\setlength{\fboxsep}{0pt}%
\setlength{\fboxrule}{1pt}%
\fbox{\includegraphics[page=2, scale=0.24]{CARLA_J1753+6310_614.pdf} \hfill \includegraphics[page=1, scale=0.20]{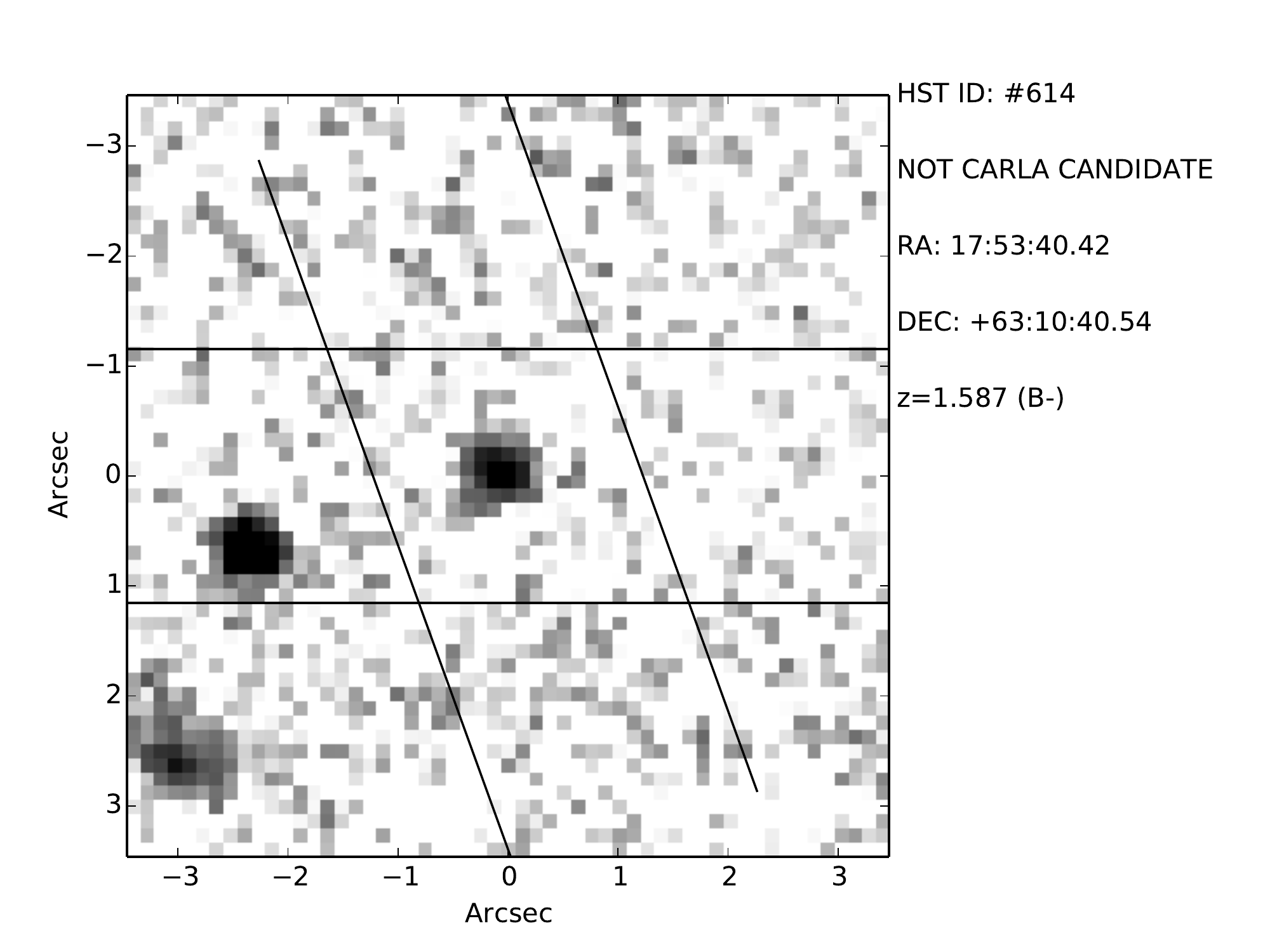} \mbox{(d)}}%
}\\%
{%
\setlength{\fboxsep}{0pt}%
\setlength{\fboxrule}{1pt}%
\fbox{\includegraphics[page=2, scale=0.24]{CARLA_J1753+6310_619.pdf} \hfill \includegraphics[page=1, scale=0.20]{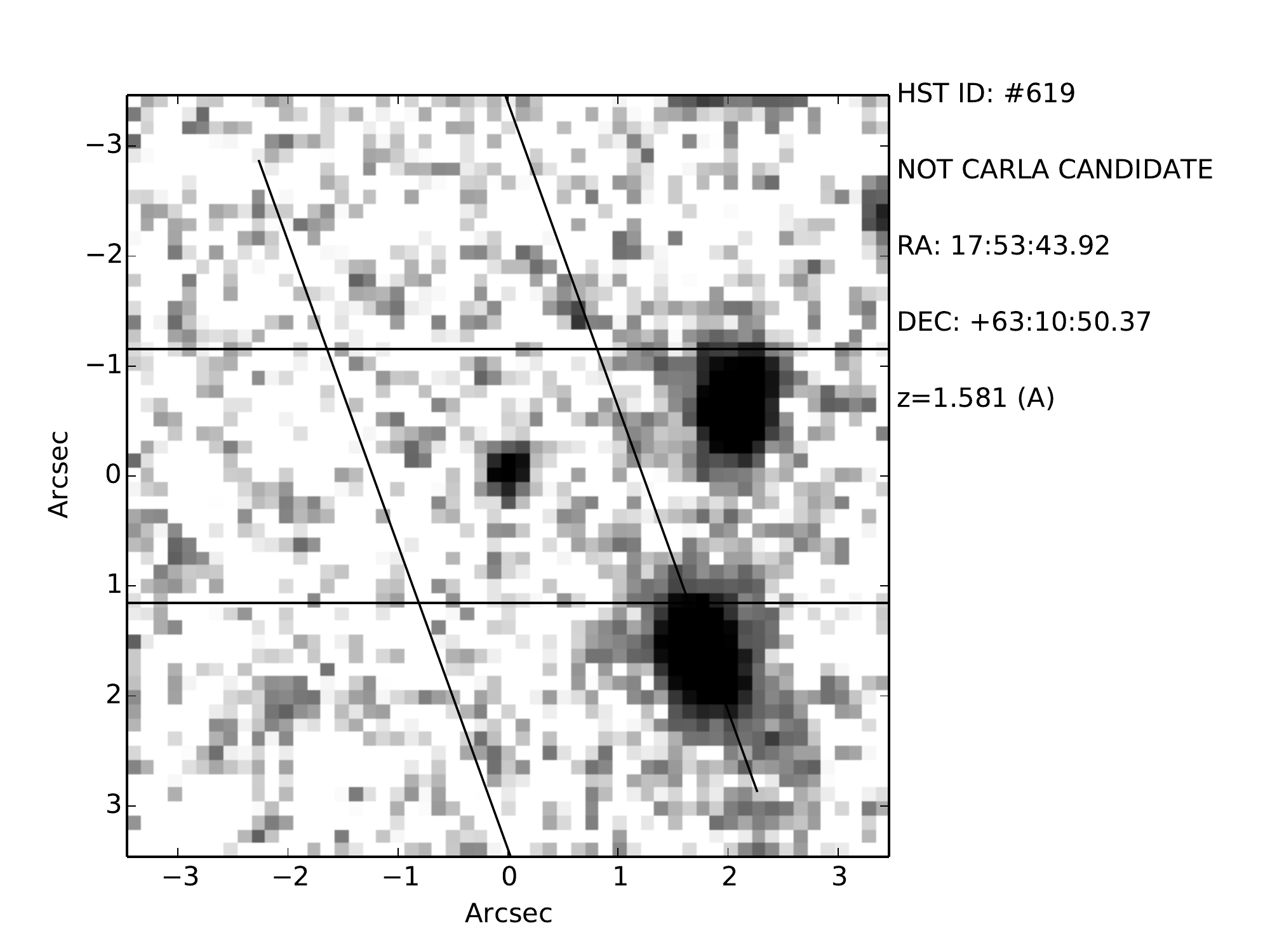} \mbox{(e)}}%
}%
\caption[CARLA~J1753+6310 member spectra]{CARLA~J1753+6310 member spectra.}
\label{fig:J1753+6310spectra}
\end{figure*}

\begin{figure*}[!ht]
{%
\setlength{\fboxsep}{0pt}%
\setlength{\fboxrule}{1pt}%
\fbox{\includegraphics[page=2, scale=0.24]{CARLA_J2227-2705_268.pdf} \hfill \includegraphics[page=1, scale=0.20]{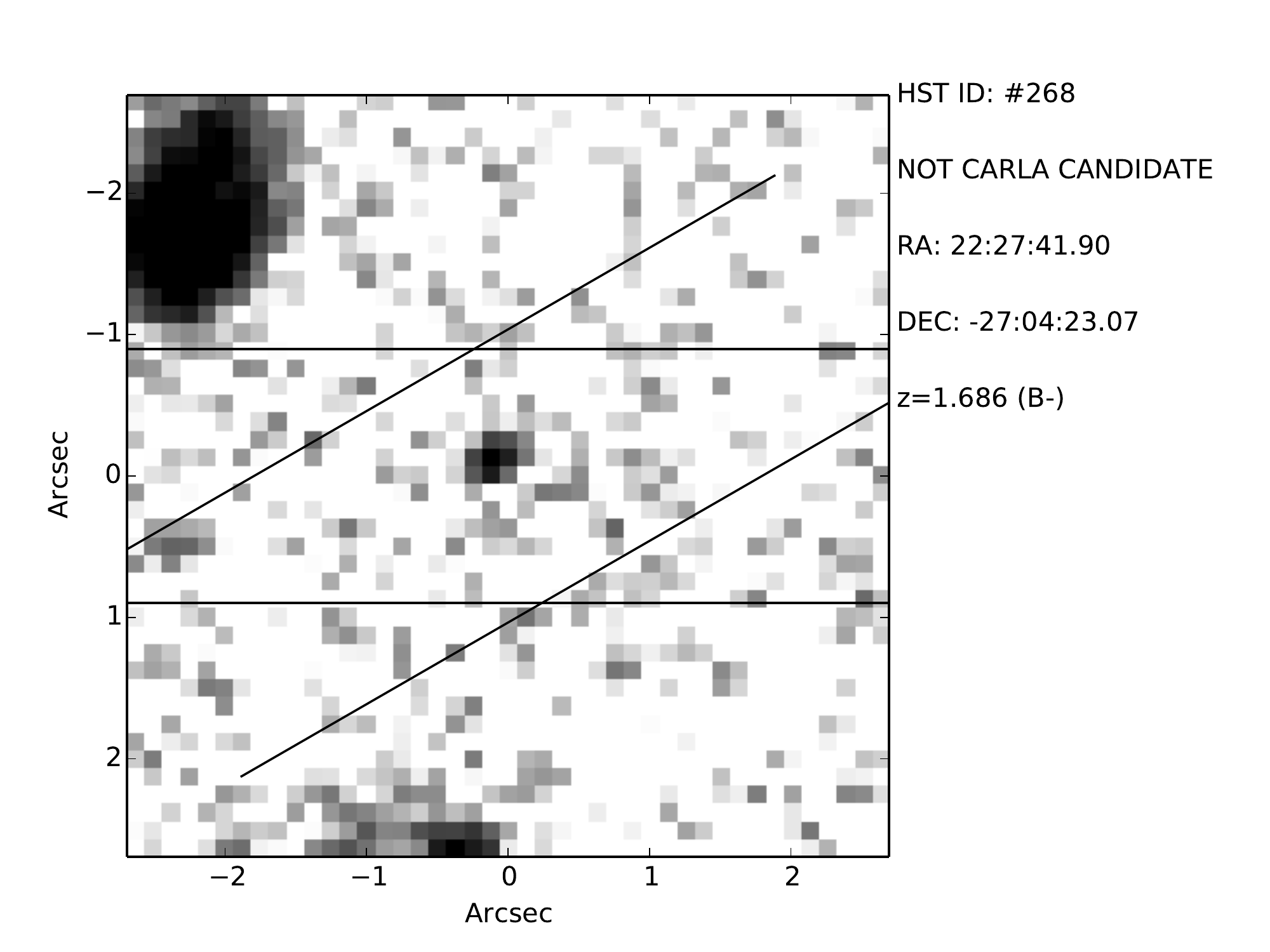} \mbox{(a)}}%
}%
{%
\setlength{\fboxsep}{0pt}%
\setlength{\fboxrule}{1pt}%
\fbox{\includegraphics[page=2, scale=0.24]{CARLA_J2227-2705_406.pdf} \hfill \includegraphics[page=1, scale=0.20]{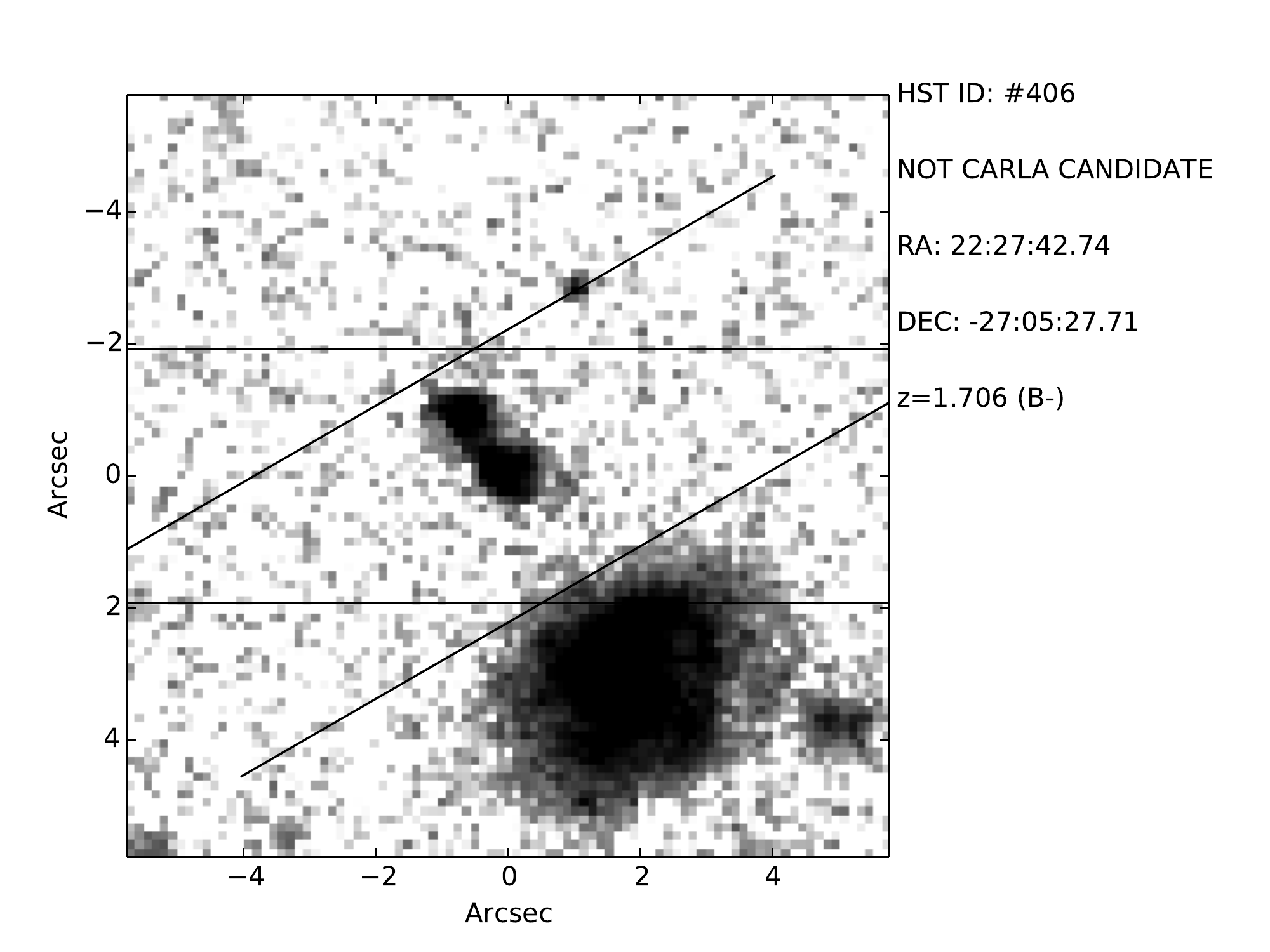} \mbox{(b)}}%
}\\%
{%
\setlength{\fboxsep}{0pt}%
\setlength{\fboxrule}{1pt}%
\fbox{\includegraphics[page=2, scale=0.24]{CARLA_J2227-2705_420.pdf} \hfill \includegraphics[page=1, scale=0.20]{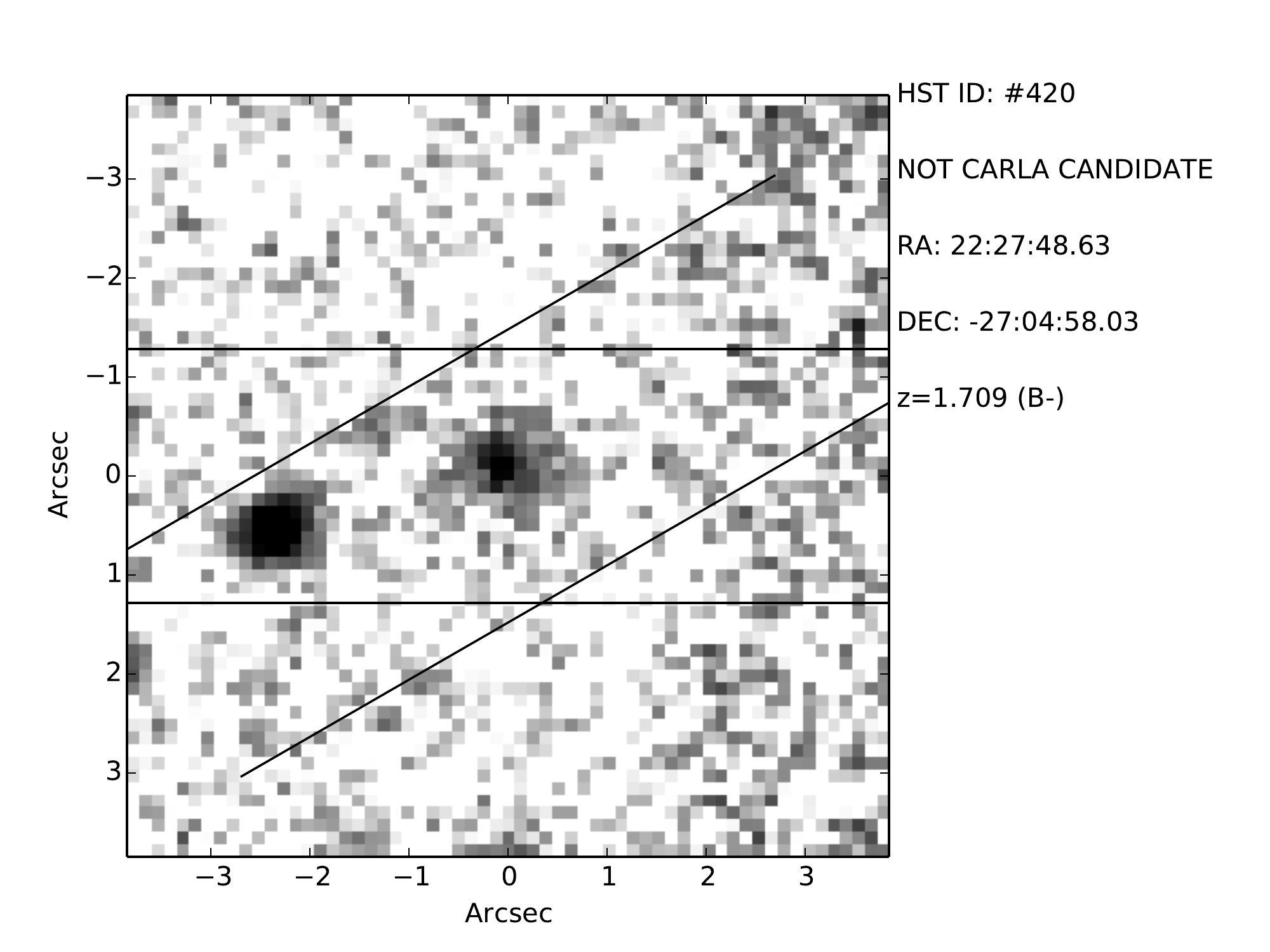} \mbox{(c)}}%
}%
{%
\setlength{\fboxsep}{0pt}%
\setlength{\fboxrule}{1pt}%
\fbox{\includegraphics[page=2, scale=0.24]{CARLA_J2227-2705_525.pdf} \hfill \includegraphics[page=1, scale=0.20]{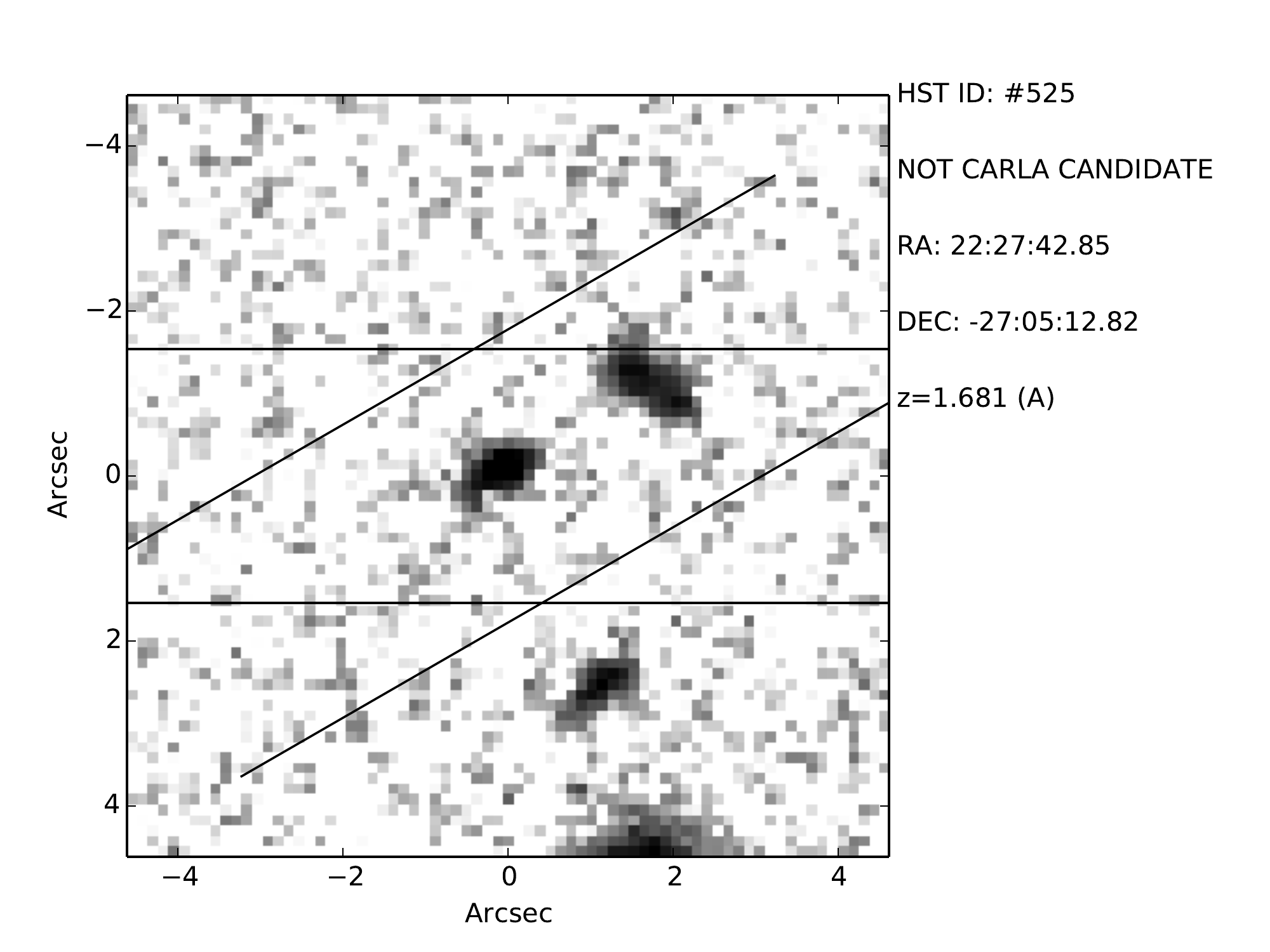} \mbox{(d)}}%
}\\%
{%
\setlength{\fboxsep}{0pt}%
\setlength{\fboxrule}{1pt}%
\fbox{\includegraphics[page=2, scale=0.24]{CARLA_J2227-2705_590.pdf} \hfill \includegraphics[page=1, scale=0.20]{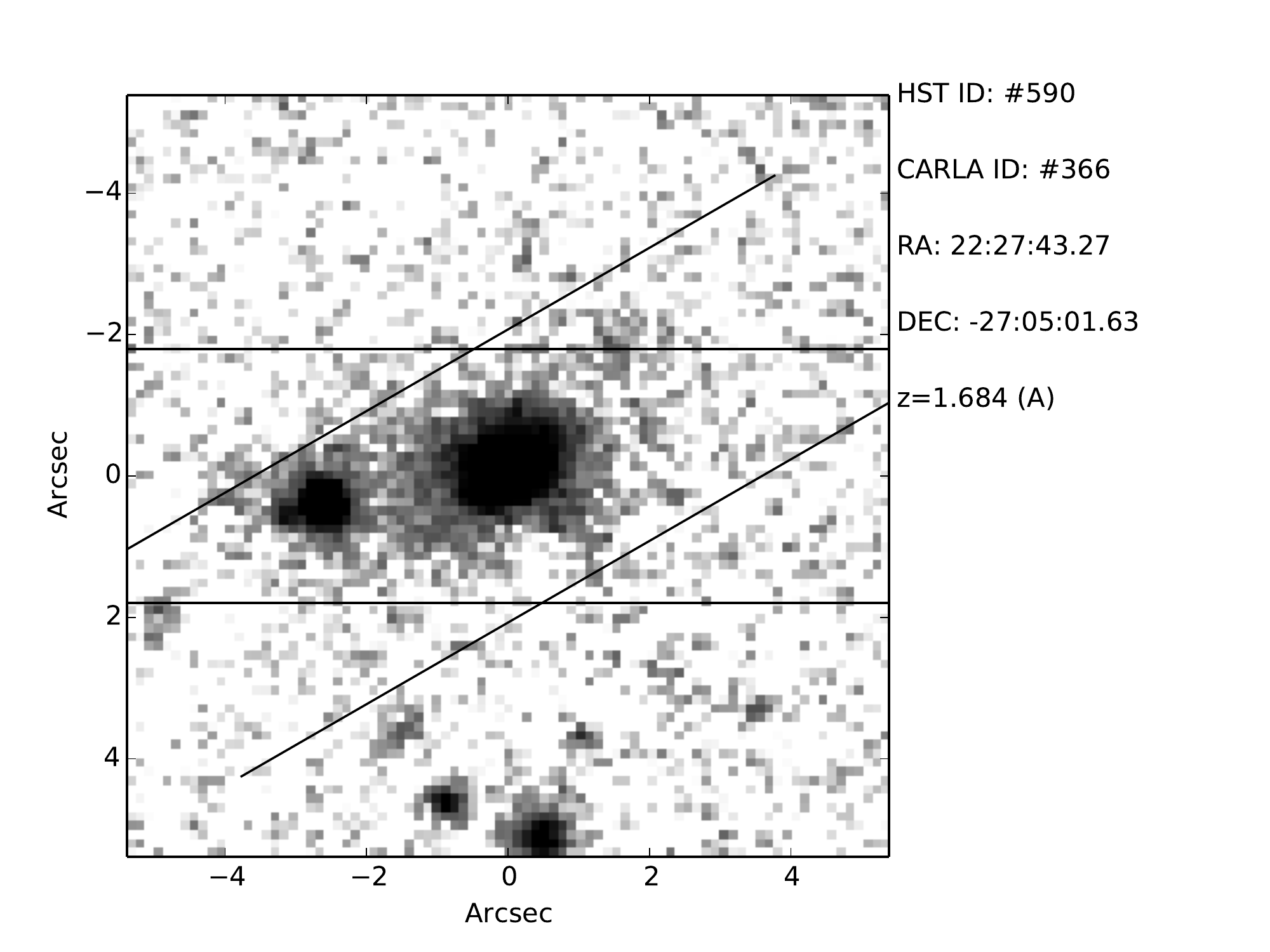} \mbox{(e)}}%
}%
{%
\setlength{\fboxsep}{0pt}%
\setlength{\fboxrule}{1pt}%
\fbox{\includegraphics[page=2, scale=0.24]{CARLA_J2227-2705_762.pdf} \hfill \includegraphics[page=1, scale=0.20]{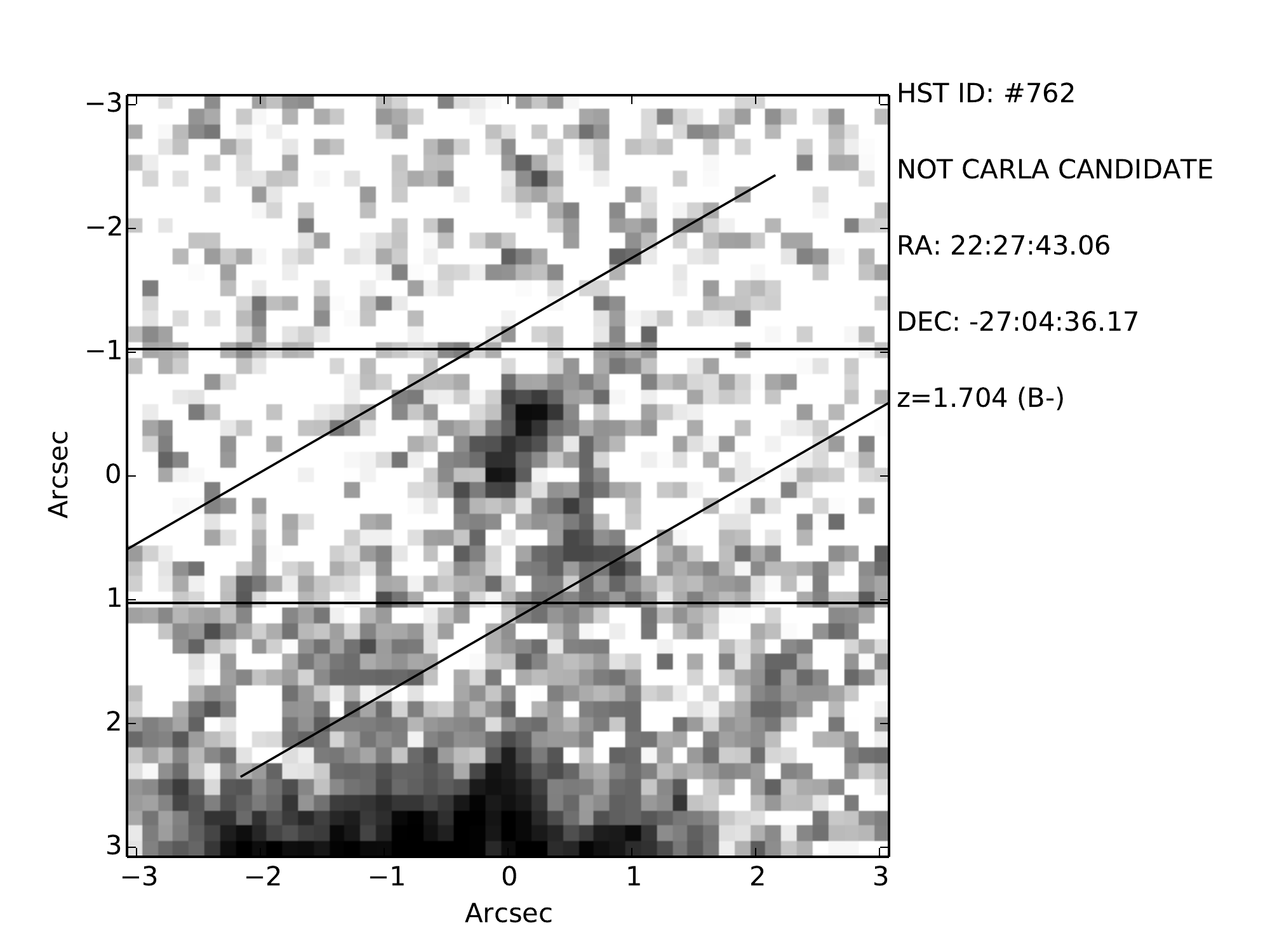} \mbox{(f)}}%
}\\%
{%
\setlength{\fboxsep}{0pt}%
\setlength{\fboxrule}{1pt}%
\fbox{\includegraphics[page=2, scale=0.24]{CARLA_J2227-2705_1022.pdf} \hfill \includegraphics[page=1, scale=0.20]{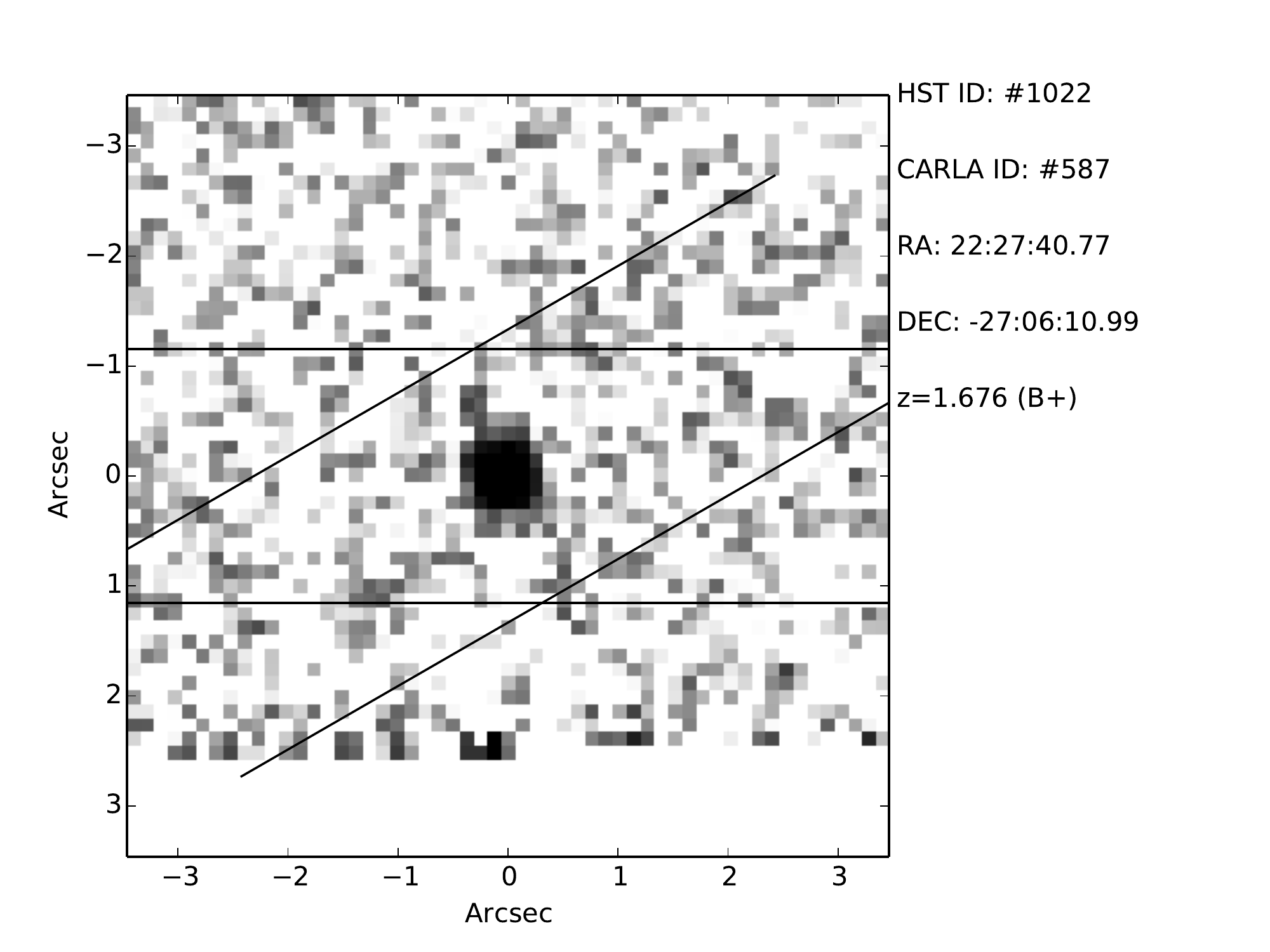} \mbox{(g)}}%
}%
\caption[CARLA~J2227-2705 member spectra]{CARLA~J2227$-$2705 member spectra.}
\label{fig:J2227-2705spectra}
\mbox{}\\
\end{figure*}

\begin{figure*}[!ht]
{%
\setlength{\fboxsep}{0pt}%
\setlength{\fboxrule}{1pt}%
\fbox{\includegraphics[page=2, scale=0.24]{CARLA_J2355-0002_213.pdf} \hfill \includegraphics[page=1, scale=0.20]{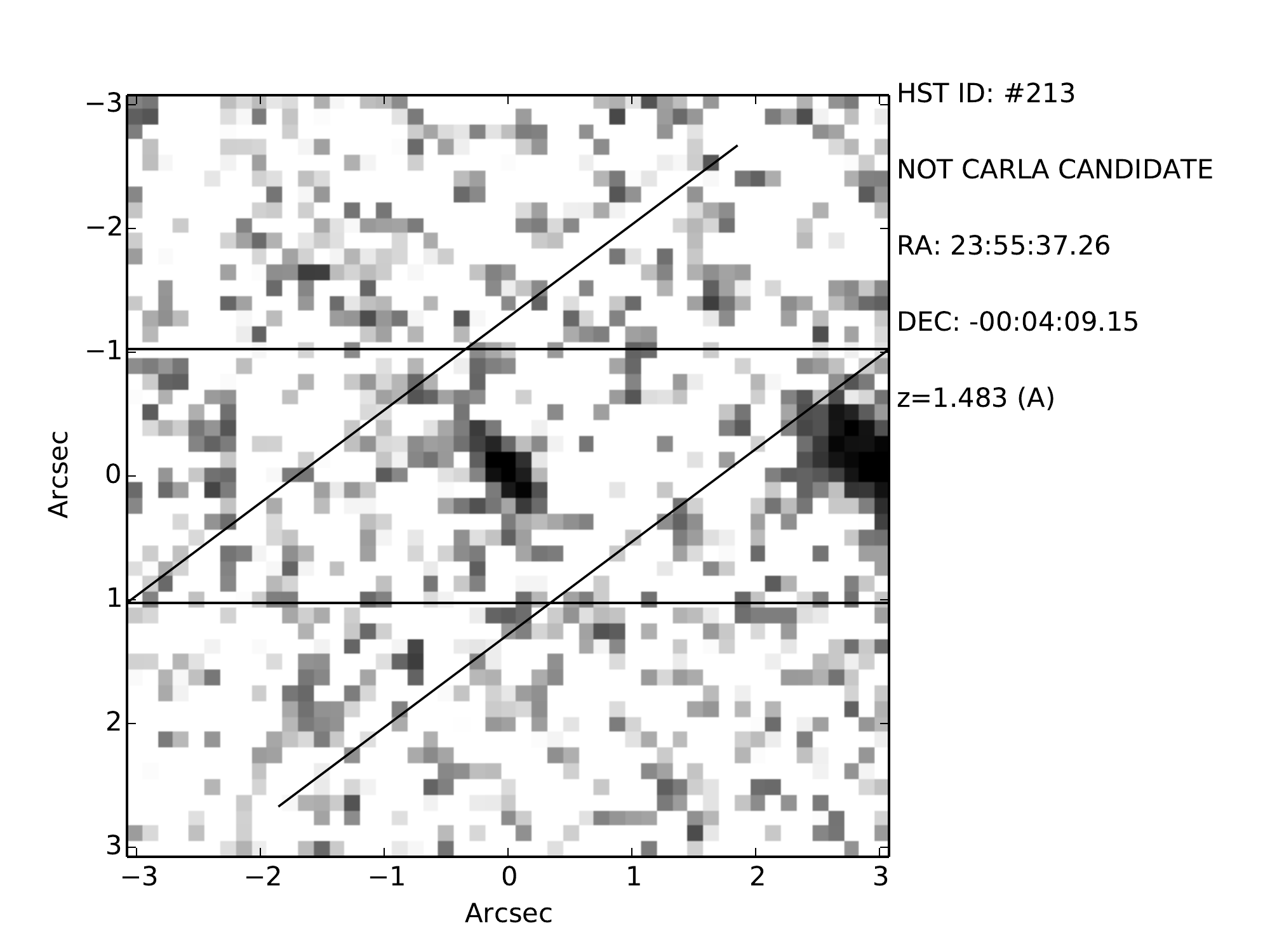} \mbox{(a)}}%
}%
{%
\setlength{\fboxsep}{0pt}%
\setlength{\fboxrule}{1pt}%
\fbox{\includegraphics[page=2, scale=0.24]{CARLA_J2355-0002_334.pdf} \hfill \includegraphics[page=1, scale=0.20]{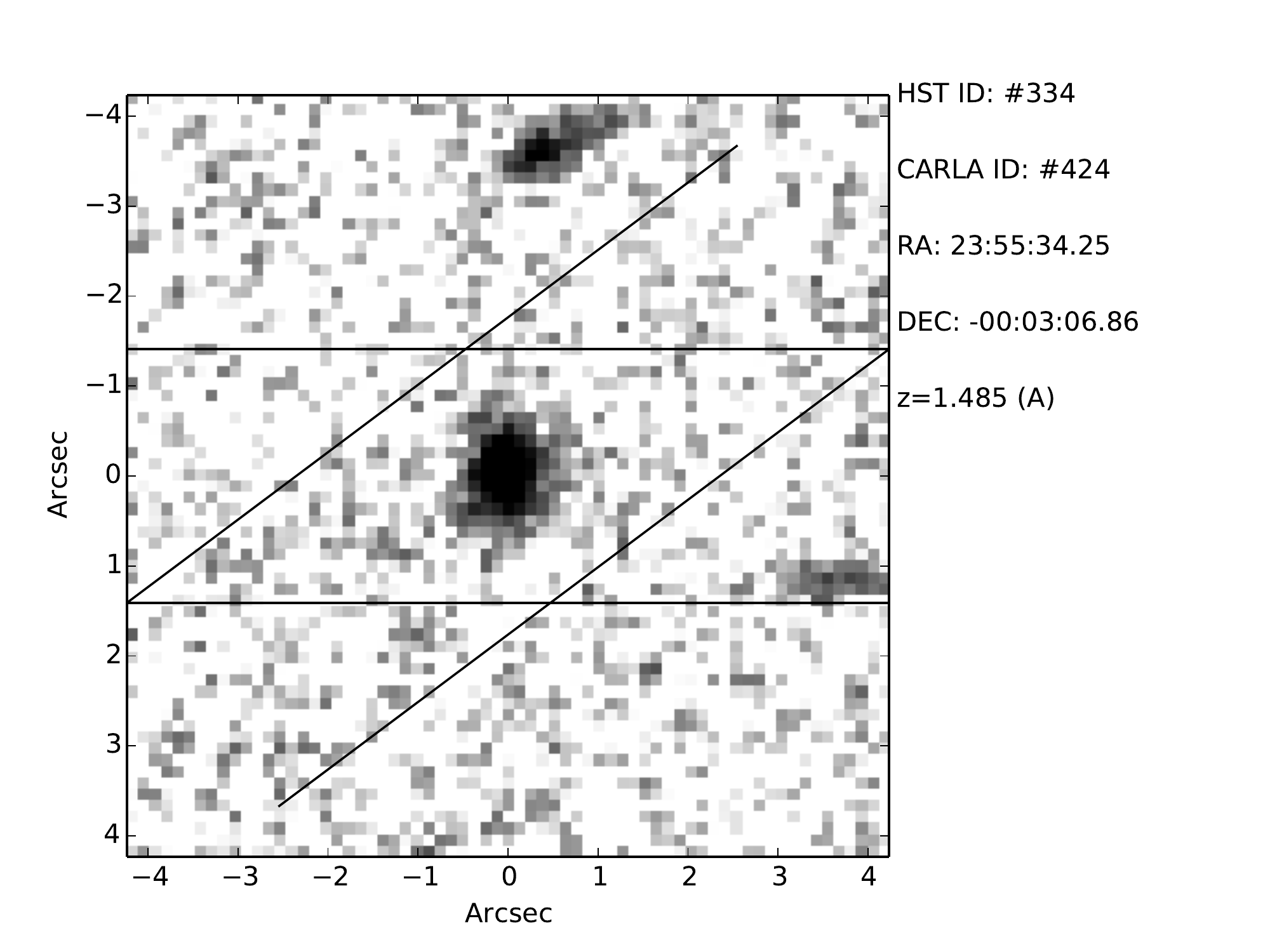} \mbox{(b)}}%
}\\%
{%
\setlength{\fboxsep}{0pt}%
\setlength{\fboxrule}{1pt}%
\fbox{\includegraphics[page=2, scale=0.24]{CARLA_J2355-0002_337.pdf} \hfill \includegraphics[page=1, scale=0.20]{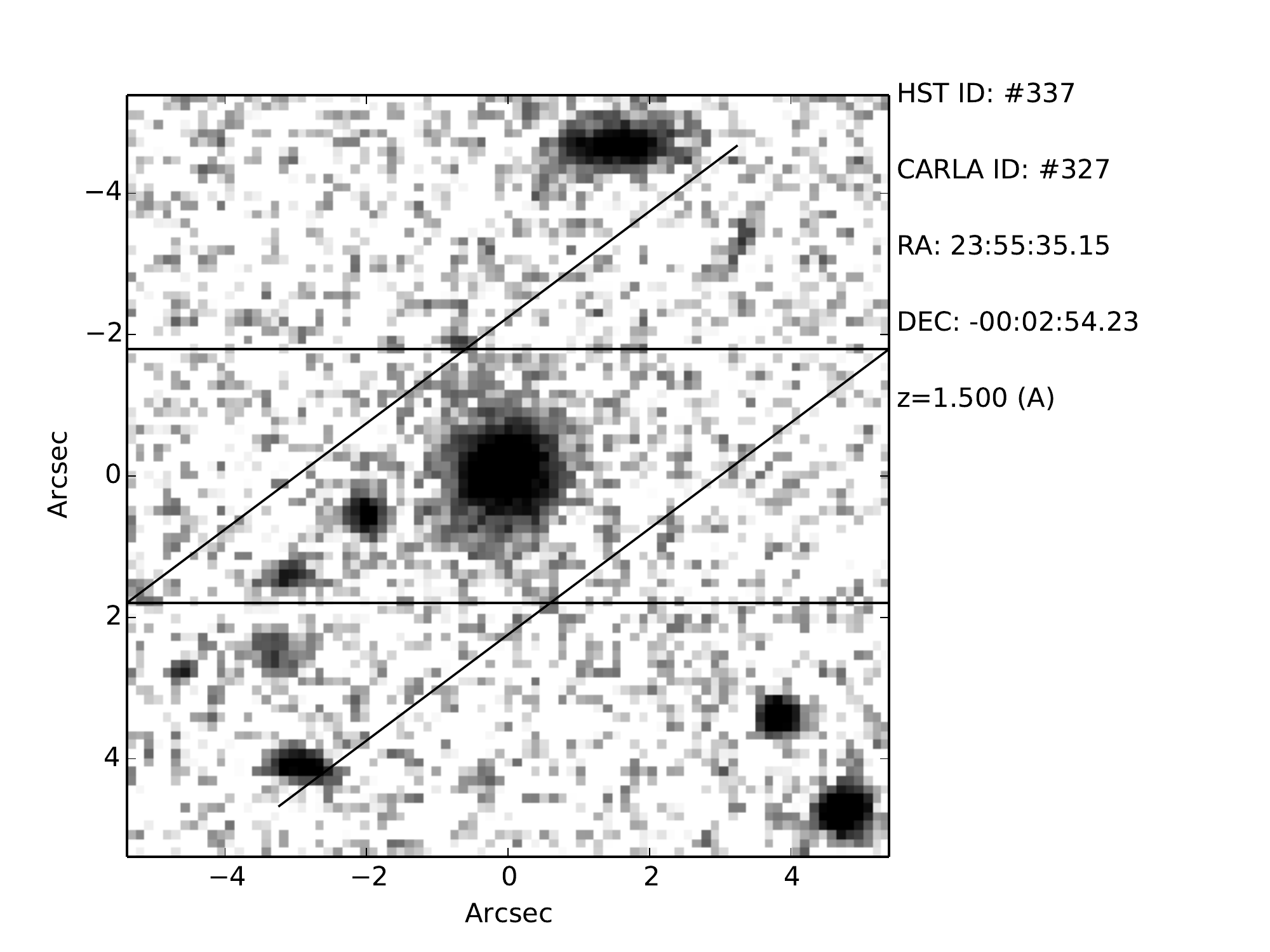} \mbox{(c)}}%
}%
{%
\setlength{\fboxsep}{0pt}%
\setlength{\fboxrule}{1pt}%
\fbox{\includegraphics[page=2, scale=0.24]{CARLA_J2355-0002_351.pdf} \hfill \includegraphics[page=1, scale=0.20]{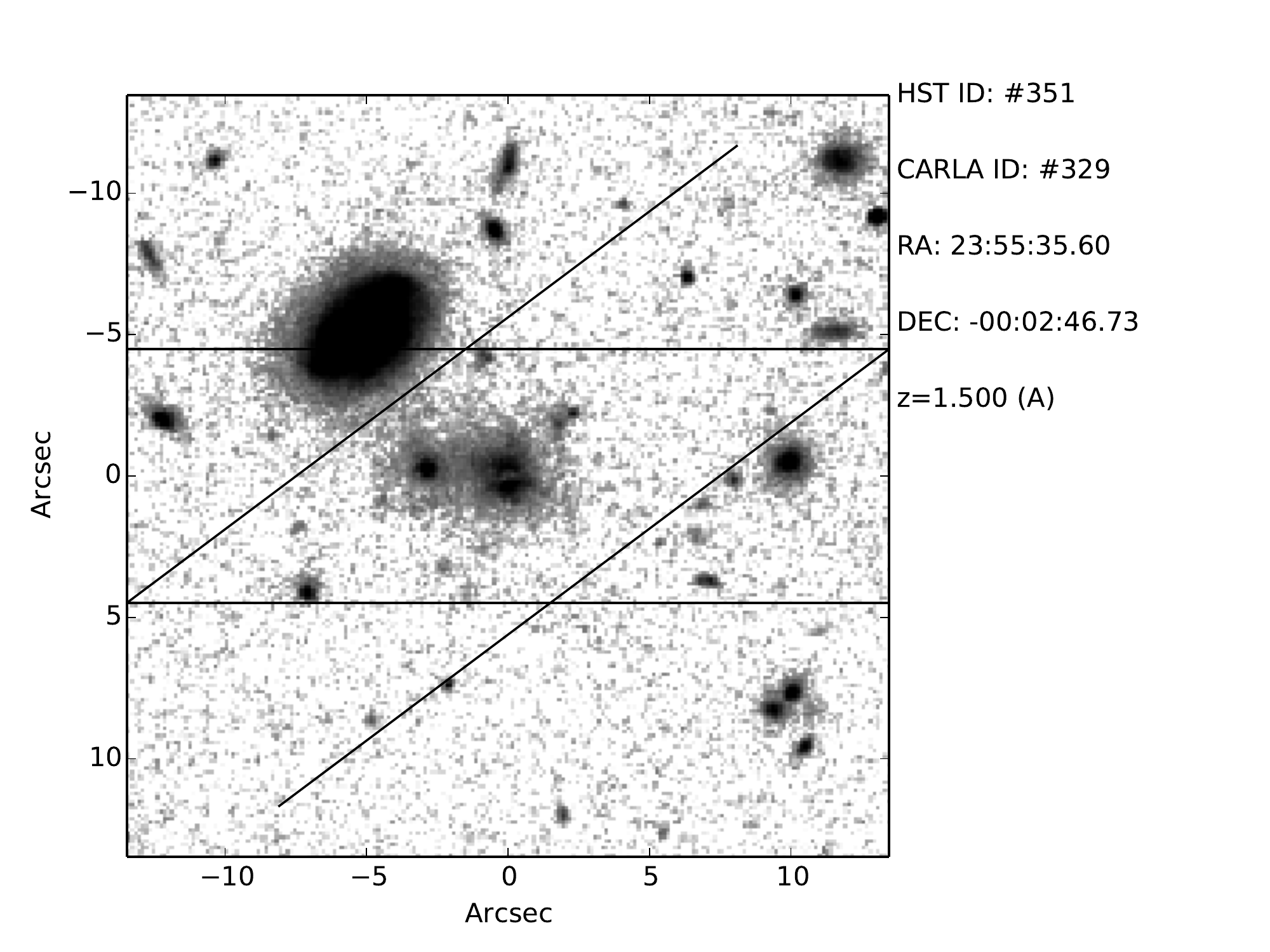} \mbox{(d)}}%
}\\%
{%
\setlength{\fboxsep}{0pt}%
\setlength{\fboxrule}{1pt}%
\fbox{\includegraphics[page=2, scale=0.24]{CARLA_J2355-0002_478.pdf} \hfill \includegraphics[page=1, scale=0.20]{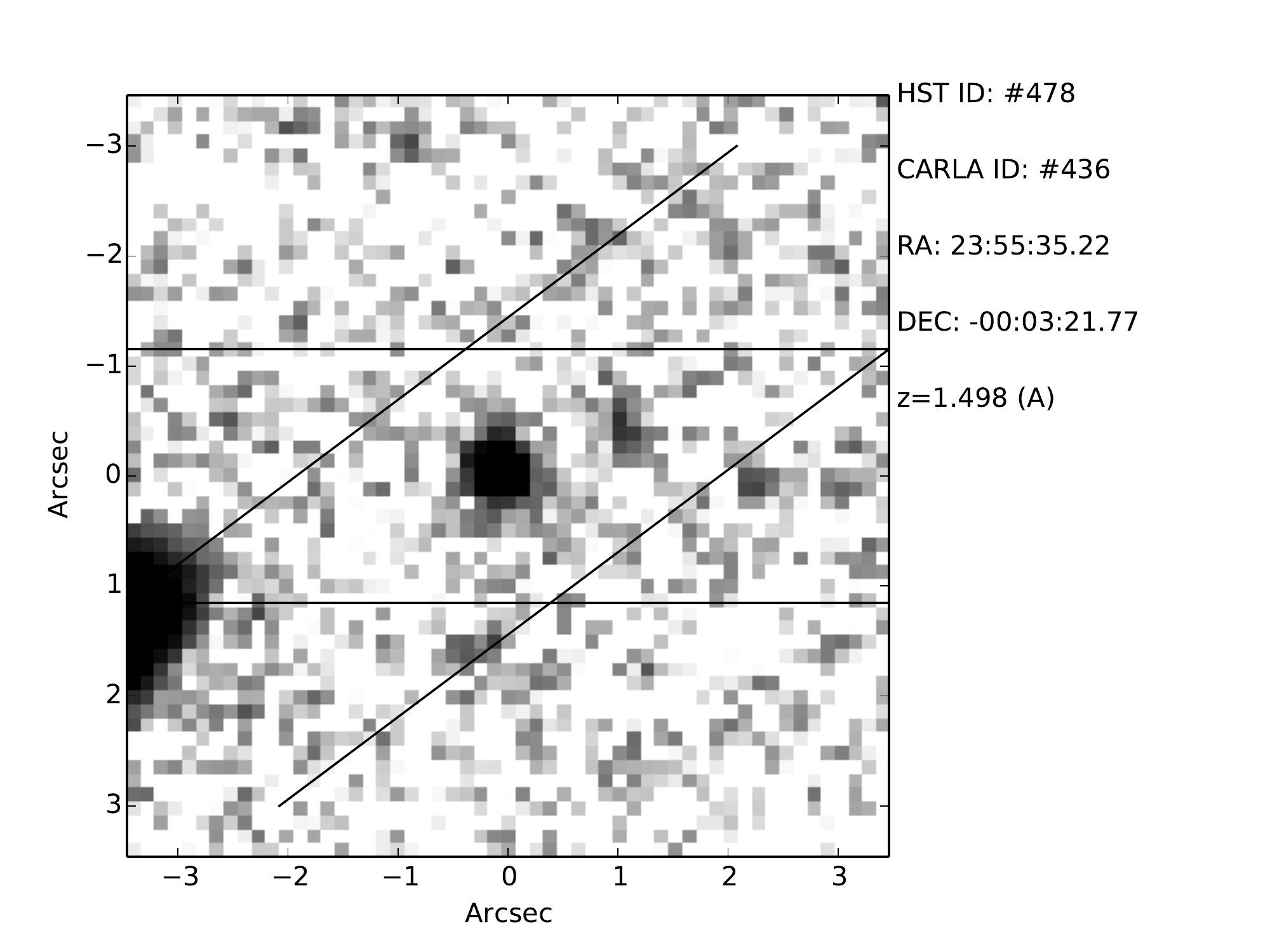} \mbox{(e)}}%
}%
{%
\setlength{\fboxsep}{0pt}%
\setlength{\fboxrule}{1pt}%
\fbox{\includegraphics[page=2, scale=0.24]{CARLA_J2355-0002_676.pdf} \hfill \includegraphics[page=1, scale=0.20]{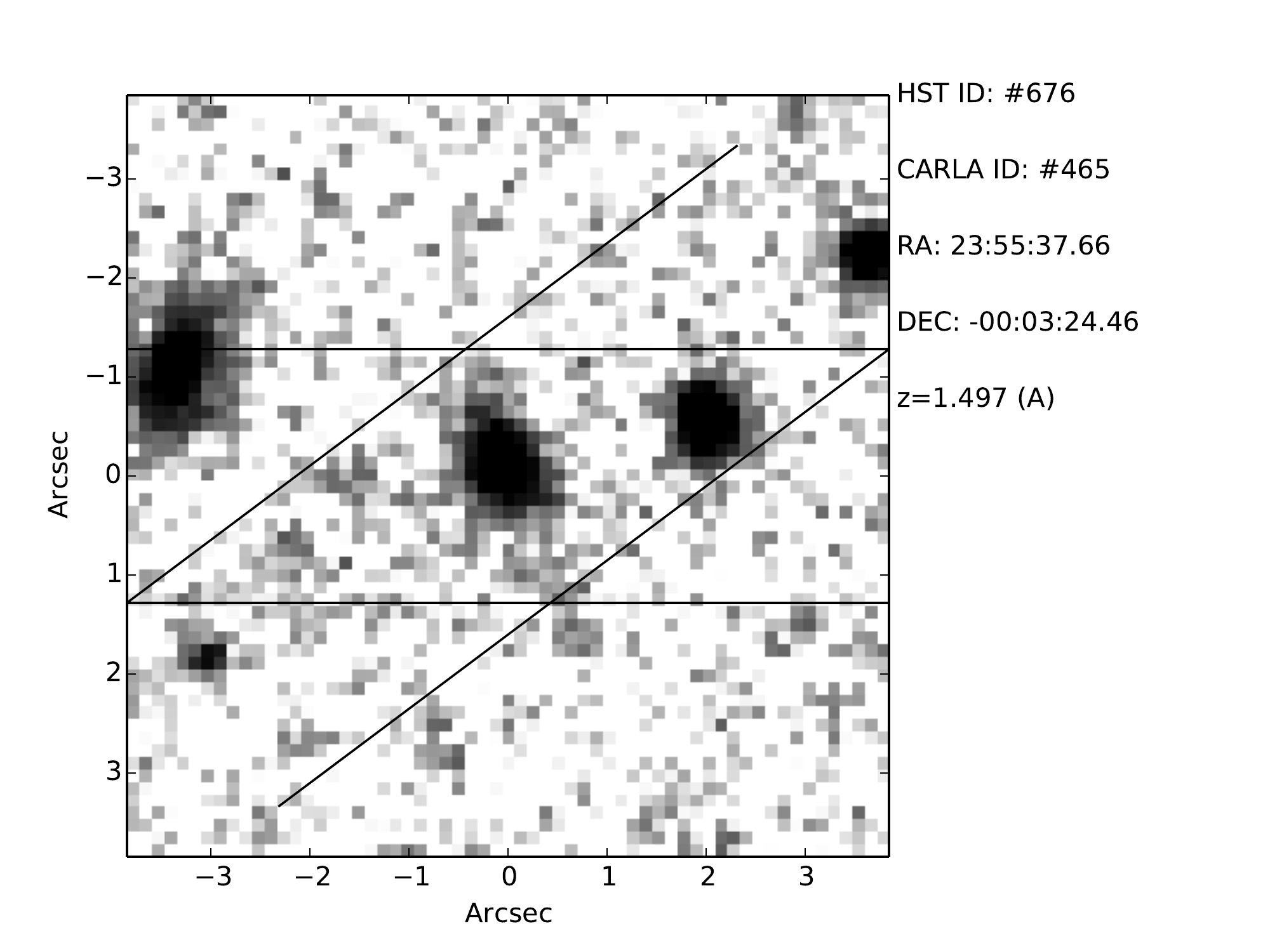} \mbox{(f)}}%
}\\%
{%
\setlength{\fboxsep}{0pt}%
\setlength{\fboxrule}{1pt}%
\fbox{\includegraphics[page=2, scale=0.24]{CARLA_J2355-0002_736.pdf} \hfill \includegraphics[page=1, scale=0.20]{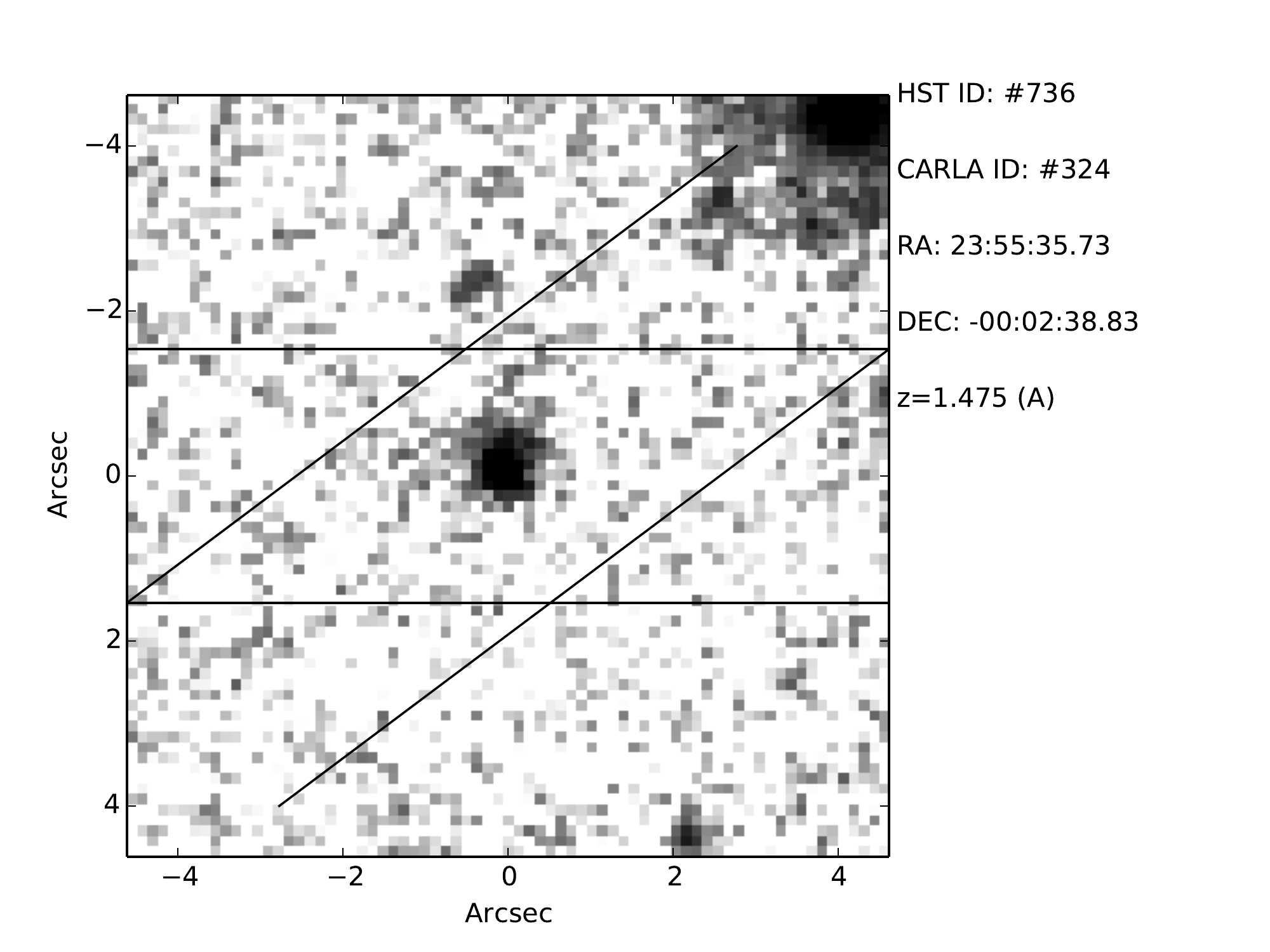} \mbox{(g)}}%
}%
{%
\setlength{\fboxsep}{0pt}%
\setlength{\fboxrule}{1pt}%
\fbox{\includegraphics[page=2, scale=0.24]{CARLA_J2355-0002_751.pdf} \hfill \includegraphics[page=1, scale=0.20]{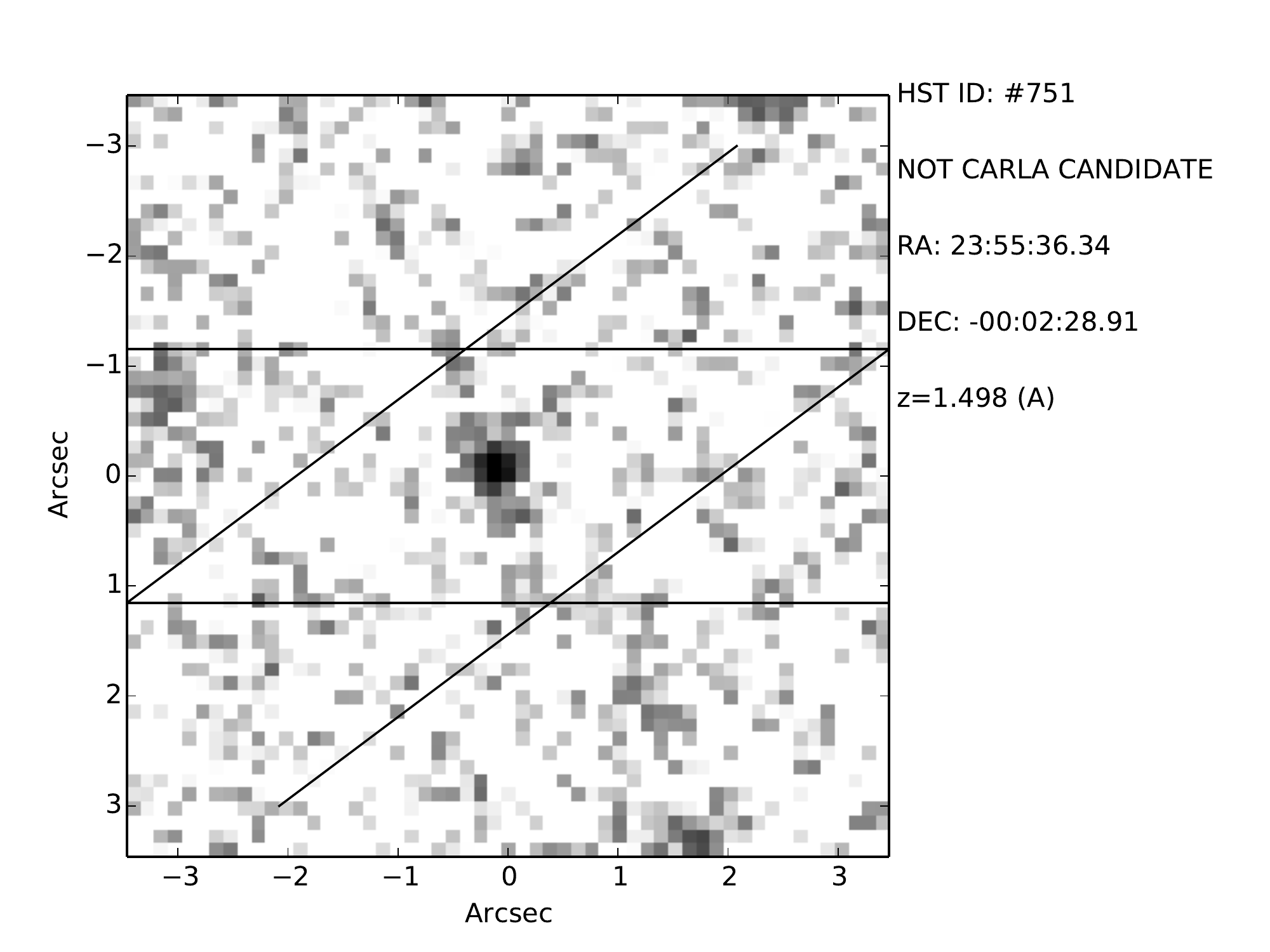} \mbox{(h)}}%
}\\%
{%
\setlength{\fboxsep}{0pt}%
\setlength{\fboxrule}{1pt}%
\fbox{\includegraphics[page=2, scale=0.24]{CARLA_J2355-0002_762.pdf} \hfill \includegraphics[page=1, scale=0.20]{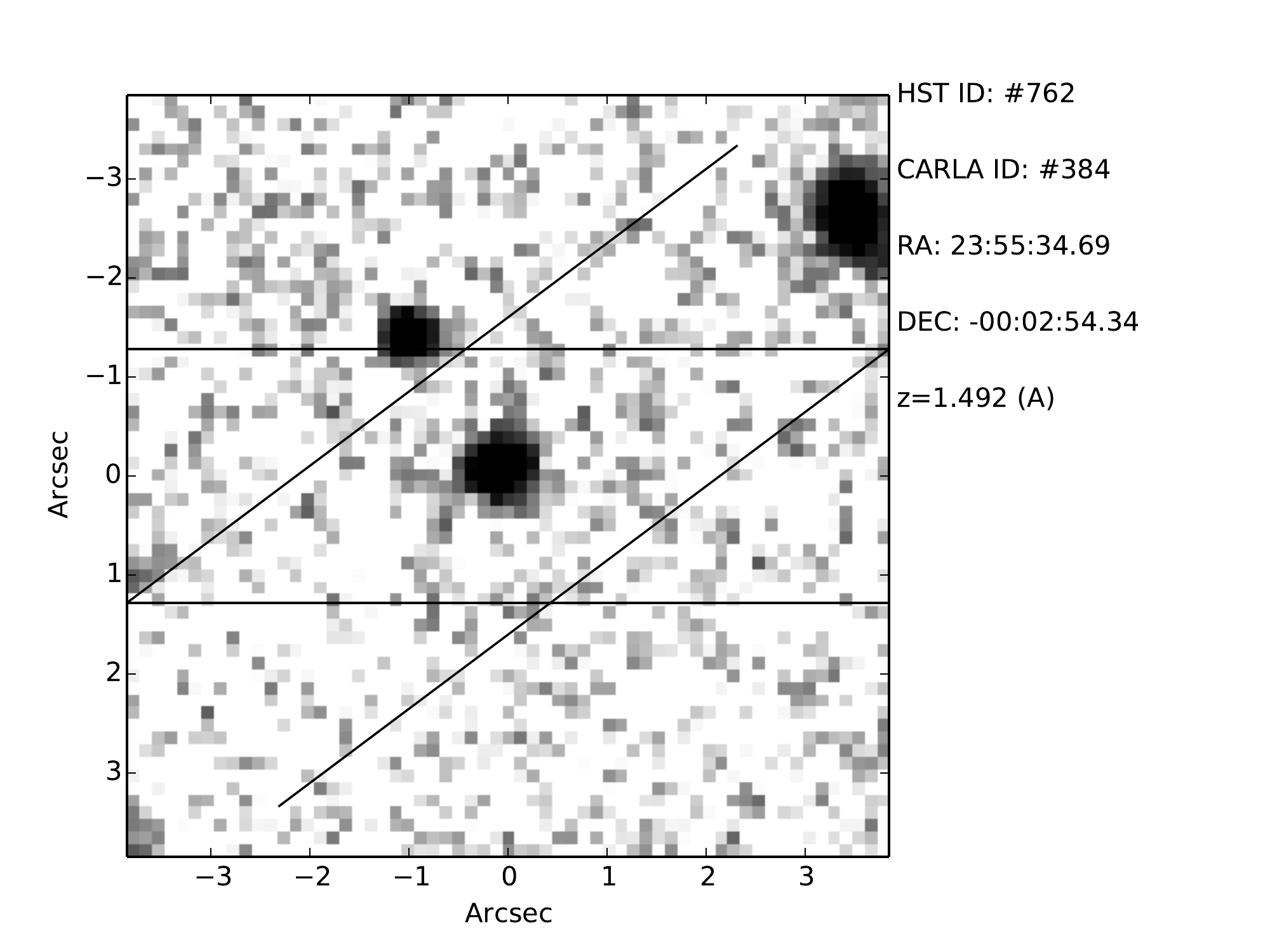} \mbox{(i)}}%
}%
{%
\setlength{\fboxsep}{0pt}%
\setlength{\fboxrule}{1pt}%
\fbox{\includegraphics[page=2, scale=0.24]{CARLA_J2355-0002_855.pdf} \hfill \includegraphics[page=1, scale=0.20]{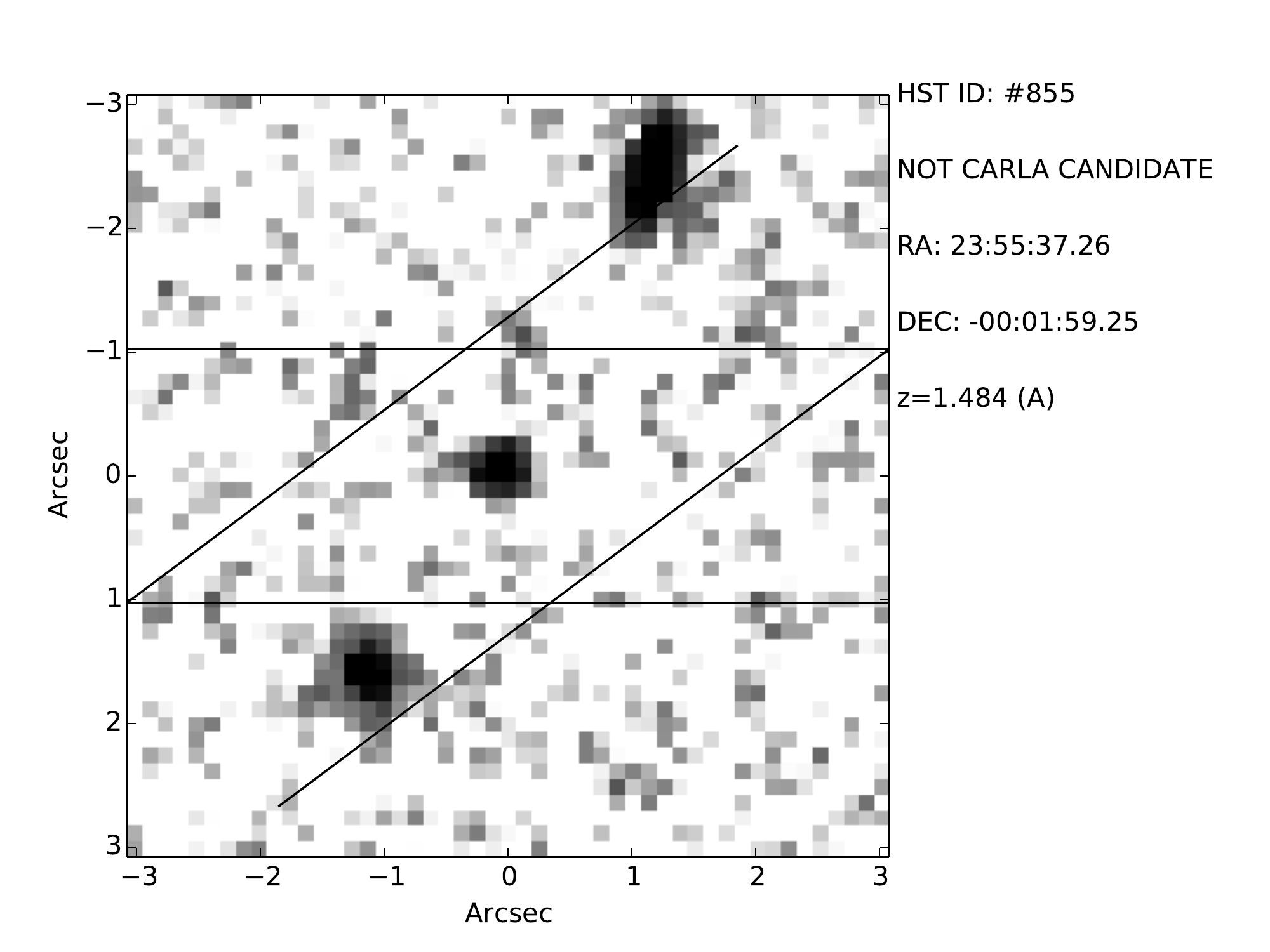} \mbox{(j)}}%
}\\%
{%
\setlength{\fboxsep}{0pt}%
\setlength{\fboxrule}{1pt}%
\fbox{\includegraphics[page=2, scale=0.24]{CARLA_J2355-0002_879.pdf} \hfill \includegraphics[page=1, scale=0.20]{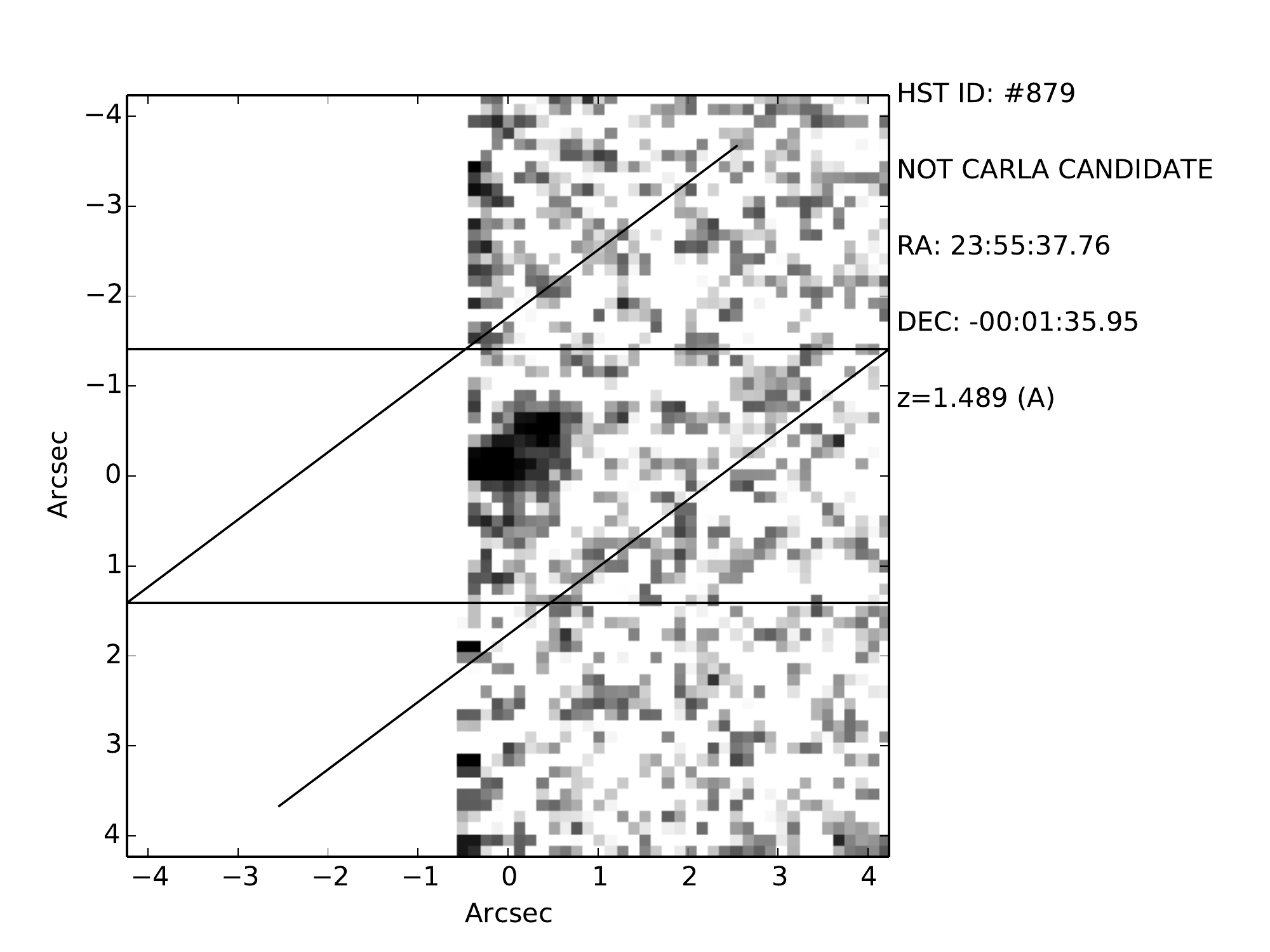} \mbox{(k)}}%
}%
{%
\setlength{\fboxsep}{0pt}%
\setlength{\fboxrule}{1pt}%
\fbox{\includegraphics[page=2, scale=0.24]{CARLA_J2355-0002_893.pdf} \hfill \includegraphics[page=1, scale=0.20]{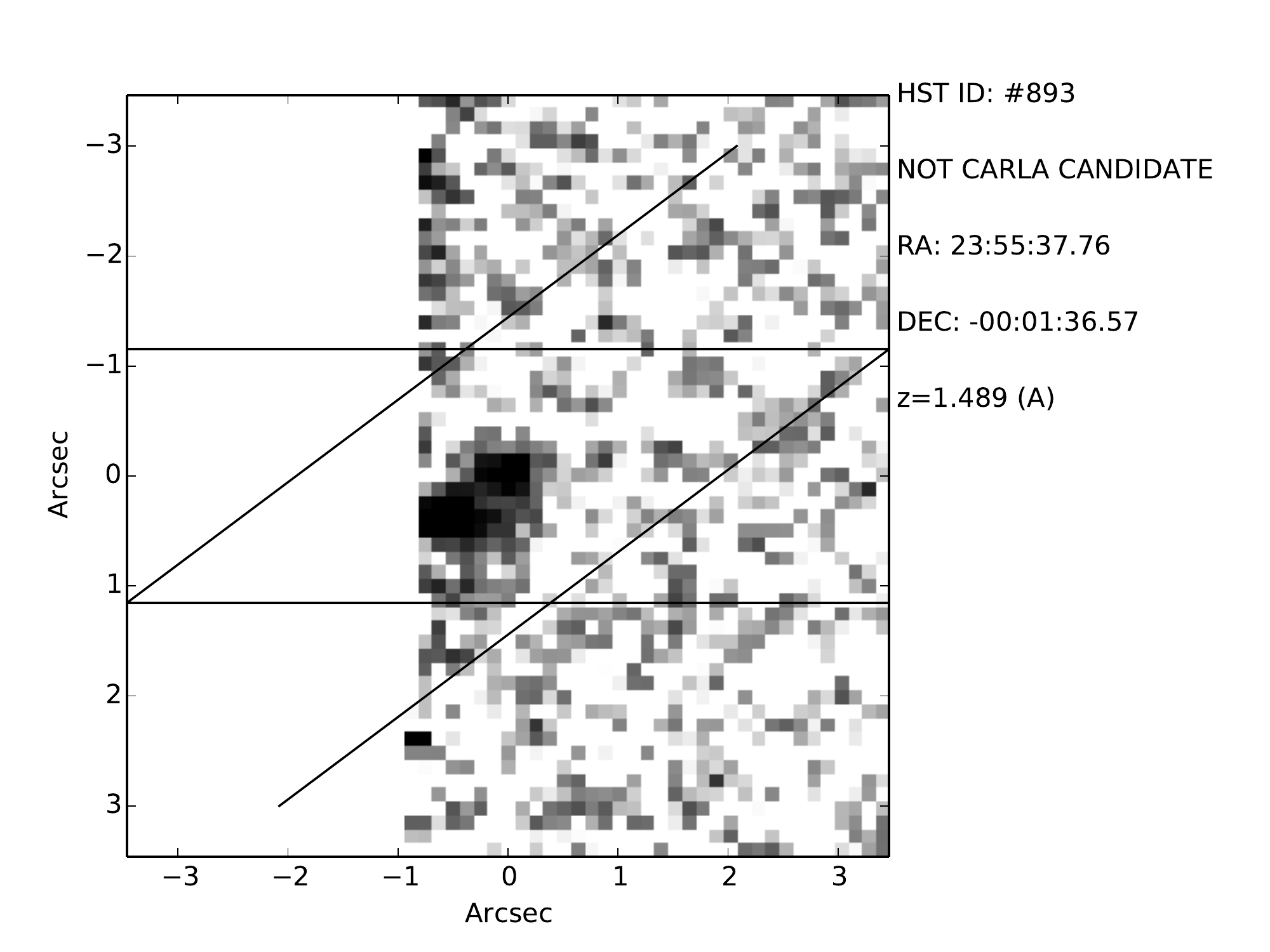} \mbox{(l)}}%
}\\%
\caption[CARLA~J2355-0002 member spectra]{CARLA~J2355-0002 member spectra.}
\label{fig:J2355-0002spectra}
\mbox{}\\
\end{figure*}

In the following appendix (Figs.~\ref{fig:J0116-2052spectra}--\ref{fig:J2355-0002spectra}, in R.A. order), we show member spectra of all confirmed structure members, except CARLA~J2039$-$2514 and CARLA~J0800+4029 which were published in N16.\\


\begin{figure*}[]
{%
\setlength{\fboxsep}{0pt}%
\setlength{\fboxrule}{1pt}%
\fbox{\includegraphics[page=2, scale=0.24]{CARLA_J1358+5752_917.pdf} \hfill \includegraphics[page=1, scale=0.20]{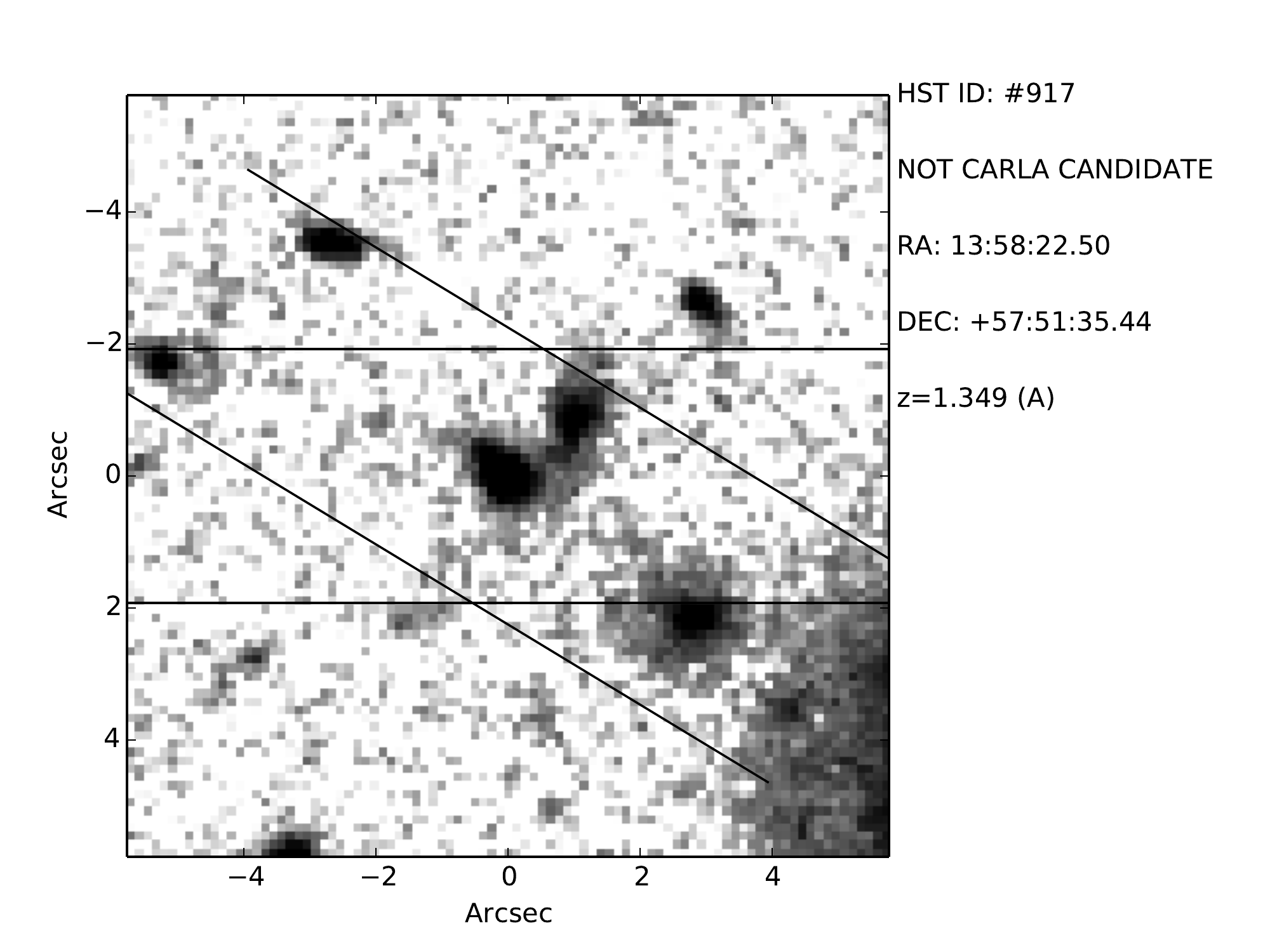} \mbox{(m)}}%
}%
{%
\setlength{\fboxsep}{0pt}%
\setlength{\fboxrule}{1pt}%
\fbox{\includegraphics[page=2, scale=0.24]{CARLA_J1358+5752_965.pdf} \hfill \includegraphics[page=1, scale=0.20]{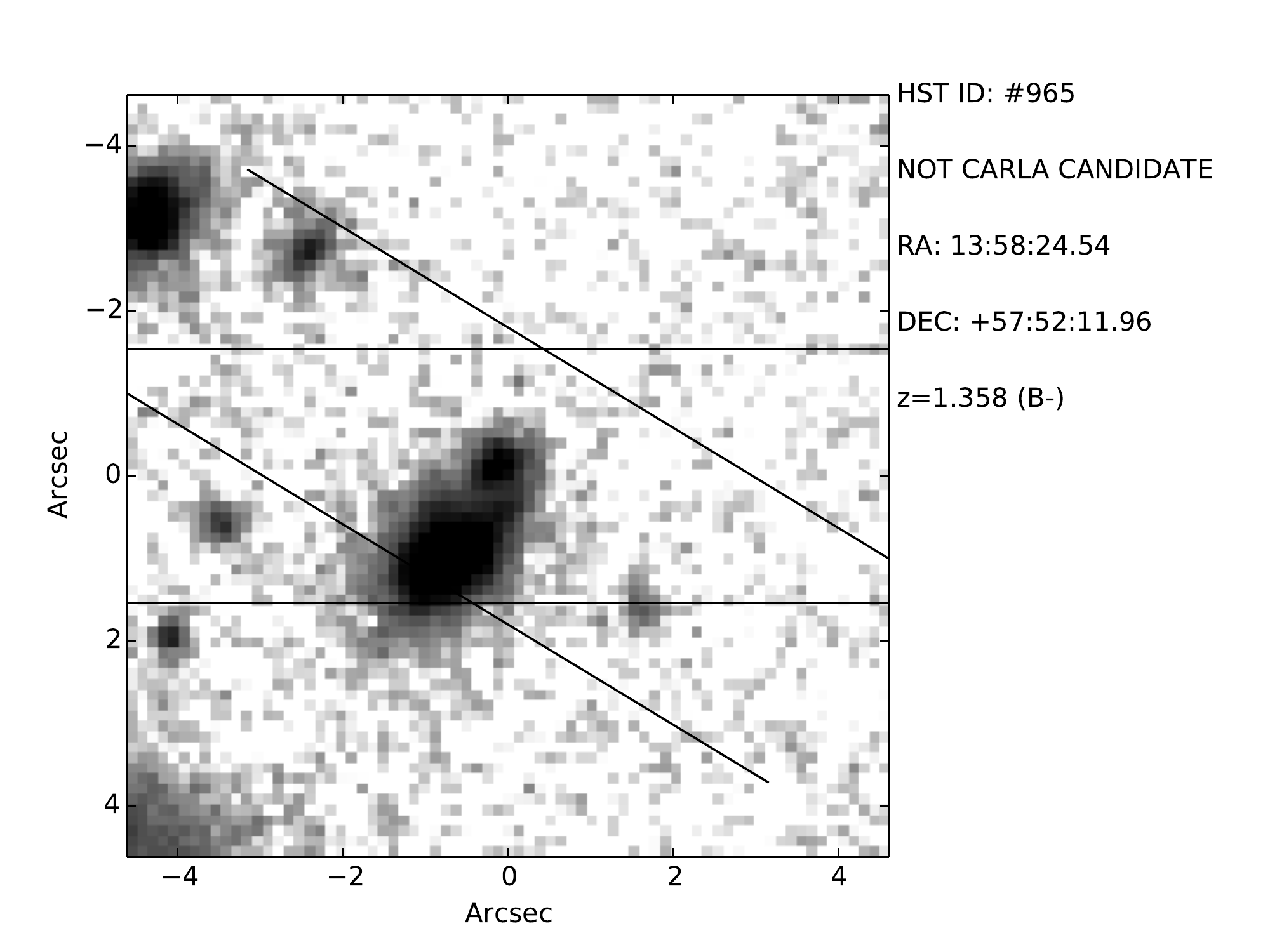} \mbox{(n)}}%
}\\%
\textbf{\mbox{}\\ Figure \ref{fig:J1358+5752spectra}} --- Continued.
\end{figure*}


\mbox{}\\

\end{document}